\documentclass[twocolumn,aps,prd,showpacs,superscriptaddress,floats,floatfix,showkeys,notitlepage,longbibliography]{revtex4-1}

\usepackage{lipsum}
\usepackage{graphicx}
\usepackage{subfigure}
\usepackage{palatino}
\usepackage{changes}
\usepackage{hyperref}
\hypersetup{colorlinks=true,linkcolor=blue,urlcolor=blue,citecolor=blue}
\usepackage[toc,page]{appendix}
\usepackage[normalem]{ulem}
\usepackage{adjustbox}
\usepackage{latexsym}
\usepackage{amsmath}
\usepackage{amssymb}
\usepackage{amsfonts}
\usepackage{times}
\usepackage{dcolumn}
\usepackage{bm}
\usepackage{tikz}
\usepackage{bigints}
\usepackage{array,tabularx,multirow}
\usepackage[tracking=true]{microtype}
\SetTracking{}{500}
\SetTracking{encoding={*}, shape=sc}{40}
\UseRawInputEncoding 
\allowdisplaybreaks

\usepackage{braket}
\usepackage{tensor}
\usepackage{slashed}
\usepackage{booktabs} 
\usepackage{float}

\newcommand{\rp}[1]{\textcolor{black}{ #1}}

\newcommand{\AO}[1]{\textcolor{black}{ #1}}

\newcommand{\AU}[1]{\textcolor{black}{ #1}}

\begin{document} \sloppy

\title{Probing a non-linear electrodynamics black hole with thin accretion disk, shadow, and deflection angle with M87* and Sgr A* from EHT}

\author{Akhil Uniyal}
\email{akhil\_uniyal@iitg.ac.in}
\affiliation{ Department of Physics, Indian Institute of Technology, Guwahati 781039, India}

\author{Reggie C. Pantig}
\email{rcpantig@mapua.edu.ph}
\affiliation{Physics Department, Map\'ua University, 658 Muralla St., Intramuros, Manila 1002, Philippines}

\author{Ali \"Ovg\"un}
\email{ali.ovgun@emu.edu.tr}
\homepage{https://www.aovgun.com}
\affiliation{Physics Department, Eastern Mediterranean University, Famagusta, 99628 North Cyprus via Mersin 10, Turkey}
\date{\today}

\begin{abstract}
Non-linear interaction between the electromagnetic fields (EMF) occurs when vacuum polarization in quantum electrodynamics (QED) happens. The field of non-linear electrodynamics, which may result from this interaction, could have important effects on black hole physics. This paper considers the asymptotically flat black hole solution in  Einstein-nonlinear electrodynamics (NLE) fields. We study the effect of the NLE parameters on the black hole deflection angle using the Gauss-Bonnet theorem in weak field limits, shadow cast using the null geodesics method, and thin accretion disk using the Novikov-Thorne model. In particular, we studied the time-averaged energy flux, the disk temperature, the differential luminosity, the different emission profiles, and infalling spherical accretion. Then we show how the physical quantities depend on $\beta$ and $C$ parameters of NLE and provide some constraints on the NLE parameters using the observations of M87* and Sgr A* from EHT.

\end{abstract}

\maketitle

\date{\today}


\section{Introduction}\label{sec:intro}
Black holes are one of the solutions to Einstein's equation, known as cast-iron predictions of General Relativity(GR). There is historical debate on the black hole's existence, and several observatories have been made to detect such a compact object. In the recent year 2015 \cite{LIGOScientific:2016aoc}, Laser Interferometer Gravitational-Wave Observatory (LIGO) detected the first gravitational waves (GWs), which were emitted from the black hole merger. Along with that recently in 2019, Event Horizon Telescope (EHT) showed the shadow of the black hole in the center of the M87* galaxy \cite{EventHorizonTelescope:2019dse,EventHorizonTelescope:2019uob,EventHorizonTelescope:2019jan,EventHorizonTelescope:2019ths,EventHorizonTelescope:2019pgp,EventHorizonTelescope:2019ggy}. \AO{When a black hole is surrounded by a bright source of light, such as a disk of hot gas, the light can be bent by the strong gravitational field of the black hole. This creates a region around the black hole called the "shadow" where no light can reach an observer. The size and shape of this shadow can provide important information about the black hole's mass, spin, and other properties. For example, a black hole that is spinning rapidly will have a shadow that is slightly distorted from a circle, due to the frame-dragging effect of its spin. The shadow of a black hole can also be used to test the predictions of general relativity and other theories of gravity. By comparing the observed shape of the shadow with theoretical predictions, scientists can learn more about the nature of black holes and the behavior of gravity in extreme conditions.} Later on EHT also came up with the polarisation image of the M87$*$ which gives a indication of the presence of the magnetic field and a possibility to explain the jets coming out of the black hole \cite{EventHorizonTelescope:2021bee,EventHorizonTelescope:2021srq,EventHorizonTelescope:2021dqv}. Recently, EHT announced observing the shadow cast of SgrA$*$ black hole in the center of Milky Way galaxy \cite{EventHorizonTelescope:2022xnr}. \AO{ The shadow image may give us valuable information about the geometry around the black hole \cite{Cunha:2018gql},} especially in the vicinity of the horizon, which can essentially give us information about the black hole's mass and spin \cite{Takahashi:2004xh}. All these observations and theoretical predictions lead us to test the GR and modified gravity theories in these regimes \cite{Volkel:2020xlc}.

GR has given very accurate results so far \cite{Will:2014kxa}, but it also has some limitations in the cosmological framework, such as explaining the accelerated expansion of the universe, initial singularities, missing mass problem, Magnitude Problem and Coincidence Problem. The testing of GR in the strong gravity regime, such as near black hole horizon, is still a big challenge \cite{Berti:2004bd}. Therefore, people try to modify the theory of gravity to \AU{look beyond the GR and understand the near horizon physics and Universe at the large-scale structure} \cite{Easson:2004fq,Nojiri:2006ri,Trodden:2006qk}. \AO{GR also suffers from the singularity problem. Since non-linear electrodynamics (NLE) can remove the singularity at the classical level, exploring the NLE corrections near the black hole region becomes very important. The NLE theory of Born-Infeld is a modification of classical electrodynamics that was proposed in the 1930s by physicist Max Born and his collaborator, physicist Leopold Infeld. The theory was developed in an attempt to eliminate certain infinities that arose in the equations of classical electrodynamics when dealing with point charges. One of the key features of the Born-Infeld theory is that it introduces a maximum value for the electric field, which helps to avoid the singularities that plague classical electrodynamics. The theory also predicts that the electromagnetic force between two charged particles should become weaker at very short distances, which is a behavior that is not observed in classical electrodynamics. Today, the Born-Infeld theory is of interest to physicists because it arises naturally in the context of certain theories of quantum gravity, such as string theory. It is also related to the study of black holes and other exotic objects in astrophysics. In these contexts, the Born-Infeld theory provides a framework for understanding the behavior of electromagnetism under extreme conditions, and it has the potential to shed light on some of the most fundamental mysteries of the universe.} Some NLE models with cosmological models can solve the initial Big Bang singularity, and early inflation \cite{Garcia-Salcedo:2000ujn,Camara:2004ap}. The gravity with non-linear electrodynamics produces the negative pressures that cause accelerated expansion \cite{Novello:2003kh,Novello:2006ng,Vollick:2008dx}. Therefore, some models of NLE have been considered to explain the Universe's accelerated expansion. \AU{Hence, because of removing the singularity and taking care of the quantum corrections which is the biggest problem in GR, NLE models attracts the community in a large scale.}

After releasing the black hole shadow image from the EHT, the scientific community got interested in understanding the black hole's features (such as mass and spin) by matching the data from EHT with the theoretical models. People have studied the shadow in the modified gravity and deformed structure of the space-time near the black hole region \cite{Okyay:2021nnh,Allahyari:2019jqz,Chen:2022nbb,Roy:2021uye,Khodadi:2020jij,Vagnozzi:2022moj,Wang:2018prk,Cunha:2019hzj,Pantig:2020odu,Pantig:2020uhp,Pantig:2021zqe,Pantig:2022toh,Pantig:2022whj,Pantig:2022sjb,Pantig:2022gih,Pantig:2022ely,physics4040084,Zuluaga:2021vjc,Li:2021ypw,Rahaman:2021kge,Stashko:2021lad,Liu:2021yev,Gyulchev:2021dvt,Guerrero:2021ues,Gan:2021xdl,Heydari-Fard:2020iiu,Heydari-Fard:2021ljh,Kazempour:2022asl,He:2022yse,Bisnovatyi-Kogan:2022ujt,Bauer:2021atk,Li:2021riw,Ovgun:2020gjz,Ovgun:2019jdo,Ovgun:2018tua,Ovgun:2021ttv,Ling:2021vgk,Belhaj:2020okh,Belhaj:2020rdb,Abdikamalov:2019ztb,Abdujabbarov:2016efm,Atamurotov:2015nra,Papnoi:2014aaa,Abdujabbarov:2012bn,Atamurotov:2013sca,Cunha:2018acu,Perlick:2015vta,Nedkova:2013msa,Li:2013jra,Cunha:2016wzk,Johannsen:2015hib,Johannsen:2015mdd,Shaikh:2019fpu,Yumoto:2012kz,Cunha:2016bpi,Moffat:2015kva,Giddings:2016btb,Cunha:2016bjh,Zakharov:2014lqa,Tsukamoto:2017fxq,Hennigar:2018hza,Kumar:2020hgm,Li2020,Cimdiker:2021cpz,Hu:2020usx,Zhong:2021mty} \AU{but having a clear picture of how NLE field effects the photon trajectory around the black hole is still an open field and in this paper we are addressing one such case.} Therefore, in order to understand the black hole physics by coupling the NLE with gravity, we choose to study the accretion disk radiation and shadow of the black hole. In 1979, Luminet explicitly derived the expression for the radiation coming from the thin accretion disk and also the accretion disk image around the black hole \cite{Luminet:1979nyg}. Cunningham and Bardeen studied the first visual appearance of the black hole (1973) \cite{1973Apx} for the case of a star orbiting around a black hole and other related scenarios. All these models of the accretion disk are based on the prescription of the disk described by Shakura and Sunyaev 1973 \cite{1973A}; Novikov and Thorne 1973 \cite{NovikovThorne}; Page and Thorne 1974 \cite{1974ApJ...191..499P}. The disk is to be assumed geometrically thin and optically thick. The disk is considered to be optically thick when the photon's mean free path is much less than the height of the accretion disk ($H$) and it is considered to be geometrically thin when its characteristic length is much greater than the height of the disk.

\AO{Gravitational lensing is also an important tool for the study of gravity \cite{Virbhadra:1999nm,Virbhadra:2002ju,Virbhadra:1998dy,Virbhadra:2007kw,Virbhadra:2008ws,Adler:2022qtb,Virbhadra:2022ybp,Virbhadra:2022iiy,Bozza:2001xd,Bozza:2002zj,Hasse:2001by,Perlick:2003vg,He:2020eah}. Weak gravitational lensing is important because it allows us to study the distribution of matter in the universe, including both visible and dark matter. By measuring the distortion of background galaxies caused by the gravitational field of foreground mass, scientists can map the distribution of mass in the universe and use this information to test theories about the evolution and structure of the universe. In addition, weak gravitational lensing is a powerful tool for studying the properties of dark energy, which is thought to be driving the acceleration of the expansion of the universe. Gibbons and Werner have proposed an alternative method for calculating the deflection angle in weak field limits for the asymptotically flat black holes \cite{Gibbons:2008rj}. These methods use the Gauss-Bonnet theorem (GBT) on the optical metric. The Gauss-Bonnet theorem is a mathematical theorem that relates the curvature of a surface to its topological features. It states that the integral of the Gaussian curvature over the surface of a closed 2D manifold is equal to a topological invariant, the Euler characteristic of the surface. This theorem has important applications in mathematics and physics, particularly in the study of black holes and the behavior of spacetime in general relativity. In physics, the Gauss-Bonnet theorem is often used to study the topology and geometry of spacetime, and it has played a role in the development of string theory and other theories of quantum gravity. Afterward, this method is extended for stationary spacetimes by Werner, in which the deflection angle of the Kerr black hole is calculated in weak fields \cite{Werner_2012}. Since then, there are many applications of the method in the literature \cite{Ovgun:2018fnk,Ovgun:2019wej,Ovgun:2018oxk,Javed:2019kon,Javed:2019rrg,Javed:2019ynm,Javed:2020lsg,Javed:2019qyg,Ovgun:2018fte,Javed:2019jag,Ishihara:2016vdc,Takizawa:2020egm,Ono:2019hkw,Ishihara:2016sfv,Ono:2017pie,Li:2020dln,Li:2020wvn,Javed:2022fsn,doi:10.1142/S0219887823500408}.  In the framework of NLE we study the weak deflection angle of both massive and null particles by exploiting the Gauss-bonnet theorem.}

It is well studied that the accretion process takes place around compact objects such as black holes and neutrons where the strong gravity effect is dominant \cite{Tucker_2018, Corral-Santana:2015fud}. In this process, gases accrete around the compact object, and due to the gravitational potential, it releases energy in the form of radiation. The emission spectra obtained from such radiations depend upon the particles' geodesic motion and the compact object's structure. Therefore, the disk properties can be studied to test the GR and deviation from the GR, such as modified gravity ($f(R)$ gravity models, Horava-Lifshitz gravity, and other possible modifications) \cite{Perez:2012bx, Harko:2009rp, Pun:2008ua, Heydari-Fard:2010agr}. Some techniques, such as continuum-fitting method and analysis of relativistic iron line profiles, can be used to distinguish the different astrophysical compact objects by studying the accretion disk properties \cite{Bambi:2013jda, Bambi:2015ldr, Jiang:2016bdj}. Therefore, this paper studies the thin accretion disk properties when GR is coupled with the NLE. We study the effect of the coupling parameter on the weak deflection angle, radiation coming from the accretion disk, and shadow of the black hole using the thin accretion disk model in the framework of NLE. In particular, we will study the time-averaged energy flux $(F)$, the disk temperature $(T)$, the differential luminosity$(dL_{\infty})$, and the different emission profiles to study the shadow of the black hole. We also study the infalling spherical accretion. We will show how the physical quantities change by incorporating the NLE with gravity. Later on, \AU{we have constraint the coupling parameter by the help of data released} by EHT for M87* and Sagittarius A*. The paper has been written with the unit system $G=M=c=1$, and metric signature $(-,+,+,+)$.

\section{Nonlinear electrodynamics Black Hole}\label{sec:formal}
Let us consider the coupling of non-linear electromagnetic field \cite{Kruglov_2015} through the GR action as
\begin{equation} \label{e1}
S=\int d^4x \sqrt{-g} [\frac{1}{2 \kappa^2} R +\mathcal{L}_\text{em}].
\end{equation}
Here, $R$ is the Ricci scalar, $\kappa^{-1}=M_\text{pl}$ where $M_\text{pl}$ is the reduced Planck mass and the EM Lagrangian is
\begin{equation} \label{e2}
\mathcal{L}_\text{em}=\frac{-\mathcal{F}}{2\beta \mathcal{F}+1},
\end{equation}
where $\beta$ \AU{has a dimension (length)$^4$ with $\beta \mathcal{F}$ dimensionless}, and $\mathcal{F}=(1/4)F_{\mu \nu}F^{\mu \nu}$. \AU{It is well known that Linear Maxwell’s electrodynamics and the Born-Infeld (BI) electrodynamics do not
produce the birefringence phenomenon \cite{Born1934410ï12437} but all other non-linear electrodynamics theory produces the birefringence. Therefore, people have tried to constrain the non-linear electrodynamics parameter using the birefringence. In our case} the constraint on the beta with an experimental setup like birefringence is not straightforward, as mentioned in \cite{kruglov2015nonlinear} by the fact that in vacuum birefringence, higher order parameter contributes. Therefore, one needs a new kind of experiment to find the bound for the coupling constant ($\beta$). Here, we try to find the bound of $\beta$ within the astrophysical regime.
Varying the action with respect to the metric and electromagnetic field will give the following field equations:
\begin{equation} \label{e3}
R_{\mu \nu}-\frac{1}{2}g_{\mu \nu}R=\kappa^2 T_{\mu \nu},
\end{equation}
\begin{equation} \label{e4}
\partial_\mu \left( \frac{ \sqrt{-g} F^{\mu \nu}}{(2 \beta \mathcal{F} + 1)^2} \right)=0.
\end{equation}
By considering non-vanishing component of the vector-potential $A_0=h(r)$, we can define
\begin{equation} \label{e5}
    \mathcal{F}=-(h'(r))^2/2,
\end{equation}
where prime denotes the derivative with respect $r$. Hence, from Eq. \eqref{e4}, we can write
\begin{equation} \label{e6}
\partial_r \left[ \frac{ r^2 h'(r)}{(1-\beta (h'(r))^2)^2} \right]=0,
\end{equation}
and
\begin{equation} \label{e7}
r^2 h'(r)=C[1-\beta (h'(r))^2]^2.
\end{equation}
Here, $C$ is a dimensionless constant of integration. We can introduce a new dimensionless variable to simplify the calculations further:
\begin{equation} \label{e8} 
y=\frac{r \sqrt{h'(r)}}{\sqrt{C}},x=\frac{r}{\sqrt{C}\beta^{1/4}},
\end{equation}
As it enables the simplification of Eq. \eqref{e7} in the following algebraic equation:
\begin{equation} \label{e9}
y^4+(y-1)x^4=0.
\end{equation}

We now use  Cardano's formula to write the solution of Eq. \eqref{e9} analytically. The only solution that gives physical meaning is
\begin{equation} \label{e10}
    y=\frac{12\sqrt{3}\sqrt{\beta}r^{2}-\beta C\lambda^{3/4}}{12\beta C\sqrt[4]{\lambda}},
\end{equation}
where
\begin{align} \label{e11}
    \lambda &= \frac{6\sqrt[3]{6}r^{2}\left(\sqrt[3]{2}\beta^{2/3}\sqrt[3]{C}\gamma^{2/3}-8\sqrt[3]{3}\beta C\right)}{\beta^{4/3}C^{5/3}\sqrt[3]{\gamma}} \nonumber \\
    \gamma &= \sqrt{3}\sqrt{256\beta C^{2}+27r^{4}}+9r^{2}.
\end{align}

One can find out the component of the vector potential by integrating Eq. \eqref{e8} as
\begin{equation}  \label{e12}
    h(r)=C \int \frac{y^2}{r^2} dr,
\end{equation}
which is
\begin{equation} \label{e13}
    h(r)=\frac{\sqrt{C}}{4 \beta^{1/4}} \int \frac{4-3y}{(1-y)^{3/4}}dy=\frac{\sqrt{C}}{5 \beta^{1/4}}(3y-8)(1-y)^{1/4},
\end{equation}
where, $y$ is the solution of the Eq. \eqref{e8}. In the region where $r \to 0$ and
\begin{equation} \label{e14}
    h(r) \to \frac{-8\sqrt{C}}{5 \beta^{1/4}}, \quad \quad h'(r) \to \frac{1}{\sqrt{\beta}}.
\end{equation}
We know that in electrostatics $E=h'(r)$, the maximum electric field we can acquire here is $E_{max}=1/\sqrt{\beta}$. Thus, the theory has no singularity at $r=0$ as the electric field has a finite value. 

Let us consider the general static and spherically symmetric spacetime in the $4$D:
\begin{equation} \label{e15}
ds^2=-f(r) dt^2+\frac{1}{f(r)} dr^2+r^2 (d\theta^2+\sin^2\theta d\phi^2),
\end{equation}
where the component of the metric can be found by the relation \cite{Kruglov_2015}
\begin{equation} \label{e16}
    f(r)=1+\frac{k_1}{r}+\frac{k_2}{r^2}+\frac{1}{r^2} \int dr \left[ \int r^2 R(r) dr \right].
\end{equation}
Here, $k_1$ and $k_2$ are the integration constants. We use the following equation to calculate the Ricci scalar
\begin{equation} \label{e17}
    R=-\kappa^2 T, T=g^{\mu \nu} T_{\mu \nu},
\end{equation}
where $T_{\mu \nu}$ is the energy-momentum tensor given by
\begin{equation} \label{e18}
    T_{\mu \nu}=-\frac{1}{(2\beta \mathcal{F}+1)^2} \left[F_{\mu}^{\alpha} F_{\nu \alpha}-g_{\mu \nu} \mathcal{F} (2\beta \mathcal{F}+1) \right].
\end{equation}
The trace of the energy-momentum tensor is
\begin{equation} \label{e19}
    T=\frac{8\beta \mathcal{F}^2}{(1+2\beta \mathcal{F})^2},
\end{equation}
which simplifies to
\begin{equation} \label{e20}
    T=\frac{2\beta (h'(r))^4}{(1-\beta(h'(r))^2)^2}.
\end{equation}
Therefore, the Ricci scalar can be written using the Eq. \eqref{e6} as
\begin{equation} \label{e21}
    R=-\frac{2C \kappa^2 \beta (h'(r))^3}{r^2}.
\end{equation}

\begin{figure}
    \centering
    \includegraphics[width=\linewidth]{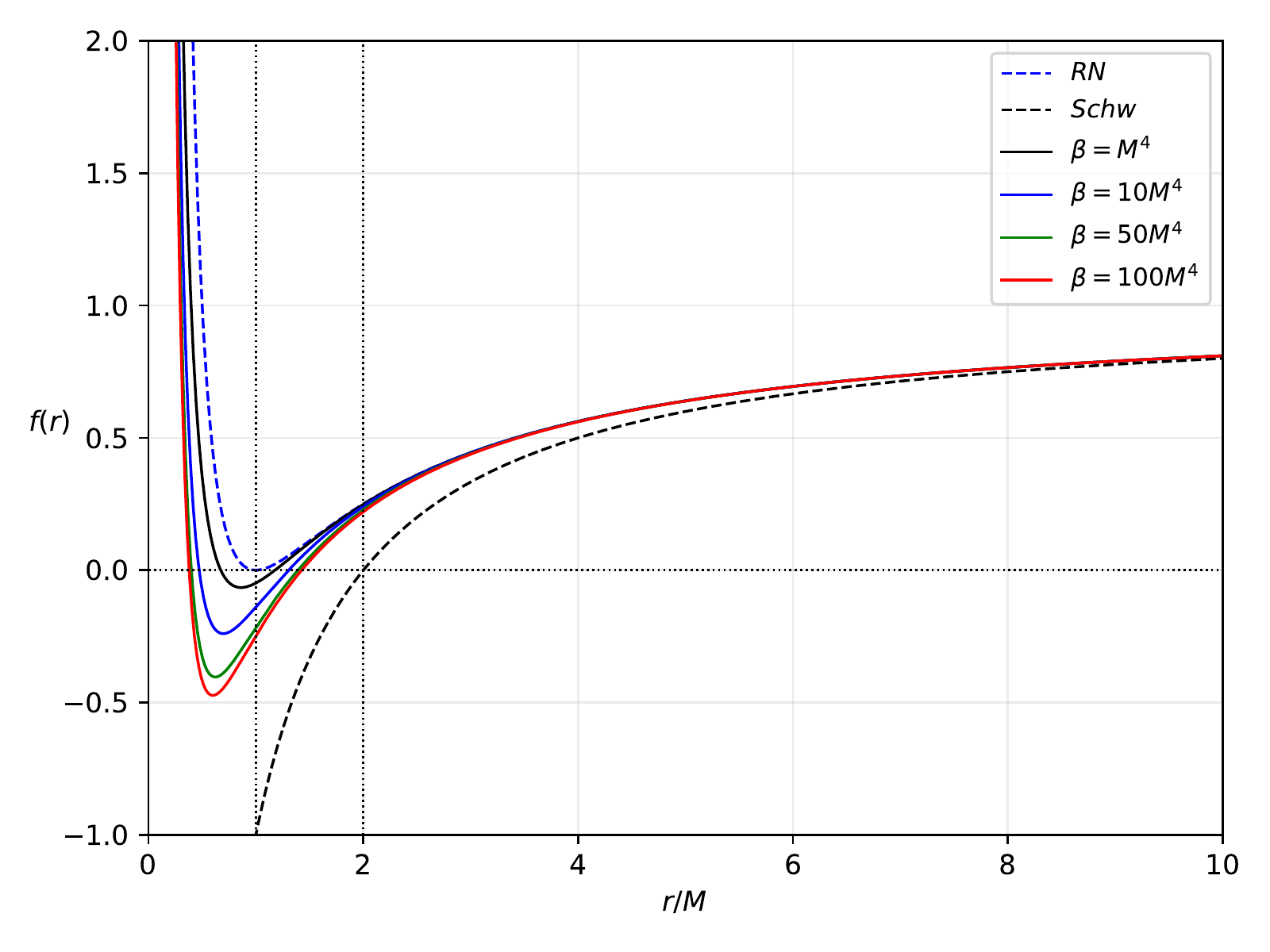}
    \caption{Plot of the lapse function $f(r)$ for different values of $\beta$. Here, $Q = C =\kappa = 1$.}
    \label{fig:c}
\end{figure}

Thus, the final and complete form of the metric component is
\begin{equation} \label{e22}
    f(r)=1-\frac{2M}{r}+\frac{Q^2}{r^2}-\frac{C^2\kappa^2}{2r^2}+\frac{C^2 \kappa^2}{30r^2}(5y^3-22y^2+32y),
\end{equation}
where $y$ is given by Eq. \eqref{e10}. With the complexity of Eq. \eqref{e10}, it is useful to find the simplified form of the metric function in Eq. \eqref{e22} in the strong and weak field limit. An implementation of series expansion for $r \to 0$ gives
\begin{equation} \label{e23}
	f(r)=1-\frac{2M}{r}+\frac{Q^2}{r^2}+\frac{16C^{3/2}\kappa^{2}}{15\beta^{1/4}r}-\frac{C^{2}\kappa^{2}}{2r^{2}},
\end{equation}
and for $r \to \infty$ we have
\begin{equation} \label{e24}
    f(r)=1-\frac{2M}{r}+\frac{Q^2}{r^2}-\frac{\beta C^{4}\kappa^{2}}{10r^{6}}.
\end{equation}
We can see in Eq. \eqref{e24} that the last term is proportional to $r^{-6}$, which is consistent to the plot of Eq. \eqref{e22} (See Fig. \ref{fig:c}) as $r$ tends to get larger. Indeed, as $r \to \infty$, the metric function reduces to the Reissner-Nordstr\"om solution, and closer to the black hole \AU{see Fig. \ref{fig:c}}, the effects of the non-linear EM become evident.

For simplicity, in all the plots, we have taken $M=C=\kappa=1$. Fig. \ref{fig:c} shows the plot of the metric function as $\beta$ varies.

\section{Null geodesics and the shadow} \label{sec:level3}
\begin{figure}
    \centering
    \includegraphics[width=\linewidth]{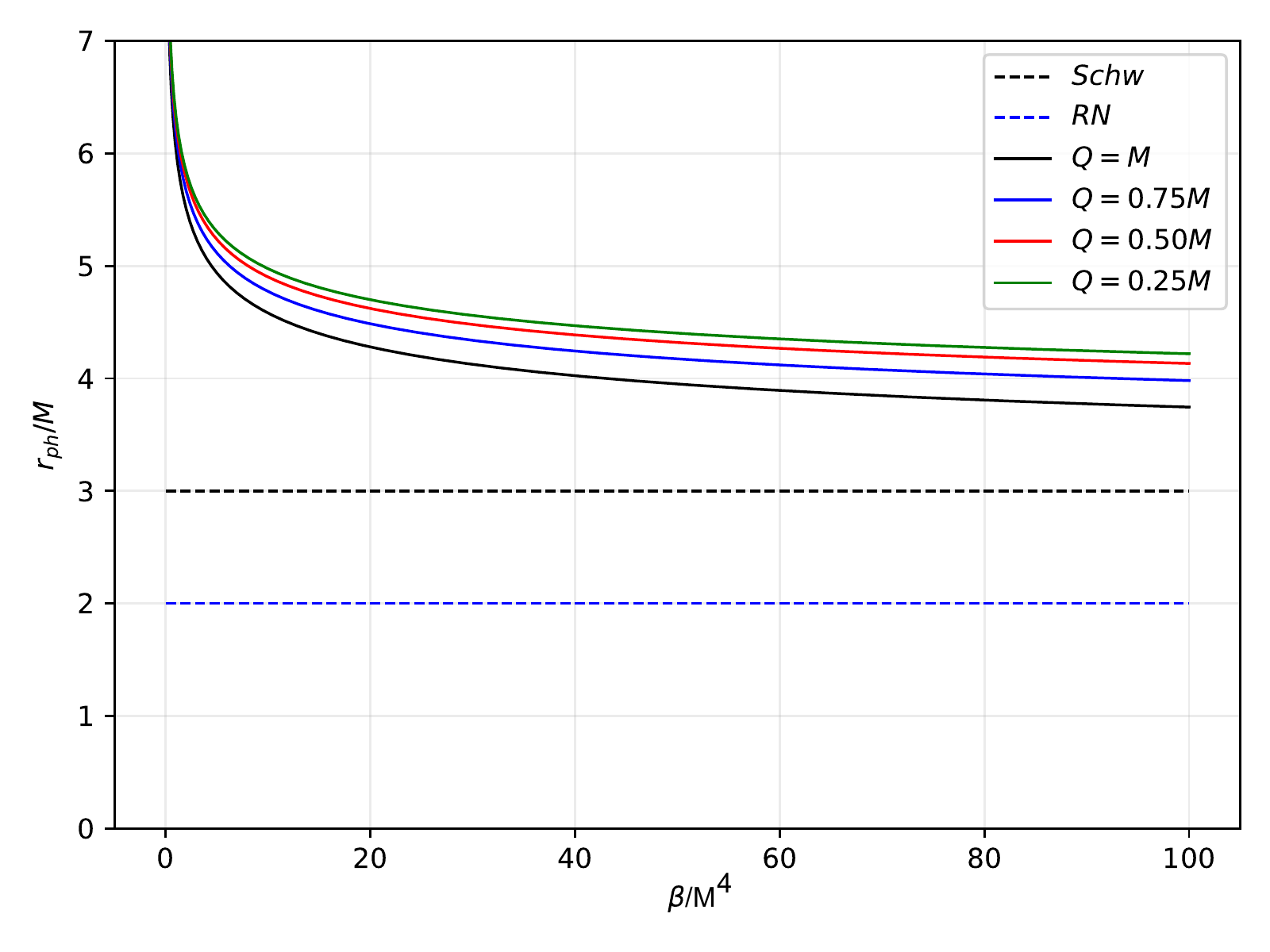}
    \caption{The photonsphere radius under the effect of the non-linear electromagnetic parameter $\beta$ perceived by an observer near the black hole. The dashed black and blue lines represent the Schwarzschild and Reissner-Nordstr\"om ($Q = M$) cases, respectively.}
    \label{fig:rphe}
\end{figure}

\rp{In this section, we will first explore the behavior of the shadow radius theoretically, and then find constraints to the NLE parameter using the data from EHT. To do so, we will consider the methodology presented in Perlick et al. \cite{Perlick:2015vta} for the calculation of the photonsphere radii as well as the shadow radius.  To begin with, consider a static, spherically symmetric (SSS) spacetime given by}
\begin{align}  \label{e25}
    ds^2&=g_{\mu \nu}dx^{\mu}dx^{\nu} \nonumber \\
    &=-A(r)dt^2+B(r)dr^2+C(r)d\Omega^2,
\end{align}
\rp{where $d\Omega^2=d\theta^2+r^2\sin^2\theta \, d\phi^2$ is the line element of the unit two-spheres. Without loss of generality, we analyze the null geodesic in the equatorial plane only such that the polar angle is fixed to $\theta = \pi/2$. As a result, $C(r) = r^2$.} Then, the Hamiltonian for light rays is given by
\begin{equation} \label{e41}
    H = \frac{1}{2} g^{ik} p_{i} p_{k} = \frac{1}{2} \left( -\frac{p_{t}^{2}}{A(r)} + \frac{p_{r}^{2}}{B(r)} + \frac{p_{\phi }^{2}}{C(r)} \right),
\end{equation}
where $A(r)=f(r)$, $B(r)=f(r)^{-1}$, and $C(r)=r^2$ wherein the metric function $f(r)$ can be represented by either Eq. \eqref{e23} or \eqref{e24}. The equations of motion for null particles are then
\begin{equation} \label{e42}
    \dot{x}^{i} = \frac{\partial H}{\partial p_{i}}, \quad \quad \dot{p}_{i} = -\frac{\partial H}{\partial x^{i}}.
\end{equation}
Here, $\dot{x}=dx/d\lambda$ and $\dot{p}$ represents the conjugate momenta. Eq. \eqref{e42} gives
\begin{equation} \label{e43}
    \dot{t} = -\frac{p_{t}}{f(r)}, \quad \quad \dot{\phi } = \frac{p_{\phi }}{r^2}, \quad \quad \dot{r} = -\frac{p_{r}}{f(r)^{-1}},
\end{equation}
$$\dot{p}_{t} = 0, \quad \quad \dot{p}_{\phi } = 0,$$
\begin{equation} \label{e44}
     \dot{p}_{r} = \frac{1}{2} \left( -\frac{p_{t}^{2} f'(r)}{f(r)^{2}} + \frac{p_{r}^{2} f'(r)^{-1}}{f(r)^{-2}} + \frac{2p_{\phi }^{2}}{r^{3}} \right).
\end{equation}
Setting $ H = 0 $, we have
\begin{equation} \label{e45}
    -\frac{p_{t}^{2}}{f(r)} + \frac{p_{r}^{2}}{f(r)^{-1}} + \frac{p_{\phi }^{2}}{r^2} = 0,
\end{equation}
and it now follows that
\begin{equation} \label{e46}
    \frac{dr}{d\phi } = \frac{\dot{r}}{\dot{\phi }} = \frac{r^2} {f(r)^{-1}}\frac{p_r}{p_{\phi}}.
\end{equation}
Setting $p_t=-\omega_{o}$, and using $p_r$, we can get the relation how $r$ changes with $\phi$:
\begin{equation} \label{e47}
    \frac{dr}{d\phi } = \pm \frac{r}{f(r)^{-1/2}} \sqrt{\frac{\omega _{o}^{2}}{p_{\phi }^{2}} h(r)^{2} - 1},
\end{equation}
where 
\begin{equation} \label{e48}
    h(r)^{2} = \frac{r^2}{f(r)}
\end{equation}
is defined. For a circular light orbit, the radial velocity and acceleration should be $ \dot{r} = 0 $ and $ \ddot{r} = 0 $ respectively, and hence, $ p_{r} = 0 $. Eq. \eqref{e44} then becomes
\begin{equation} \label{e49}
    0 = -\frac{\omega _{o}^{2}}{f(r)} + \frac{p_{\phi }^{2}}{r^2}.
\end{equation}
Since $ \dot{p}_{r} = 0 $, Eq. \eqref{e43} can be rewritten as
\begin{equation} \label{e50}
    \dot{p}_{r} = 0 = -\frac{\omega _{o}^{2} f'(r)}{f(r)^{2}} + \frac{2p_{\phi }^{2}}{r^3}.
\end{equation}
Using Eqs. \eqref{e48} and \eqref{e49}, we find
\begin{equation} \label{e51}
    p_{\phi }^{2} = r^2  \frac{\omega _{o}^{2}}{f(r)},
\end{equation}
\begin{equation} \label{e52}
    p_{\phi }^{2} = r^3  \frac{\omega _{o}^{2} f'(r)}{f(r)^{2}}.
\end{equation}
The implication of subtracting Eqs. \eqref{e50} and \eqref{e51} give the information on how to find the radius of the photon sphere:
\begin{equation} \label{e53}
    0 = \frac{d}{dr} h(r)^{2}.
\end{equation}
Let us start with the photonsphere perceived by someone near the black hole. To simplify the calculation, let us define
\begin{equation} \label{e54}
    m = M + \frac{16C^{3/2}\kappa^2}{15\beta^{1/4}}, \quad q^2 = Q^2-\frac{C^2\kappa^2}{2}.
\end{equation}
These definitions will enable us to write Eq. \eqref{e23} as
\begin{equation} \label{e55}
    f(r) = 1 - \frac{2m}{r} + \frac{q^2}{r^2}.
\end{equation}
Following Eq. \eqref{e53}, the analytical form of the photonsphere radius $r_\text{ph}$ is
\begin{align} \label{e56}
    r_\text{ph} &= \frac{3M}{2}+\frac{8\kappa^{2}C^{3/2}}{5\beta^{1/4}} \nonumber\\
    &\pm\frac{1}{2}\sqrt{9\left(\frac{16\kappa^{2}C^{3/2}}{15\beta^{1/4}}+M\right)^{2}+4\kappa^{2}C^{2}-8Q^{2}},
\end{align}
where we take the positive root since it is the one that gives physical results. Also, note that $\beta \ne 0$ in this case. We plot the above equation numerically (See Fig. \ref{fig:rphe}), where it shows the values of $\beta$ for the photonsphere to exist.
As the parameter $\beta$ increases, the photonsphere radius essentially decreases. An asymptotic increase in $r_\text{ph}$ is seen when $\beta \to 0$. Then, for a given value of $\beta$, we observe an increase in photonsphere radius as the black hole charge $Q$ decreases.

\begin{figure}
    \centering
    \includegraphics[width=\linewidth]{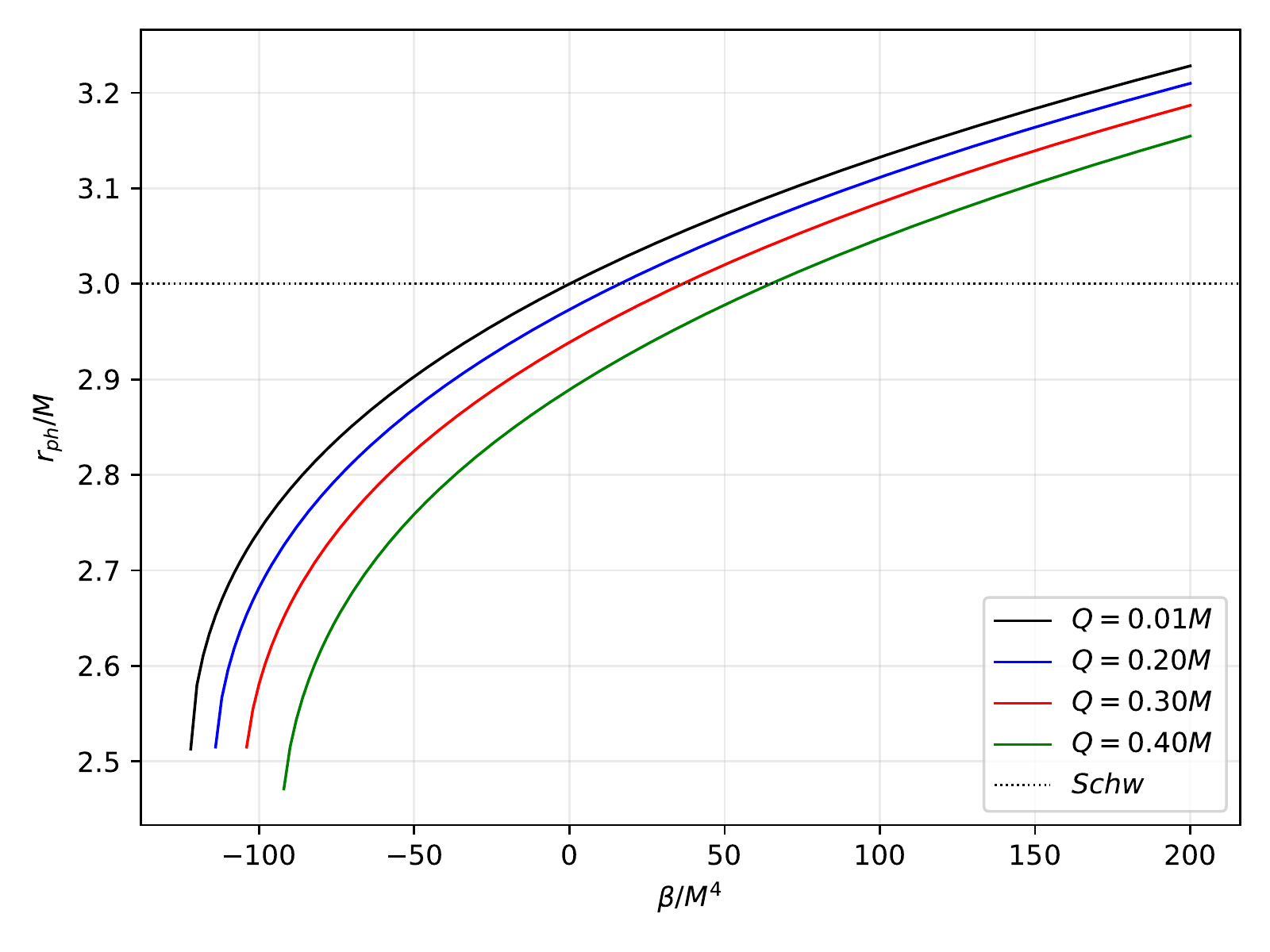}
    \caption{The photonsphere radius under the effect of the non-linear electromagnetic parameter $\beta$ perceived by a remote observer at $r_\text{o} \to \infty$.}
    \label{fig:rpha}
\end{figure}

For a remote observer, using Eq. \eqref{e24} in finding the photonsphere gives
\begin{equation} \label{e57}
    3Mr^{5}-2Q^{2}r^{4}-r^{6}+\frac{2\kappa^2 \beta C^4}{5}=0,
\end{equation}
where the plot is shown in Fig. \ref{fig:rpha}.
For an observer in an asymptotically flat region of spacetime, negative values of $\beta$ are permitted as they influence the photonsphere radius. The Schwarzschild case is at $r = 3M$. It can be observed that as $\beta$ decreases, the photonsphere radius also decreases. Also, in the plot, we compared several values for the black hole charge $Q$, including $Q = 0.01M$. We see that $\beta$ can still have a strong influence in causing deviation from the Schwarzschild or RN case.

\begin{figure}
    \centering
    \includegraphics[width=\linewidth]{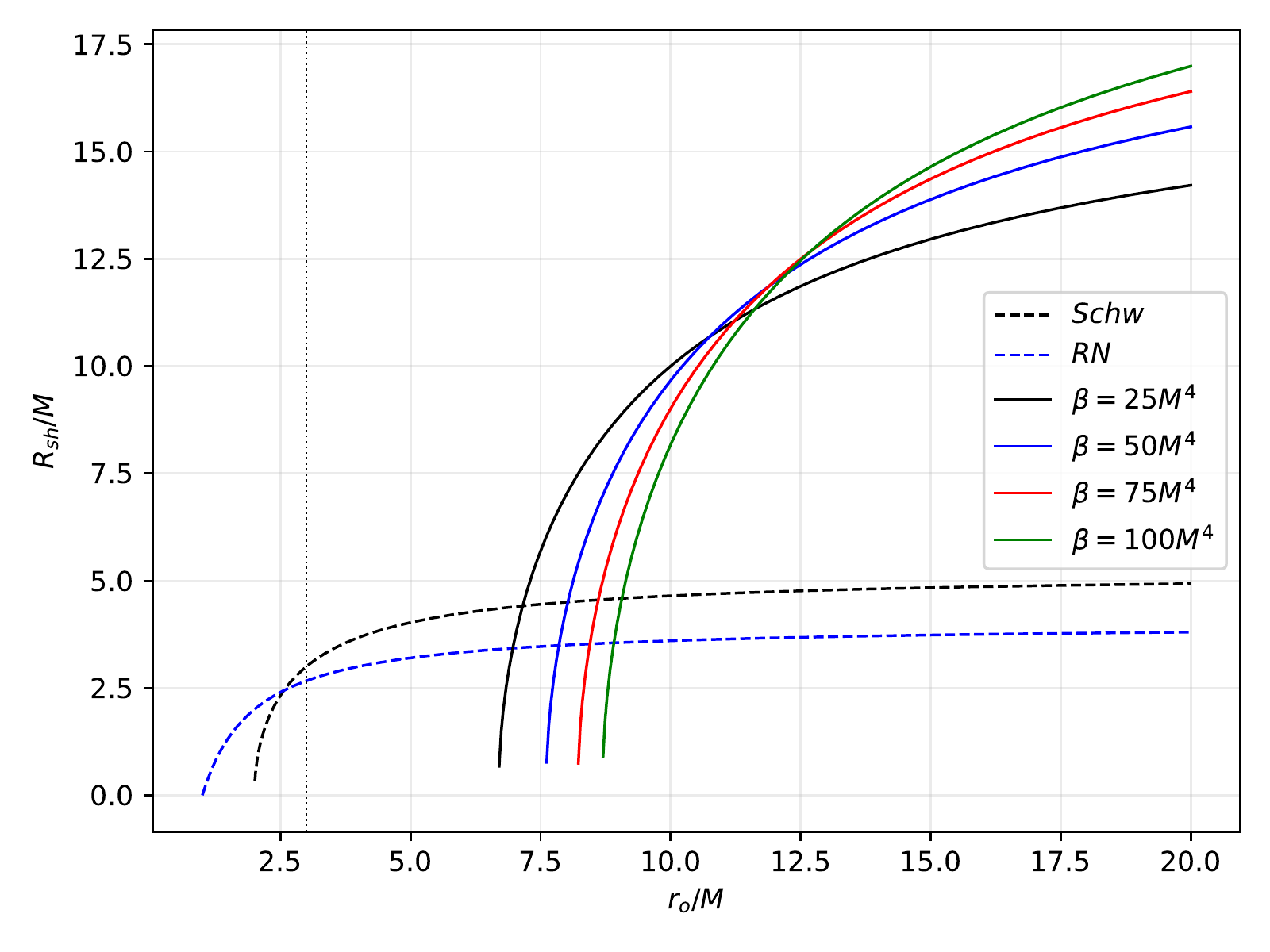}
    \caption{The shadow radius under the effect of the non-linear electromagnetic parameter $\beta$ perceived by an observer near the black hole. The dashed black and blue lines represent the Schwarzschild and Reissner-Nordstr\"om cases, respectively. The values of $Q$ are the same as in Fig. \ref{fig:rphe}.}
    \label{fig:rshc}
\end{figure}

Let us now determine the behavior of the shadow radius. It depends on the initial direction of light rays that spiral towards the outermost photon sphere. The angular radius of the shadow $ \alpha _{\text{sh}} $ is defined by
\begin{equation} \label{e58}
    \cot(\alpha_{\text{sh}}) = \frac{f(r)^{-1/2}}{r} \left. \frac{dr}{d\phi } \right|_{r=r_\text{o}}.
\end{equation}
If the light ray goes out again after reaching $ r_{\text{ph}} $, the orbit equation in Eq. \eqref{e47} can be rewritten as
\begin{equation} \label{e59}
    \frac{dr}{d\phi } = \pm \frac{r}{f(r)^{-1/2}} \sqrt{\frac{h(r)^{2}}{h(r_{\text{ph}})^{2}} - 1}.
\end{equation}
Thus, the angular radius of the shadow becomes
\begin{equation} \label{e60}
    \cot^{2}(\alpha _\text{sh}) = \frac{h(r_\text{o})^{2}}{h(r_{\text{ph}})^{2}} - 1,
\end{equation}
and by using a trigonometric identity, $ 1 + \cot^{2}(\alpha_{\text{sh}}) = \frac{1}{\sin^{2}(\alpha_{\text{sh}})} $, it can be rewritten as
\begin{equation} \label{e61}
    \sin^{2}(\alpha _\text{sh}) = \frac{h(r_{\text{ph}})^{2}}{h(r_\text{o})^{2}},
\end{equation}
where $ h(r_{\text{ph}})$ must be evaluated using the value (or expression) photonsphere radius, and $ r_\text{o} $ specifies the location of the observer. Keeping in mind Eqs. \eqref{e54} and \eqref{e56}, the expression of the shadow radius as measured by an observer near the black hole is
\begin{equation}
    R_\text{sh}=\frac{\left(\sqrt{9m^{2}-8q^{2}}+3m\right)^{2}\sqrt{r_\text{o}^{2}-2r_\text{o}m+q^{2}}}{2\sqrt{2}r_\text{o}\sqrt{m\sqrt{9m^{2}-8q^{2}}+3m^{2}-2q^{2}}}.
\end{equation}
In Fig. \ref{fig:rshc}, we plot how the shadow radius behaves for different values of $\beta$ where maximal black hole charge is assumed ($Q=1$).
We observe that there are points where the solid lines intersect the Schwarzschild and Reissner-Nordstr\"om shadow radius. It means that even if there is an influence of $\beta$, there are radial points where the shadow radius becomes identical to the Schwarzschild and Reissner-Nordstr\"om cases. In these points, the effect of $\beta$ increases the radial position where this similarity occurs. Another observation is that as the observer gets a little far from the black hole, a drastic increase in the shadow radius can be seen relative to the known cases.

\begin{figure}
    \centering
    \includegraphics[width=\linewidth]{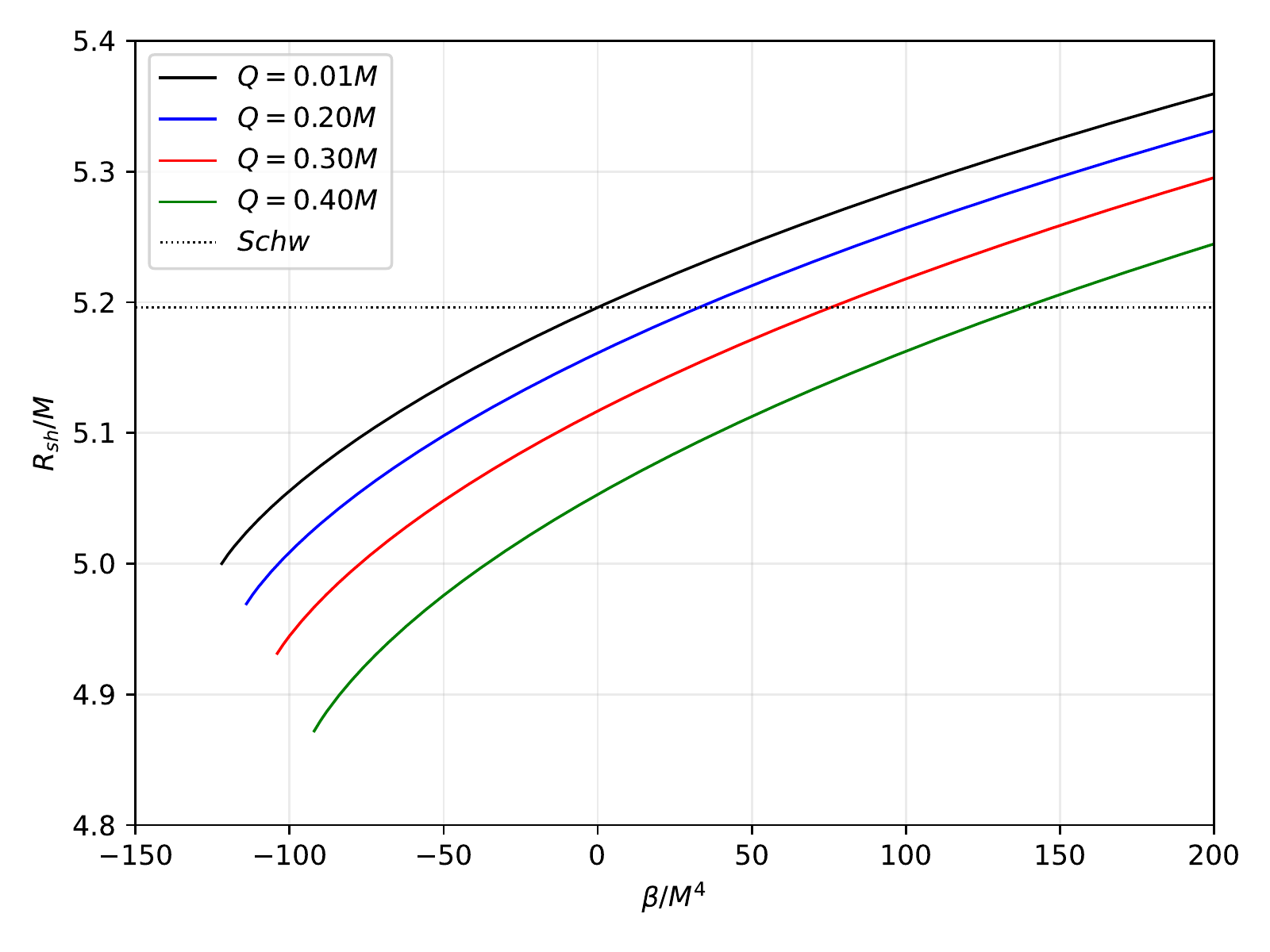}
    \caption{The shadow radius under the effect of the non-linear electromagnetic parameter $\beta$ perceived by a remote observer at $r_\text{o} \to \infty$. The black dotted line represents the shadow radius for the Schwarzschild black hole.}
    \label{fig:rshf}
\end{figure}

For the remote observer, and using Eq. \eqref{e24}, the shadow radius is approximated as
\begin{equation} \label{e48}
    R_\text{sh} \sim \frac{r_\text{ph}^{4}(r_\text{o}-M)}{r_\text{o}\sqrt{-\frac{\beta\kappa^{2}C^{4}}{10}-2Mr_\text{ph}^{5}+Q^{2}r_\text{ph}^{4}+r_\text{ph}^{6}}},
\end{equation}
where $r_\text{ph}$ is the solution to Eq. \eqref{e57} for a certain value for $\beta$ and $Q$. In Fig. \ref{fig:rshf}, we plot the behavior of $R_\text{sh}$ at large distance from the black hole.
We included a low charge case $Q = 0.01M$ for comparison. The shadow radius decreases in value for the remote case as $\beta$ continues to become negative. It is the same behavior as the photonsphere radius, as seen in Fig. \ref{fig:rpha}. We can also tell how the shadow radius is sensitive to the effects of $\beta$ since even when $Q = 0.01M$, a slight decrease in $\beta$ results to a noticeable decrease in $R_\text{sh}$.

\subsection{Constraints on NED parameters of black hole with the EHT observations of M87* and Sgr A*}
In this section, we find constraints to the coupling parameter $\beta$ using the observation data provided by the EHT for M87* and Sagittarius A* from their shadow images. We only focused on the non-rotating case since the rotation parameter $a$ of Sgr. A* is small enough to have considerable deviation to the shadow radius \cite{Vagnozzi:2022moj}. Furthermore, it has also been concluded for M87* that it is difficult to distinguish between a Kerr BH ($a = 0.60M$) and a dilaton BH (non-rotating) based on BH shadow images alone using general-relativistic magnetohydrodynamical simulations and radiative-transfer calculations to generate synthetic shadow images in comparison to the present observation from the EHT Ref. \cite{Mizuno:2018lxz},. Also remarked in Ref. \cite{EventHorizonTelescope:2021dqv} that the shadow size of M87* lies within the range of $3\sqrt{3}(1\pm0.17) M$, whether their model is spherically symmetric or axisymmetric.

As reported in Ref. \cite{EventHorizonTelescope:2019dse}, for the M87*, the angular diameter of the shadow is $\theta_\text{M87*} = 42 \pm 3 \:\mu$as, the distance of the M87* from the Earth is $D = 16.8$ Mpc, and the mass of the M87* is $M_\text{M87*} = 6.5 \pm 0.90$x$10^9 \: M_\odot$. Similarly, for Sagittarius A* the data is provided in recent EHT paper \cite{EventHorizonTelescope:2022xnr}. The angular diameter of the shadow is $\theta_\text{Sgr.A*} = 48.7 \pm 7 \:\mu$as (EHT), the distance of the Sgr. A* from the Earth is $D = 8277\pm33$ pc and mass of the black hole is $M_\text{Sgr. A*} = 4.3 \pm 0.013$x$10^6 \: M_\odot$ (VLTI) \cite{Wang2022,EventHorizonTelescope:2022xnr}. Now, once we have the above data about the black hole, we can calculate the diameter of the shadow size in units of mass by using the following expression \cite{Bambi:2019tjh},
\begin{equation}
    d_\text{sh} = \frac{D \theta}{M}
\end{equation}
Hence, the theoretical shadow diameter, however, can be obtained via $d_\text{sh}^\text{theo} = 2R_\text{sh}$. Therefore, by using the above expression, we get the diameter of the shadow image of M87* $d^\text{M87*}_\text{sh} = (11 \pm 1.5)M$ and for Sgr. A* $d^\text{Sgr.A*}_\text{sh} = (9.5 \pm 1.4)M$.

\begin{figure}
    \centering
    \includegraphics[width=\linewidth]{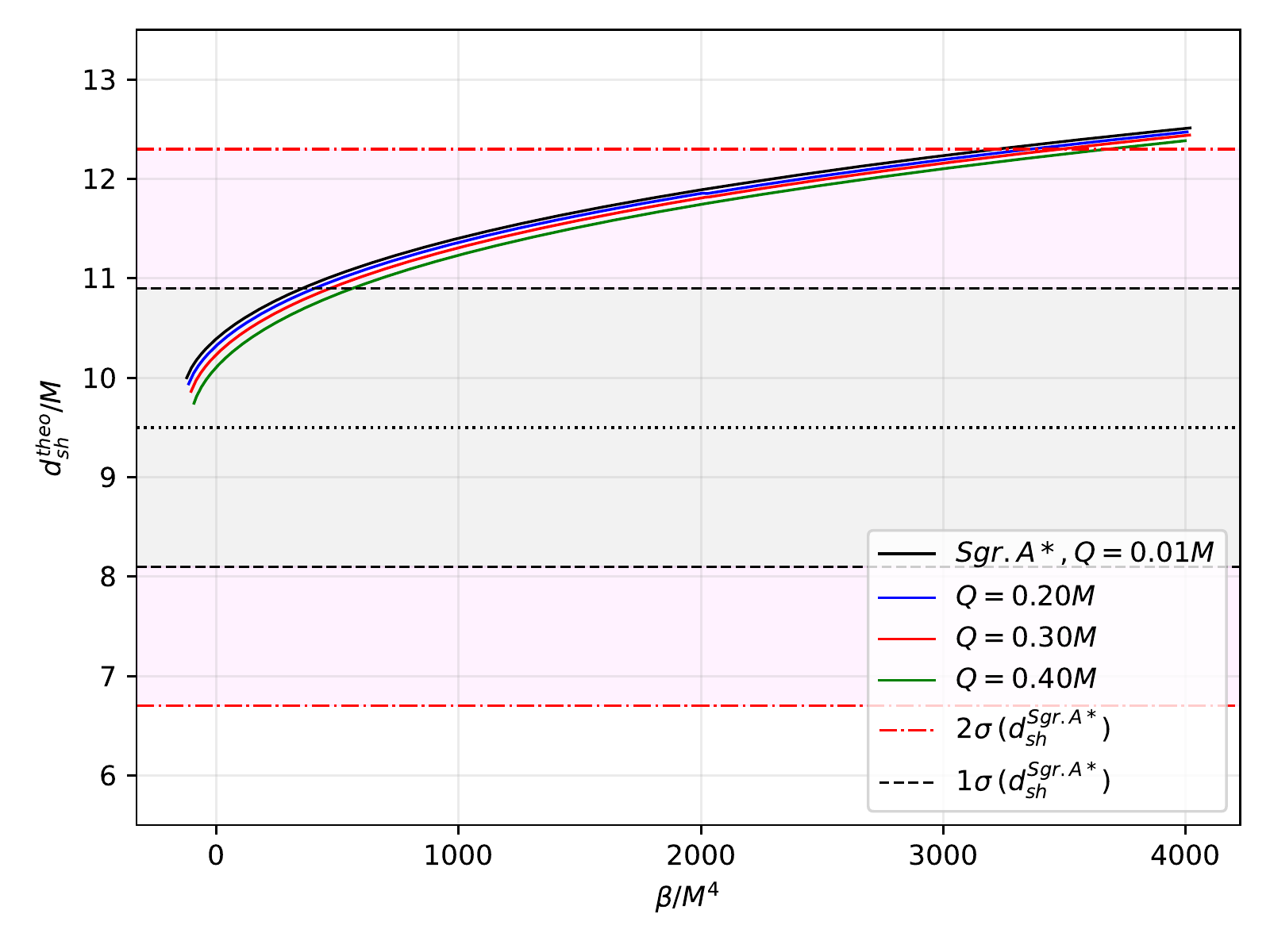}
    \includegraphics[width=\linewidth]{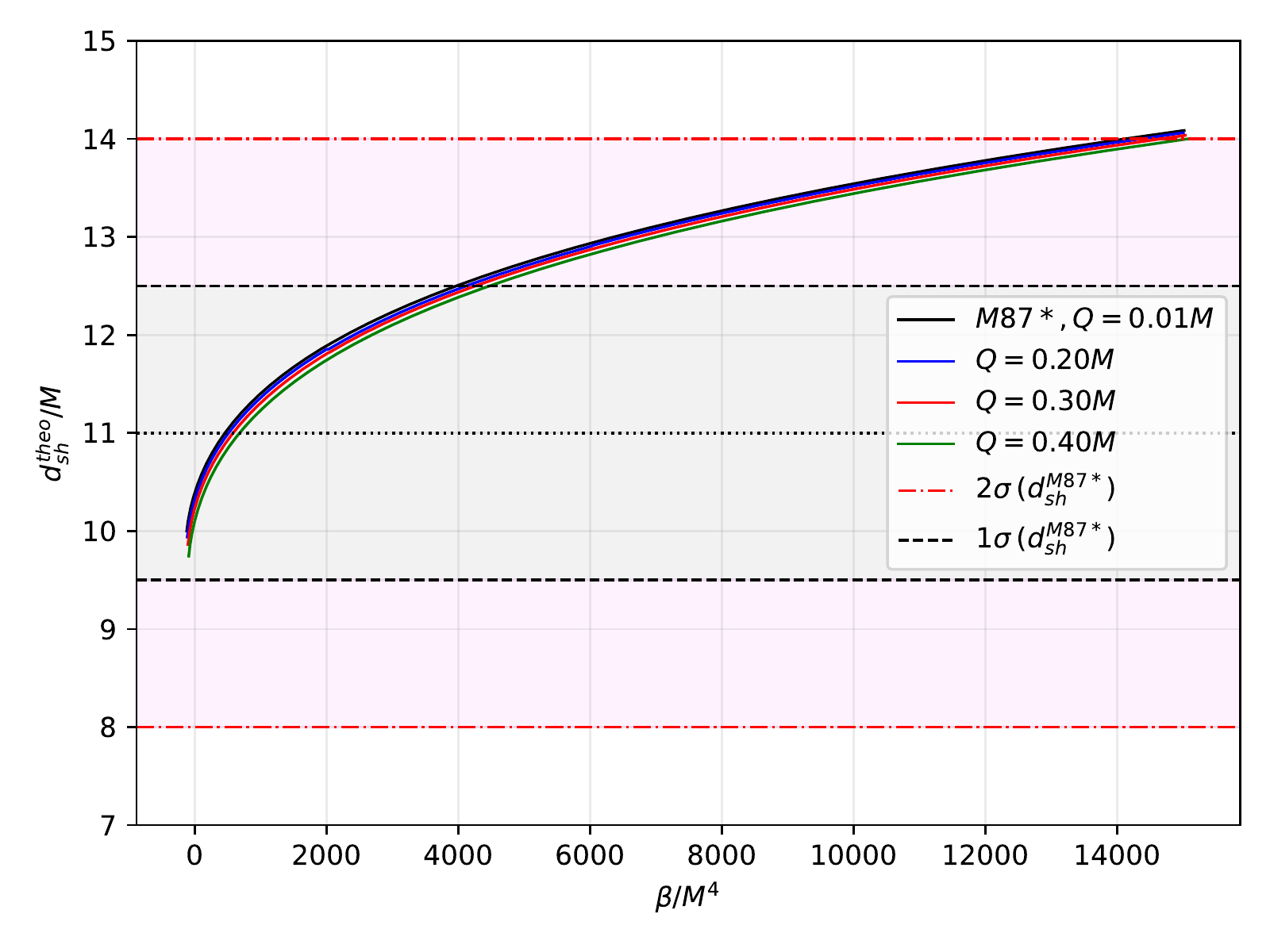}
    \caption{Plot showing the constraints for coupling parameter $\beta$. For the values of $\beta$ at the mean, $1\sigma$ and $2\sigma$ levels, see Table \ref{Tab1}.}
    \label{fig:cons}
\end{figure}

The variation of the diameter of the shadow image with coupling parameter $\beta$ for M87* and for Sgr. A* is shown in Fig. \ref{fig:cons}, showing uncertainties at $1\sigma$ and $2\sigma$ levels. The numerical values for the upper bounds in $\beta$ is found in Table \ref{Tab1}. Note that due to Eq. \eqref{e57}, there is an inner value for $\beta$, which are all negative, for some value of $Q$. For $Q = 0.01M, 0.20M, 0.30M$, and $0.40M$, these are $\beta/M^4 = -122, -114, -104$, and $-92$, respectively. We observed no lower bounds for $\beta$. Furthermore, all the values for $\beta$ were chosen in Figs. \ref{fig:c}-\ref{fig:rshf} fall within the uncertainty levels.
\begin{table}[!ht]
    \centering
    \begin{tabular}{ c c c c c c }
    \hline
    \hline
      Sgr. A* &  \multicolumn{2}{c}{$2\sigma$} & \multicolumn{2}{c}{$1\sigma$} & \multicolumn{1}{c}{observed $R_\text{sh}$} \\
    \hline
    charge $Q$ & upper & lower & upper & lower  & mean   \\
    \hline
    $0.01M$ & 357 & - & 3223 & -  & -    \\
    $0.20M$ & 406 & - & 3360 & -  & -   \\
    $0.30M$ & 471 & - & 3485 & - & - \\
    $0.40M$ & 563 & - & 3679 & - & - \\
    \hline
    \end{tabular}
    \quad
    \begin{tabular}{ c c c c c c }
    \hline
    \hline
      M87* &  \multicolumn{2}{c}{$2\sigma$} & \multicolumn{2}{c}{$1\sigma$} & \multicolumn{1}{c}{observed $R_\text{sh}$} \\
    \hline
    charge $Q$ & upper & lower & upper & lower  & mean   \\
    \hline
    $0.01M$ & 14118 & - & 3966 & -  & 457    \\
    $0.20M$ & 14354 & - & 4118 & -  & 512   \\
    $0.30M$ & 14655 & - & 4257 & - & 580 \\
    $0.40M$ & 15072 & - & 4468 & - & 679 \\
    \hline
    \end{tabular}
    \caption{Values of $\beta/M^4$ as generated in Fig. \ref{fig:cons}.}
    \label{Tab1}
\end{table}

\rp{Fig. \ref{fig:cons} also shows how the shadow radius behaves as the coupling parameter varies while $r_\text{o}$, which is the observer's radial distance from the black hole, remains fixed. Note that such a behavior will be identical since we have used Eq. \eqref{e48}. However, it turns out that the data for M87* gives a better constraints for the coupling parameter since there are points which crosses the mean of the shadow diameter. It would mean that there is a certain value for the coupling parameter that gives the observed value of M87*'s shadow. Note that the NLE coupling parameter is strong near the black hole, thus it can greatly affect the photonsphere behavior. As a consequence, the shadow cast can also be affected even if perceived by a remote observer. By examining Eq. \eqref{e48}, we can see that $\beta$ have greater influence than that of the black hole charge $Q$. Thus, the curves we see in Fig. \ref{fig:cons} is mainly due to $\beta$, and the charge's effect is to decrease slightly the value of the shadow diameter.}

\section{Deflection Angle Using Gauss-Bonnet Theorem In Weak Field Limits} \label{sec:level2}
In this section, we explore another black hole phenomenon which is weak deflection angle and consider the effect of the NLE parameter. Using Eq. \eqref{e25}, the Jacobi metric reads
\begin{align} \label{e26}
    dl^2&=g_{ij}dx^{i}dx^{j} \nonumber \\
    &=(E^2-m^2A(r))\left(\frac{B(r)}{A(r)}dr^2+\frac{C(r)}{A(r)}d\Omega^2\right).
\end{align}
Here, $E$ is the particle's energy per unit mass $m$. It is easy to see how the Jacobi metric $dl^2$ reduces to the optical metric $dt^2$ for null particles as $m=0$ and $E=1$. The energy $E$ is one of the constants of motion, and asymptotic observers perceive the energy of a particle far from the black hole as
\begin{equation} \label{e27}
    E = \frac{m}{\sqrt{1-v^2}},
\end{equation}
where $v$ is the particle's velocity as a fraction of the speed of light $c$. \rp{By restricting light rays along the equatorial plane only, the Jacobi metric can then be rewritten as}
\begin{equation} \label{e28}
    dl^2=m^2[(1-v^2)^{-1}-A(r)]\left(\frac{B(r)}{A(r)}dr^2+\frac{C(r)}{A(r)}d\phi^2\right)
\end{equation}
without loss of generality. The determinant of the Jacobi metric above can also be easily calculated as
\begin{equation} \label{e29}
    g=m^4B(r)C(r)\frac{[A(r)(v^2-1)+1]^2}{A(r)^2(v^2-1)^2}.
\end{equation}
Next, we will use these equations to find the weak deflection angle using the Gauss-Bonnet theorem (GBT), originally stated as \cite{Carmo2016,Klingenberg2013}
\begin{equation} \label{e30}
    \iint_DKdS+\sum\limits_{a=1}^N \int_{\partial D_{a}} \kappa_{\text{g}} d\ell+ \sum\limits_{a=1}^N \theta_{a} = 2\pi\chi(D).
\end{equation}
Here, the Gaussian curvature $K$ describing the domain $D$ is a freely orientable $2D$ curved surface $S$ with an infinitesimal area element is $dS$. The boundary of $D$ are given by $\partial D_{\text{a}}$ (a=$1,2,..,N$), and the geodesic curvature $\kappa_{\text{g}}$ is integrated over the path $d\ell$ along a positive convention. Also, $\theta_\text{a}$ is the jump angle, which $\chi(D)$ is the Euler characteristic, which in our case is equal to $1$ since $D$ is in a non-singular region.

It was shown by \cite{Ishihara:2016vdc} that in a SSS spacetime admitting asymptotic flatness, Eq. \eqref{e30} can be written as
\begin{equation} \label{e31}
    \hat{\alpha} = -\iint_{_{\text{R}}^{\infty }\square _{\text{S}}^{\infty}}KdS.
\end{equation}
With $K$ being related to the $2D$ Riemann tensor given by
\begin{equation} \label{e32}
    K=\frac{R_{r\phi r\phi }}{g},
\end{equation}
and by construction
\begin{equation} \label{e33}
    dS = \sqrt{g}dr d\phi,
\end{equation}
then we can write Eq. \eqref{e31} as
\begin{equation} \label{e34}
\hat{\alpha}=\int_{0}^{\pi}\int_{r_\text{o}}^{\infty}\mathcal{K}drd\phi,
\end{equation}
where
\begin{equation} \label{e35}
    \mathcal{K}=-\frac{\partial}{\partial r}\left[\frac{1}{2}\sqrt{\frac{A(r)^{2}}{B(r)C(r)}}\frac{\partial}{\partial r}\left(\frac{D(r)}{C(r)}\right)\right].
\end{equation}
Here, $r_\text{o}$ is the distance of the closest approach, which can be found by solving the orbit equation ($u=1/r$)
\begin{equation} \label{e36}
    \left(\frac{du}{d\phi}\right)^2 = \frac{C(u)^2u^4}{A(u)B(u)}\left[\frac{1}{b^2v^2}-A(u)\left(\frac{1-v^2}{b^2v^2}+\frac{1}{C(u)}\right)\right].
\end{equation}
One can verify that the leading order is $u_\text{o}=\sin\phi/b$, which is sufficient. With $u=1/r$, we can finally write Eq. \eqref{e34} as
\begin{equation} \label{e37}
    \hat{\alpha}=-\int_{0}^{\pi}\int_{0}^{u_{o}}\frac{\mathcal{K}}{u^2}dud\phi.
\end{equation}
In leading orders, we found $\mathcal{K}$ as
\begin{align} \label{e38}
    &\mathcal{K}=\frac{M\left(-v^{2}-1\right)}{r^{2}v^{2}}+\frac{Q^{2}\left(v^{2}+2\right)}{r^{3}v^{2}}-\frac{3\beta\kappa^{2}C^{4}\left(v^{2}+6\right)}{10r^{7}v^{2}} \nonumber \\
    &+\frac{3MQ^{2}\left(v^{4}+9v^{2}-6\right)}{2r^{4}v^{4}}+\frac{Q^{2}\beta\kappa^{2}C^{4}\left(v^{4}+24v^{2}-16\right)}{5r^{9}v^{4}} \nonumber \\
    &-\frac{7M\beta\kappa^{2}C^{4}\left(v^{4}+21v^{2}-14\right)}{20r^{8}v^{4}} \nonumber \\
    &+\frac{27MQ^{2}\beta\kappa^{2}C^{4}\left(v^{6}+45v^{4}-60v^{2}+24\right)}{40r^{10}v^{6}}.
\end{align}
Hereafter, evaluating Eq. \eqref{e38}, the weak deflection angle of a massive particle is now
\begin{align} \label{e39}
    &\hat{\alpha}= \frac{2M\left(v^{2}+1\right)}{bv^{2}}-\frac{\pi Q^{2}\left(v^{2}+2\right)}{4b^{2}v^{2}}+\frac{\pi\beta \kappa^{2}C^{4}\left(v^{2}+6\right)}{64b^{6}v^{2}} \nonumber \\
    &-\frac{2MQ^{2}\left(v^{4}+9v^{2}-6\right)}{3b^{3}v^{4}}-\frac{7\pi\beta \kappa^{2}C^{4}Q^{2}\left(v^{4}+24v^{2}-16\right)}{1024b^{8}v^{4}} \nonumber \\
    &+\frac{8\beta \kappa^{2}MC^{4}\left(v^{4}+21v^{2}-14\right)}{175b^{7}v^{4}} \nonumber \\
    &-\frac{32\beta \kappa^{2}MC^{4}Q^{2}\left(v^{6}+45v^{4}-60v^{2}+24\right)}{525b^{9}v^{6}}.
\end{align}
For null particles where $v=1$, we find
\begin{align} \label{e40}
    &\hat{\alpha}=\frac{4M}{b}-\frac{3\pi Q^{2}}{4b^{2}}+\frac{7\pi\beta \kappa^{2}C^{4}}{64b^{6}}-\frac{8MQ^{2}}{3b^{3}} \nonumber \\
    &-\frac{63\pi\beta \kappa^{2}C^{4}Q^{2}}{1024b^{8}}+\frac{64\beta \kappa^{2}MC^{4}}{175b^{7}}-\frac{64\beta \kappa^{2}MC^{4}Q^{2}}{105b^{9}}.
\end{align}
It is clear that without the influence of the non-linear electromagnetic field, Eq. \eqref{e40} agrees with the known expression for $\hat{\alpha}$ in Reissner-Nordstr\"om black hole. The dominating contribution of the non-linear electromagnetic field can be seen in the third term, and the effect is to increase the value of the weak deflection angle. Furthermore, we can already discern that as the impact parameter $b/M$ increases, the value of $\hat{\alpha}$ indeed approaches immediately that of a Reissner-Nordstr\"om black hole. \rp{Nevertheless, when $b/M$ considered is near the photonsphere radius, the deviation die to $\beta$ is noticeable for a given value of $Q$, which makes it a better probe than the shadow radius deviation.}
\begin{figure}
    \centering
    \includegraphics[width=\linewidth]{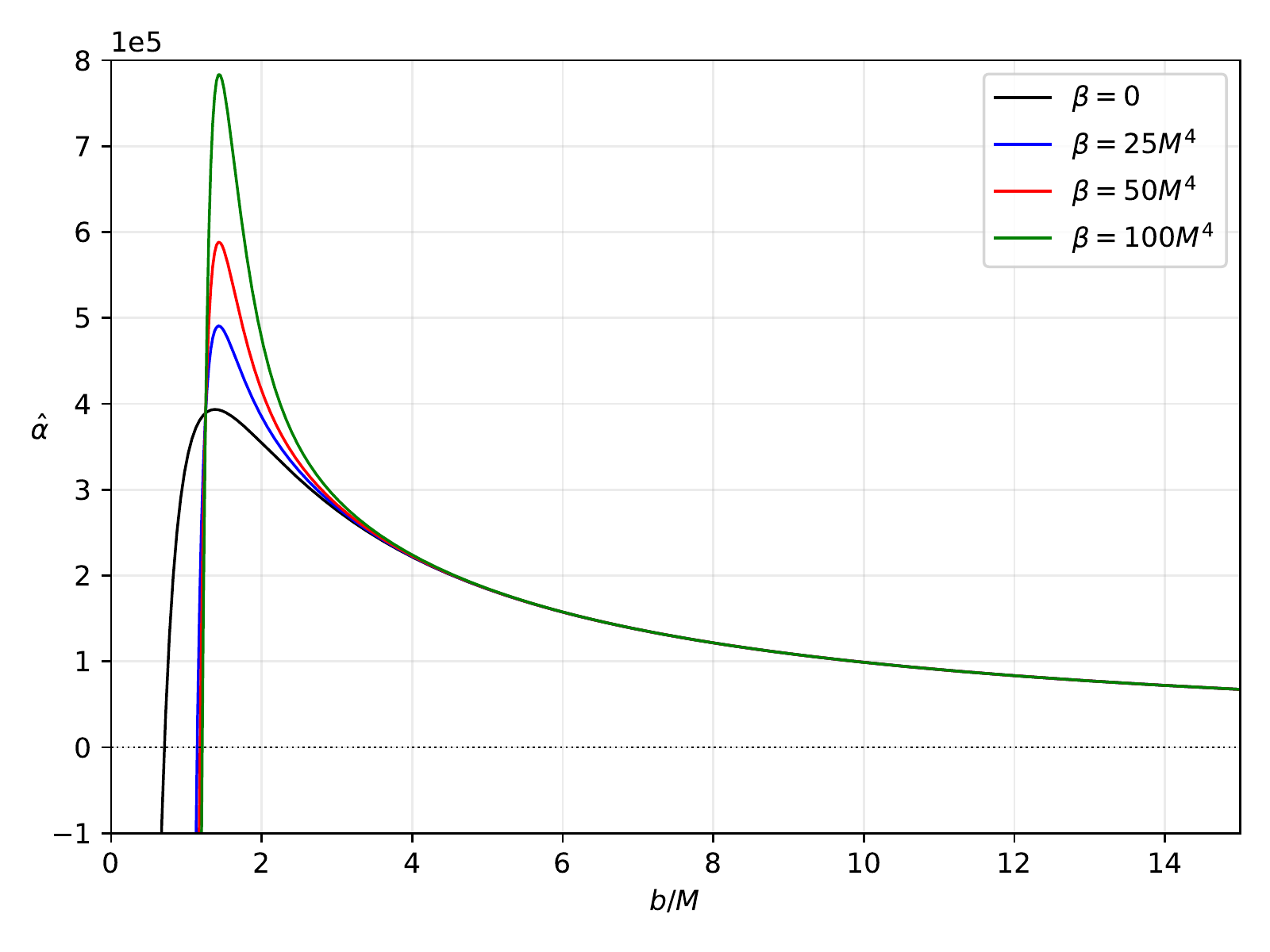}
    \caption{Plot of the weak deflection angle in $\mu$as. Here, $Q=k=M=1$ and $v=0.80$ are chosen to see how $\hat{\alpha}$ varies with $\beta$.}
    \label{fig:wdat}
\end{figure}
\begin{figure}
    \centering
    \includegraphics[width=\linewidth]{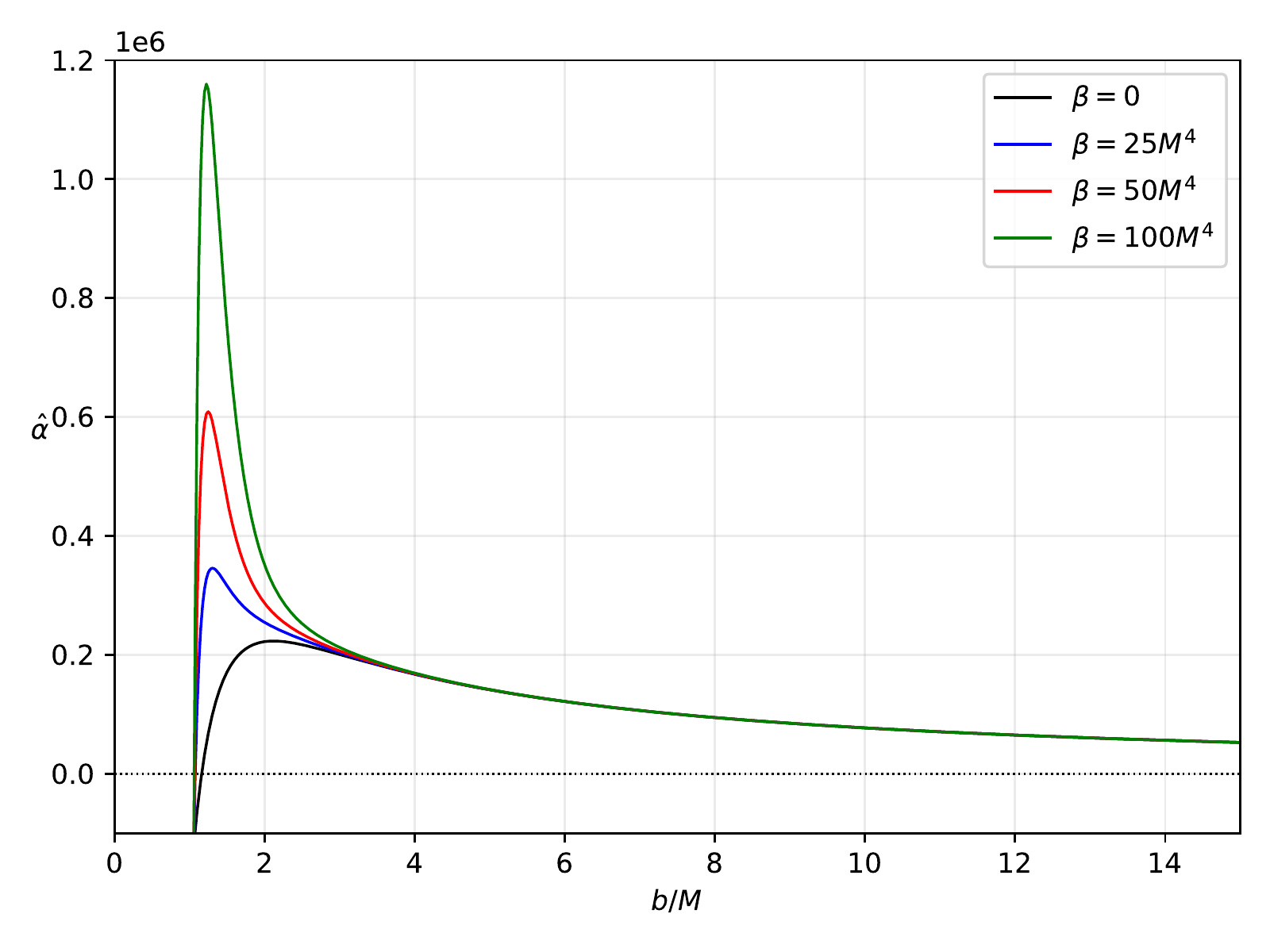}
    \caption{Plot of the weak deflection angle in $\mu$as. Here, $Q=k=M=1$ are chosen to see how $\hat{\alpha}$ varies with $\beta$. We also set $v=1$, which represents null particles.}
    \label{fig:wdan}
\end{figure}

\section{Thin Accretion Disk Model}
In this section, we will study the particle trajectory in the equatorial plane of the thin accretion disk around the black hole using different models.

\subsection{Novikov-Thorne accretion model}
\begin{figure}
    \centering
    \includegraphics[width=\linewidth]{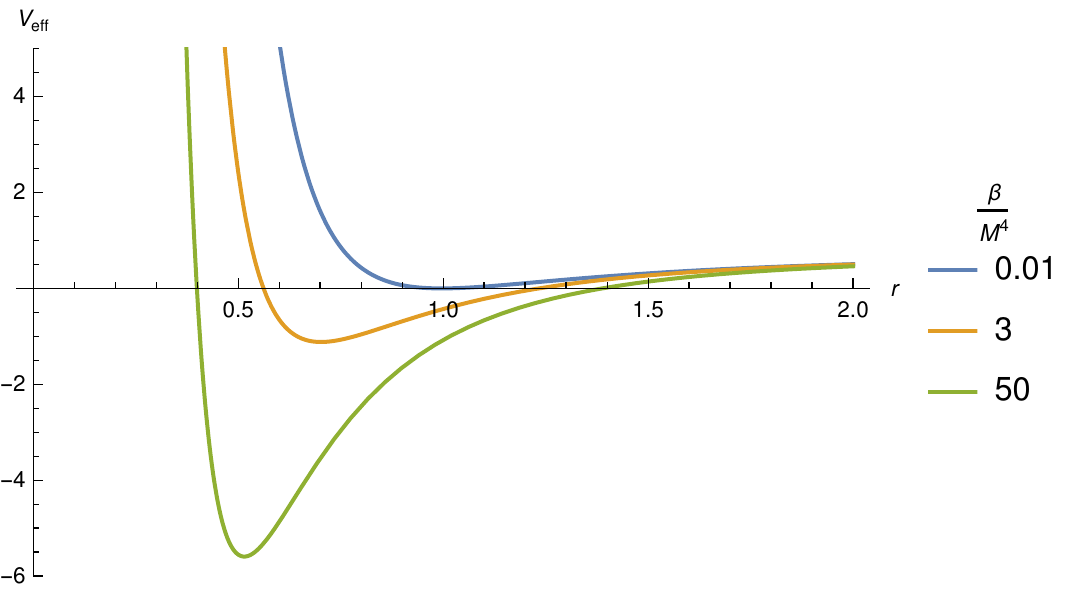}
    \caption{The effective potential for different values of $\beta=0.01 M^4 (\text{blue})$, $3 M^4 (\text{yellow})$ and $50 M^4 (\text{green})$. } \label{fig:e}
\end{figure}

First, we use the standard framework to explain thin accretion disk processes known as Novikov-Thorne model \cite{NovikovThorne}, which is a generalization of the Shakura-Sunyaev \cite{1973A}. For that, let us consider the Lagrangian for the test particle, which is moving around a compact object \cite{Heydari-Fard:2020iiu}
\begin{equation}
    \mathcal{L}=\frac{1}{2}g_{\mu \nu}\dot{x}^\mu \dot{x}^{\nu},
\end{equation}
where $g_{\mu \nu}$ is known as the spacetime metric, and a dot in the equation denotes differentiation with respect to the affine parameter. Now let us consider a general form of static and spherically symmetric spacetime metric
\begin{equation}
    ds^2=g_{tt}dt^2+g_{rr}dr^2+g_{\theta \theta}d\theta^2+g_{\phi \phi}d\phi^2,
\end{equation}
where we considered that the metric components $g_{tt}$ , $g_{rr}$ , $g_{\theta \theta}$ and $g_{\phi \phi}$ only depend on the radial coordinate $r$ if we consider the equatorial plane. Using the well-known, Euler-Lagrange equations, in the equatorial plane of the disk $\theta = \pi/2$
\begin{equation}
    \dot{t}=-\frac{E}{g_{tt}},
\end{equation}
\begin{equation}
    \dot{\phi}=\frac{L}{g_{\phi \phi}},
\end{equation}
where $E$ and $L$ are known as the particle's speciﬁc energy and angular momentum moving in the equatorial plane. Now, by considering $2\mathcal{L}=-1$ for a test particle and using Eqs. $(22)$ and $(23)$ to get the following equations
\begin{equation}
    -g_{tt}g_{rr}\dot{r}^2+V_\text{eff}=E^2,
\end{equation}
where the $V_\text{eff}$, effective potential is deﬁned as
\begin{equation}
    V_\text{eff}=-g_{tt}\left(1+\frac{L^2}{g_{\phi \phi}} \right).
\end{equation}
The effective potential for the different values of the parameter $\beta=0.01M^4,3M^4,50M^4$ is shown in the Fig. ~\ref{fig:e}. It has been observed that as we increase the coupling constant $\beta$, the dip in the effective potential increases, affecting the accretion disk properties. 

\begin{figure}
    \centering
    \includegraphics[width=\linewidth]{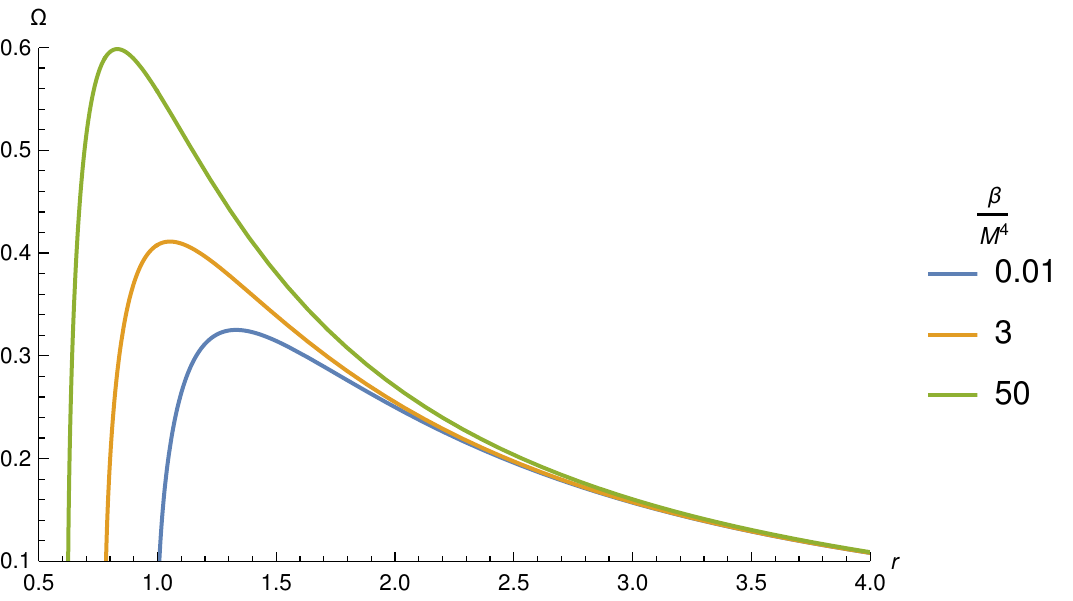}
    \caption{Angular velocity for different values of $\beta=0.01M^4 (\text{blue})$, $3M^4 (\text{yellow})$ and $50M^4 (\text{green})$. } \label{fig:g}
\end{figure}

\begin{figure}
    \centering
    \includegraphics[width=\linewidth]{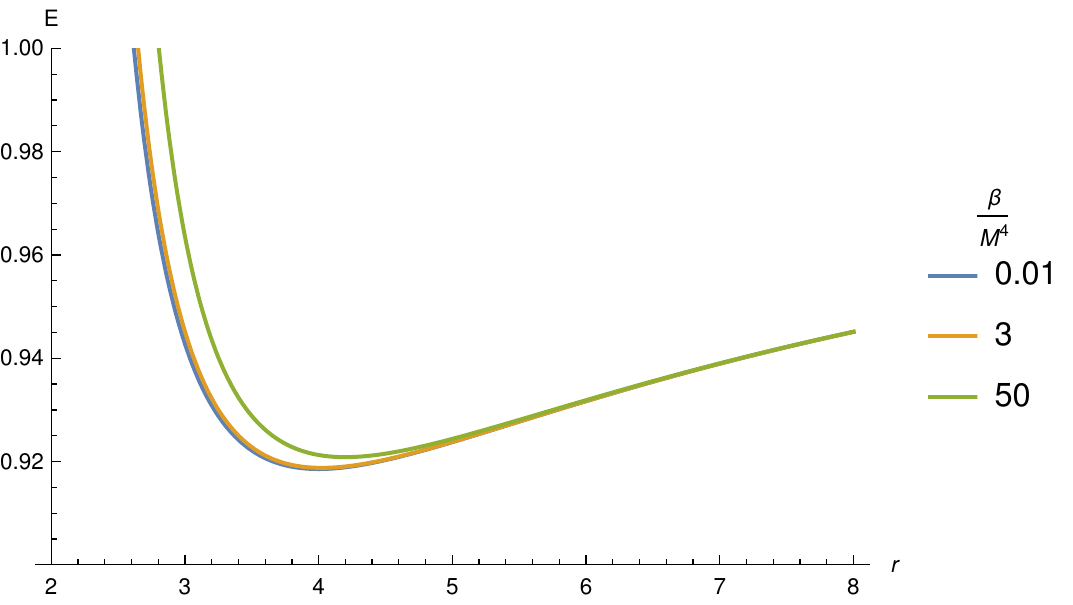}
    \caption{Specific Energy for different values of $\beta=0.01M^4 (\text{blue})$, $3M^4 (\text{yellow})$ and $50M^4 (\text{green})$. } \label{fig:h}
\end{figure}

\begin{figure}
    \centering
    \includegraphics[width=\linewidth]{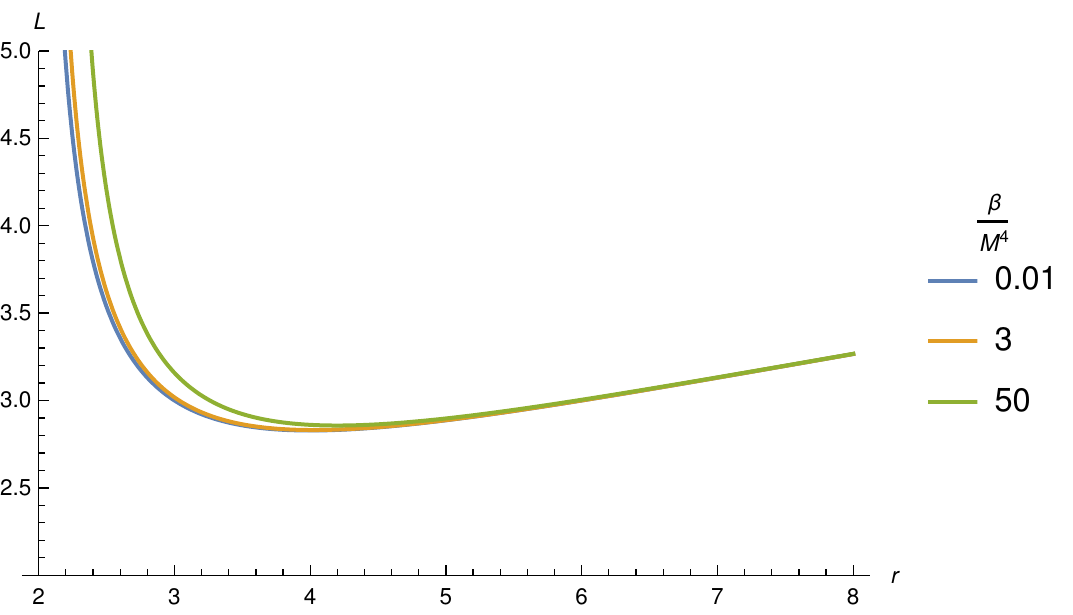}
    \caption{Specific angular momentum for different values of $\beta=0.01M^4 (\text{blue})$, $3M^4 (\text{yellow})$ and $50M^4 (\text{green})$.} \label{fig:i}
\end{figure}

Now for the stable circular orbits, we use the required conditions $V_\text{eff}=0$ and $V_{\text{eff},r}=0$, and obtain the speciﬁc energy $(E)$, speciﬁc angular momentum $(L)$ and angular velocity $(\Omega)$ of particles moving in the equatorial plane and in the presence of the gravitational potential of the black hole as follows:
\begin{equation}
    E=-\frac{g_{tt}}{\sqrt{-g_{tt}-g_{\phi \phi}\Omega^2}}
\end{equation}
\begin{equation}
    L=-\frac{g_{\phi \phi}\Omega}{\sqrt{-g_{tt}-g_{\phi \phi}\Omega^2}}
\end{equation}
\begin{equation}
    \Omega=\frac{d\phi}{dt}=\sqrt{\frac{-g_{tt,r}}{g_{\phi \phi,r}}}
\end{equation}
The plots for the angular velocity, specific energy and specific angular momentum has been shown in Fig. ~\ref{fig:g}, ~\ref{fig:h} and ~\ref{fig:i} for the different values of coupling parameter $\beta$. As we increase $\beta$, all the quantities increase, which is governed by the change in the effective potential with $\beta$. 

Now, to obtain the innermost stable circular orbit $r_\text{ISCO}$ of the NLE black hole we use the following required condition $V_{\text{eff},rr}=0$,
\begin{equation}
    E^2g_{\phi \phi,rr}+L^2g_{tt,rr}+(g_{tt}g_{\phi \phi})_{rr}=0
\end{equation}

We showed the horizon position, marginally bound orbit, and ISCO position for different values of $\beta=0.01M^4,3M^4,50M^4$ in \AU{Table} ~\ref{Tab:I}. It has been observed that as the coupling parameter increase, the horizon position, marginally stable orbit, and ISCO position also increase. The change is much larger than in the Schwarzchild case; hence one can distinguish the NLE from the Schwarzchild by using the shadow image and other accretion disk properties.

\begin{table*}
\centering
\begin{tabular}{ c c c c }
\hline
\hline
 Parameter & $\beta=0.01M^4$ & $\beta=3M^4$ & $\beta=50M^4$ \\
\hline
Horizon ($r_h$)   & $1.02961$    & $1.23722$ &   $1.39125$ \\
 \\ marginally bound orbit ($r_\text{mb}$) &   $2.81661$  & $2.84512$   & $2.98107$ \\
\\ Innermost stable circular orbit ($r_\text{ISCO}$) & $4.00008$ &  $4.02260$ & $4.19824$\\
\hline
\end{tabular}
\caption{The position of the horizon, marginally bound orbit, and ISCO for different coupling parameter values $\beta$.}
\label{Tab:I}
\end{table*}

Now, for a thin accretion disk, we consider that the vertical disk height ($H$) is much smaller than the disk's characteristic radius ($r$). The whole disk can be considered as in the local hydrodynamical equilibrium at each point; the pressure gradient and vertical entropy gradient are negligible in the disk. It is assumed that the cooling in the disk is efficient enough to prevent the disk from cumulating the heat generated by the stresses and dynamical friction within the disk; hence this cooling \AU{helps the disk to stabilize its vertically thin characteristic.} We also consider the disk to be in a steady state, which means that the mass accretion rate ($\dot{M}$), is constant with time. The disk's inner edge is at ISCO and the matter far from the black hole is considered to be following Keplerian motion. The stress energy-momentum tensor for the matter which is accreting around the compact object can be written in the form \cite{NovikovThorne,1974ApJ...191..499P}
\begin{equation}
    T^{\mu \nu}=\rho_0 u^\mu u^\nu+2u^{(\mu}q^{\nu)}+t^{\mu \nu},
\end{equation}
where $u_\mu q^\mu=0$ and $u_\mu t^{\mu \nu}=0$. Here, $u^\mu$ is the four-velocity of the orbiting particles, and $\rho_0$, $q^\mu$ and $t^{\mu \nu}$ are respectively the rest mass density, energy flow vector, and stress tensor of the accreting matter. Now using the rest-mass conservation, $\triangledown_\mu (\rho_0 u^\mu)=0$, we calculated the time-averaged mass accretion rate, $\dot{M}$ which comes out to be independent of the accretion disk radius
\begin{equation}
    \dot{M}=-2\pi \sqrt{-g}\Sigma u^r=\text{const},
\end{equation}
where $\Sigma$ is the time-averaged surface density and is deﬁned as
\begin{equation}
    \Sigma= \int_{-h}^{+h} <\rho_0>dz.
\end{equation}
Here, $z$ is in the cylindrical coordinate system. Now using the conservation laws for the energy, $\triangledown_\mu E^\mu=0$, and angular momentum $(L)$, $\triangledown_\mu J^\mu=0$, we obtained the following equations,
\begin{equation} \label{e77}
    [\dot{M}E-2\pi \sqrt{-g} \Omega W^r_\phi]_{,r}=4\pi rF(r)E,
\end{equation}
and
\begin{equation} \label{e78}
    [\dot{M}L-2\pi \sqrt{-g} \Omega W^r_\phi]_{,r}=4\pi rF(r)L,
\end{equation}
where $W^r_\phi$ is called the averaged torque and given by
\begin{equation}
    W^r_\phi= \int_{-h}^{h} <t^r_\phi> dz,
\end{equation}
and $<t^r_\phi>$ is the $(\phi,r)$ component of the stress-energy tensor which is calculated averaged over the time scale $\delta t$ and angle $\delta \phi=2\phi$ . The Eq. \eqref{e77} shows an energy balanced equation such that the rest mass energy of the disk ($\dot{M} E$) and the energy due to the torques in the disk ($2 \pi \sqrt{-g} \Omega W^r_\phi$) are well balanced by the radiated energy ($4 \pi r F(r) E$) from the surface of the disk. 

Similarly, Eq. \eqref{e78} shows angular momentum balanced equation such that the angular momentum transferred by the disk's rest mass ($\dot{M} L$) and the angular momentum due to the torques in the disk ($2\pi \sqrt{-g} \Omega W^r_\phi $) are well balanced by the angular momentum transferred by the surface of the disk by the outgoing radiation ($4 \pi r F(r) E$).

Now, by using the energy-angular momentum relation $E_{,r}=\Omega L_{,r}$ and removing $W^r_\phi$ from equations $(33)$ and $(34)$, the time-averaged energy flux $F(r)$ emitted from the surface of an accretion disk around the compact object is given by
\begin{equation}
    F(r)=-\frac{\Omega_{,r}\dot{M}}{4\pi \sqrt{-g}(E-\Omega L)^2} \int_{r_\text{ISCO}}^{r}(E-\Omega L)L_{,r} dr.
\end{equation}

\begin{figure}
    \centering
    \includegraphics[width=\linewidth]{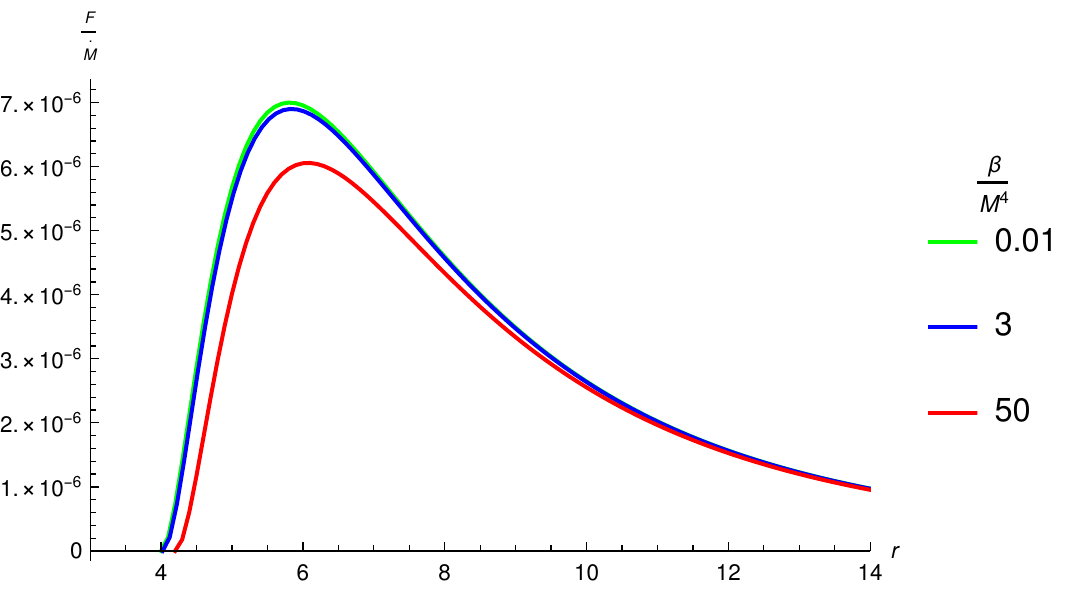}
    \caption{Energy flux for different values of $\beta=0.01M^4 (\text{green})$, $3M^4 (\text{blue})$ and $50M^4 (\text{red})$.} \label{fig:m}
\end{figure}

\begin{figure}
    \centering
    \includegraphics[width=\linewidth]{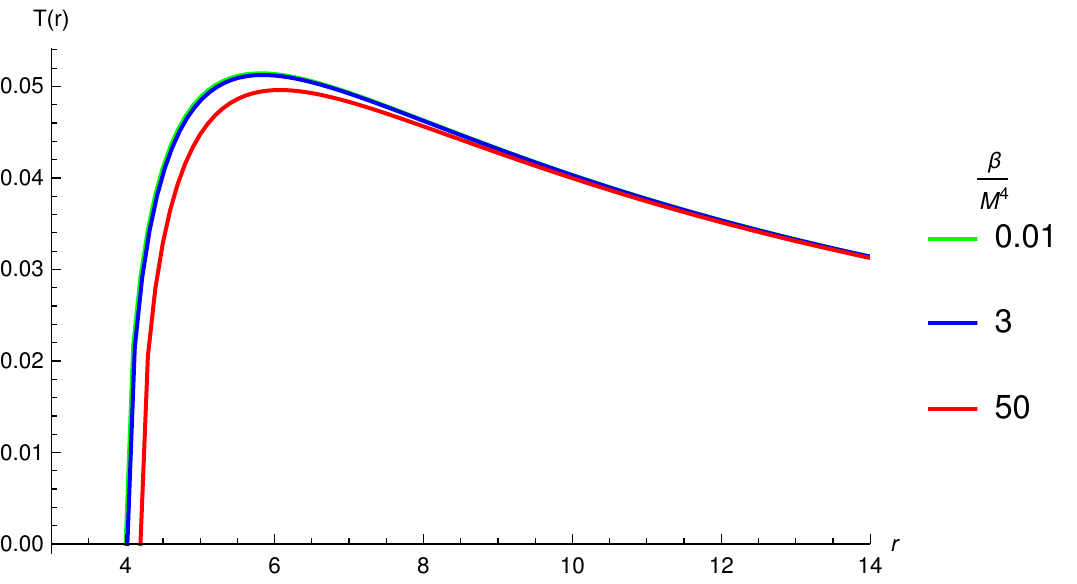}
    \caption{Temperature for different values of $\beta=0.01M^4 (\text{green})$, $3M^4 (\text{blue})$ and $50M^4 (\text{red})$. } \label{fig:n}
\end{figure}

\begin{figure}
    \centering
    \includegraphics[width=\linewidth]{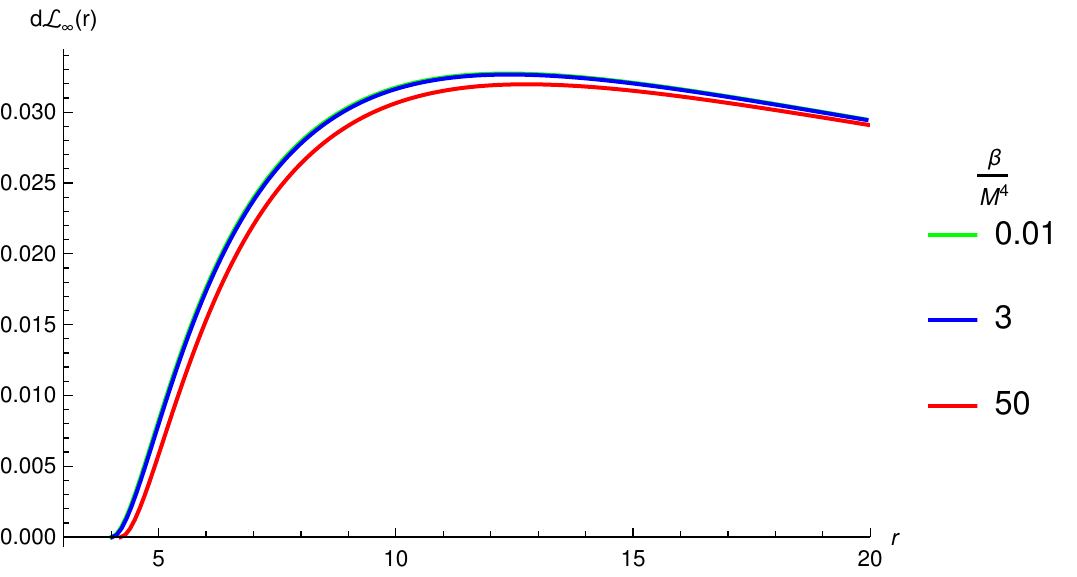}
    \caption{Differential luminosity for different values of $\beta=0.01M^4 (\text{green})$, $3M^4 (\text{blue})$ and $50M^4 (\text{red})$. } \label{fig:o}
\end{figure}

The disk temperature is related to the energy flux by the equation,
\begin{equation}
    F(r)=\sigma T^4(r),
\end{equation}
where $\sigma$ is known as the Stefan-Boltzmann constant. Now, by combining the conservation laws of energy and angular momentum, it provides us the differential of the luminosity $L_\infty$ at the infinity as \cite{1974ApJ...191..499P,Joshi:2013dva}
\begin{equation}
    \frac{d\mathcal{L_\infty}}{d\ln r}=4 \pi r \sqrt{-g} E \mathcal{F}(r).
\end{equation}
The plots for the energy flux \AU{per unit mass accretion rate, temperature of the thin accretion disk or more precisely the radial variation of $\sigma^{1/4} T/\dot{M}^{1/4}$, and differential luminosity per unit mass accretion rate} are shown in the Figs.~\ref{fig:m},~\ref{fig:n} and~\ref{fig:o} respectively. It can be observed from the figures that as we turn on the NLE effect and increase the coupling parameter $\beta$, the radiative flux, temperature, and differential luminosity decrease, which means that for the larger value of the coupling constant, the radiation from the disk is less and the corresponding temperature of the disk is lesser compare to the lower value of the $\beta$. 

\subsection{Direct Emission, Lensing Ring And Photon Ring}
In this section, we classify the incoming rays coming from the observer sky as in Wald et al. \cite{Gralla:2019xty} by defining the number of orbits $n(\gamma)=\phi/2 \pi$, where, $\phi$ is known as the final azimuthal coordinate of the ray once it is completely free from the gravitational lensing. The number of orbits essentially gives the number of times the geodesic $\gamma$ has crossed the equatorial plane of the accretion disk. We define it as the following:
(1) $n<3/4$ is direct emission, crossed the equatorial plane only once.
(2) $3/4<n<5/4$ is the lensing ring that crosses the equatorial plane twice.
(3) $n>5/4$ is a photon ring that crosses the equatorial plane more than twice.

\begin{figure*}
    \centering
    \includegraphics[width=.3\textwidth]{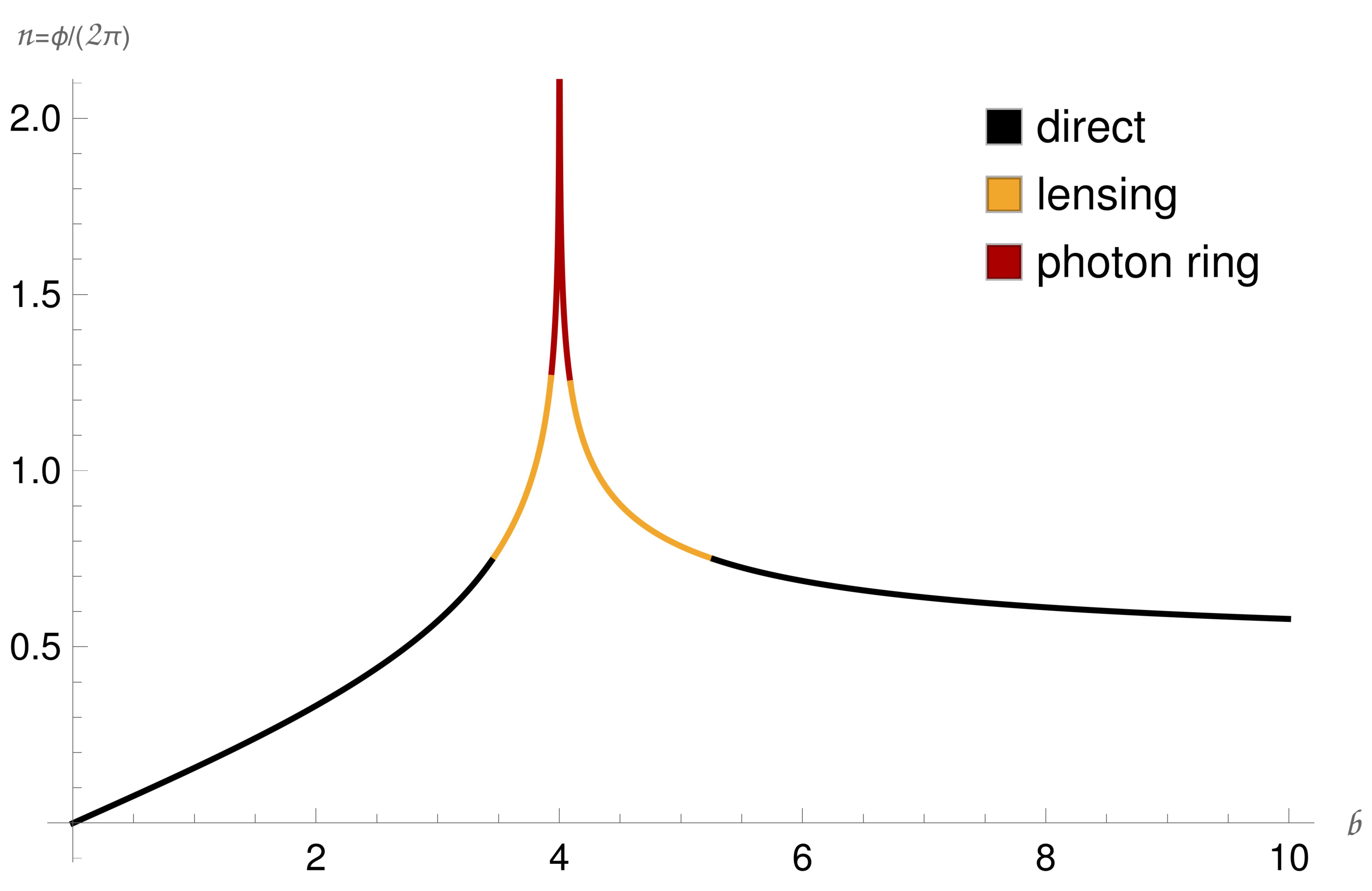} \hfill
    \includegraphics[width=.3\textwidth]{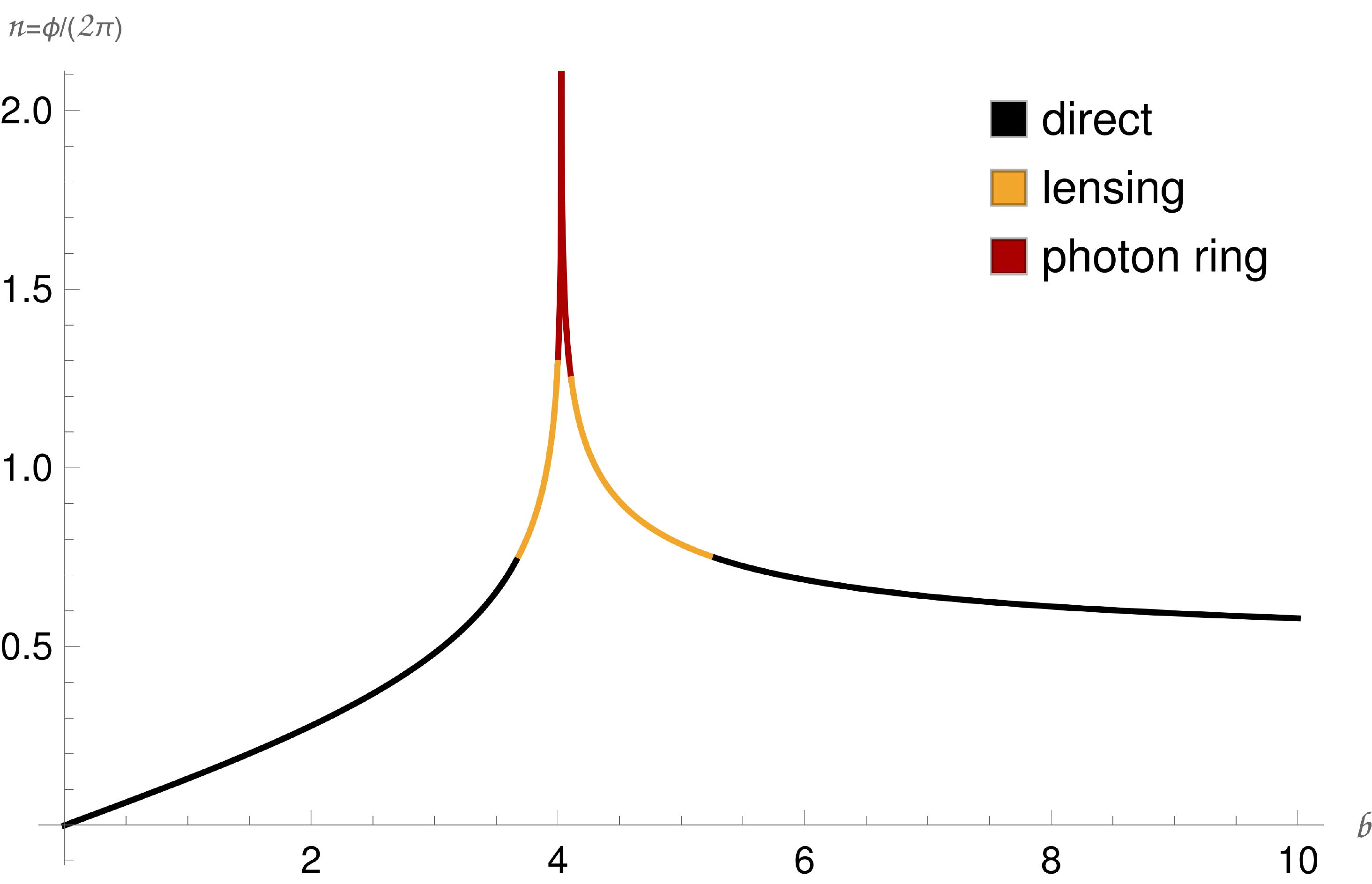} \hfill
    \includegraphics[width=.3\textwidth]{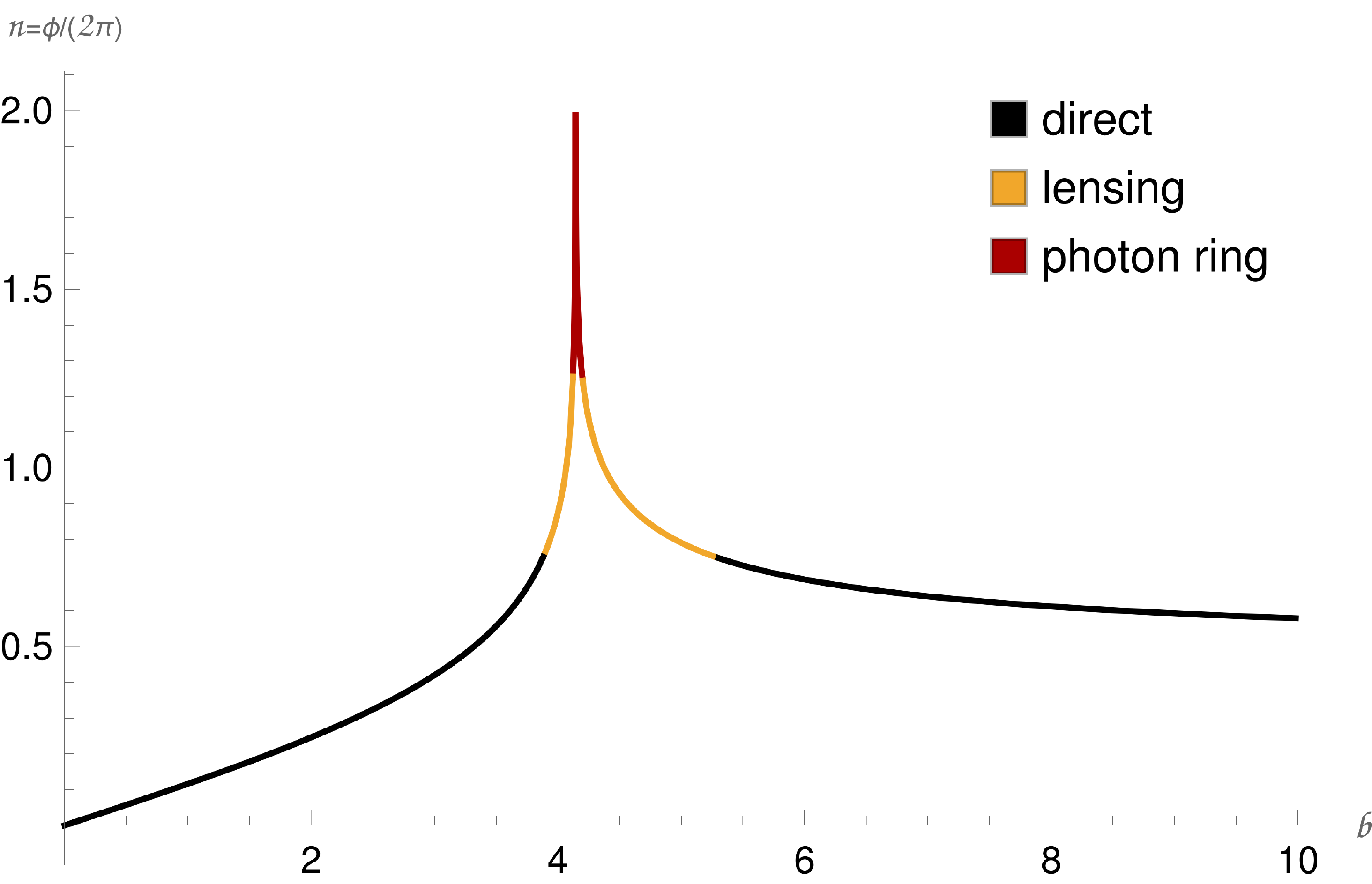} \hfill
    \vfill
    \centering
    \includegraphics[width=.3\textwidth]{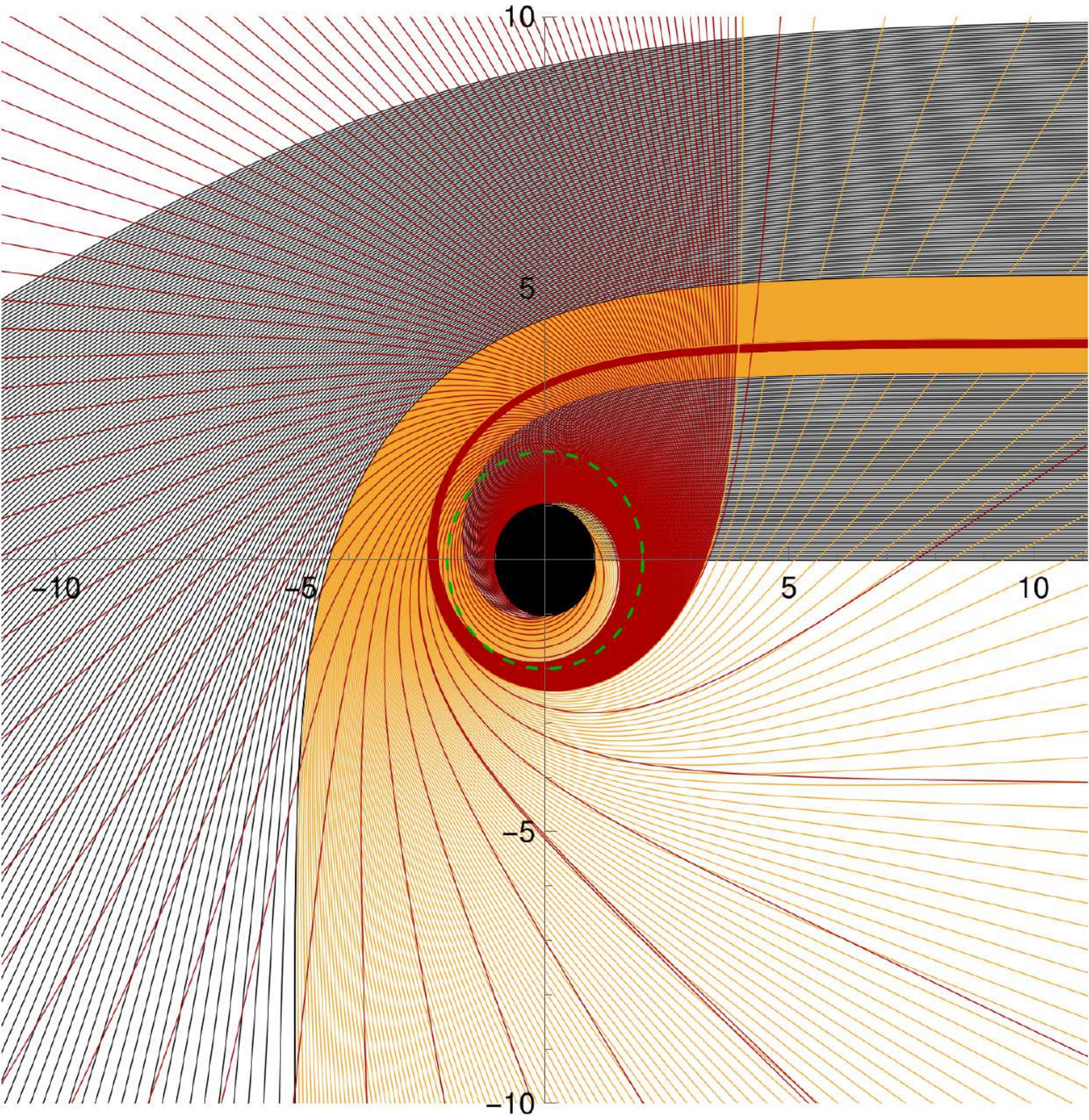} \hfill
    \includegraphics[width=.3\textwidth]{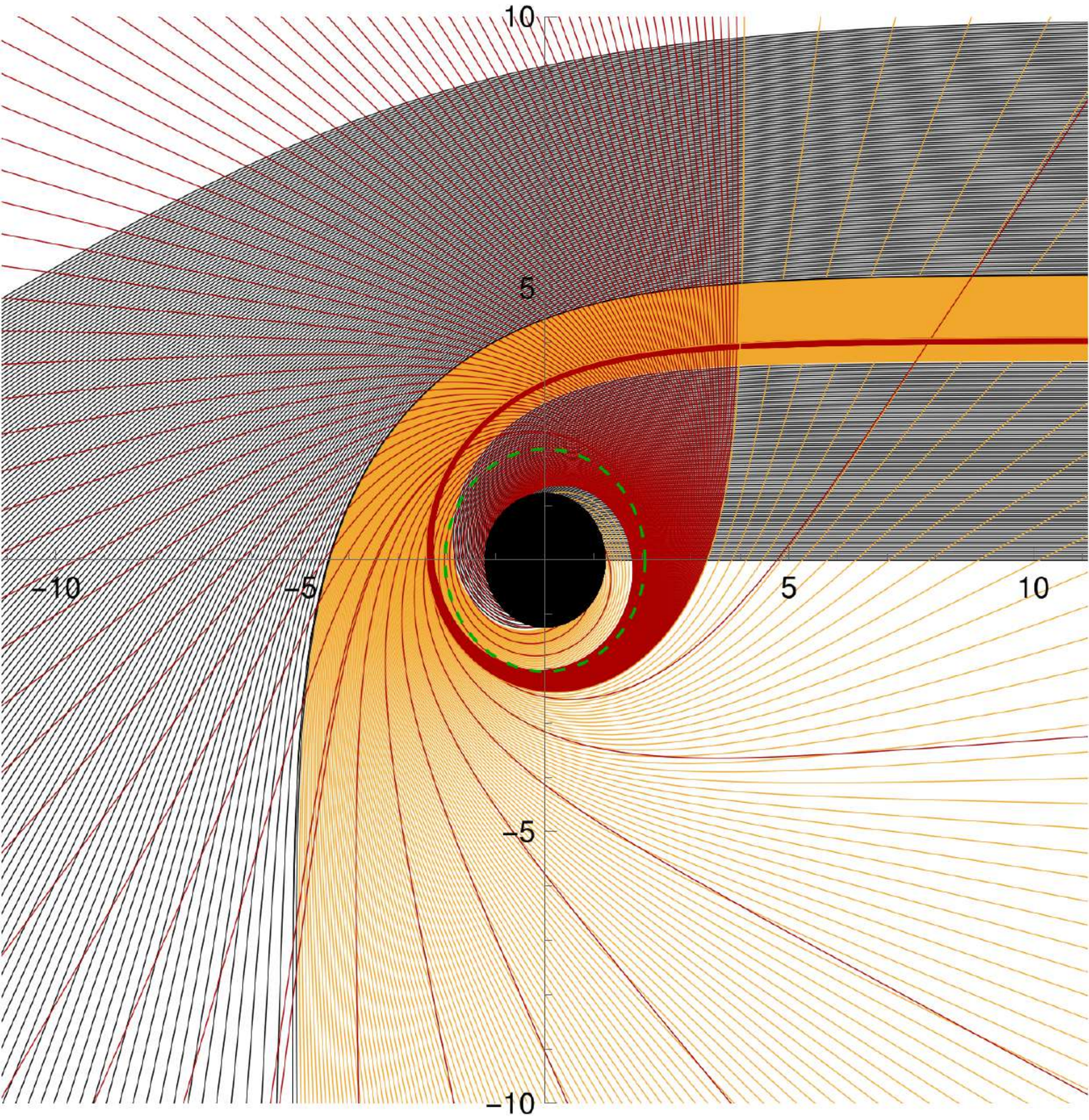} \hfill
    \includegraphics[width=.3\textwidth]{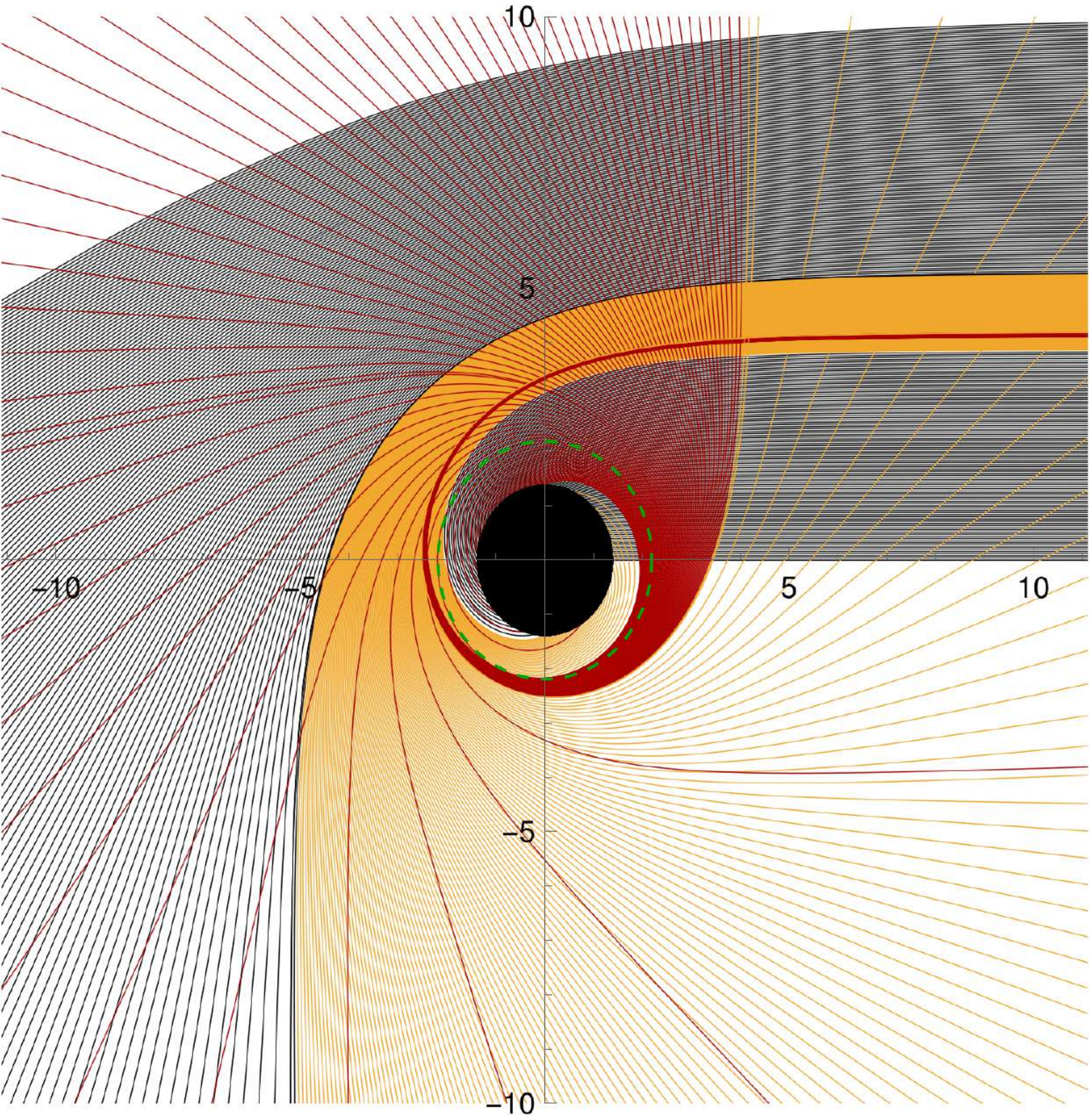} \hfill
    \caption{Behavior of the photons for different values of $\beta=0.01M^4,3M^4,50M^4$. The red, orange, and black lines represent the photon ring, lensing ring, and direct emission, respectively. The green dotted line represents the photon orbit, and the black disk represents the event horizon of the black hole \AU{In the top panel, we show the fractional number of orbits, $n = \phi/(2\pi)$, where $\phi$ is the total change in (orbit plane) azimuthal angle outside the horizon and in the bottom panel, we show a selection of associated photon trajectories, treating $r$, $\phi$ as Euclidean polar coordinates.}}
    \label{fig:15}
\end{figure*}

Here, we have shown the number of orbits $n$ vs impact parameter $b$ (See Fig. \ref{fig:15}). The black color denotes the direct emission rays, yellow represents the lensing ray, and red represents the photon ring rays. The dashed green circle in the ray tracing figure shows the photon orbit. Note that for calculating thin-accretion disk, null geodesics and shadow cast, we use the Okyay-\"Ovg\"un \textit{Mathematica} notebook package \cite{Okyay:2021nnh}, (used in \cite{Chakhchi:2022fls}).
\begin{table*}
\centering
\begin{tabular}{ p{3cm} p{4cm} p{4cm} p{4cm}}
 \hline
 \hline
 Parameter & $\beta=0.01M^4$ & $\beta=3M^4$ & $\beta=50M^4$ \\
 \hline
 Direct Emission \newline $n<3/4$   & $b<3.46012$ \newline $b>5.27012$    & $b<3.67745$ \newline $b>5.27745$ &   $b<3.88071$ \newline $b>5.30071$ \\
 \\ Lensing Ring \newline $3/4<n<5/4$ &   $3.46012<b<3.93012$ \newline $4.09012<b<5.27012$  & $3.67745<b<3.99745$ \newline $4.10745<b<5.27745$   & $3.88071<b<4.12071$ \newline $4.20071<b<5.30071$ \\
\\ Photon Ring \newline $n>5/4$ & $3.93012<b<4.09012$ & $3.99745<b<4.10745$ &  $4.12071<b<4.20071$\\
\hline
\end{tabular}
\caption{The region of direct emission, lensing ring, and photon ring for the different parameter $\beta$.}
\label{Tab:II}
\end{table*}
   
We can see the variation of the black hole shadow with increasing the $\beta$ in Table \ref{Tab:II}. We find out that the shadow also increases its angular area as the horizon shifts. Therefore, the behavior of the photons also changes once it crosses the equatorial plane. Also, from the table and figures, we noticed that the lensing and photon rings' range decreases with the increasing coupling constant $\beta$, if we fix $C=1$. It shows that as we increase the coupling parameter $\beta$, the contribution in the brightness of the lensing and photon rings will decrease. It can also be seen that when the impact parameter is very close to the critical impact parameter $b \pm b_\text{c}$, the photon orbit shows the narrow peak in $(b,\phi)$ plane, and after that, as $b$ increases, the photon trajectories are always direct emission in all the cases. In the next section, we will study the observed emission intensity of the accretion disk within the framework of the NLE model.

\subsection{Transfer functions and observed specific intensities}
In this section, we will investigate the emitted intensity from the black hole. We assume that the disk will emit the radiation isotropically in the rest frame of the static worldlines of the observer. Now, from Liouville's theorem, $I^{em}_{\nu}/\nu_e^3$ is conserved along the path of the light ray,  we can write the observed intensity as
\begin{equation}
    I_{\nu'}^\text{obs}=g^3 I_\nu^\text{em},
\end{equation}
where $g=\sqrt{f(r)}$ and $I_{\nu'}^{obs}$  is the observed intensity at the frequency $\nu'$. We can integrate over all the frequencies, as $\nu'=g d\nu$ and $I_\text{em}=\int I_\nu^{em} d\nu $. Therefore, the observed frequency will be given by
\begin{equation}
    I^\text{obs}=g^4 I_\text{em}.
\end{equation}
$I_\text{em}$ here is the total emitted specific intensity from the accretion disk. Therefore, the total intensity received by the observer will be,
\begin{equation}
    I(r)=\sum_{n} I^\text{obs}(r)|_{r=r_m(b)},
\end{equation}
where $r_{m}(b)$ is the $m^{th}$ intersection outside the horizon in the equatorial plane, which we call the transfer function. The transfer function gives the relationship between the radial coordinates and the impact parameter of the photon. Therefore, the slope of the transfer function tells us about the demagnified scale of the transfer function \cite{Gralla:2019xty,Zeng:2020vsj}. Hence, it is called the demagnification factor. We have shown the transfer functions for the different values of the coupling constant $\beta$.
\begin{figure*}
    \centering
    \includegraphics[width=.32\textwidth]{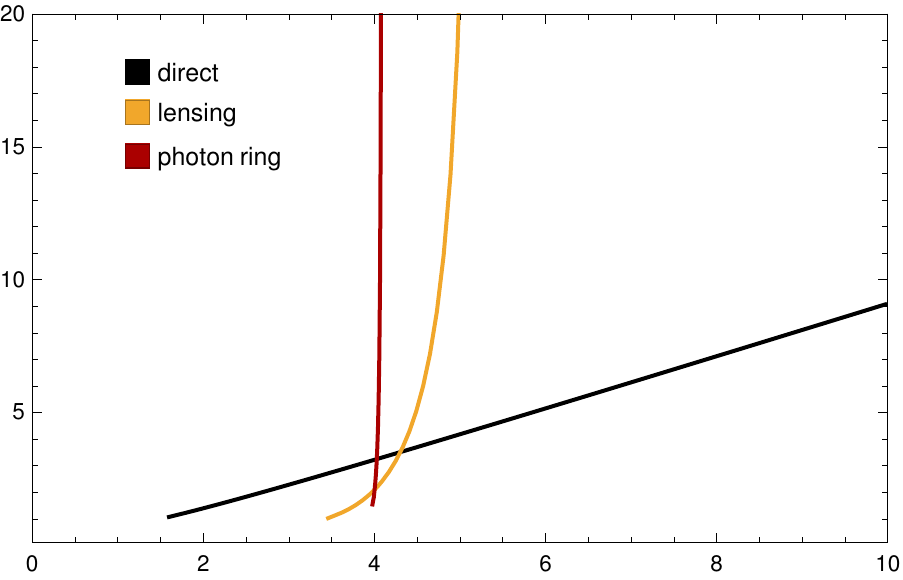}
    \includegraphics[width=.32\textwidth]{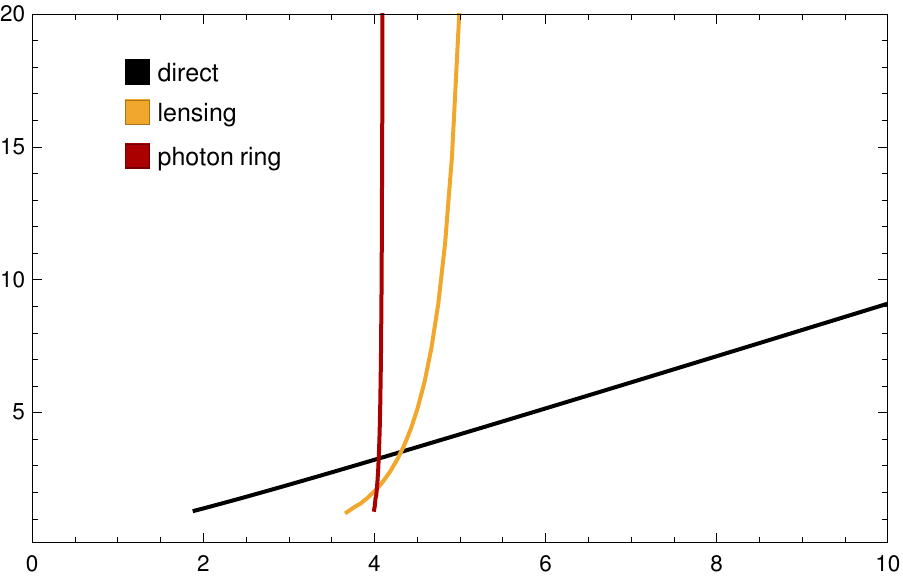}
    \includegraphics[width=.32\textwidth]{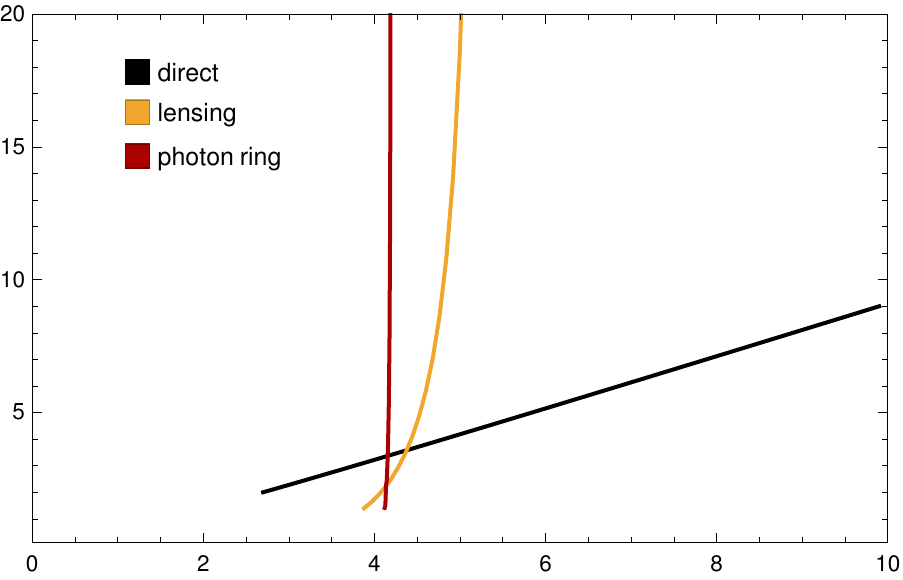}
    \caption{The first three transfer function of a NLE black hole for different values of $\beta=0.01M^4,3M^4,50M^4$. Here, $y$ axis is the transfer function ($r_m(b)$) and $x$ axis is impact parameter ($b$).} \label{fig:16}
\end{figure*}

In Fig. \ref{fig:16}, it is clear that the black dots $(m=1)$, which indicates the transfer function for the direct emission, have an almost constant slope. Since the slope of the direct emission profile is one, it indicates the redshifted source profile. The yellow dots $(m=2)$ represent the transfer function for the lensing ring. We can see in Fig. that for $m=2$, when $b$ approaches $b_\text{c}$, the slope is small, but as $b$ increases, the slope increases rapidly. In other words, the image of the back side of the accretion disk will be demagnified in this case as the slope is much greater than one. Now, the red dots $(m=3)$, correspond to the photon ring. The slope is very close to infinity for the photon ring, indicating that the accretion disk's front side image will be extremely demagnified. Hence, the total observed flux's contribution mainly comes from the direct emission. Since the higher $m>4$ will have much less contribution to the observed flux, we are only interested till $m=3$ transfer function.\\

\subsection{Observational features of direct emission, photon and lensing rings}
As we discussed in the previous section that the specific brightness only depends on $r$ for the observer at infinity, we have considered three toy models for intensity profile $I_\text{em}$,
\begin{itemize}
  \item Model 1:\[
    I_\text{EM}^1(r)= 
\begin{cases}
    \left( \frac{1}{r-(r_\text{ISCO}-1)} \right)^2   ,&  r\geq r_\text{ISCO}\\
    0,              & r \leq r_\text{ISCO}
\end{cases}
\]
  \item Model 2:\[   I_\text{EM}^2(r)= 
\begin{cases}
    \left( \frac{1}{r-(r_\text{ph}-1)} \right)^3   ,&  r\geq r_\text{ph}\\
    0,              & r \leq r_\text{ph}
\end{cases}
\]
   \item Model 3:\[   I_\text{EM}^3(r)= 
\begin{cases}
    \frac{1-\arctan(r-(r_\text{ISCO}-1))}{1-\arctan(r_\text{ph})}    ,&  r\geq r_\text{h}\\
    0,              & r \leq r_\text{h}
\end{cases}
\]
\end{itemize}
where $r_\text{ISCO}$,$r_\text{ph}$ and $r_\text{h}$ is the innermost stable orbit, photon sphere and event horizon respectively. These three models have their properties, such as the decay rate being very large in the second model and very slow in the third model. In the third model we have considered, the emission starts directly from the horizon; in the second model, it starts from the photon sphere; in the first model, it starts from the ISCO \AU{Although these three models seems to be highly idealised case but they can give a qualitative idea about the photons behaviour around the black hole}.
\begin{figure*}
    \centering
    \includegraphics[width=.3\textwidth]{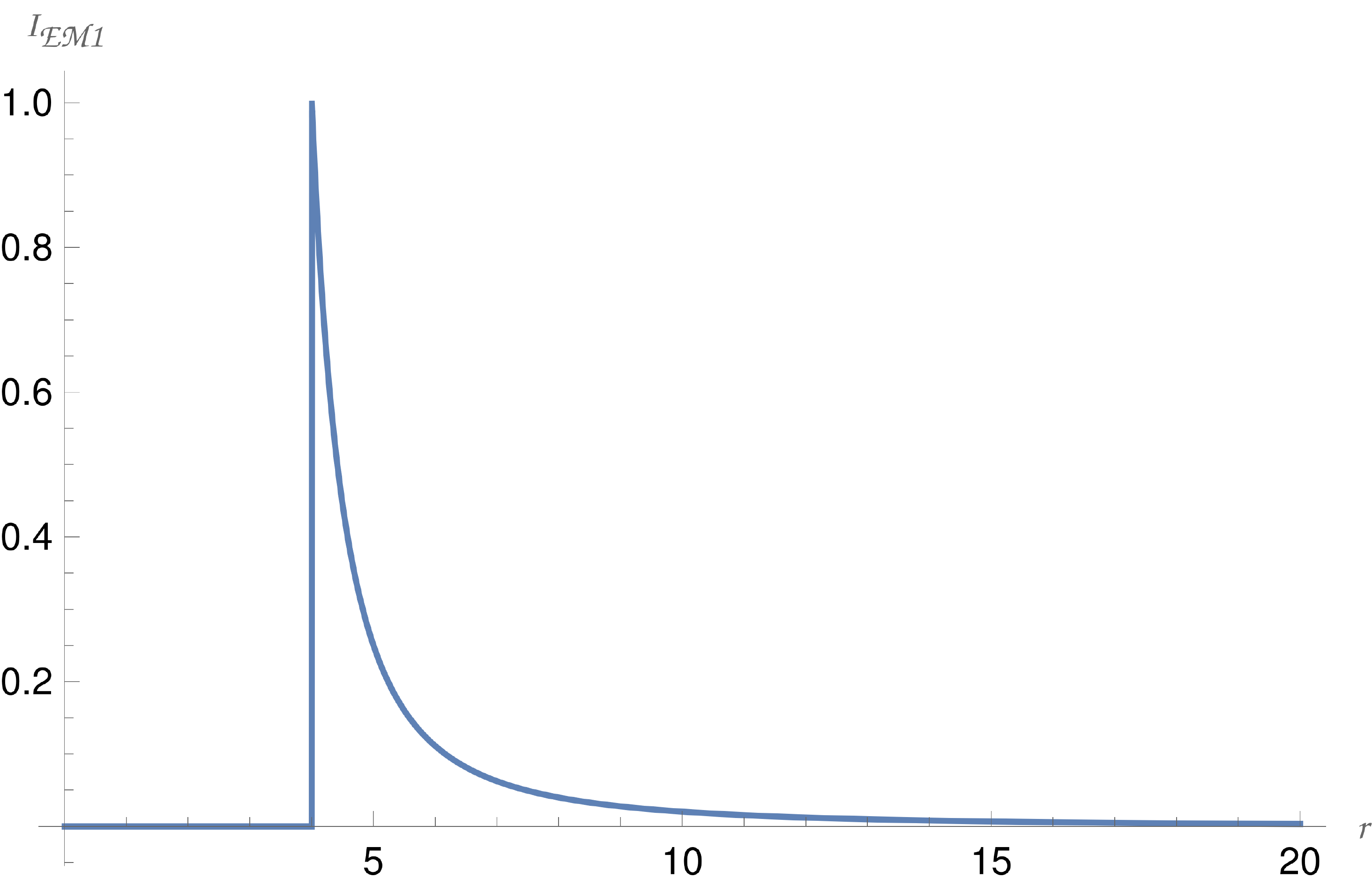}
    \includegraphics[width=.3\textwidth]{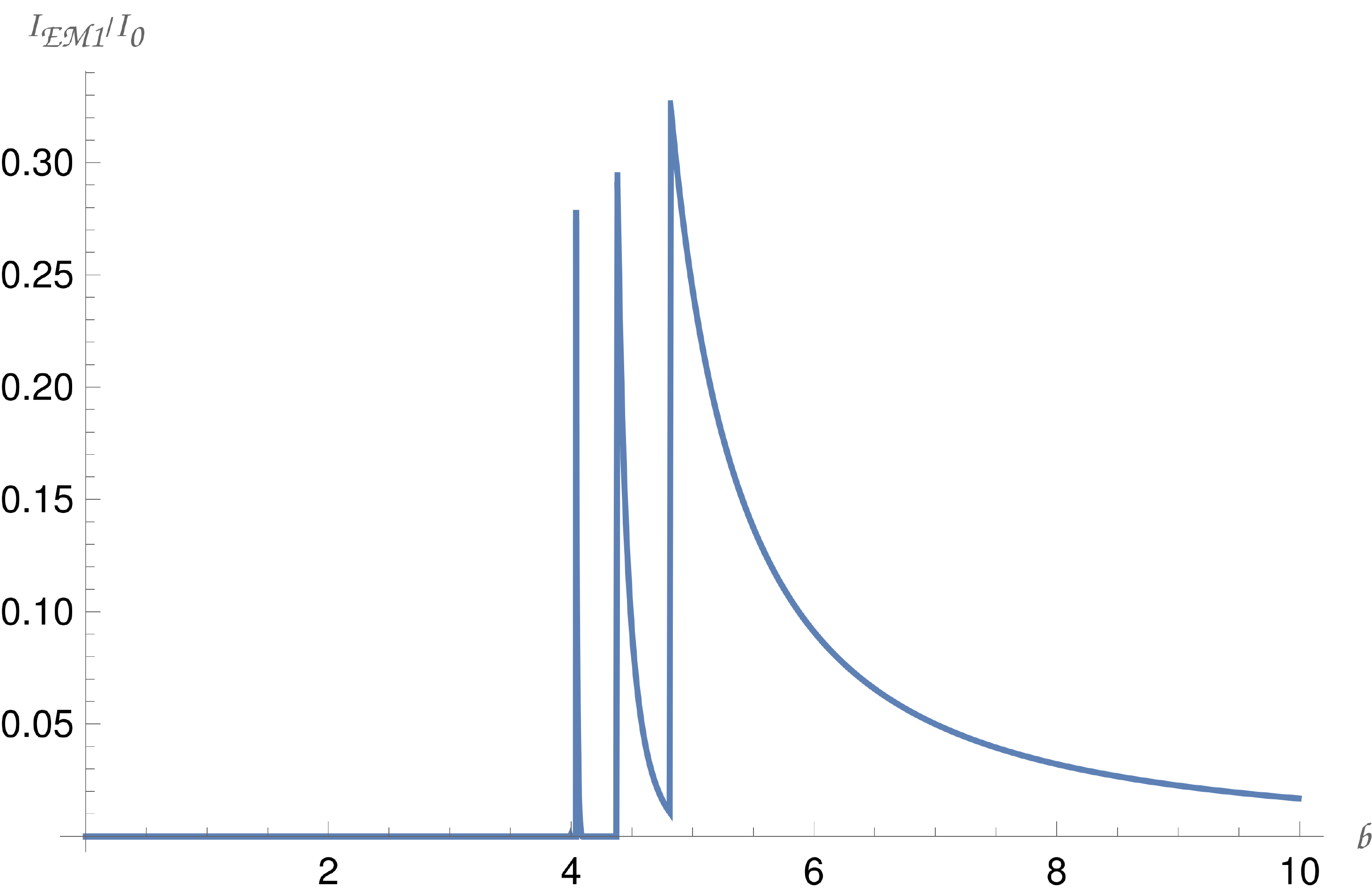}
    \includegraphics[width=.3\textwidth]{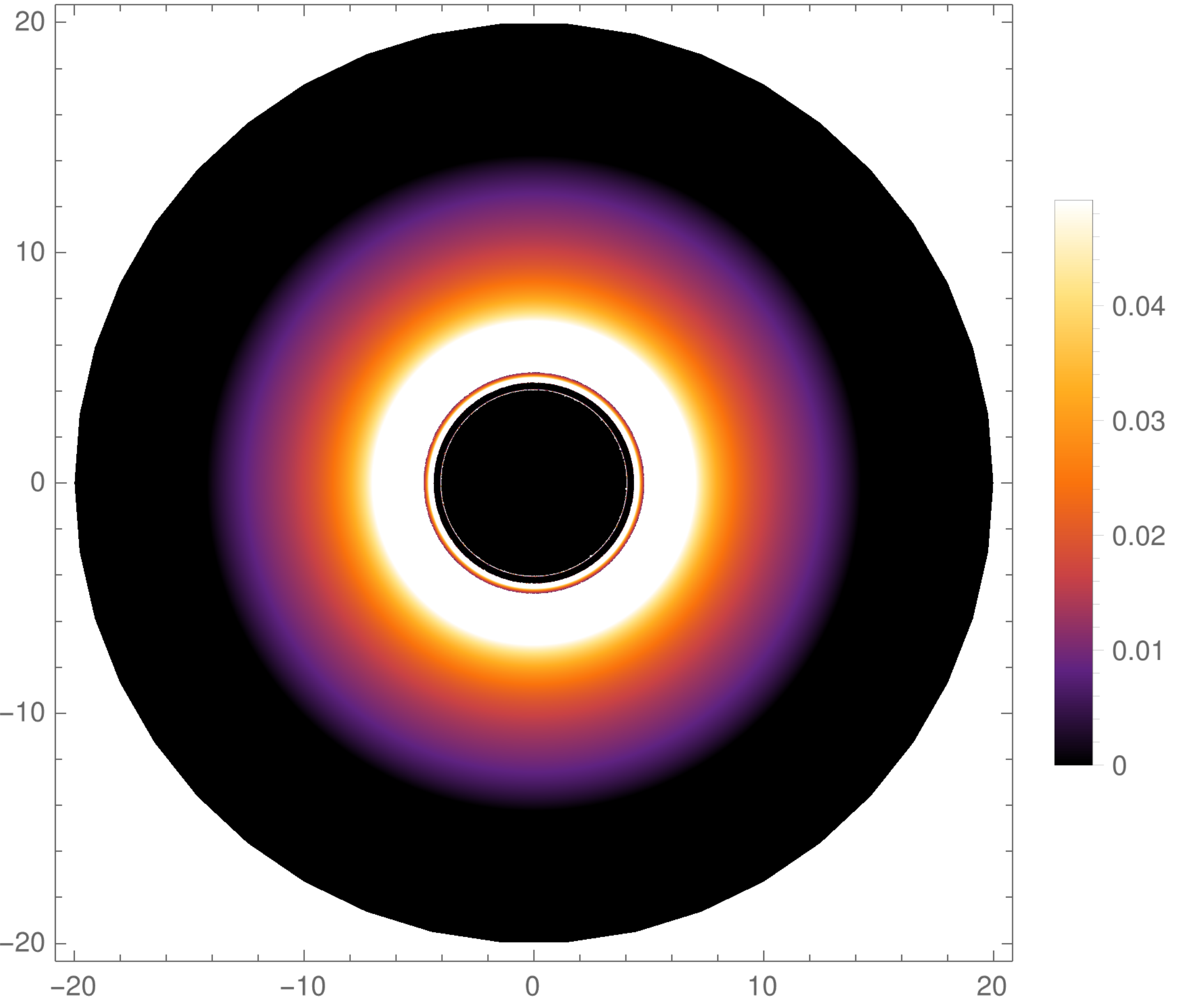}
    \vfill
    \centering
    \includegraphics[width=.3\textwidth]{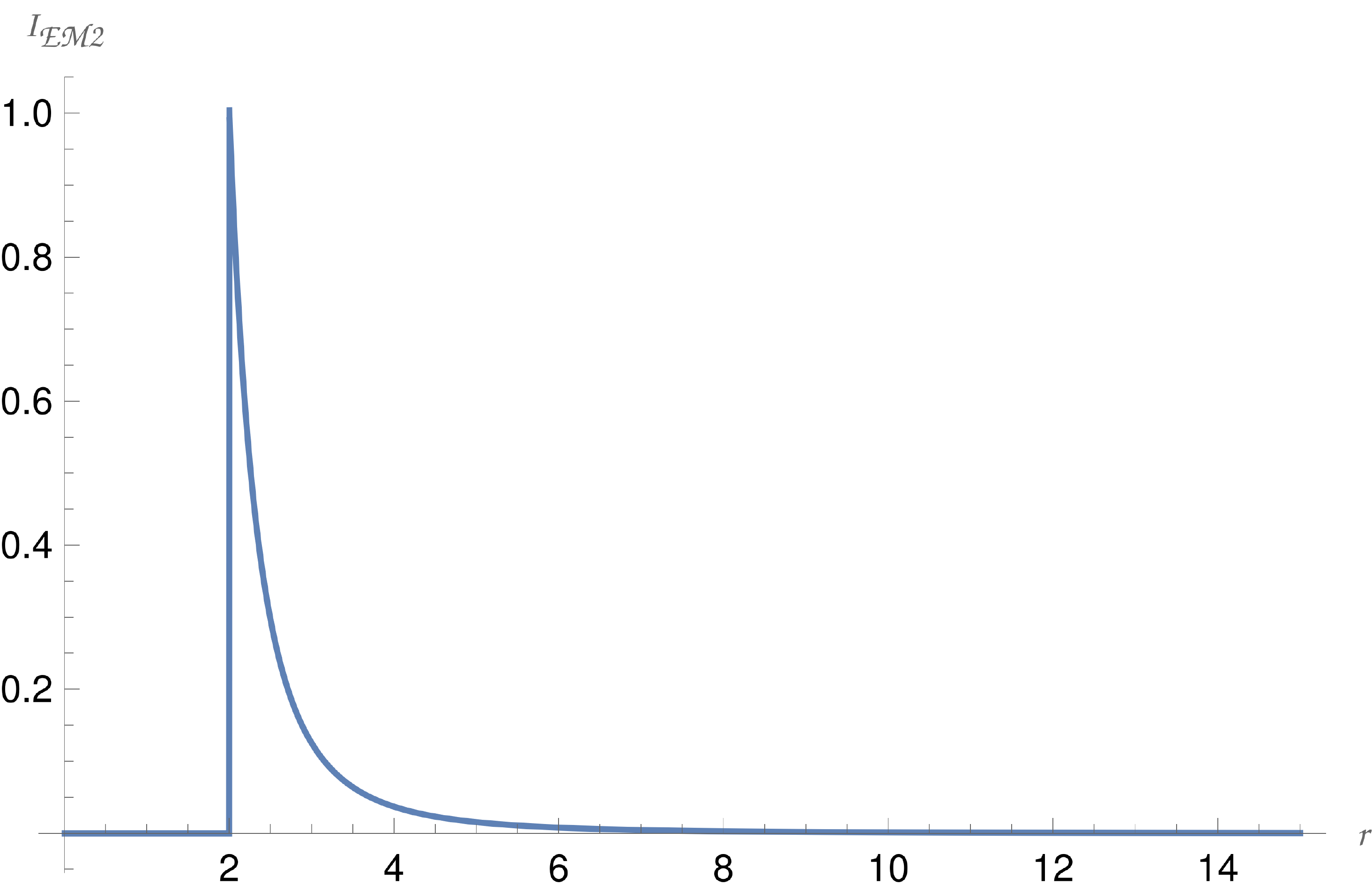}
    \includegraphics[width=.3\textwidth]{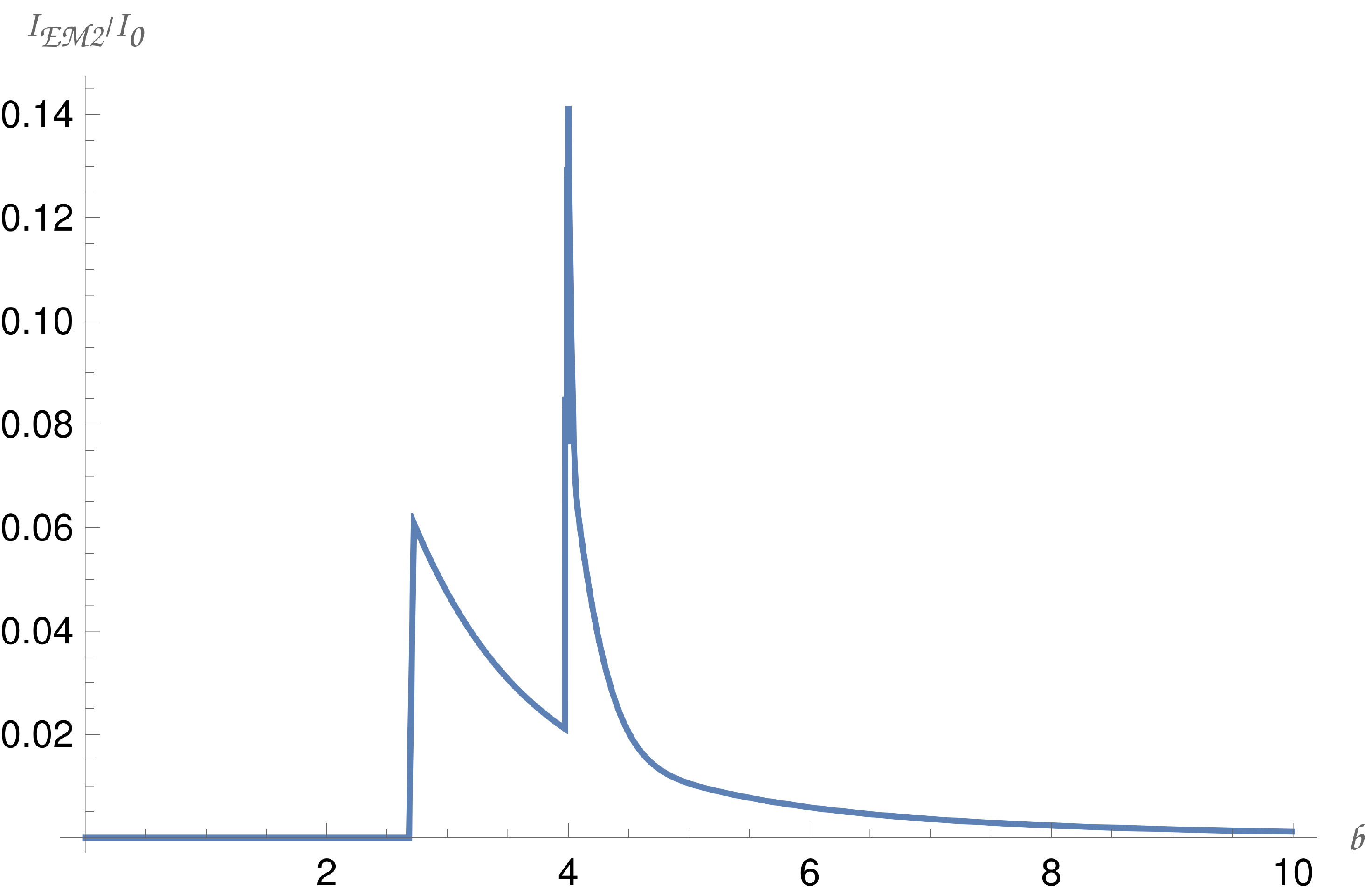}
    \includegraphics[width=.3\textwidth]{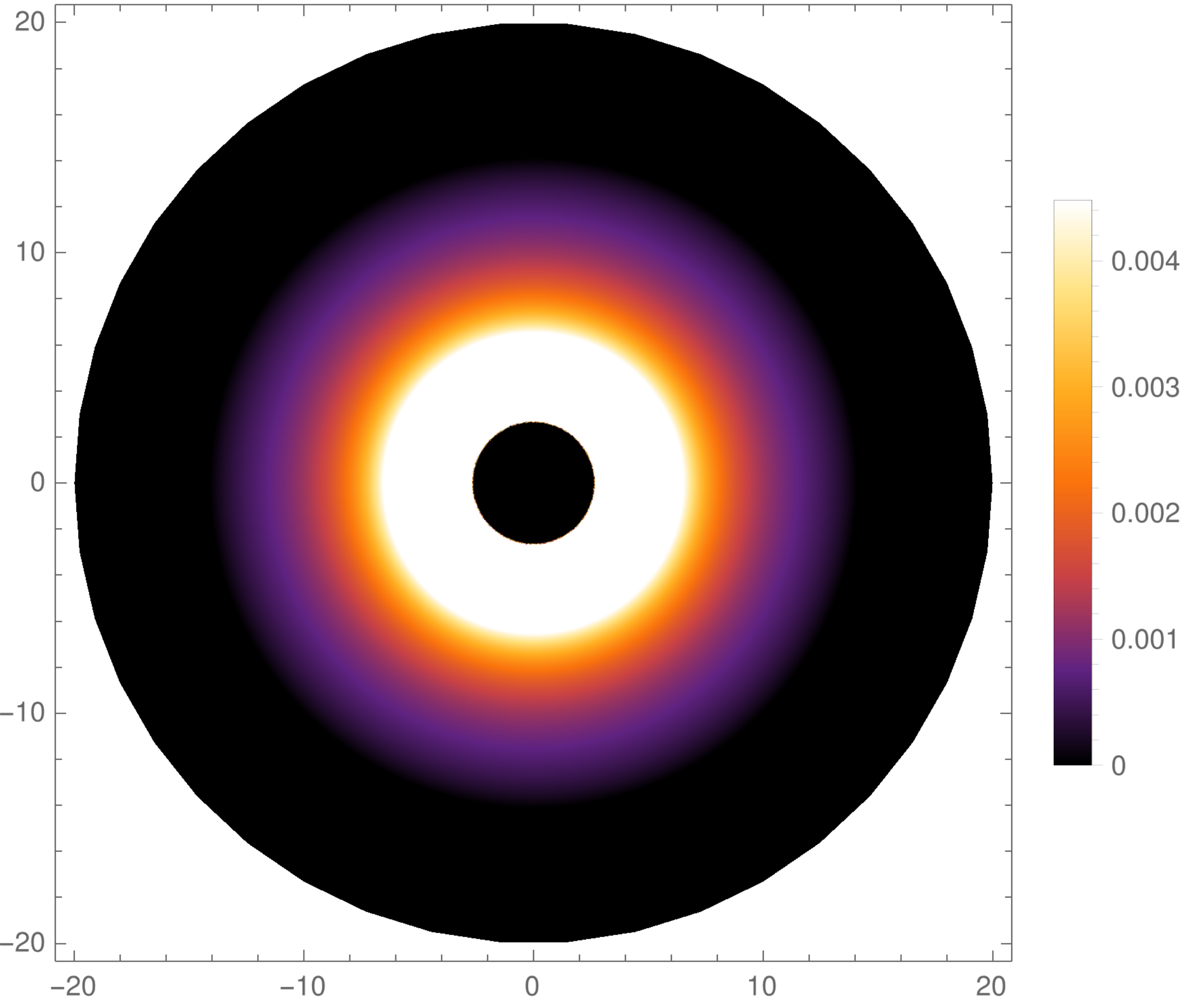}
    \vfill
    \centering
    \includegraphics[width=.3\textwidth]{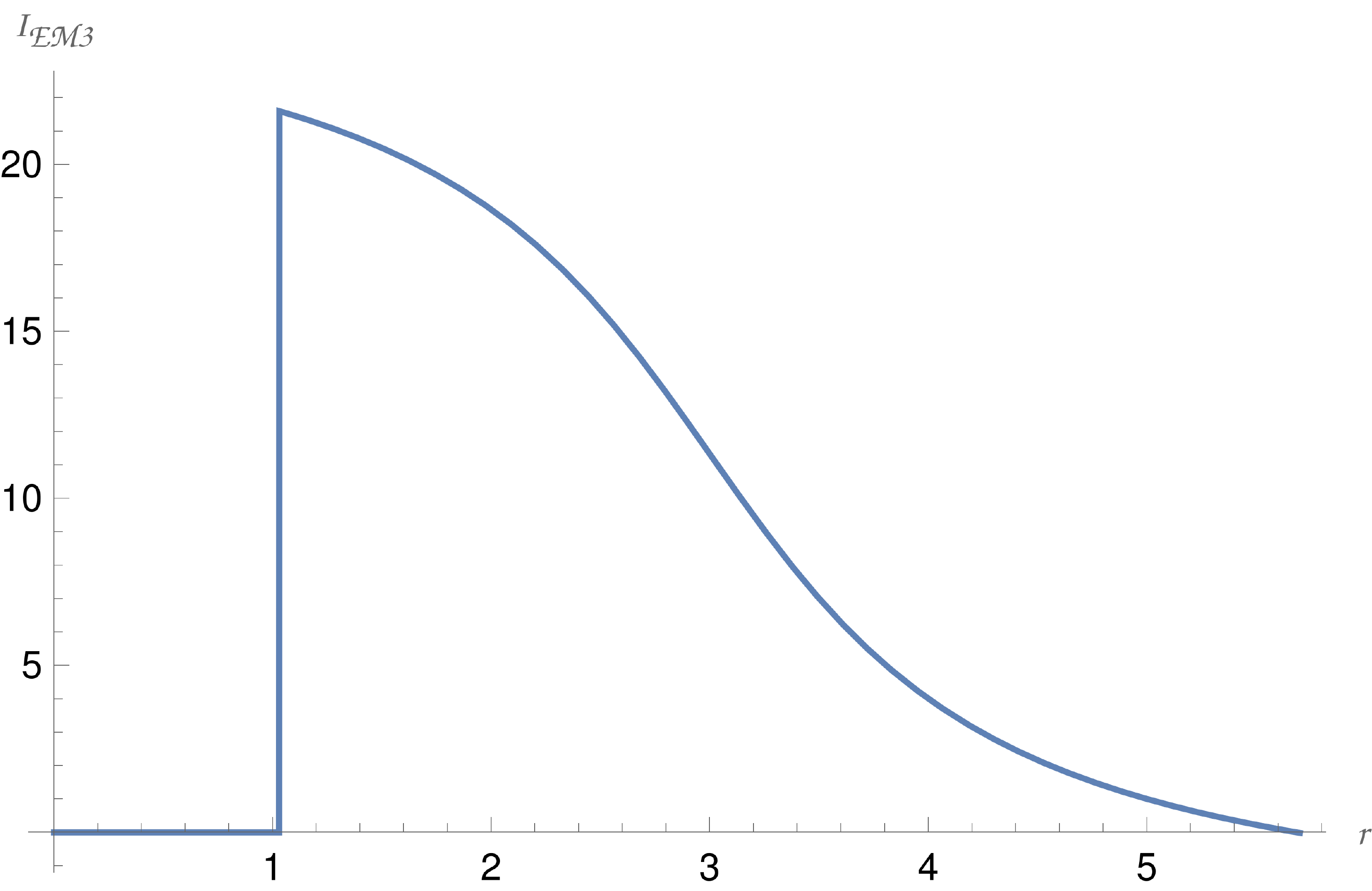}
    \includegraphics[width=.3\textwidth]{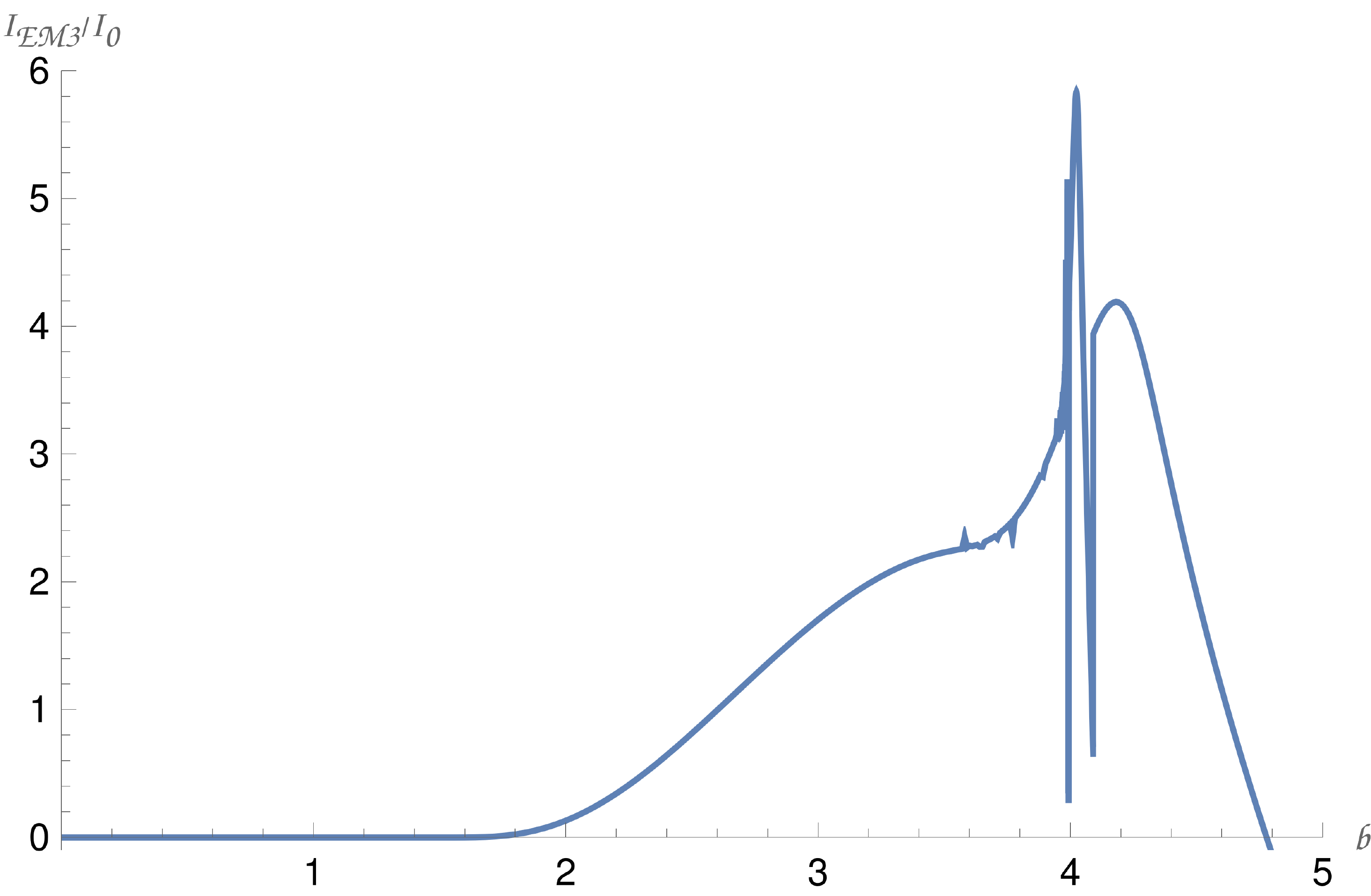}
    \includegraphics[width=.3\textwidth]{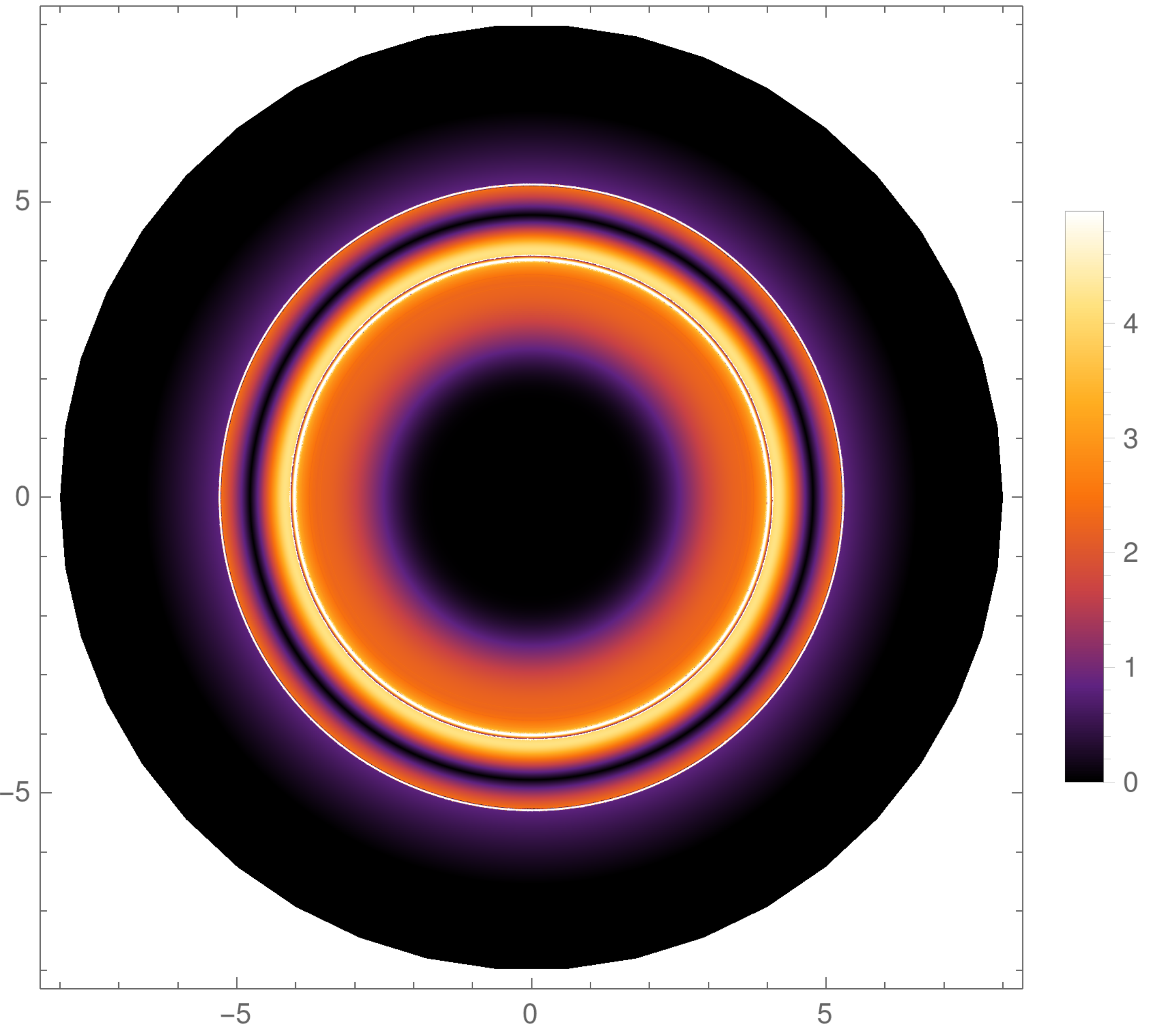}
    \caption{Observational appearance of the thin disk with different emission profile for $\beta=0.01M^4$, viewed from a face-on orientation. The upper row is for the emission profile intensity given by  model $1$, the second row for model $2$, and the third row is for model $3$ described in section C. \AU{In the plots the emitted and observed intensities $I_{EM}$ and $I_{obs}$ are normalized to the maximum value $I_0$ of the emitted intensity outside the horizon.} }
    \label{fig:17}
\end{figure*}
\begin{figure*}
    \centering
    \includegraphics[width=.3\textwidth]{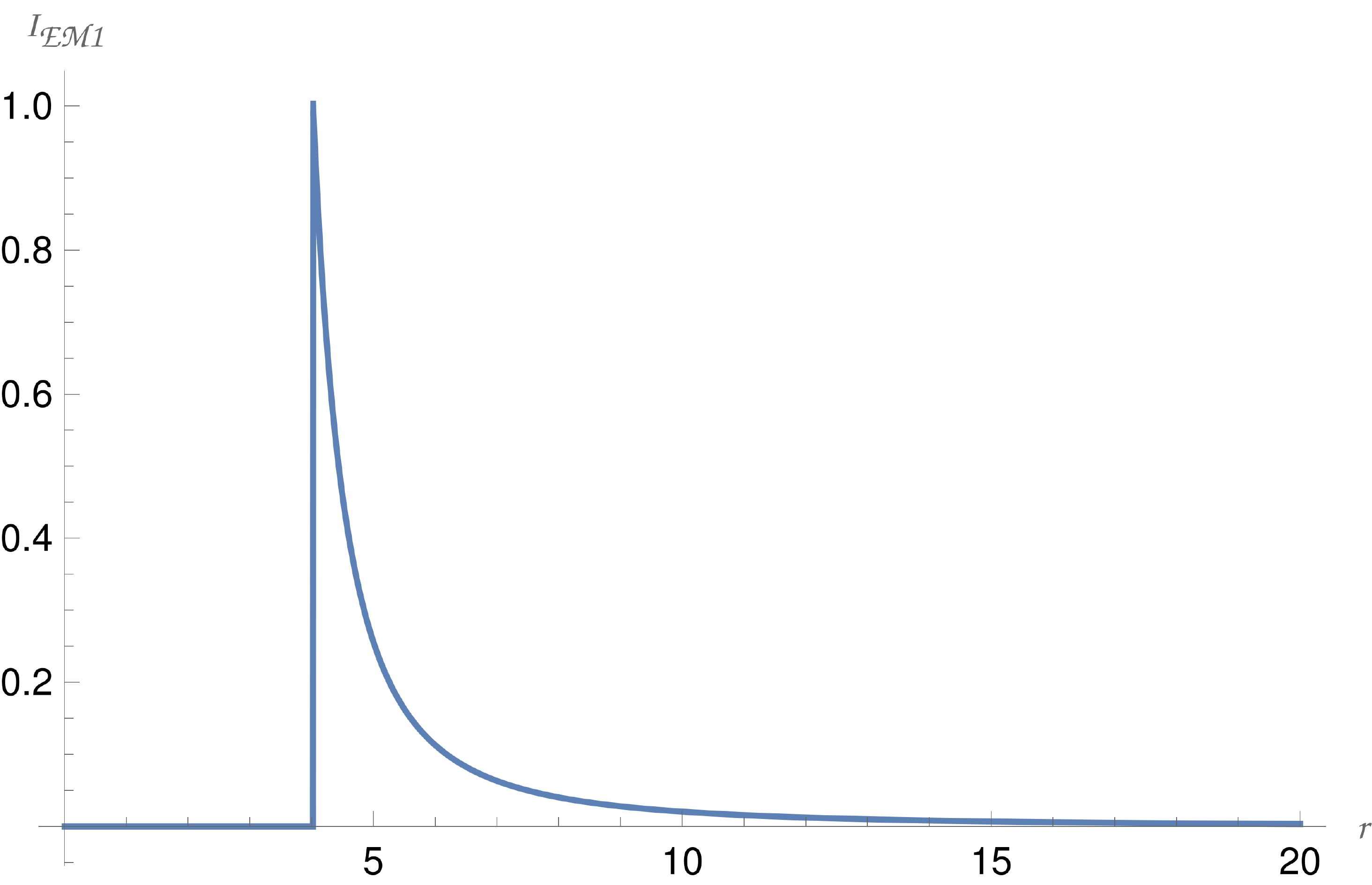}
    \includegraphics[width=.3\textwidth]{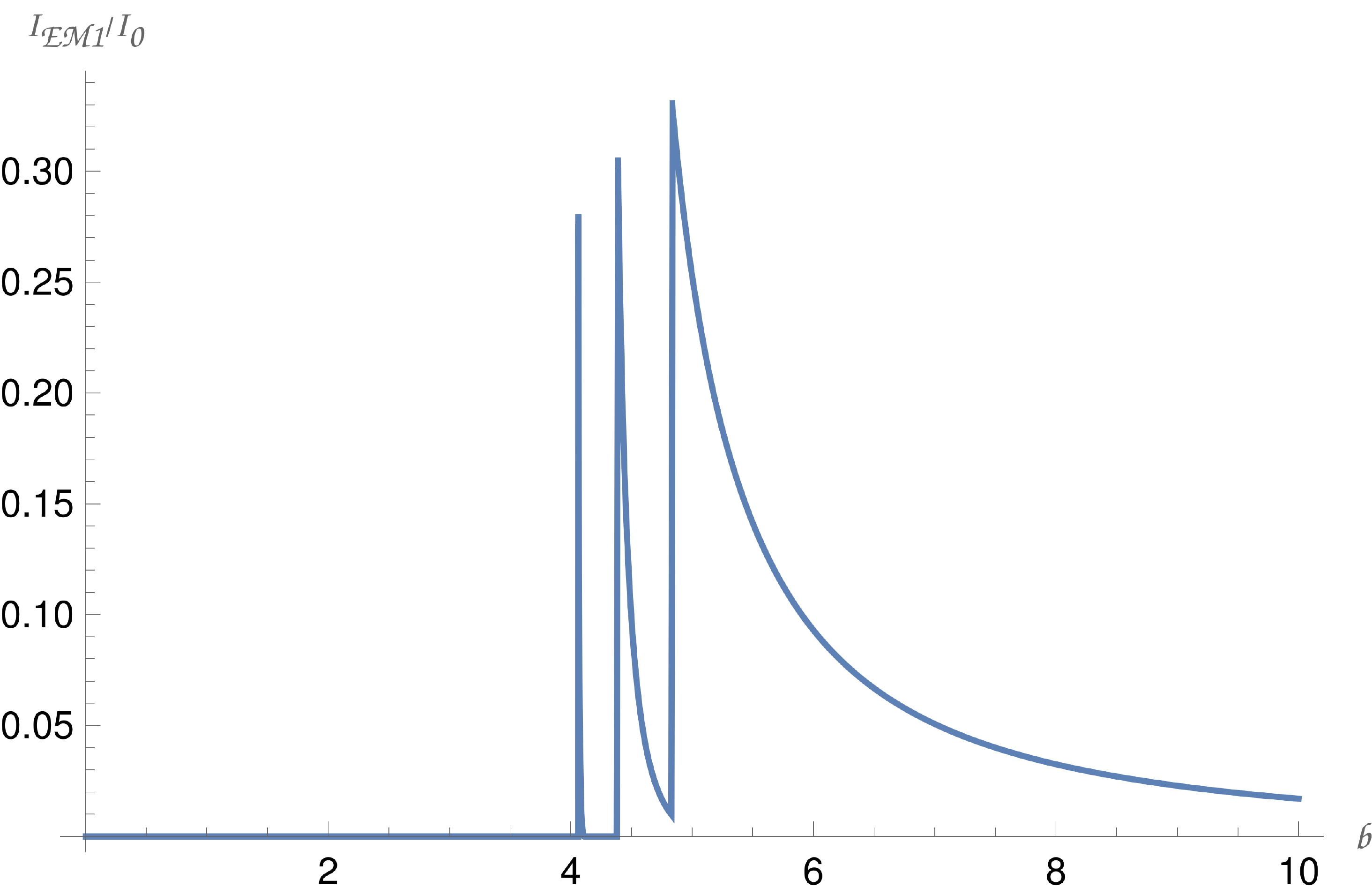}
    \includegraphics[width=.3\textwidth]{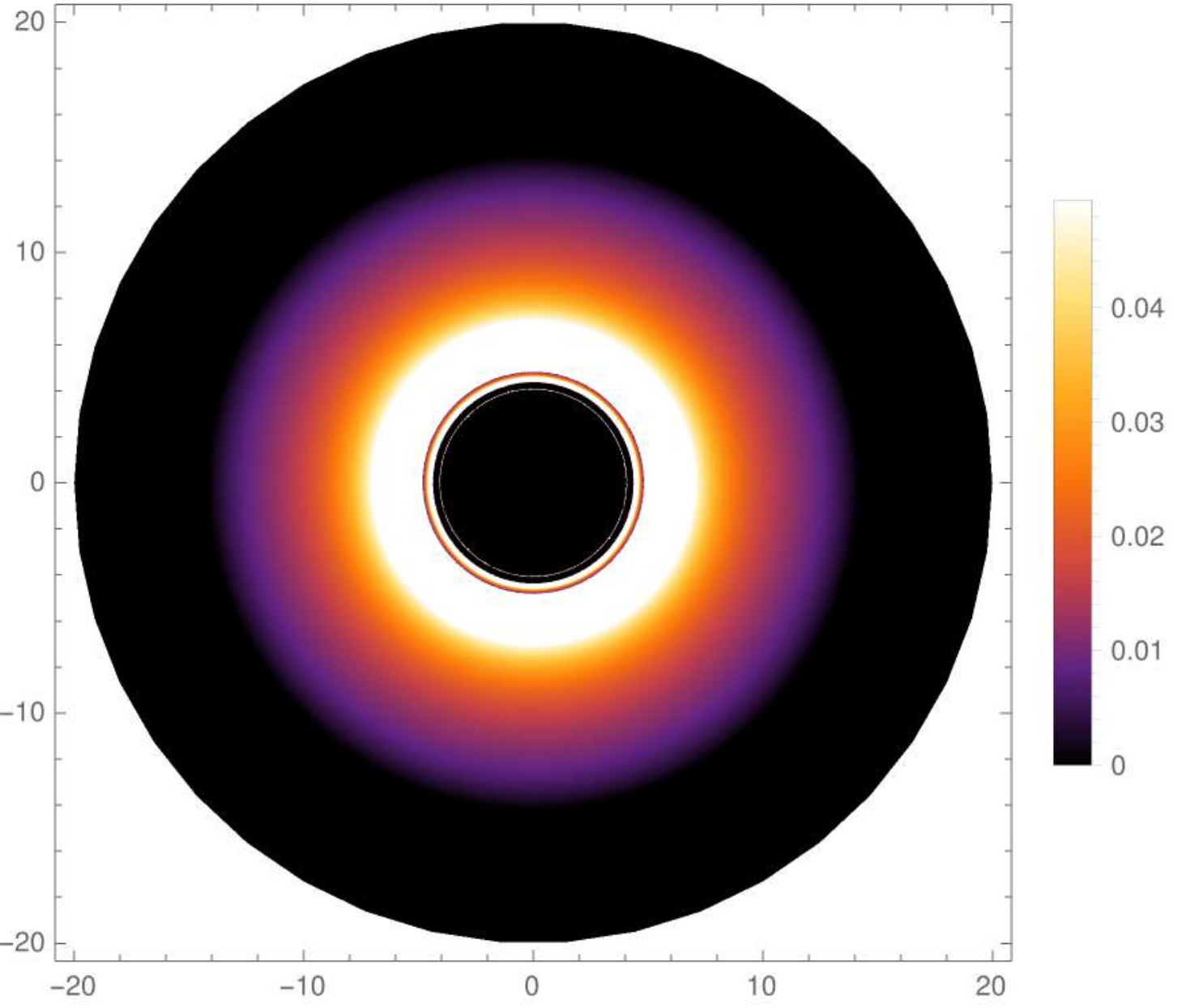}
    \vfill
    \centering
    \includegraphics[width=.3\textwidth]{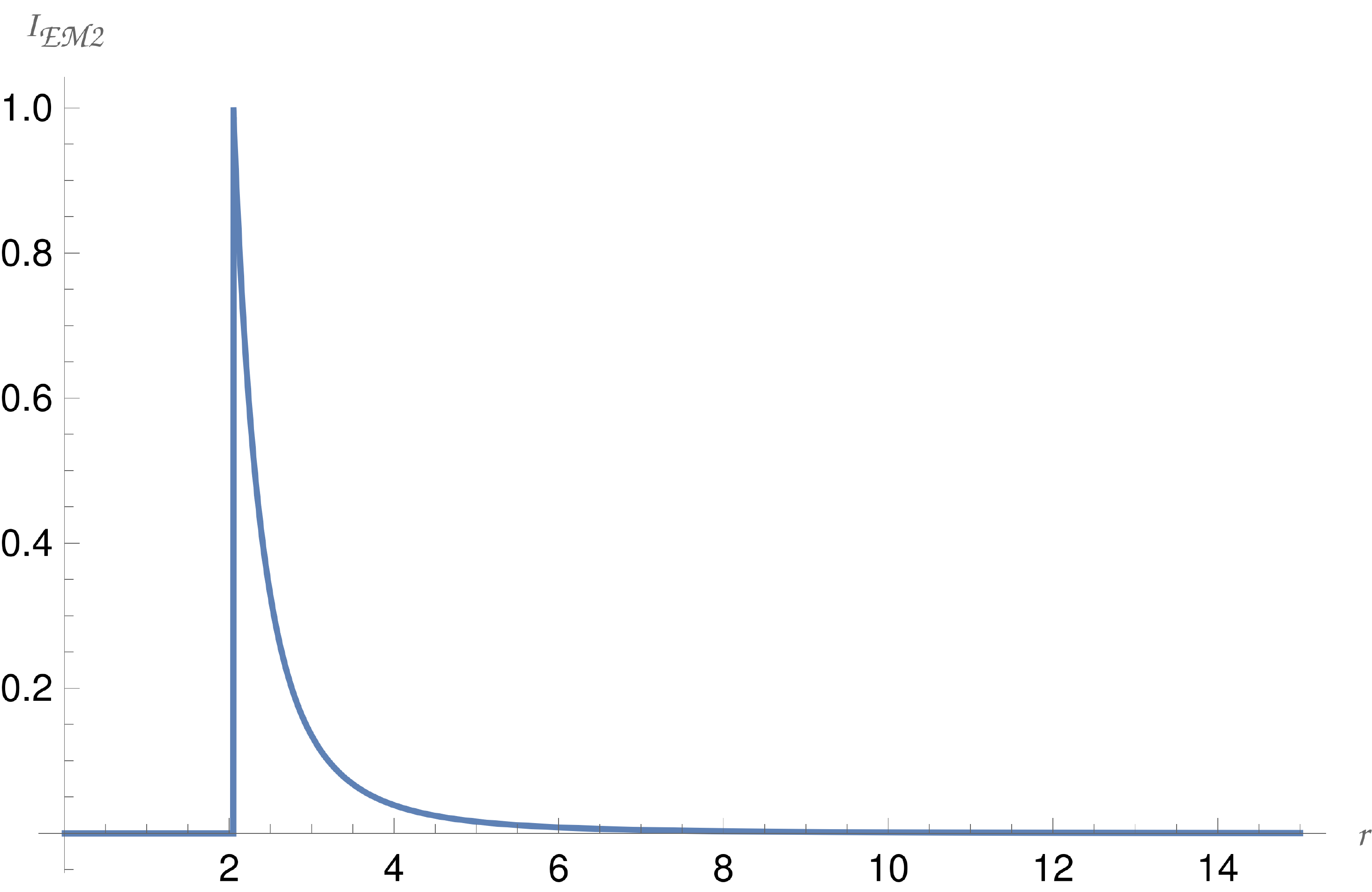}
    \includegraphics[width=.3\textwidth]{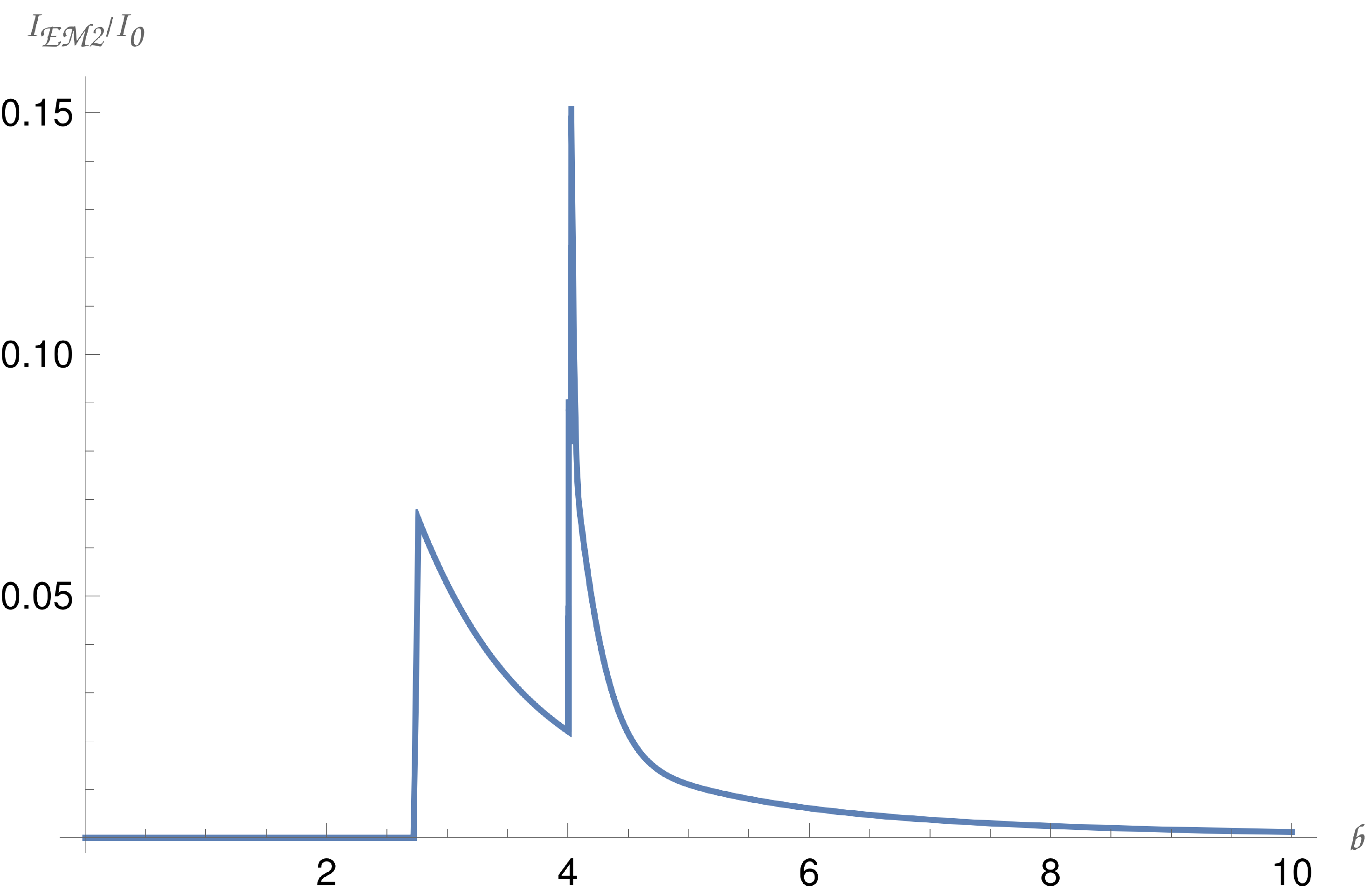}
    \includegraphics[width=.3\textwidth]{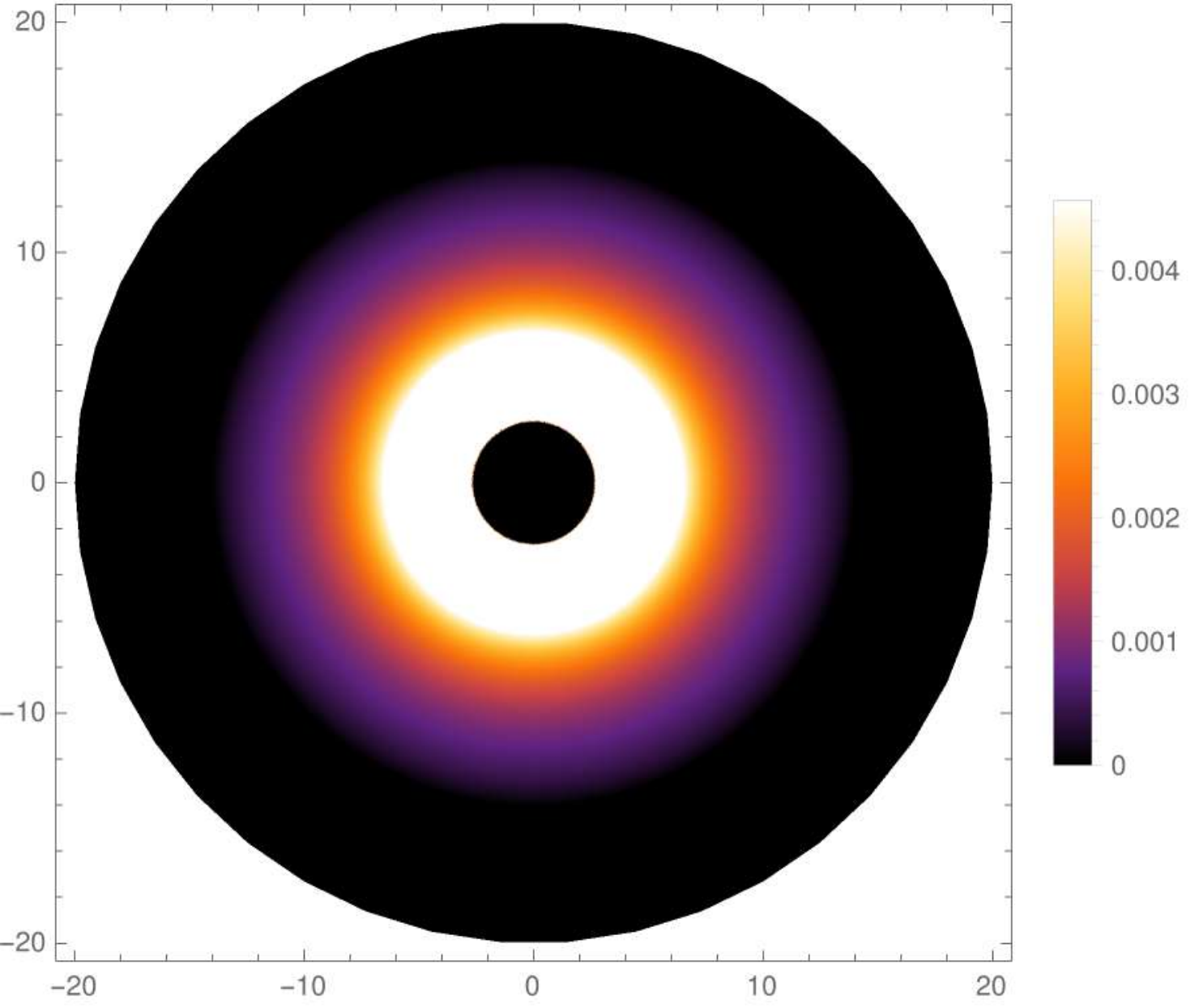}
    \vfill
    \centering
    \includegraphics[width=.3\textwidth]{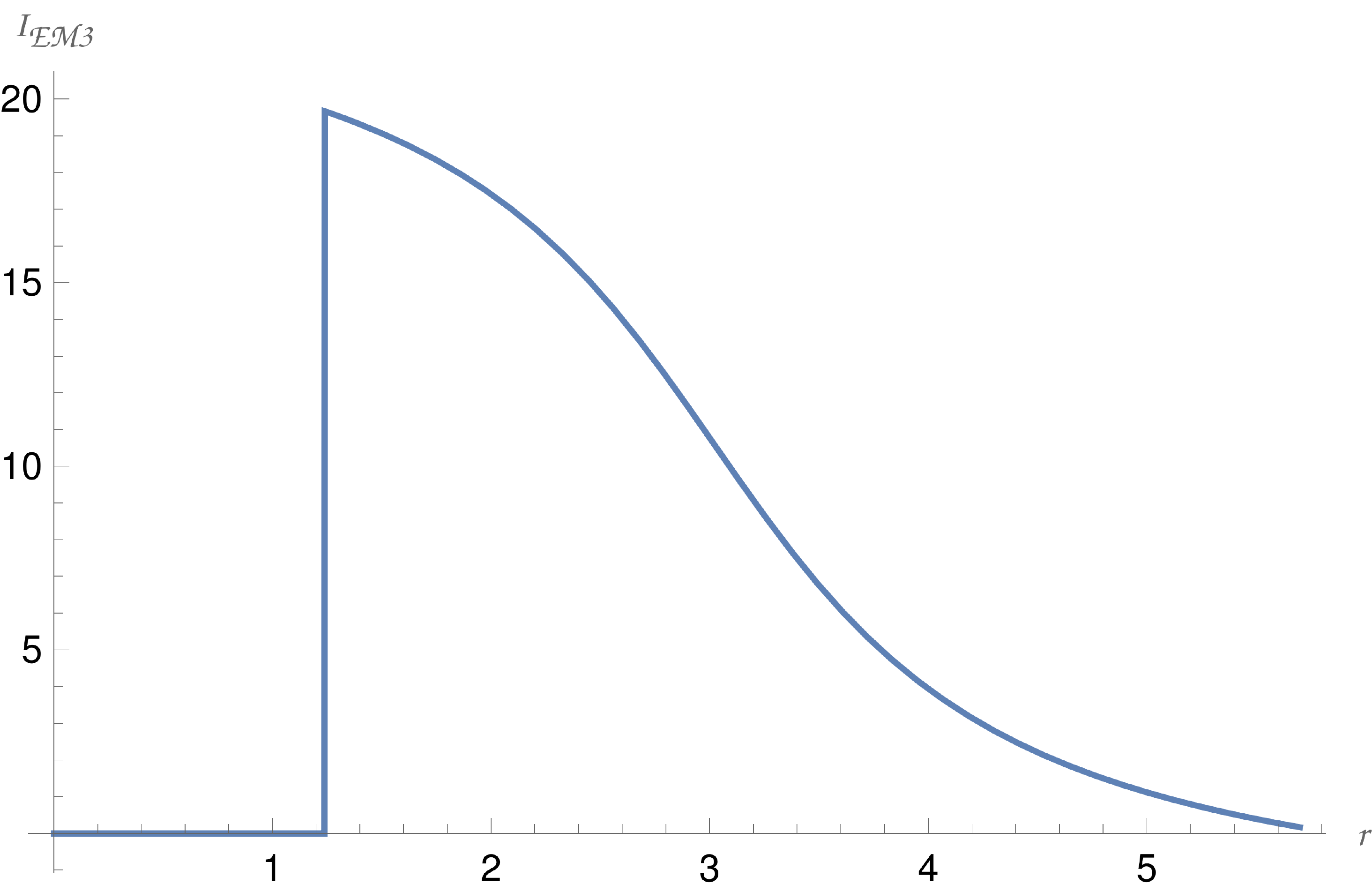}
    \includegraphics[width=.3\textwidth]{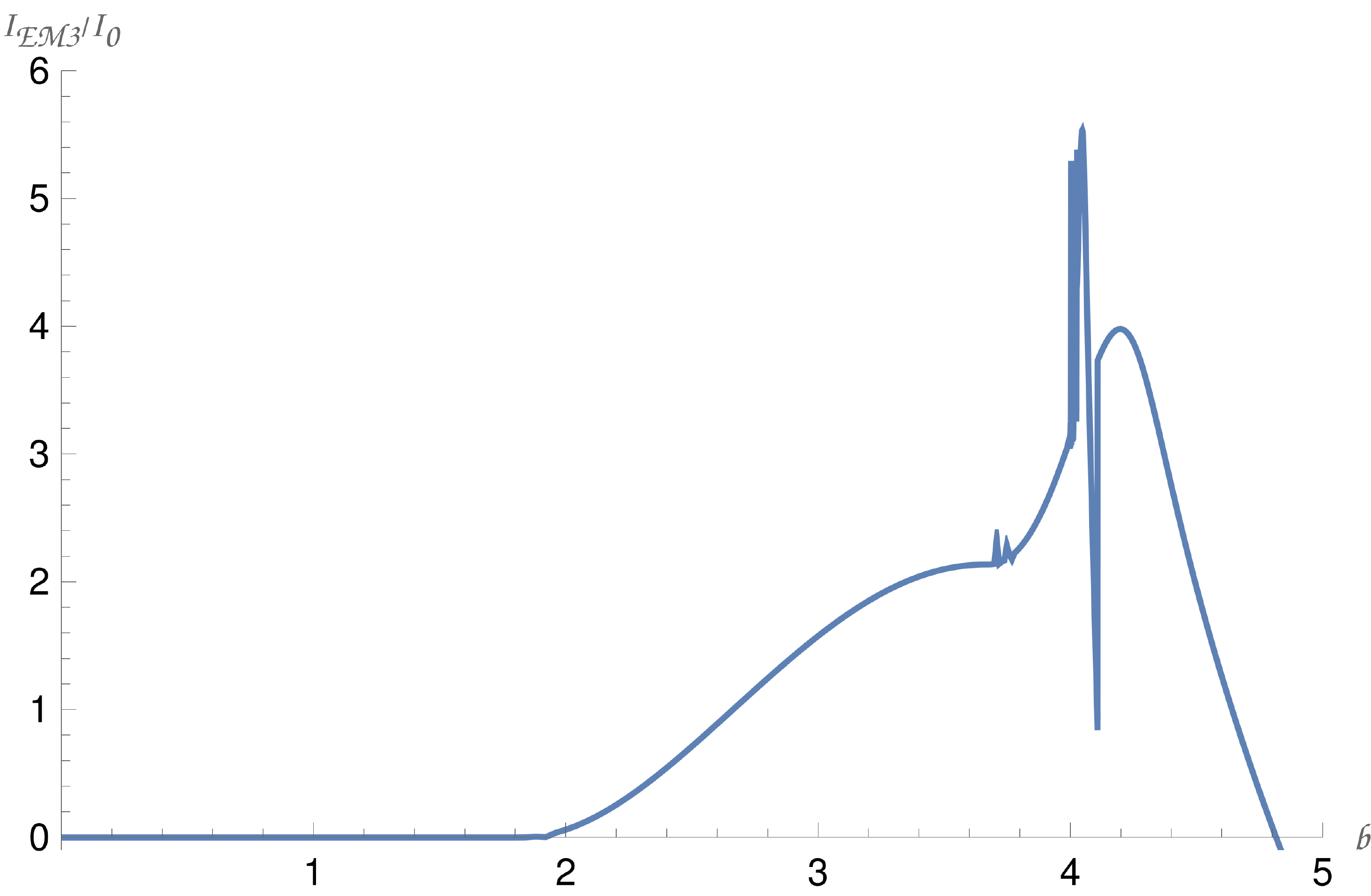}
    \includegraphics[width=.3\textwidth]{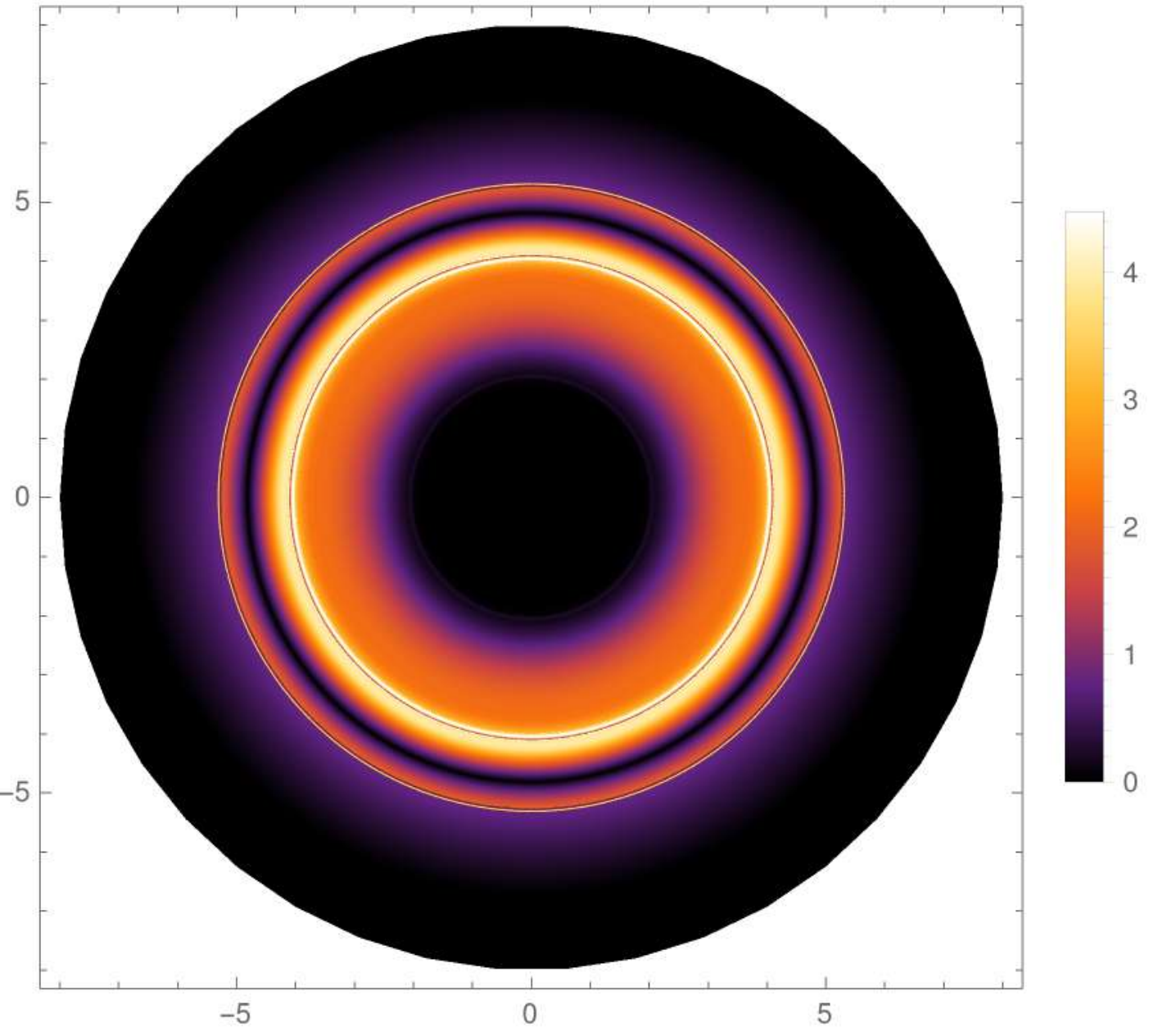}
    \caption{Observational appearance of the thin disk with different emission profile for $\beta=3M^4$, viewed from a face-on orientation. The upper row is for the emission profile intensity given by  model $1$, the second row for model $2$, and the third row is for model $3$ described in section C. \AU{In the plots the emitted and observed intensities $I_{EM}$ and $I_{obs}$ are normalized to the maximum value $I_0$ of the emitted intensity outside the horizon.}}
    \label{fig:18}
\end{figure*}
\begin{figure*}
    \centering
    \includegraphics[width=.3\textwidth]{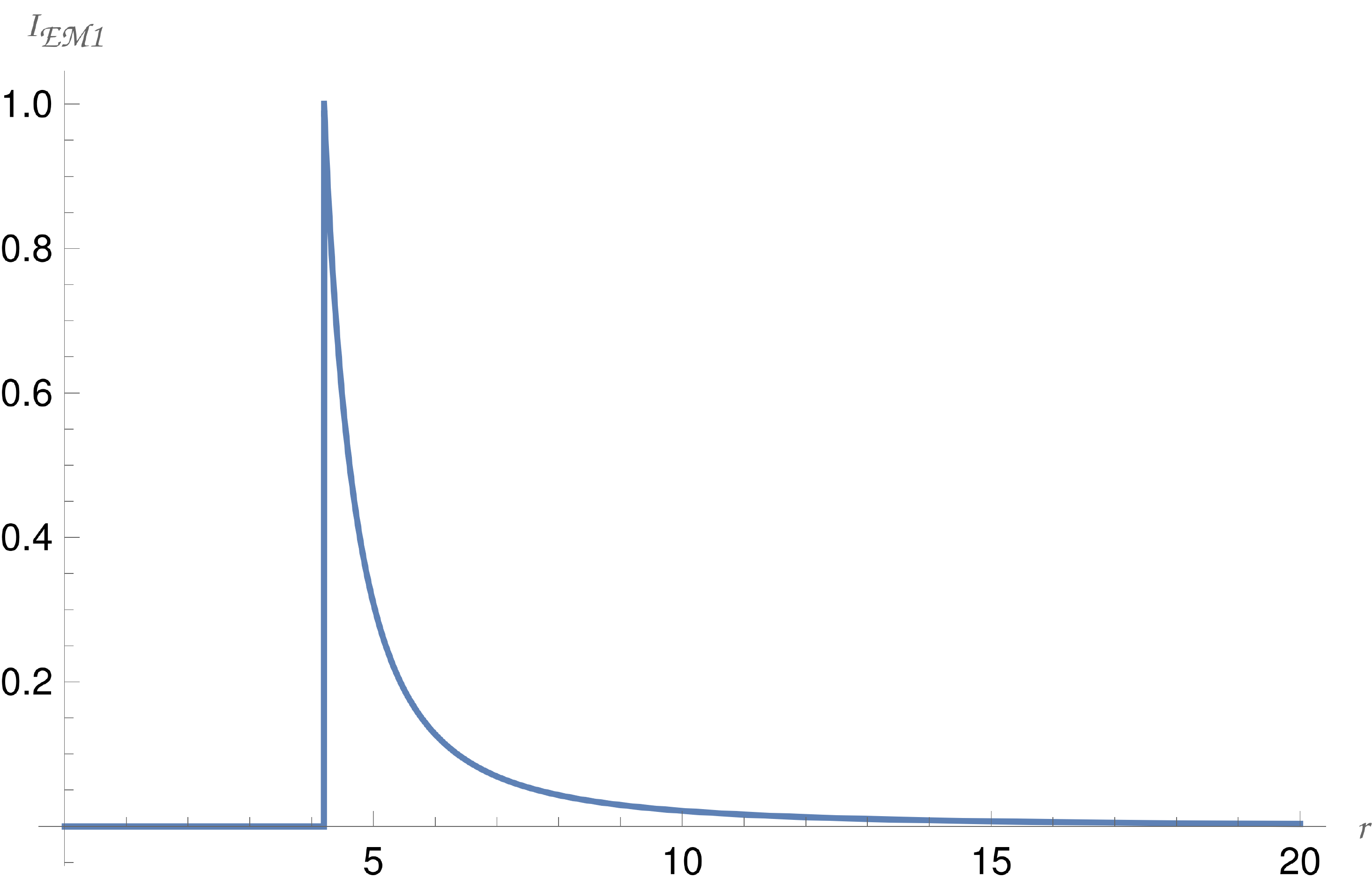}
    \includegraphics[width=.3\textwidth]{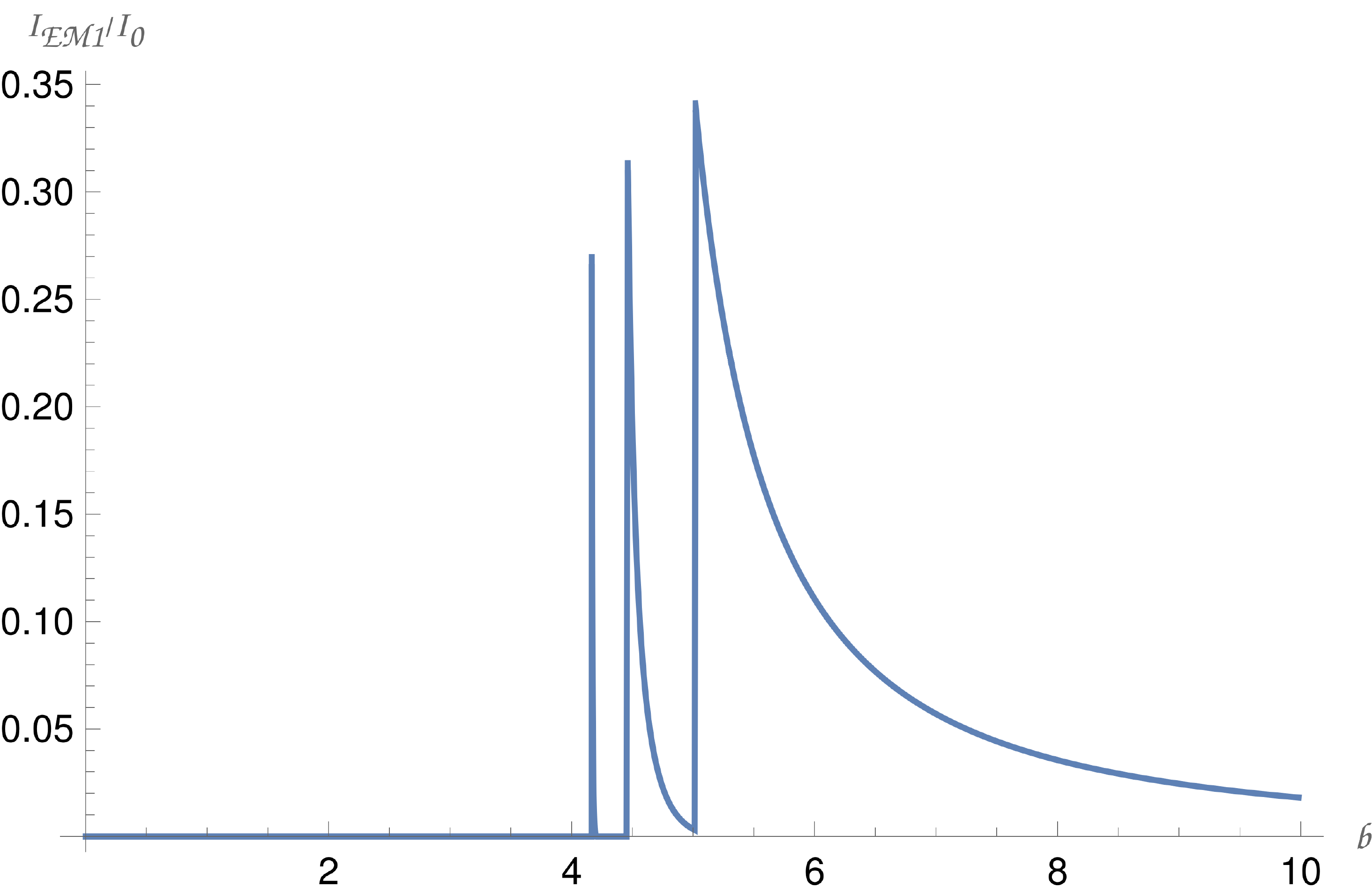}
    \includegraphics[width=.3\textwidth]{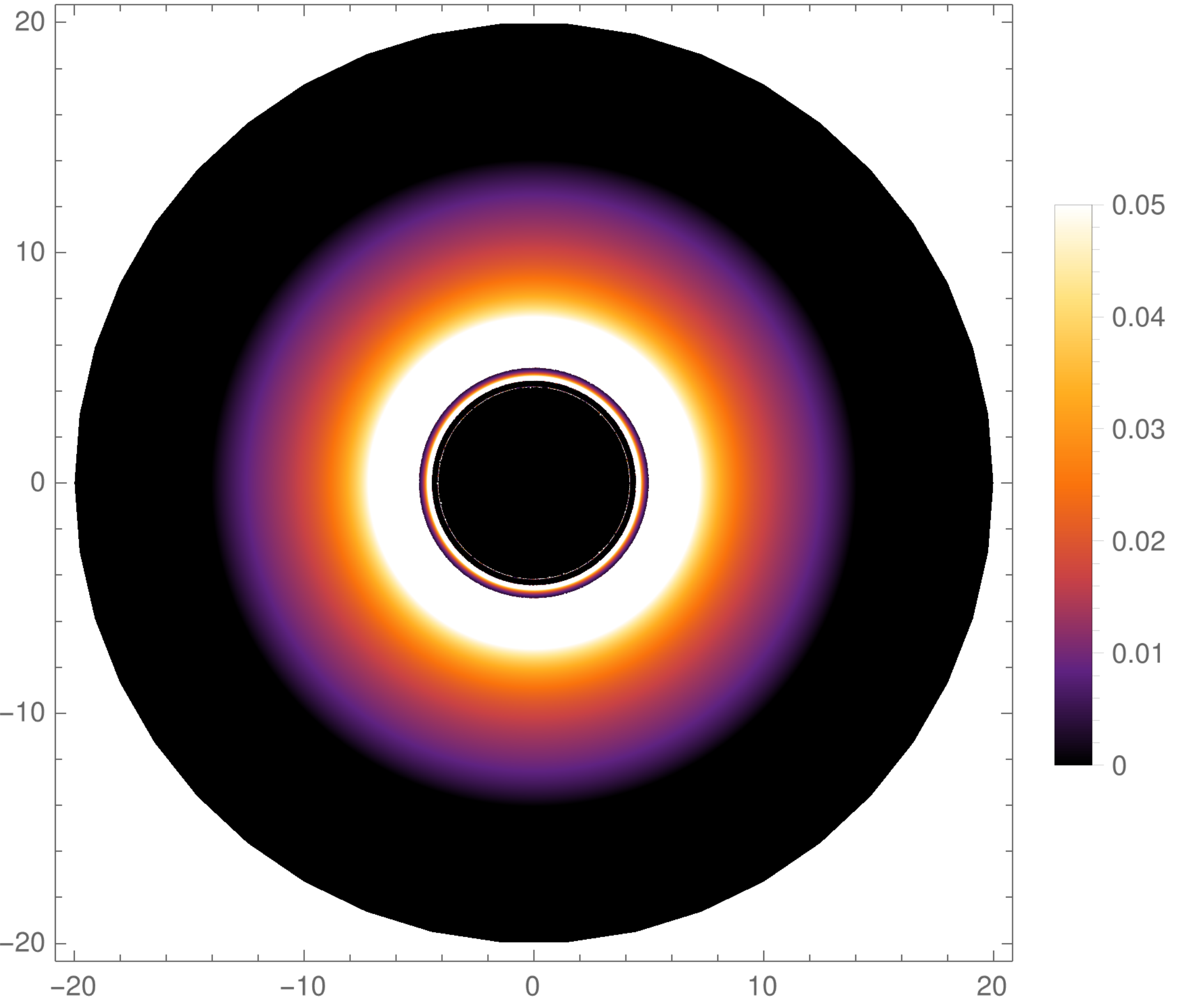}
    \vfill
    \centering
    \includegraphics[width=.3\textwidth]{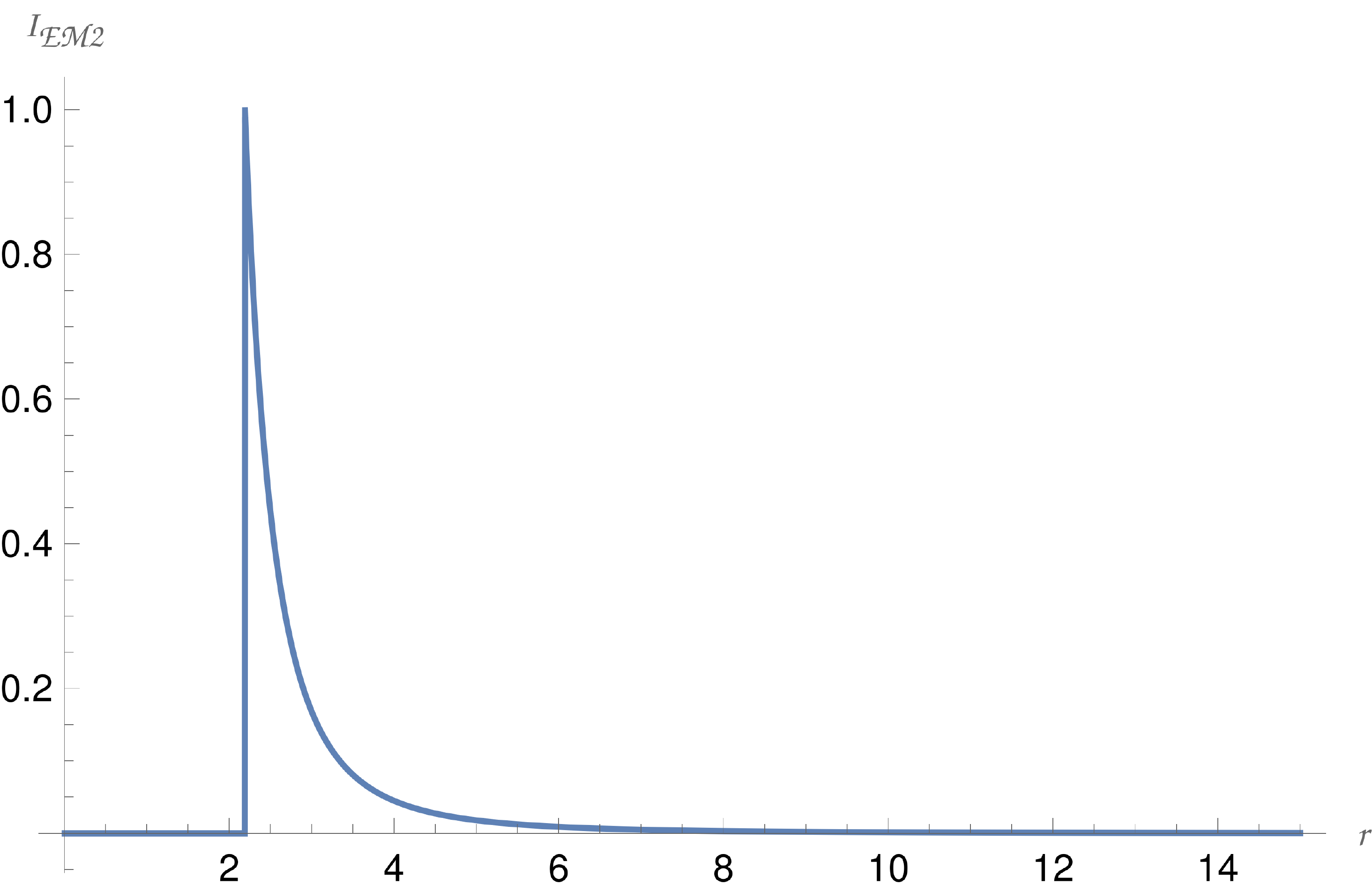}
    \includegraphics[width=.3\textwidth]{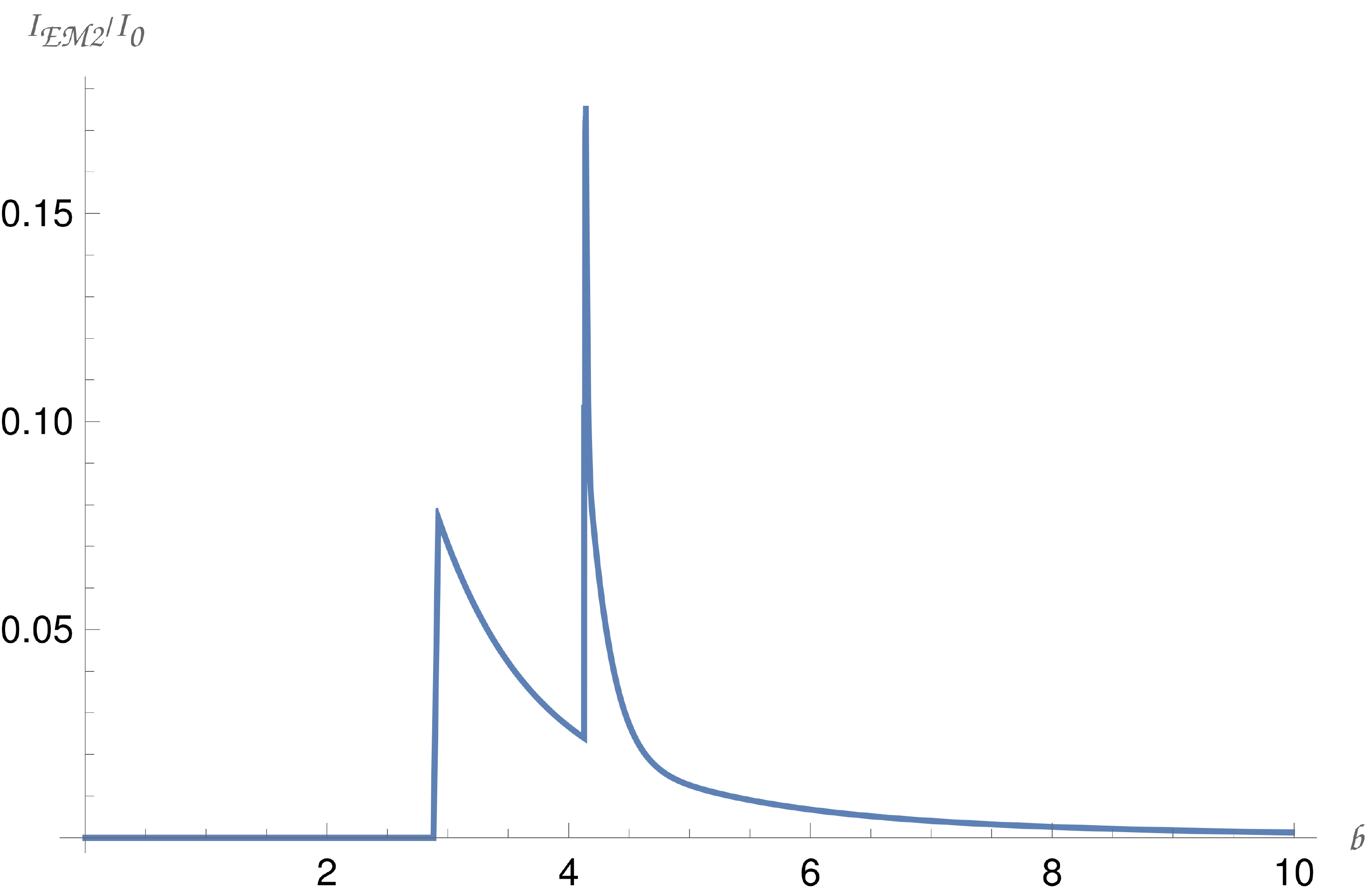}
    \includegraphics[width=.3\textwidth]{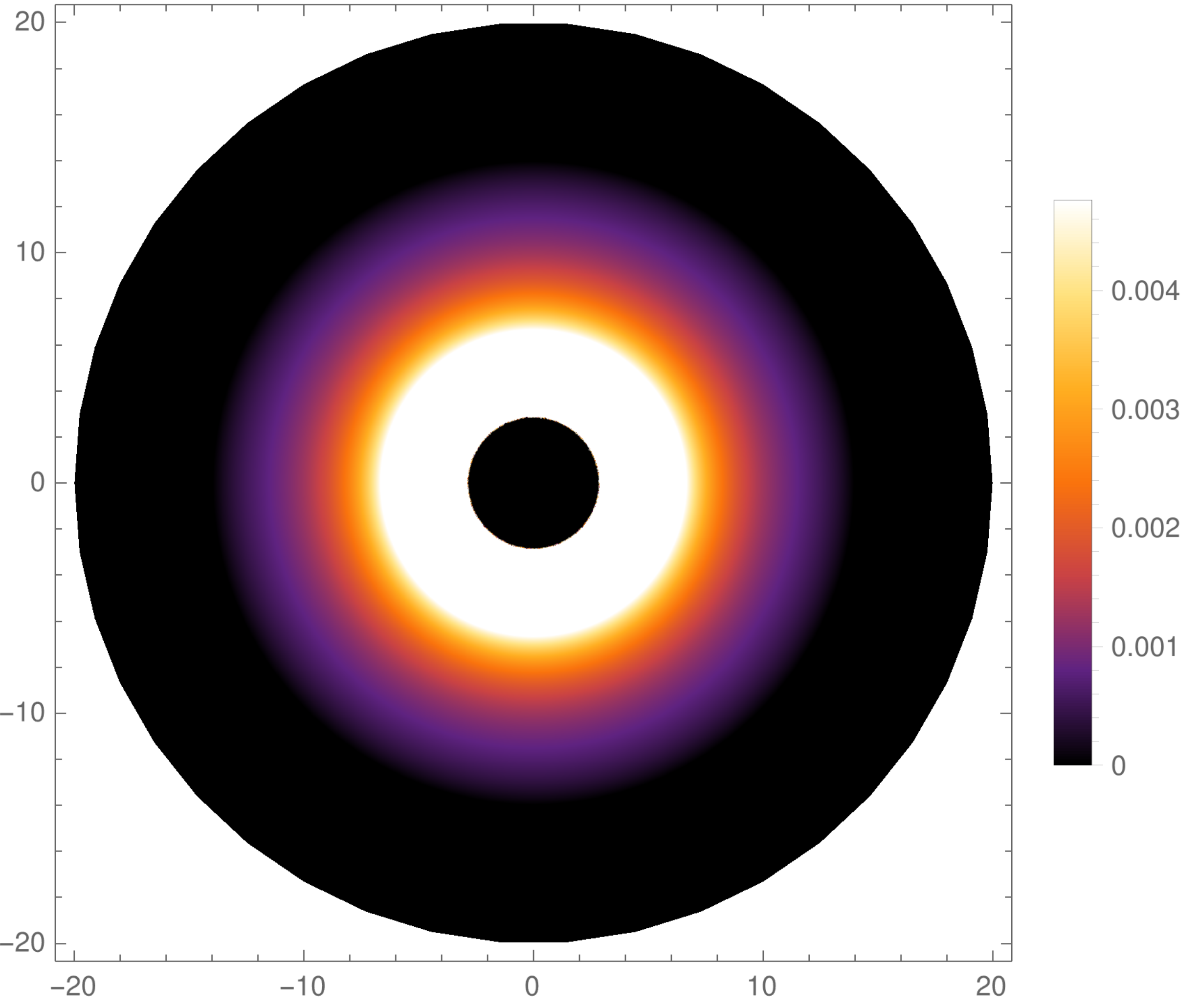}
    \vfill
    \centering
    \includegraphics[width=.3\textwidth]{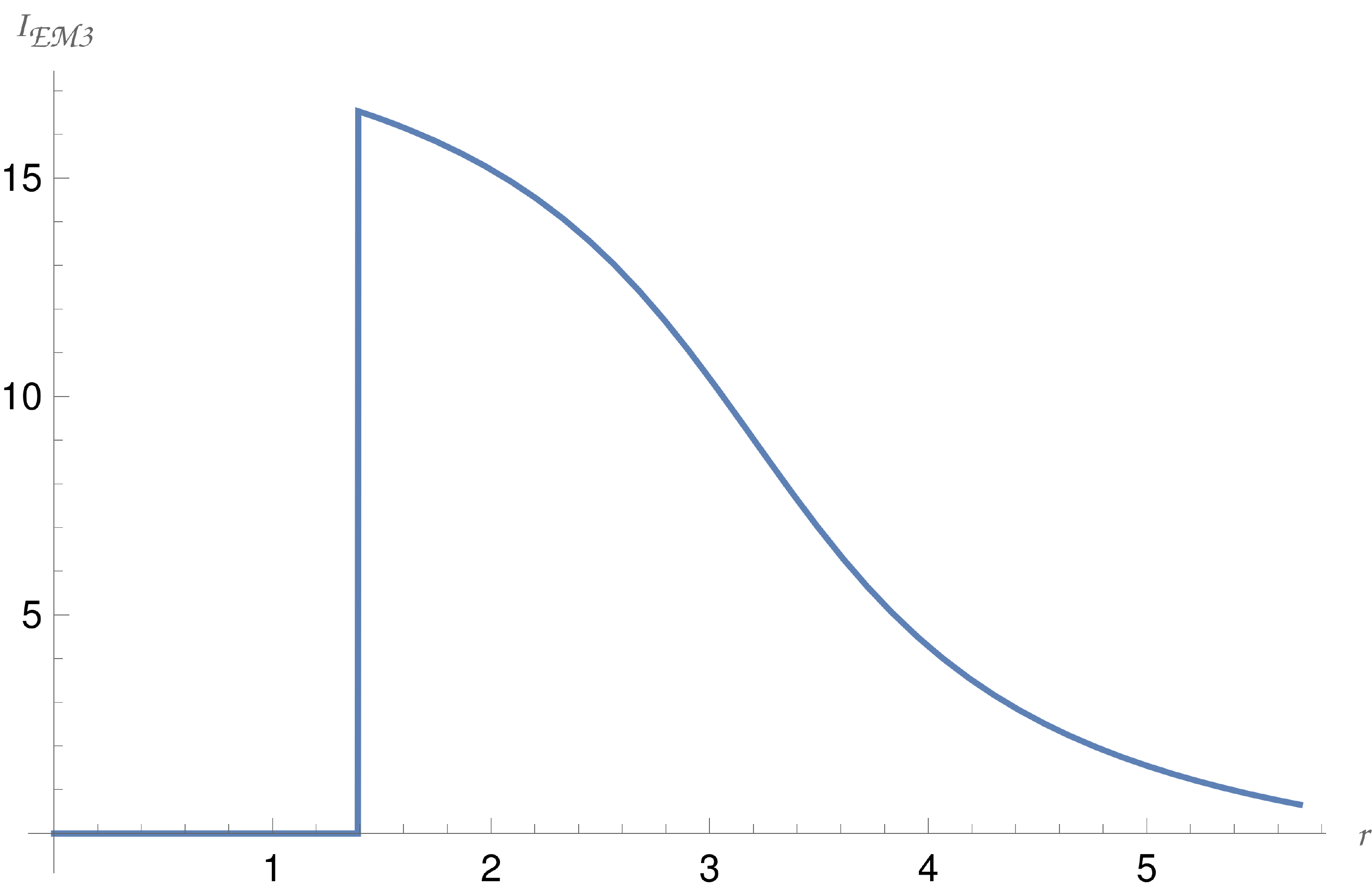}
    \includegraphics[width=.3\textwidth]{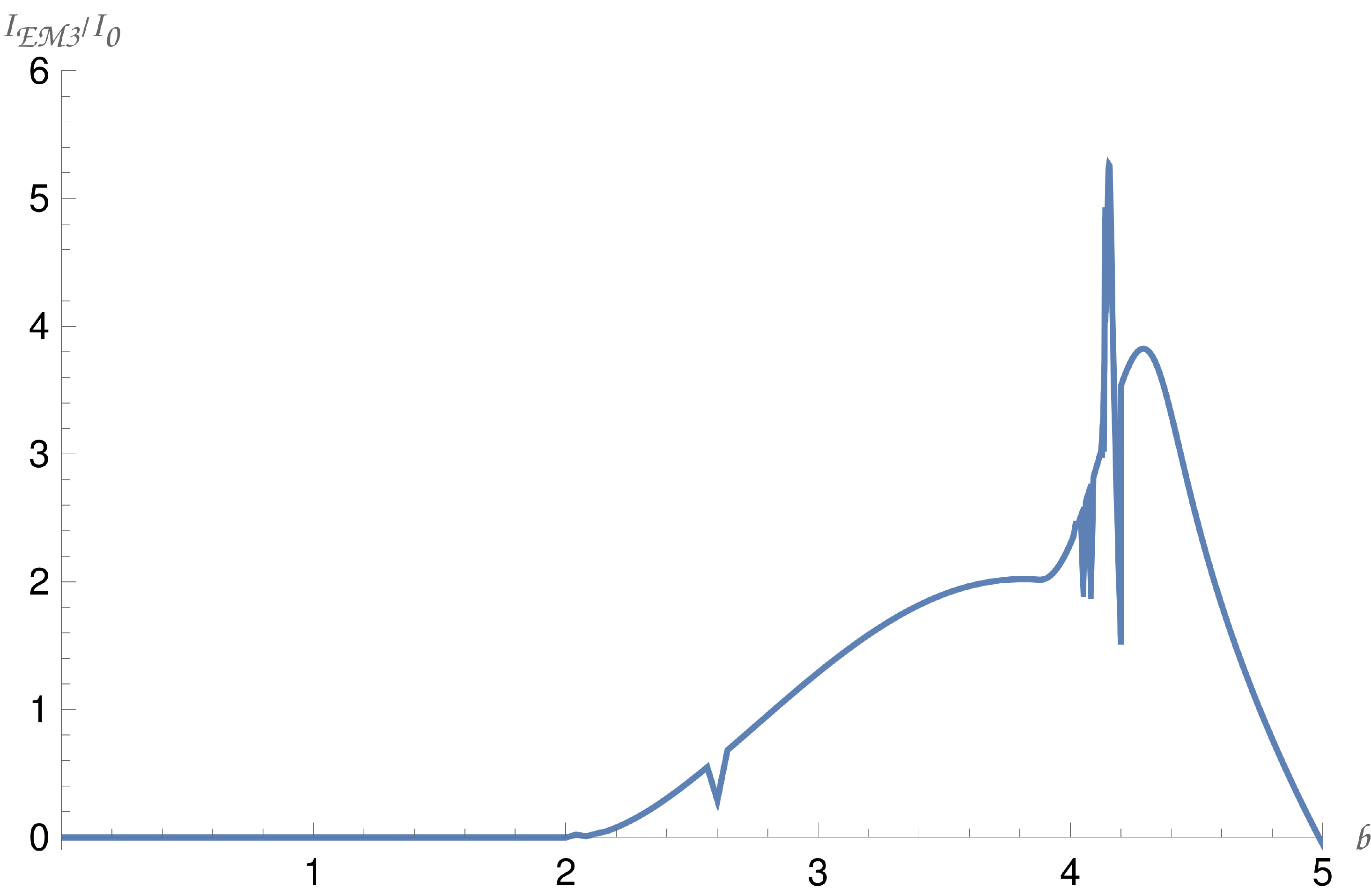}
    \includegraphics[width=.3\textwidth]{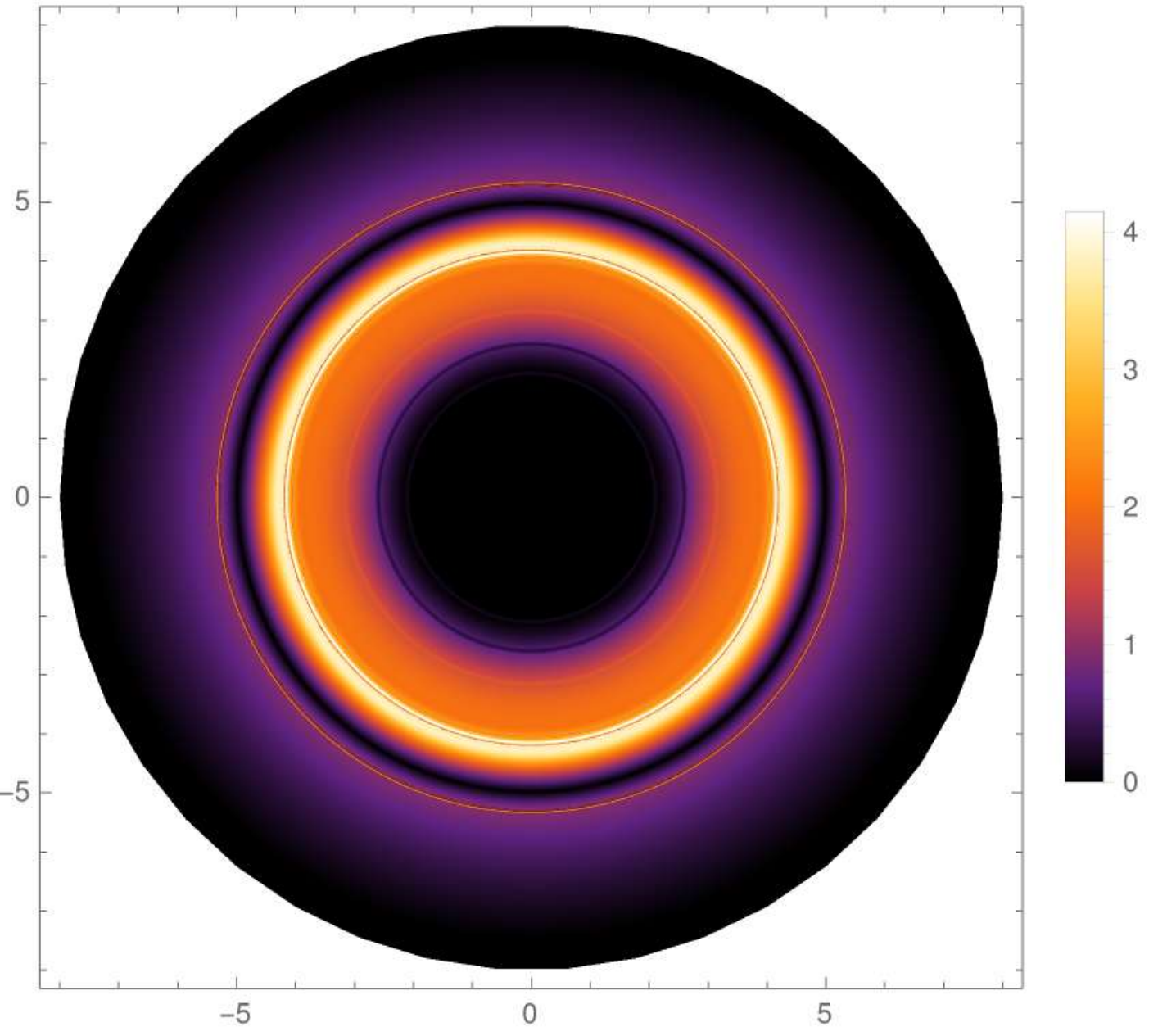}
    \caption{Observational appearance of the thin disk with different emission profile for $\beta=50M^4$, viewed from a face-on orientation. The upper row is for the emission profile intensity given by  model $1$, the second row for model $2$, and the third row is for model $3$ described in section C. \AU{In the plots the emitted and observed intensities $I_{EM}$ and $I_{obs}$ are normalized to the maximum value $I_0$ of the emitted intensity outside the horizon.}}
    \label{fig:19}
\end{figure*}

In the Figs. \ref{fig:17},\ref{fig:18} and \ref{fig:19}, we have shown the observation appearance of the thin accretion disk with the observed intensity and the impact factor corresponding to the three models for different coupling constant parameter $\beta=0.01M^4,3M^4,50M^4$ respectively. In all the figures, we see that for the first model (top row), the emission intensity (first row, first column) has a peak near the critical impact parameter $r\sim b_\text{c}$ and then it decreases as radial distance increases and becomes zero. In this case, the photon sphere lies in the interior region of the emission part of the disk. Also, we observed that due to the gravitational lensing, we have two independently separated peaks within the region of lensing and photon rings of the emission intensity (first row, second column). However, we noticed that the peak of the photon and lensing rings are not only smaller than the direct emission but also have a narrow observational area. Hence, we observed that the observed intensity has a huge contribution coming from the direct emission, a small contribution from the lensing rings, and very little contribution from the photon rings. It is also clear from the $2D$ shadow image (first row, third column) that the direct emission and photon ring mostly dominate the optical appearance and is not visible.

For the second model (second row), the intensity of the emission has a peak at photon sphere $r\sim r_\text{ph}$ and then decreases as radial distance increases (second row, first column). In this case, the observed intensity profile (second row, second column) peaks due to the direct emission, and then it shows an attenuation with increasing $r$. We observed that the photon and lensing rings are coinciding, which improves the total intensity of this particular area, and therefore, we get a new peak due to the photon ring, lensing ring, and direct emission. However, as discussed in the last section, the photon and lensing rings are highly demagnetized and have a very narrow area in the observed intensity. Therefore, we still have a dominant contribution coming from the direct emission in the observed intensity which can be seen in the $2D$ shadow image (second row, third column).

For the third model (third row), the peak in the intensity starts from the horizon ($r_h$) and then decreases with increasing $r$ (third row, first column). In this case, the region of the photon ring, lensing ring, and direct emission coincide for a large range of $r$, which can be seen in the observed intensity (third row, second column). We observed here that the intensity gently increases from a slightly larger than the horizon of the black hole and then suddenly increases sharply to a peak value in the photon ring region. After that, due to the contribution from the lensing ring, the observed intensity shows a higher peak followed by a sudden decrease, then decreases and goes to zero. As a result, we observed multiple rings in the $2D$ shadow image coming from the contribution of the photon ring, lensing ring, and direct emission (third row, third column).

We can compare all the emission profiles with the observed intensity and optical appearance of the disk for the different values of the coupling constant $\beta=0.01M^4,3M^4,50M^4$ when taking $C=1$. For all three models, as we increase the coupling constant $\beta$, the observed intensity increases, but it is much less than what we get for the Schwarzschild spacetime \cite{Gralla:2019xty}. Therefore, we observed significant changes in the observed intensities when we include the NLE effect in theory, and we can easily distinguish the NLE black hole from the Schwarzschild black hole.

\subsection{Shadow With Infalling Spherical Accretion}
In this section, we will study the image of the NLE black hole with an infalling spherical accretion by considering the optically thin disk \cite{Bambi:2013nla}. Here we consider an infalling spherical accretion model known as the dynamical model, which can describe the radiative gas moving around the black hole forming the accretion disk. We will investigate the parameter $\beta$ to understand the observational characteristics. For the observer sitting at infinity, the specific intensity is given by
\begin{equation}
 I_\text{obs}= \int_\Gamma g^3 j(\nu_e) dl_\text{prop},
\end{equation}
where $\nu_e$, $g$, and $\nu_{obs}$ are known as the photon frequency, redshift factor, and observed photon frequency, respectively. Now, let us consider the rest frame of the emitter, where, $j(\nu_e) \propto \frac{\delta (\nu_e-\nu_f)}{r^2}$ is known as the emissivity per unit volume, which is having $1/r^2$ radial profile. Here, $\delta$ is the well-known delta function, $\nu_f$ is called the frequency of the light radiation, which is considered monochromatic, and $dl_\text{prop}$ is the infinitesimal proper length. Now, the redshift factor for the black hole spacetime is given by
\begin{equation}
g=\frac{\mathcal{K_{\rho}}u_{o}^{\rho}}{\mathcal{K_{\sigma}}u_{e}^{\sigma}},
\end{equation}
where $\mathcal{K^{\mu}}=\dot{x_{\mu}}$ is known as the four-velocity of the photon, and $u_o^{\mu}=(1,0,0,0)$ is known as the four velocities of the static observer at infinity. Now, the four-velocity $(u_e^{\mu})$ of the infalling accretion is given by
\begin{equation}
u^t_e=f(r)^{-1},u^r_e=-\sqrt{1-f(r)},u^{\theta}_e=u^\phi_e=0.
\end{equation}
Hence, we can write the four velocities of the photon by using the null geodesic
\begin{equation}
\mathcal{K}_t=\frac{1}{b},\frac{\mathcal{K}_r}{\mathcal{K}_t}=\pm \frac{1}{f(r)}\sqrt{1-f(r)\frac{b^2}{r^2}},
\end{equation}
where the sign $\pm$ shows that the photon goes or away from the black hole. Therefore, we can write the redshift factor for the infalling accretion
\begin{equation}
g=\left[ u_e^t + \left( \frac{\mathcal{K}_r}{\mathcal{K}_e} \right) u^r_e \right]^{-1},
\end{equation}
and the proper distance will b modified as
\begin{equation}
dl_\text{prop}=\mathcal{K}_{\mu}u^{\mu}_e d\lambda = \frac{\mathcal{K}_t}{g |\mathcal{K}_r |}dr.
\end{equation}
Hence, now we can integrate equation $(38)$ over all the frequencies and get the expression observed intensity with infalling spherical accretion
\begin{equation}
I_\text{obs} \propto \int_{\Gamma} \frac{g^3 \mathcal{K}_t dr}{r^2 |\mathcal{K}_r |}.
\end{equation}
Based on the above equation, we explore the influence of the NLE on the near horizon of the black hole shadow and its brightness distribution.

\begin{figure}
    \centering
    \includegraphics[width=\linewidth]{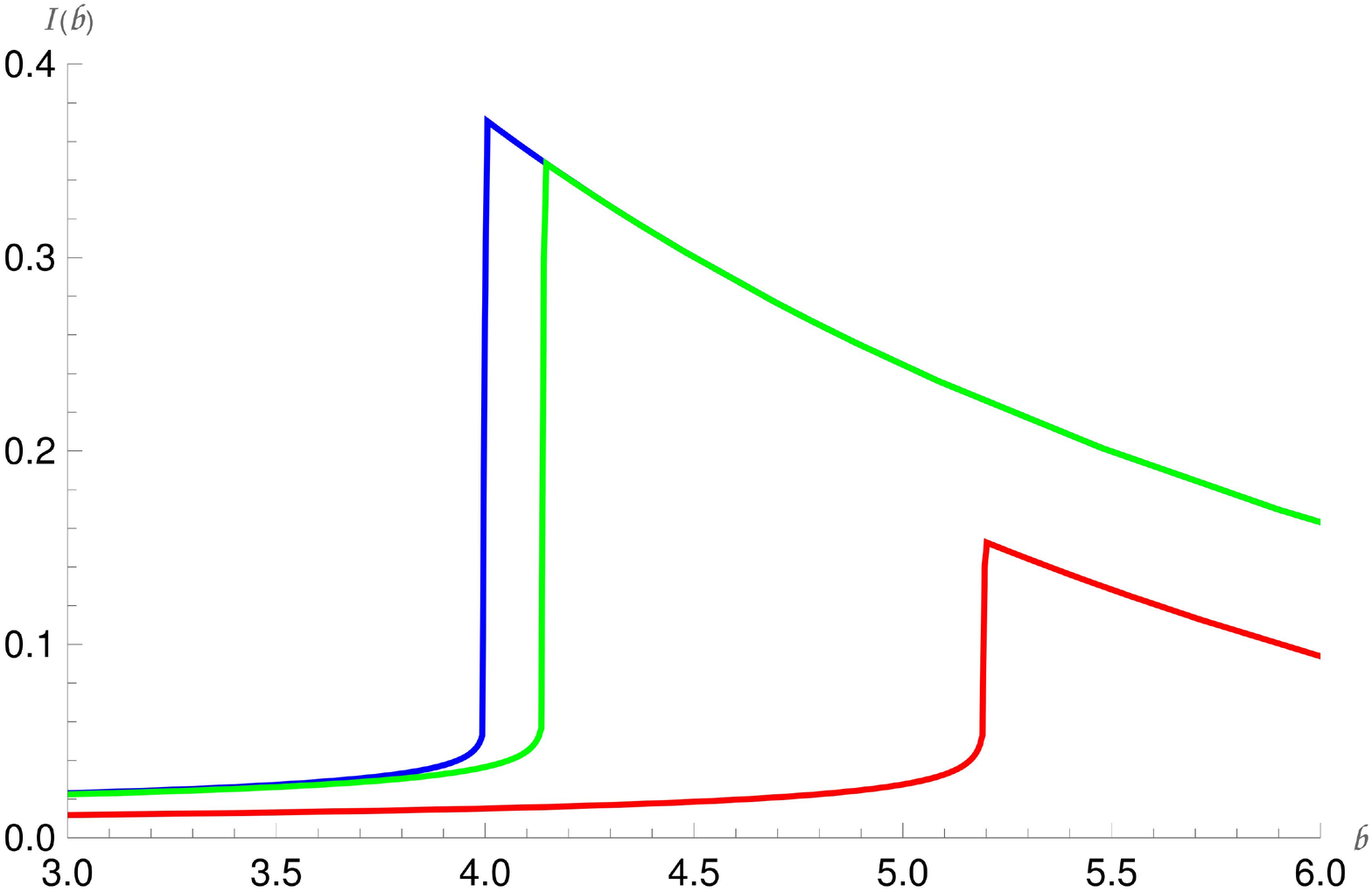}
    \caption{Observed intensity of the infalling spherical accreting matter for Schwarzschild  (\text{red}) and for different values of $\beta=0.01M^4 (\text{green}),50M^4 (\text{blue})$ respectively.} \label{fig:compp}
\end{figure}

\begin{figure}
    \centering
    \includegraphics[width=.3\textwidth]{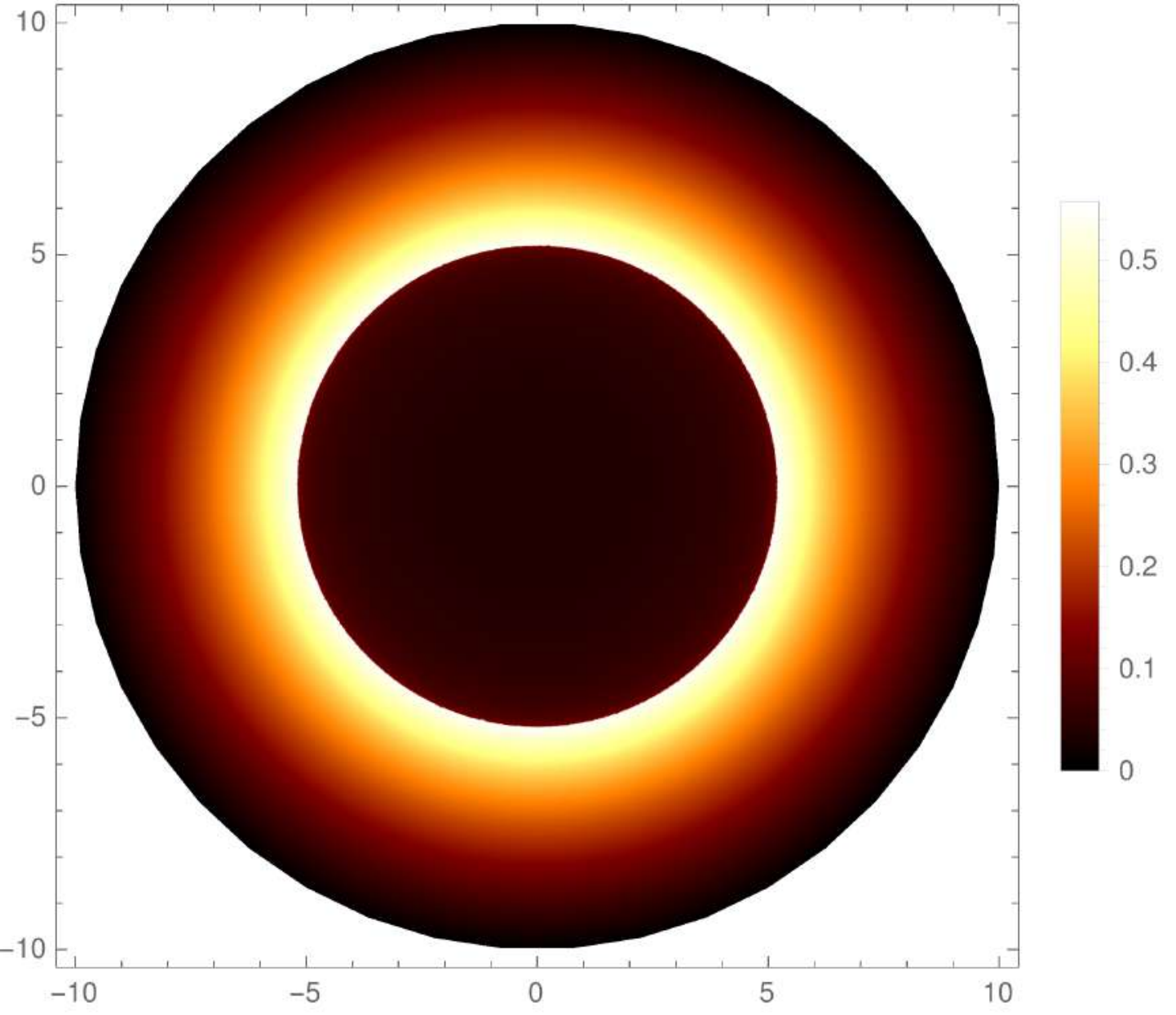}
    \includegraphics[width=.3\textwidth]{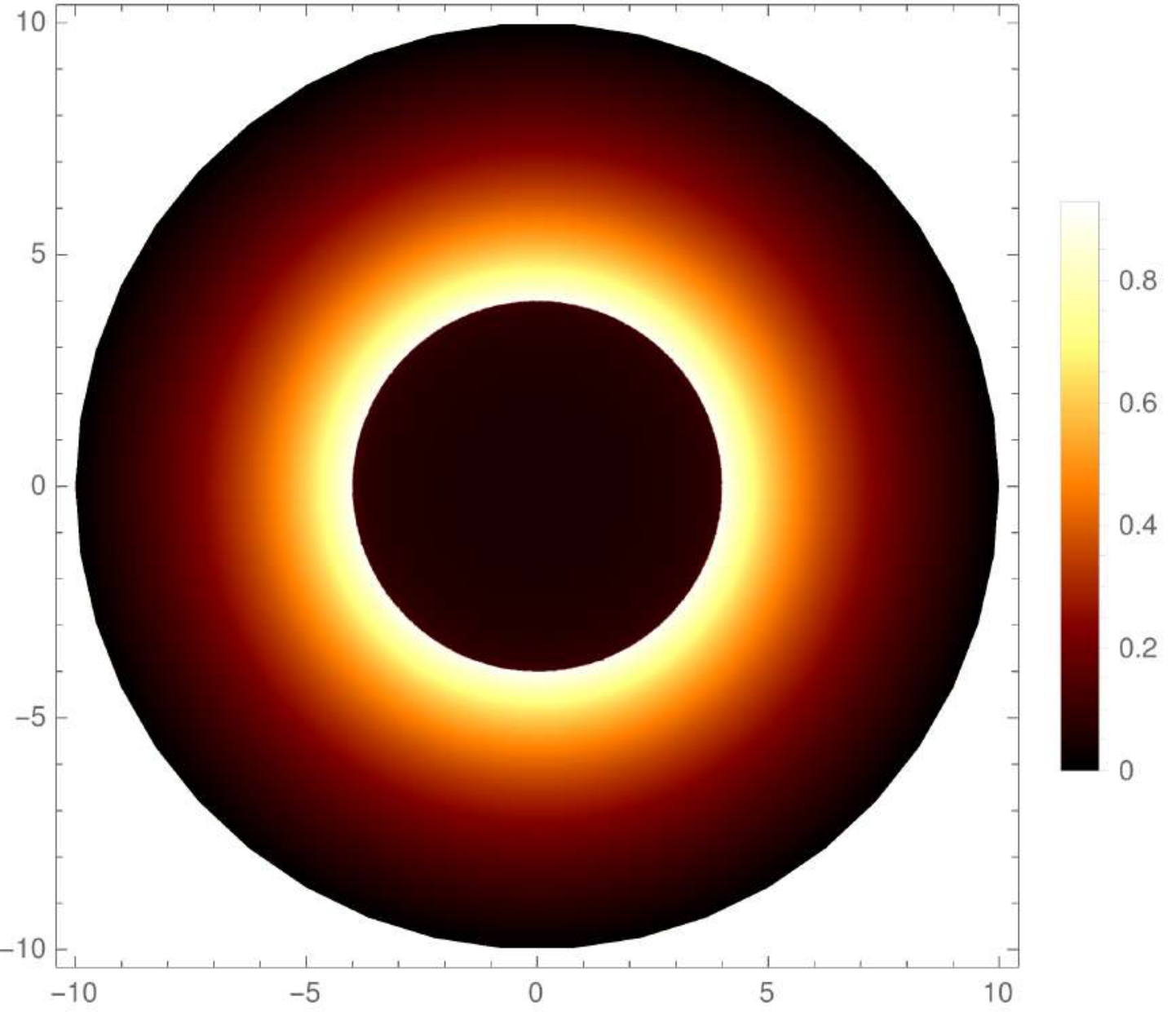}
    \includegraphics[width=.3\textwidth]{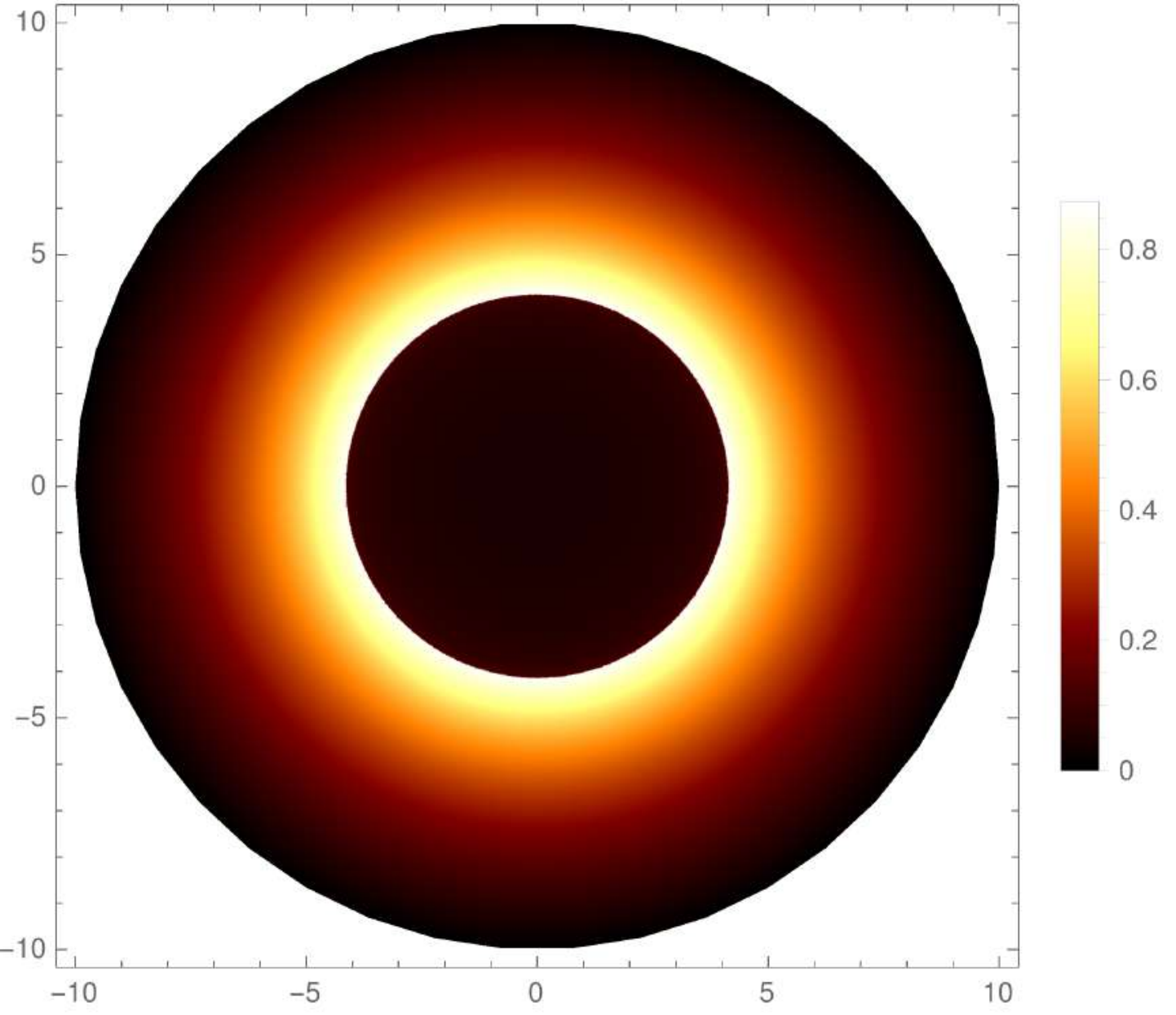}
    \caption{Image of the black hole shadow with infalling spherical accretion for Schwarzschild (top) and different values of $\beta=0.01M^4,50M^4$ respectively.} \label{fig:21}
\end{figure}

We can see from the Fig.\ref{fig:compp} that the specific intensity $I_\text{obs}$ increases sharply as the impact parameter $b$ increases and it shows a peak at $b \sim b_{ph}$. The region where $b>b_\text{c}$, the specific intensity $I_\text{obs}$, decreases and in the limit of $b \to \infty$, it goes to zero. For Schwarzschild (red), the intensity is less than the NLE black hole, and as we increase the coupling parameter $\beta$, the peak value increases but decreases at the same rate. For comparison we have shown $\beta=0.01M^4$(green) and $\beta=50M^4$(blue). A similar effect can be seen in the $2D$ shadow image (see Fig. \ref{fig:21}). 


\section{Conclusions}\label{sec:conclude}
In this paper, we studied the observational characteristic of the thin accretion disk around the black hole by coupling gravity with non-linear electrodynamics. We have found two free parameters $\beta$ and $C$, whose effects approach the Reisnner-Nordstr\"om case in the asymptotically flat regions. Nonetheless, we examined the effect of these parameters through the deviations in the black hole shadow, weak deflection angle, and radiation flux.

\rp{To properly choose values of the NLE parameter $\beta$, we initially find constraints through the EHT data on the shadow diameter. We found a lower range for $\beta$ in Sgr. A*, while a higher one in M87*. Theoretically, we also have analyzed the NLE effect on the photonsphere and shadow radius for observer very close and remote to the black hole.} We find that both radius increases due to the effect of $\beta$ for both observers. However, while the increase in shadow radius is abnormally large for observers close to the black hole, observers at $r\to \infty$ see deviations close to the Reissner-Nordstr\"om case. These deviations can be detected with future sophisticated astronomical tools since we represent the observer at $r\to \infty$, and $\beta$ is seen to give considerable deviation even at a low black hole charge ($Q = 0.01M$).

For the weak deflection angle, the deviation to the known case is noticeable when the impact parameter is comparable to the value of the photonsphere radius. The effect of $\beta$ increases the value of $\hat{\alpha}$ (in $\mu$as) for the timelike particles. However, a much higher value is seen for null particles ($v=1$). It is observed that the deviation becomes so small as $b/M$ further increases. Thus, to detect the effects of $\beta$, one must only consider those particles grazing somewhat near the photonsphere. \rp{With how sensitive the weak deflection angle is on these regions, the deviation caused by the NLE parameter $\beta$ can be possibly detected by future space technologies such as the EHT ($10\mu$as), ESA GAIA mission ($20\mu$as-$7\mu$as) \cite{Liu_2017}, and the futuristic VLBI RadioAstron that can achieve at around $1-10\mu$as \cite{Kardashev2013}.}

Furthermore, we studied the properties of the thin accretion disk such as time-averaged energy flux $(F)$, the disk temperature $(T)$, the differential luminosity$(dL_{\infty})$ within the framework of NLE space-time. We have observed that as you increase the coupling parameter $\beta$, the dip in the potential increases which has a direct effect on the physical quantities such as it decreases the peak value of radiative flux and differential luminosity.

A similar kind of effect has been seen using the Okyay-\"Ovg\"un \textit{Mathematica} notebook package \cite{Okyay:2021nnh}, (used in also \cite{Chakhchi:2022fls}), when we investigated the black hole shadow and rings with three toy models of a thin accretion disk. We find that increasing the parameter decreases the first peak in intensity while the last peak increases and the corresponding critical impact parameter increases. Also, we found that in each case, the direct emission played the dominant role in the shadow image, but for the third model of the emission profile, the lensing ring played a significant role because of NLE coupling.
 
In the case of the spherical accretion flow, we noticed that the peak value of the observed intensity increases with increasing the coupling parameter $\beta$. For the case when we turn on the NLE effect, the shadow size increases in comparison with the Schwarzschild. We have shown two cases ($\beta=0.01M^4,50M^4$) and observed that the $2D$ depiction of the intensity plot looks very similar, except the higher beta will have higher brightness near the horizon.

Since no prior bound exists on the coupling constant $\beta$ from the laboratory experiments as mentioned in \cite{kruglov2015nonlinear}, we tried to put the constraint on $\beta$ by using the astrophysical environment. We have shown the bounds for the coupling parameter $\beta$ by using the data provided by EHT for M87* and Sgr. A* for their shadow diameter. We have observed that there are no lower bounds for $\beta$, and M87* is still a good candidate for constraining $\beta$ since there are particular values of $\beta$ intersecting the mean of the shadow diameter. For the upper bounds of beta, see Table \ref{Tab1}.
 
With regards to future work, it would be interesting to see whether this approach could be extended to the other stationary/static compact objects in plasma medium and whether there is an effect of magnetic plasma medium on the shadow. In the near future, we are hopeful that the EHT will give hints about the features of the black hole's shadow. If this observation occurs, we expect to see some of the experimental results of this theoretical work.

\begin{acknowledgements} A. {\"O}. and R. P. would like to acknowledge networking support by the COST Action CA18108 - Quantum gravity phenomenology in the multi-messenger approach (QG-MM).
\end{acknowledgements}

\bibliography{references}

\begin{thebibliography}{153}%
\makeatletter
\providecommand \@ifxundefined [1]{%
 \@ifx{#1\undefined}
}%
\providecommand \@ifnum [1]{%
 \ifnum #1\expandafter \@firstoftwo
 \else \expandafter \@secondoftwo
 \fi
}%
\providecommand \@ifx [1]{%
 \ifx #1\expandafter \@firstoftwo
 \else \expandafter \@secondoftwo
 \fi
}%
\providecommand \natexlab [1]{#1}%
\providecommand \enquote  [1]{``#1''}%
\providecommand \bibnamefont  [1]{#1}%
\providecommand \bibfnamefont [1]{#1}%
\providecommand \citenamefont [1]{#1}%
\providecommand \href@noop [0]{\@secondoftwo}%
\providecommand \href [0]{\begingroup \@sanitize@url \@href}%
\providecommand \@href[1]{\@@startlink{#1}\@@href}%
\providecommand \@@href[1]{\endgroup#1\@@endlink}%
\providecommand \@sanitize@url [0]{\catcode `\\12\catcode `\$12\catcode
  `\&12\catcode `\#12\catcode `\^12\catcode `\_12\catcode `\%12\relax}%
\providecommand \@@startlink[1]{}%
\providecommand \@@endlink[0]{}%
\providecommand \url  [0]{\begingroup\@sanitize@url \@url }%
\providecommand \@url [1]{\endgroup\@href {#1}{\urlprefix }}%
\providecommand \urlprefix  [0]{URL }%
\providecommand \Eprint [0]{\href }%
\providecommand \doibase [0]{http://dx.doi.org/}%
\providecommand \selectlanguage [0]{\@gobble}%
\providecommand \bibinfo  [0]{\@secondoftwo}%
\providecommand \bibfield  [0]{\@secondoftwo}%
\providecommand \translation [1]{[#1]}%
\providecommand \BibitemOpen [0]{}%
\providecommand \bibitemStop [0]{}%
\providecommand \bibitemNoStop [0]{.\EOS\space}%
\providecommand \EOS [0]{\spacefactor3000\relax}%
\providecommand \BibitemShut  [1]{\csname bibitem#1\endcsname}%
\let\auto@bib@innerbib\@empty
\bibitem [{\citenamefont {Abbott}\ \emph {et~al.}(2016)\citenamefont {Abbott}
  \emph {et~al.}}]{LIGOScientific:2016aoc}%
  \BibitemOpen
  \bibfield  {author} {\bibinfo {author} {\bibfnamefont {B.~P.}\ \bibnamefont
  {Abbott}} \emph {et~al.} (\bibinfo {collaboration} {LIGO Scientific,
  Virgo}),\ }\bibfield  {title} {\enquote {\bibinfo {title} {{Observation of
  Gravitational Waves from a Binary Black Hole Merger}},}\ }\href {\doibase
  10.1103/PhysRevLett.116.061102} {\bibfield  {journal} {\bibinfo  {journal}
  {Phys. Rev. Lett.}\ }\textbf {\bibinfo {volume} {116}},\ \bibinfo {pages}
  {061102} (\bibinfo {year} {2016})},\ \Eprint
  {http://arxiv.org/abs/1602.03837} {arXiv:1602.03837 [gr-qc]} \BibitemShut
  {NoStop}%
\bibitem [{\citenamefont {Akiyama}\ \emph
  {et~al.}(2019{\natexlab{a}})\citenamefont {Akiyama} \emph
  {et~al.}}]{EventHorizonTelescope:2019dse}%
  \BibitemOpen
  \bibfield  {author} {\bibinfo {author} {\bibfnamefont {Kazunori}\
  \bibnamefont {Akiyama}} \emph {et~al.} (\bibinfo {collaboration} {Event
  Horizon Telescope}),\ }\bibfield  {title} {\enquote {\bibinfo {title} {{First
  M87 Event Horizon Telescope Results. I. The Shadow of the Supermassive Black
  Hole}},}\ }\href {\doibase 10.3847/2041-8213/ab0ec7} {\bibfield  {journal}
  {\bibinfo  {journal} {Astrophys. J. Lett.}\ }\textbf {\bibinfo {volume}
  {875}},\ \bibinfo {pages} {L1} (\bibinfo {year} {2019}{\natexlab{a}})},\
  \Eprint {http://arxiv.org/abs/1906.11238} {arXiv:1906.11238 [astro-ph.GA]}
  \BibitemShut {NoStop}%
\bibitem [{\citenamefont {Akiyama}\ \emph
  {et~al.}(2019{\natexlab{b}})\citenamefont {Akiyama} \emph
  {et~al.}}]{EventHorizonTelescope:2019uob}%
  \BibitemOpen
  \bibfield  {author} {\bibinfo {author} {\bibfnamefont {Kazunori}\
  \bibnamefont {Akiyama}} \emph {et~al.} (\bibinfo {collaboration} {Event
  Horizon Telescope}),\ }\bibfield  {title} {\enquote {\bibinfo {title} {{First
  M87 Event Horizon Telescope Results. II. Array and Instrumentation}},}\
  }\href {\doibase 10.3847/2041-8213/ab0c96} {\bibfield  {journal} {\bibinfo
  {journal} {Astrophys. J. Lett.}\ }\textbf {\bibinfo {volume} {875}},\
  \bibinfo {pages} {L2} (\bibinfo {year} {2019}{\natexlab{b}})},\ \Eprint
  {http://arxiv.org/abs/1906.11239} {arXiv:1906.11239 [astro-ph.IM]}
  \BibitemShut {NoStop}%
\bibitem [{\citenamefont {Akiyama}\ \emph
  {et~al.}(2019{\natexlab{c}})\citenamefont {Akiyama} \emph
  {et~al.}}]{EventHorizonTelescope:2019jan}%
  \BibitemOpen
  \bibfield  {author} {\bibinfo {author} {\bibfnamefont {Kazunori}\
  \bibnamefont {Akiyama}} \emph {et~al.} (\bibinfo {collaboration} {Event
  Horizon Telescope}),\ }\bibfield  {title} {\enquote {\bibinfo {title} {{First
  M87 Event Horizon Telescope Results. III. Data Processing and
  Calibration}},}\ }\href {\doibase 10.3847/2041-8213/ab0c57} {\bibfield
  {journal} {\bibinfo  {journal} {Astrophys. J. Lett.}\ }\textbf {\bibinfo
  {volume} {875}},\ \bibinfo {pages} {L3} (\bibinfo {year}
  {2019}{\natexlab{c}})},\ \Eprint {http://arxiv.org/abs/1906.11240}
  {arXiv:1906.11240 [astro-ph.GA]} \BibitemShut {NoStop}%
\bibitem [{\citenamefont {Akiyama}\ \emph
  {et~al.}(2019{\natexlab{d}})\citenamefont {Akiyama} \emph
  {et~al.}}]{EventHorizonTelescope:2019ths}%
  \BibitemOpen
  \bibfield  {author} {\bibinfo {author} {\bibfnamefont {Kazunori}\
  \bibnamefont {Akiyama}} \emph {et~al.} (\bibinfo {collaboration} {Event
  Horizon Telescope}),\ }\bibfield  {title} {\enquote {\bibinfo {title} {{First
  M87 Event Horizon Telescope Results. IV. Imaging the Central Supermassive
  Black Hole}},}\ }\href {\doibase 10.3847/2041-8213/ab0e85} {\bibfield
  {journal} {\bibinfo  {journal} {Astrophys. J. Lett.}\ }\textbf {\bibinfo
  {volume} {875}},\ \bibinfo {pages} {L4} (\bibinfo {year}
  {2019}{\natexlab{d}})},\ \Eprint {http://arxiv.org/abs/1906.11241}
  {arXiv:1906.11241 [astro-ph.GA]} \BibitemShut {NoStop}%
\bibitem [{\citenamefont {Akiyama}\ \emph
  {et~al.}(2019{\natexlab{e}})\citenamefont {Akiyama} \emph
  {et~al.}}]{EventHorizonTelescope:2019pgp}%
  \BibitemOpen
  \bibfield  {author} {\bibinfo {author} {\bibfnamefont {Kazunori}\
  \bibnamefont {Akiyama}} \emph {et~al.} (\bibinfo {collaboration} {Event
  Horizon Telescope}),\ }\bibfield  {title} {\enquote {\bibinfo {title} {{First
  M87 Event Horizon Telescope Results. V. Physical Origin of the Asymmetric
  Ring}},}\ }\href {\doibase 10.3847/2041-8213/ab0f43} {\bibfield  {journal}
  {\bibinfo  {journal} {Astrophys. J. Lett.}\ }\textbf {\bibinfo {volume}
  {875}},\ \bibinfo {pages} {L5} (\bibinfo {year} {2019}{\natexlab{e}})},\
  \Eprint {http://arxiv.org/abs/1906.11242} {arXiv:1906.11242 [astro-ph.GA]}
  \BibitemShut {NoStop}%
\bibitem [{\citenamefont {Akiyama}\ \emph
  {et~al.}(2019{\natexlab{f}})\citenamefont {Akiyama} \emph
  {et~al.}}]{EventHorizonTelescope:2019ggy}%
  \BibitemOpen
  \bibfield  {author} {\bibinfo {author} {\bibfnamefont {Kazunori}\
  \bibnamefont {Akiyama}} \emph {et~al.} (\bibinfo {collaboration} {Event
  Horizon Telescope}),\ }\bibfield  {title} {\enquote {\bibinfo {title} {{First
  M87 Event Horizon Telescope Results. VI. The Shadow and Mass of the Central
  Black Hole}},}\ }\href {\doibase 10.3847/2041-8213/ab1141} {\bibfield
  {journal} {\bibinfo  {journal} {Astrophys. J. Lett.}\ }\textbf {\bibinfo
  {volume} {875}},\ \bibinfo {pages} {L6} (\bibinfo {year}
  {2019}{\natexlab{f}})},\ \Eprint {http://arxiv.org/abs/1906.11243}
  {arXiv:1906.11243 [astro-ph.GA]} \BibitemShut {NoStop}%
\bibitem [{\citenamefont {Akiyama}\ \emph
  {et~al.}(2021{\natexlab{a}})\citenamefont {Akiyama} \emph
  {et~al.}}]{EventHorizonTelescope:2021bee}%
  \BibitemOpen
  \bibfield  {author} {\bibinfo {author} {\bibfnamefont {Kazunori}\
  \bibnamefont {Akiyama}} \emph {et~al.} (\bibinfo {collaboration} {Event
  Horizon Telescope}),\ }\bibfield  {title} {\enquote {\bibinfo {title} {{First
  M87 Event Horizon Telescope Results. VII. Polarization of the Ring}},}\
  }\href {\doibase 10.3847/2041-8213/abe71d} {\bibfield  {journal} {\bibinfo
  {journal} {Astrophys. J. Lett.}\ }\textbf {\bibinfo {volume} {910}},\
  \bibinfo {pages} {L12} (\bibinfo {year} {2021}{\natexlab{a}})},\ \Eprint
  {http://arxiv.org/abs/2105.01169} {arXiv:2105.01169 [astro-ph.HE]}
  \BibitemShut {NoStop}%
\bibitem [{\citenamefont {Akiyama}\ \emph
  {et~al.}(2021{\natexlab{b}})\citenamefont {Akiyama} \emph
  {et~al.}}]{EventHorizonTelescope:2021srq}%
  \BibitemOpen
  \bibfield  {author} {\bibinfo {author} {\bibfnamefont {Kazunori}\
  \bibnamefont {Akiyama}} \emph {et~al.} (\bibinfo {collaboration} {Event
  Horizon Telescope}),\ }\bibfield  {title} {\enquote {\bibinfo {title} {{First
  M87 Event Horizon Telescope Results. VIII. Magnetic Field Structure near The
  Event Horizon}},}\ }\href {\doibase 10.3847/2041-8213/abe4de} {\bibfield
  {journal} {\bibinfo  {journal} {Astrophys. J. Lett.}\ }\textbf {\bibinfo
  {volume} {910}},\ \bibinfo {pages} {L13} (\bibinfo {year}
  {2021}{\natexlab{b}})},\ \Eprint {http://arxiv.org/abs/2105.01173}
  {arXiv:2105.01173 [astro-ph.HE]} \BibitemShut {NoStop}%
\bibitem [{\citenamefont {Kocherlakota}\ \emph {et~al.}(2021)\citenamefont
  {Kocherlakota} \emph {et~al.}}]{EventHorizonTelescope:2021dqv}%
  \BibitemOpen
  \bibfield  {author} {\bibinfo {author} {\bibfnamefont {Prashant}\
  \bibnamefont {Kocherlakota}} \emph {et~al.} (\bibinfo {collaboration} {Event
  Horizon Telescope}),\ }\bibfield  {title} {\enquote {\bibinfo {title}
  {{Constraints on black-hole charges with the 2017 EHT observations of
  M87*}},}\ }\href {\doibase 10.1103/PhysRevD.103.104047} {\bibfield  {journal}
  {\bibinfo  {journal} {Phys. Rev. D}\ }\textbf {\bibinfo {volume} {103}},\
  \bibinfo {pages} {104047} (\bibinfo {year} {2021})},\ \Eprint
  {http://arxiv.org/abs/2105.09343} {arXiv:2105.09343 [gr-qc]} \BibitemShut
  {NoStop}%
\bibitem [{\citenamefont {Akiyama}\ \emph {et~al.}(2022)\citenamefont {Akiyama}
  \emph {et~al.}}]{EventHorizonTelescope:2022xnr}%
  \BibitemOpen
  \bibfield  {author} {\bibinfo {author} {\bibfnamefont {Kazunori}\
  \bibnamefont {Akiyama}} \emph {et~al.} (\bibinfo {collaboration} {Event
  Horizon Telescope}),\ }\bibfield  {title} {\enquote {\bibinfo {title} {{First
  Sagittarius A* Event Horizon Telescope Results. I. The Shadow of the
  Supermassive Black Hole in the Center of the Milky Way}},}\ }\href {\doibase
  10.3847/2041-8213/ac6674} {\bibfield  {journal} {\bibinfo  {journal}
  {Astrophys. J. Lett.}\ }\textbf {\bibinfo {volume} {930}},\ \bibinfo {pages}
  {L12} (\bibinfo {year} {2022})}\BibitemShut {NoStop}%
\bibitem [{\citenamefont {Cunha}\ \emph {et~al.}(2018)\citenamefont {Cunha},
  \citenamefont {Herdeiro},\ and\ \citenamefont {Rodriguez}}]{Cunha:2018gql}%
  \BibitemOpen
  \bibfield  {author} {\bibinfo {author} {\bibfnamefont {Pedro V.~P.}\
  \bibnamefont {Cunha}}, \bibinfo {author} {\bibfnamefont {Carlos A.~R.}\
  \bibnamefont {Herdeiro}}, \ and\ \bibinfo {author} {\bibfnamefont {Maria~J.}\
  \bibnamefont {Rodriguez}},\ }\bibfield  {title} {\enquote {\bibinfo {title}
  {{Does the black hole shadow probe the event horizon geometry?}}}\ }\href
  {\doibase 10.1103/PhysRevD.97.084020} {\bibfield  {journal} {\bibinfo
  {journal} {Phys. Rev. D}\ }\textbf {\bibinfo {volume} {97}},\ \bibinfo
  {pages} {084020} (\bibinfo {year} {2018})},\ \Eprint
  {http://arxiv.org/abs/1802.02675} {arXiv:1802.02675 [gr-qc]} \BibitemShut
  {NoStop}%
\bibitem [{\citenamefont {Takahashi}(2004)}]{Takahashi:2004xh}%
  \BibitemOpen
  \bibfield  {author} {\bibinfo {author} {\bibfnamefont {Rohta}\ \bibnamefont
  {Takahashi}},\ }\bibfield  {title} {\enquote {\bibinfo {title} {{Shapes and
  positions of black hole shadows in accretion disks and spin parameters of
  black holes}},}\ }\href {\doibase 10.1086/422403} {\bibfield  {journal}
  {\bibinfo  {journal} {J. Korean Phys. Soc.}\ }\textbf {\bibinfo {volume}
  {45}},\ \bibinfo {pages} {S1808--S1812} (\bibinfo {year} {2004})},\ \Eprint
  {http://arxiv.org/abs/astro-ph/0405099} {arXiv:astro-ph/0405099} \BibitemShut
  {NoStop}%
\bibitem [{\citenamefont {V\"olkel}\ \emph {et~al.}(2021)\citenamefont
  {V\"olkel}, \citenamefont {Barausse}, \citenamefont {Franchini},\ and\
  \citenamefont {Broderick}}]{Volkel:2020xlc}%
  \BibitemOpen
  \bibfield  {author} {\bibinfo {author} {\bibfnamefont {Sebastian~H.}\
  \bibnamefont {V\"olkel}}, \bibinfo {author} {\bibfnamefont {Enrico}\
  \bibnamefont {Barausse}}, \bibinfo {author} {\bibfnamefont {Nicola}\
  \bibnamefont {Franchini}}, \ and\ \bibinfo {author} {\bibfnamefont
  {Avery~E.}\ \bibnamefont {Broderick}},\ }\bibfield  {title} {\enquote
  {\bibinfo {title} {{EHT tests of the strong-field regime of general
  relativity}},}\ }\href {\doibase 10.1088/1361-6382/ac27ed} {\bibfield
  {journal} {\bibinfo  {journal} {Class. Quant. Grav.}\ }\textbf {\bibinfo
  {volume} {38}},\ \bibinfo {pages} {21LT01} (\bibinfo {year} {2021})},\
  \Eprint {http://arxiv.org/abs/2011.06812} {arXiv:2011.06812 [gr-qc]}
  \BibitemShut {NoStop}%
\bibitem [{\citenamefont {Will}(2014)}]{Will:2014kxa}%
  \BibitemOpen
  \bibfield  {author} {\bibinfo {author} {\bibfnamefont {Clifford~M.}\
  \bibnamefont {Will}},\ }\bibfield  {title} {\enquote {\bibinfo {title} {{The
  Confrontation between General Relativity and Experiment}},}\ }\href {\doibase
  10.12942/lrr-2014-4} {\bibfield  {journal} {\bibinfo  {journal} {Living Rev.
  Rel.}\ }\textbf {\bibinfo {volume} {17}},\ \bibinfo {pages} {4} (\bibinfo
  {year} {2014})},\ \Eprint {http://arxiv.org/abs/1403.7377} {arXiv:1403.7377
  [gr-qc]} \BibitemShut {NoStop}%
\bibitem [{\citenamefont {Berti}\ \emph {et~al.}(2005)\citenamefont {Berti},
  \citenamefont {Buonanno},\ and\ \citenamefont {Will}}]{Berti:2004bd}%
  \BibitemOpen
  \bibfield  {author} {\bibinfo {author} {\bibfnamefont {Emanuele}\
  \bibnamefont {Berti}}, \bibinfo {author} {\bibfnamefont {Alessandra}\
  \bibnamefont {Buonanno}}, \ and\ \bibinfo {author} {\bibfnamefont
  {Clifford~M.}\ \bibnamefont {Will}},\ }\bibfield  {title} {\enquote {\bibinfo
  {title} {{Estimating spinning binary parameters and testing alternative
  theories of gravity with LISA}},}\ }\href {\doibase
  10.1103/PhysRevD.71.084025} {\bibfield  {journal} {\bibinfo  {journal} {Phys.
  Rev. D}\ }\textbf {\bibinfo {volume} {71}},\ \bibinfo {pages} {084025}
  (\bibinfo {year} {2005})},\ \Eprint {http://arxiv.org/abs/gr-qc/0411129}
  {arXiv:gr-qc/0411129} \BibitemShut {NoStop}%
\bibitem [{\citenamefont {Easson}(2004)}]{Easson:2004fq}%
  \BibitemOpen
  \bibfield  {author} {\bibinfo {author} {\bibfnamefont {Damien~A.}\
  \bibnamefont {Easson}},\ }\bibfield  {title} {\enquote {\bibinfo {title}
  {{Cosmic acceleration and modified gravitational models}},}\ }\href {\doibase
  10.1142/S0217751X04022578} {\bibfield  {journal} {\bibinfo  {journal} {Int.
  J. Mod. Phys. A}\ }\textbf {\bibinfo {volume} {19}},\ \bibinfo {pages}
  {5343--5350} (\bibinfo {year} {2004})},\ \Eprint
  {http://arxiv.org/abs/astro-ph/0411209} {arXiv:astro-ph/0411209} \BibitemShut
  {NoStop}%
\bibitem [{\citenamefont {Nojiri}\ and\ \citenamefont
  {Odintsov}(2006)}]{Nojiri:2006ri}%
  \BibitemOpen
  \bibfield  {author} {\bibinfo {author} {\bibfnamefont {Shin'ichi}\
  \bibnamefont {Nojiri}}\ and\ \bibinfo {author} {\bibfnamefont {Sergei~D.}\
  \bibnamefont {Odintsov}},\ }\bibfield  {title} {\enquote {\bibinfo {title}
  {{Introduction to modified gravity and gravitational alternative for dark
  energy}},}\ }\href {\doibase 10.1142/S0219887807001928} {\bibfield  {journal}
  {\bibinfo  {journal} {eConf}\ }\textbf {\bibinfo {volume} {C0602061}},\
  \bibinfo {pages} {06} (\bibinfo {year} {2006})},\ \Eprint
  {http://arxiv.org/abs/hep-th/0601213} {arXiv:hep-th/0601213} \BibitemShut
  {NoStop}%
\bibitem [{\citenamefont {Trodden}(2008)}]{Trodden:2006qk}%
  \BibitemOpen
  \bibfield  {author} {\bibinfo {author} {\bibfnamefont {Mark}\ \bibnamefont
  {Trodden}},\ }\bibfield  {title} {\enquote {\bibinfo {title} {{Cosmic
  Acceleration and Modified Gravity}},}\ }\href {\doibase
  10.1142/S0218271807011723} {\bibfield  {journal} {\bibinfo  {journal} {Int.
  J. Mod. Phys. D}\ }\textbf {\bibinfo {volume} {16}},\ \bibinfo {pages}
  {2065--2074} (\bibinfo {year} {2008})},\ \Eprint
  {http://arxiv.org/abs/astro-ph/0607510} {arXiv:astro-ph/0607510} \BibitemShut
  {NoStop}%
\bibitem [{\citenamefont {Garcia-Salcedo}\ and\ \citenamefont
  {Breton}(2000)}]{Garcia-Salcedo:2000ujn}%
  \BibitemOpen
  \bibfield  {author} {\bibinfo {author} {\bibfnamefont {Ricardo}\ \bibnamefont
  {Garcia-Salcedo}}\ and\ \bibinfo {author} {\bibfnamefont {Nora}\ \bibnamefont
  {Breton}},\ }\bibfield  {title} {\enquote {\bibinfo {title} {{Born-Infeld
  cosmologies}},}\ }\href {\doibase 10.1016/S0217-751X(00)00216-9} {\bibfield
  {journal} {\bibinfo  {journal} {Int. J. Mod. Phys. A}\ }\textbf {\bibinfo
  {volume} {15}},\ \bibinfo {pages} {4341--4354} (\bibinfo {year} {2000})},\
  \Eprint {http://arxiv.org/abs/gr-qc/0004017} {arXiv:gr-qc/0004017}
  \BibitemShut {NoStop}%
\bibitem [{\citenamefont {Camara}\ \emph {et~al.}(2004)\citenamefont {Camara},
  \citenamefont {de~Garcia~Maia}, \citenamefont {Carvalho},\ and\ \citenamefont
  {Lima}}]{Camara:2004ap}%
  \BibitemOpen
  \bibfield  {author} {\bibinfo {author} {\bibfnamefont {C.~S.}\ \bibnamefont
  {Camara}}, \bibinfo {author} {\bibfnamefont {M.~R.}\ \bibnamefont
  {de~Garcia~Maia}}, \bibinfo {author} {\bibfnamefont {J.~C.}\ \bibnamefont
  {Carvalho}}, \ and\ \bibinfo {author} {\bibfnamefont {Jose Ademir~Sales}\
  \bibnamefont {Lima}},\ }\bibfield  {title} {\enquote {\bibinfo {title}
  {{Nonsingular FRW cosmology and nonlinear electrodynamics}},}\ }\href
  {\doibase 10.1103/PhysRevD.69.123504} {\bibfield  {journal} {\bibinfo
  {journal} {Phys. Rev. D}\ }\textbf {\bibinfo {volume} {69}},\ \bibinfo
  {pages} {123504} (\bibinfo {year} {2004})},\ \Eprint
  {http://arxiv.org/abs/astro-ph/0402311} {arXiv:astro-ph/0402311} \BibitemShut
  {NoStop}%
\bibitem [{\citenamefont {Novello}\ \emph {et~al.}(2004)\citenamefont
  {Novello}, \citenamefont {Perez~Bergliaffa},\ and\ \citenamefont
  {Salim}}]{Novello:2003kh}%
  \BibitemOpen
  \bibfield  {author} {\bibinfo {author} {\bibfnamefont {M.}~\bibnamefont
  {Novello}}, \bibinfo {author} {\bibfnamefont {Santiago~Esteban}\ \bibnamefont
  {Perez~Bergliaffa}}, \ and\ \bibinfo {author} {\bibfnamefont
  {J.}~\bibnamefont {Salim}},\ }\bibfield  {title} {\enquote {\bibinfo {title}
  {{Non-linear electrodynamics and the acceleration of the universe}},}\ }\href
  {\doibase 10.1103/PhysRevD.69.127301} {\bibfield  {journal} {\bibinfo
  {journal} {Phys. Rev. D}\ }\textbf {\bibinfo {volume} {69}},\ \bibinfo
  {pages} {127301} (\bibinfo {year} {2004})},\ \Eprint
  {http://arxiv.org/abs/astro-ph/0312093} {arXiv:astro-ph/0312093} \BibitemShut
  {NoStop}%
\bibitem [{\citenamefont {Novello}\ \emph {et~al.}(2007)\citenamefont
  {Novello}, \citenamefont {Goulart}, \citenamefont {Salim},\ and\
  \citenamefont {Perez~Bergliaffa}}]{Novello:2006ng}%
  \BibitemOpen
  \bibfield  {author} {\bibinfo {author} {\bibfnamefont {M.}~\bibnamefont
  {Novello}}, \bibinfo {author} {\bibfnamefont {E.}~\bibnamefont {Goulart}},
  \bibinfo {author} {\bibfnamefont {J.~M.}\ \bibnamefont {Salim}}, \ and\
  \bibinfo {author} {\bibfnamefont {S.~E.}\ \bibnamefont {Perez~Bergliaffa}},\
  }\bibfield  {title} {\enquote {\bibinfo {title} {{Cosmological Effects of
  Nonlinear Electrodynamics}},}\ }\href {\doibase 10.1088/0264-9381/24/11/015}
  {\bibfield  {journal} {\bibinfo  {journal} {Class. Quant. Grav.}\ }\textbf
  {\bibinfo {volume} {24}},\ \bibinfo {pages} {3021--3036} (\bibinfo {year}
  {2007})},\ \Eprint {http://arxiv.org/abs/gr-qc/0610043} {arXiv:gr-qc/0610043}
  \BibitemShut {NoStop}%
\bibitem [{\citenamefont {Vollick}(2008)}]{Vollick:2008dx}%
  \BibitemOpen
  \bibfield  {author} {\bibinfo {author} {\bibfnamefont {Dan~N.}\ \bibnamefont
  {Vollick}},\ }\bibfield  {title} {\enquote {\bibinfo {title} {{Homogeneous
  and isotropic cosmologies with nonlinear electromagnetic radiation}},}\
  }\href {\doibase 10.1103/PhysRevD.78.063524} {\bibfield  {journal} {\bibinfo
  {journal} {Phys. Rev. D}\ }\textbf {\bibinfo {volume} {78}},\ \bibinfo
  {pages} {063524} (\bibinfo {year} {2008})},\ \Eprint
  {http://arxiv.org/abs/0807.0448} {arXiv:0807.0448 [gr-qc]} \BibitemShut
  {NoStop}%
\bibitem [{\citenamefont {Okyay}\ and\ \citenamefont
  {\"Ovg\"un}(2022)}]{Okyay:2021nnh}%
  \BibitemOpen
  \bibfield  {author} {\bibinfo {author} {\bibfnamefont {Mert}\ \bibnamefont
  {Okyay}}\ and\ \bibinfo {author} {\bibfnamefont {Ali}\ \bibnamefont
  {\"Ovg\"un}},\ }\bibfield  {title} {\enquote {\bibinfo {title} {{Nonlinear
  electrodynamics effects on the black hole shadow, deflection angle,
  quasinormal modes and greybody factors}},}\ }\href {\doibase
  10.1088/1475-7516/2022/01/009} {\bibfield  {journal} {\bibinfo  {journal}
  {JCAP}\ }\textbf {\bibinfo {volume} {01}},\ \bibinfo {pages} {009} (\bibinfo
  {year} {2022})},\ \Eprint {http://arxiv.org/abs/2108.07766} {arXiv:2108.07766
  [gr-qc]} \BibitemShut {NoStop}%
\bibitem [{\citenamefont {Allahyari}\ \emph {et~al.}(2020)\citenamefont
  {Allahyari}, \citenamefont {Khodadi}, \citenamefont {Vagnozzi},\ and\
  \citenamefont {Mota}}]{Allahyari:2019jqz}%
  \BibitemOpen
  \bibfield  {author} {\bibinfo {author} {\bibfnamefont {Alireza}\ \bibnamefont
  {Allahyari}}, \bibinfo {author} {\bibfnamefont {Mohsen}\ \bibnamefont
  {Khodadi}}, \bibinfo {author} {\bibfnamefont {Sunny}\ \bibnamefont
  {Vagnozzi}}, \ and\ \bibinfo {author} {\bibfnamefont {David~F.}\ \bibnamefont
  {Mota}},\ }\bibfield  {title} {\enquote {\bibinfo {title} {{Magnetically
  charged black holes from non-linear electrodynamics and the Event Horizon
  Telescope}},}\ }\href {\doibase 10.1088/1475-7516/2020/02/003} {\bibfield
  {journal} {\bibinfo  {journal} {JCAP}\ }\textbf {\bibinfo {volume} {02}},\
  \bibinfo {pages} {003} (\bibinfo {year} {2020})},\ \Eprint
  {http://arxiv.org/abs/1912.08231} {arXiv:1912.08231 [gr-qc]} \BibitemShut
  {NoStop}%
\bibitem [{\citenamefont {Chen}\ \emph {et~al.}(2022)\citenamefont {Chen},
  \citenamefont {Roy}, \citenamefont {Vagnozzi},\ and\ \citenamefont
  {Visinelli}}]{Chen:2022nbb}%
  \BibitemOpen
  \bibfield  {author} {\bibinfo {author} {\bibfnamefont {Yifan}\ \bibnamefont
  {Chen}}, \bibinfo {author} {\bibfnamefont {Rittick}\ \bibnamefont {Roy}},
  \bibinfo {author} {\bibfnamefont {Sunny}\ \bibnamefont {Vagnozzi}}, \ and\
  \bibinfo {author} {\bibfnamefont {Luca}\ \bibnamefont {Visinelli}},\
  }\bibfield  {title} {\enquote {\bibinfo {title} {{Superradiant evolution of
  the shadow and photon ring of Sgr A$^\star$}},}\ }\href@noop {} {\  (\bibinfo
  {year} {2022})},\ \Eprint {http://arxiv.org/abs/2205.06238} {arXiv:2205.06238
  [astro-ph.HE]} \BibitemShut {NoStop}%
\bibitem [{\citenamefont {Roy}\ \emph {et~al.}(2022)\citenamefont {Roy},
  \citenamefont {Vagnozzi},\ and\ \citenamefont {Visinelli}}]{Roy:2021uye}%
  \BibitemOpen
  \bibfield  {author} {\bibinfo {author} {\bibfnamefont {Rittick}\ \bibnamefont
  {Roy}}, \bibinfo {author} {\bibfnamefont {Sunny}\ \bibnamefont {Vagnozzi}}, \
  and\ \bibinfo {author} {\bibfnamefont {Luca}\ \bibnamefont {Visinelli}},\
  }\bibfield  {title} {\enquote {\bibinfo {title} {{Superradiance evolution of
  black hole shadows revisited}},}\ }\href {\doibase
  10.1103/PhysRevD.105.083002} {\bibfield  {journal} {\bibinfo  {journal}
  {Phys. Rev. D}\ }\textbf {\bibinfo {volume} {105}},\ \bibinfo {pages}
  {083002} (\bibinfo {year} {2022})},\ \Eprint
  {http://arxiv.org/abs/2112.06932} {arXiv:2112.06932 [astro-ph.HE]}
  \BibitemShut {NoStop}%
\bibitem [{\citenamefont {Khodadi}\ \emph {et~al.}(2020)\citenamefont
  {Khodadi}, \citenamefont {Allahyari}, \citenamefont {Vagnozzi},\ and\
  \citenamefont {Mota}}]{Khodadi:2020jij}%
  \BibitemOpen
  \bibfield  {author} {\bibinfo {author} {\bibfnamefont {Mohsen}\ \bibnamefont
  {Khodadi}}, \bibinfo {author} {\bibfnamefont {Alireza}\ \bibnamefont
  {Allahyari}}, \bibinfo {author} {\bibfnamefont {Sunny}\ \bibnamefont
  {Vagnozzi}}, \ and\ \bibinfo {author} {\bibfnamefont {David~F.}\ \bibnamefont
  {Mota}},\ }\bibfield  {title} {\enquote {\bibinfo {title} {{Black holes with
  scalar hair in light of the Event Horizon Telescope}},}\ }\href {\doibase
  10.1088/1475-7516/2020/09/026} {\bibfield  {journal} {\bibinfo  {journal}
  {JCAP}\ }\textbf {\bibinfo {volume} {09}},\ \bibinfo {pages} {026} (\bibinfo
  {year} {2020})},\ \Eprint {http://arxiv.org/abs/2005.05992} {arXiv:2005.05992
  [gr-qc]} \BibitemShut {NoStop}%
\bibitem [{\citenamefont {Vagnozzi}\ \emph {et~al.}(2022)\citenamefont
  {Vagnozzi}, \citenamefont {Roy}, \citenamefont {Tsai},\ and\ \citenamefont
  {Visinelli}}]{Vagnozzi:2022moj}%
  \BibitemOpen
  \bibfield  {author} {\bibinfo {author} {\bibfnamefont {Sunny}\ \bibnamefont
  {Vagnozzi}}, \bibinfo {author} {\bibfnamefont {Rittick}\ \bibnamefont {Roy}},
  \bibinfo {author} {\bibfnamefont {Yu-Dai}\ \bibnamefont {Tsai}}, \ and\
  \bibinfo {author} {\bibfnamefont {Luca}\ \bibnamefont {Visinelli}},\
  }\bibfield  {title} {\enquote {\bibinfo {title} {{Horizon-scale tests of
  gravity theories and fundamental physics from the Event Horizon Telescope
  image of Sagittarius A$^*$}},}\ }\href@noop {} {\  (\bibinfo {year}
  {2022})},\ \Eprint {http://arxiv.org/abs/2205.07787} {arXiv:2205.07787
  [gr-qc]} \BibitemShut {NoStop}%
\bibitem [{\citenamefont {Wang}\ \emph {et~al.}(2019)\citenamefont {Wang},
  \citenamefont {Xu},\ and\ \citenamefont {Wei}}]{Wang:2018prk}%
  \BibitemOpen
  \bibfield  {author} {\bibinfo {author} {\bibfnamefont {Hui-Min}\ \bibnamefont
  {Wang}}, \bibinfo {author} {\bibfnamefont {Yu-Meng}\ \bibnamefont {Xu}}, \
  and\ \bibinfo {author} {\bibfnamefont {Shao-Wen}\ \bibnamefont {Wei}},\
  }\bibfield  {title} {\enquote {\bibinfo {title} {{Shadows of Kerr-like black
  holes in a modified gravity theory}},}\ }\href@noop {} {\bibfield  {journal}
  {\bibinfo  {journal} {JCAP}\ }\textbf {\bibinfo {volume} {03}},\ \bibinfo
  {pages} {046} (\bibinfo {year} {2019})},\ \Eprint
  {http://arxiv.org/abs/1810.12767} {arXiv:1810.12767 [gr-qc]} \BibitemShut
  {NoStop}%
\bibitem [{\citenamefont {Cunha}\ \emph {et~al.}(2020)\citenamefont {Cunha},
  \citenamefont {Eir\'o}, \citenamefont {Herdeiro},\ and\ \citenamefont
  {Lemos}}]{Cunha:2019hzj}%
  \BibitemOpen
  \bibfield  {author} {\bibinfo {author} {\bibfnamefont {Pedro V.~P.}\
  \bibnamefont {Cunha}}, \bibinfo {author} {\bibfnamefont {Nelson~A.}\
  \bibnamefont {Eir\'o}}, \bibinfo {author} {\bibfnamefont {Carlos A.~R.}\
  \bibnamefont {Herdeiro}}, \ and\ \bibinfo {author} {\bibfnamefont {Jos\'e
  P.~S.}\ \bibnamefont {Lemos}},\ }\bibfield  {title} {\enquote {\bibinfo
  {title} {{Lensing and shadow of a black hole surrounded by a heavy accretion
  disk}},}\ }\href@noop {} {\bibfield  {journal} {\bibinfo  {journal} {JCAP}\
  }\textbf {\bibinfo {volume} {03}},\ \bibinfo {pages} {035} (\bibinfo {year}
  {2020})},\ \Eprint {http://arxiv.org/abs/1912.08833} {arXiv:1912.08833
  [gr-qc]} \BibitemShut {NoStop}%
\bibitem [{\citenamefont {Pantig}\ and\ \citenamefont
  {Rodulfo}(2020{\natexlab{a}})}]{Pantig:2020odu}%
  \BibitemOpen
  \bibfield  {author} {\bibinfo {author} {\bibfnamefont {Reggie~C.}\
  \bibnamefont {Pantig}}\ and\ \bibinfo {author} {\bibfnamefont {Emmanuel~T.}\
  \bibnamefont {Rodulfo}},\ }\bibfield  {title} {\enquote {\bibinfo {title}
  {{Weak deflection angle of a dirty black hole}},}\ }\href {\doibase
  10.1016/j.cjph.2020.06.015} {\bibfield  {journal} {\bibinfo  {journal} {Chin.
  J. Phys.}\ }\textbf {\bibinfo {volume} {66}},\ \bibinfo {pages} {691--702}
  (\bibinfo {year} {2020}{\natexlab{a}})},\ \Eprint
  {http://arxiv.org/abs/2003.00764} {arXiv:2003.00764 [gr-qc]} \BibitemShut
  {NoStop}%
\bibitem [{\citenamefont {Pantig}\ and\ \citenamefont
  {Rodulfo}(2020{\natexlab{b}})}]{Pantig:2020uhp}%
  \BibitemOpen
  \bibfield  {author} {\bibinfo {author} {\bibfnamefont {Reggie~C.}\
  \bibnamefont {Pantig}}\ and\ \bibinfo {author} {\bibfnamefont {Emmanuel~T.}\
  \bibnamefont {Rodulfo}},\ }\bibfield  {title} {\enquote {\bibinfo {title}
  {{Rotating dirty black hole and its shadow}},}\ }\href {\doibase
  10.1016/j.cjph.2020.08.001} {\bibfield  {journal} {\bibinfo  {journal} {Chin.
  J. Phys.}\ }\textbf {\bibinfo {volume} {68}},\ \bibinfo {pages} {236--257}
  (\bibinfo {year} {2020}{\natexlab{b}})},\ \Eprint
  {http://arxiv.org/abs/2003.06829} {arXiv:2003.06829 [gr-qc]} \BibitemShut
  {NoStop}%
\bibitem [{\citenamefont {Pantig}\ \emph {et~al.}(2021)\citenamefont {Pantig},
  \citenamefont {Yu}, \citenamefont {Rodulfo},\ and\ \citenamefont
  {\"Ovg\"un}}]{Pantig:2021zqe}%
  \BibitemOpen
  \bibfield  {author} {\bibinfo {author} {\bibfnamefont {Reggie~C.}\
  \bibnamefont {Pantig}}, \bibinfo {author} {\bibfnamefont {Paul~K.}\
  \bibnamefont {Yu}}, \bibinfo {author} {\bibfnamefont {Emmanuel~T.}\
  \bibnamefont {Rodulfo}}, \ and\ \bibinfo {author} {\bibfnamefont {Ali}\
  \bibnamefont {\"Ovg\"un}},\ }\bibfield  {title} {\enquote {\bibinfo {title}
  {{Shadow and weak deflection angle of extended uncertainty principle black
  hole surrounded with dark matter}},}\ }\href@noop {} {\bibfield  {journal}
  {\bibinfo  {journal} {Annals of Physics}\ }\textbf {\bibinfo {volume}
  {436}},\ \bibinfo {pages} {168722} (\bibinfo {year} {2021})},\ \Eprint
  {http://arxiv.org/abs/2104.04304} {arXiv:2104.04304 [gr-qc]} \BibitemShut
  {NoStop}%
\bibitem [{\citenamefont {Pantig}\ and\ \citenamefont
  {\"Ovg\"un}(2022{\natexlab{a}})}]{Pantig:2022toh}%
  \BibitemOpen
  \bibfield  {author} {\bibinfo {author} {\bibfnamefont {Reggie~C.}\
  \bibnamefont {Pantig}}\ and\ \bibinfo {author} {\bibfnamefont {Ali}\
  \bibnamefont {\"Ovg\"un}},\ }\bibfield  {title} {\enquote {\bibinfo {title}
  {{Dark matter effect on the weak deflection angle by black holes at the
  center of Milky Way and M87 galaxies}},}\ }\href {\doibase
  10.1140/epjc/s10052-022-10319-8} {\bibfield  {journal} {\bibinfo  {journal}
  {Eur. Phys. J. C}\ }\textbf {\bibinfo {volume} {82}},\ \bibinfo {pages} {391}
  (\bibinfo {year} {2022}{\natexlab{a}})},\ \Eprint
  {http://arxiv.org/abs/2201.03365} {arXiv:2201.03365 [gr-qc]} \BibitemShut
  {NoStop}%
\bibitem [{\citenamefont {Pantig}\ and\ \citenamefont
  {\"Ovg\"un}(2022{\natexlab{b}})}]{Pantig:2022whj}%
  \BibitemOpen
  \bibfield  {author} {\bibinfo {author} {\bibfnamefont {Reggie~C.}\
  \bibnamefont {Pantig}}\ and\ \bibinfo {author} {\bibfnamefont {Ali}\
  \bibnamefont {\"Ovg\"un}},\ }\bibfield  {title} {\enquote {\bibinfo {title}
  {{Dehnen halo effect on a black hole in an ultra-faint dwarf galaxy}},}\
  }\href {\doibase 10.1088/1475-7516/2022/08/056} {\bibfield  {journal}
  {\bibinfo  {journal} {JCAP}\ }\textbf {\bibinfo {volume} {08}},\ \bibinfo
  {pages} {056} (\bibinfo {year} {2022}{\natexlab{b}})},\ \Eprint
  {http://arxiv.org/abs/2202.07404} {arXiv:2202.07404 [astro-ph.GA]}
  \BibitemShut {NoStop}%
\bibitem [{\citenamefont {Pantig}\ and\ \citenamefont
  {\"Ovg\"un}(2022{\natexlab{c}})}]{Pantig:2022sjb}%
  \BibitemOpen
  \bibfield  {author} {\bibinfo {author} {\bibfnamefont {Reggie~C.}\
  \bibnamefont {Pantig}}\ and\ \bibinfo {author} {\bibfnamefont {Ali}\
  \bibnamefont {\"Ovg\"un}},\ }\bibfield  {title} {\enquote {\bibinfo {title}
  {{Black hole in quantum wave dark matter}},}\ }\href {\doibase
  10.1002/prop.202200164} {\bibfield  {journal} {\bibinfo  {journal} {Fortsch.
  Phys.}\ }\textbf {\bibinfo {volume} {2022}},\ \bibinfo {pages} {2200164}
  (\bibinfo {year} {2022}{\natexlab{c}})},\ \Eprint
  {http://arxiv.org/abs/2210.00523} {arXiv:2210.00523 [gr-qc]} \BibitemShut
  {NoStop}%
\bibitem [{\citenamefont {Pantig}\ \emph {et~al.}(2022)\citenamefont {Pantig},
  \citenamefont {Mastrototaro}, \citenamefont {Lambiase},\ and\ \citenamefont
  {\"Ovg\"un}}]{Pantig:2022gih}%
  \BibitemOpen
  \bibfield  {author} {\bibinfo {author} {\bibfnamefont {Reggie~C.}\
  \bibnamefont {Pantig}}, \bibinfo {author} {\bibfnamefont {Leonardo}\
  \bibnamefont {Mastrototaro}}, \bibinfo {author} {\bibfnamefont {Gaetano}\
  \bibnamefont {Lambiase}}, \ and\ \bibinfo {author} {\bibfnamefont {Ali}\
  \bibnamefont {\"Ovg\"un}},\ }\bibfield  {title} {\enquote {\bibinfo {title}
  {{Shadow, lensing, quasinormal modes, greybody bounds and neutrino
  propagation by dyonic ModMax black holes}},}\ }\href {\doibase
  10.1140/epjc/s10052-022-11125-y} {\bibfield  {journal} {\bibinfo  {journal}
  {Eur. Phys. J. C}\ }\textbf {\bibinfo {volume} {82}},\ \bibinfo {pages}
  {1155} (\bibinfo {year} {2022})},\ \Eprint {http://arxiv.org/abs/2208.06664}
  {arXiv:2208.06664 [gr-qc]} \BibitemShut {NoStop}%
\bibitem [{\citenamefont {Pantig}\ and\ \citenamefont
  {\"Ovg\"un}(2023)}]{Pantig:2022ely}%
  \BibitemOpen
  \bibfield  {author} {\bibinfo {author} {\bibfnamefont {Reggie~C.}\
  \bibnamefont {Pantig}}\ and\ \bibinfo {author} {\bibfnamefont {Ali}\
  \bibnamefont {\"Ovg\"un}},\ }\bibfield  {title} {\enquote {\bibinfo {title}
  {{Testing dynamical torsion effects on the charged black
  hole\textquoteright{}s shadow, deflection angle and greybody with M87* and
  Sgr. A* from EHT}},}\ }\href {\doibase 10.1016/j.aop.2022.169197} {\bibfield
  {journal} {\bibinfo  {journal} {Annals Phys.}\ }\textbf {\bibinfo {volume}
  {448}},\ \bibinfo {pages} {169197} (\bibinfo {year} {2023})},\ \Eprint
  {http://arxiv.org/abs/2206.02161} {arXiv:2206.02161 [gr-qc]} \BibitemShut
  {NoStop}%
\bibitem [{\citenamefont {Lobos}\ and\ \citenamefont
  {Pantig}(2022)}]{physics4040084}%
  \BibitemOpen
  \bibfield  {author} {\bibinfo {author} {\bibfnamefont {Nikko John Leo~S.}\
  \bibnamefont {Lobos}}\ and\ \bibinfo {author} {\bibfnamefont {Reggie~C.}\
  \bibnamefont {Pantig}},\ }\bibfield  {title} {\enquote {\bibinfo {title}
  {Generalized extended uncertainty principle black holes: Shadow and lensing
  in the macro- and microscopic realms},}\ }\href {\doibase
  10.3390/physics4040084} {\bibfield  {journal} {\bibinfo  {journal} {Physics}\
  }\textbf {\bibinfo {volume} {4}},\ \bibinfo {pages} {1318--1330} (\bibinfo
  {year} {2022})}\BibitemShut {NoStop}%
\bibitem [{\citenamefont {Zuluaga}\ and\ \citenamefont
  {S\'anchez}(2021)}]{Zuluaga:2021vjc}%
  \BibitemOpen
  \bibfield  {author} {\bibinfo {author} {\bibfnamefont {Fabi\'an~H.}\
  \bibnamefont {Zuluaga}}\ and\ \bibinfo {author} {\bibfnamefont {Luis~A.}\
  \bibnamefont {S\'anchez}},\ }\bibfield  {title} {\enquote {\bibinfo {title}
  {{Accretion disk around a Schwarzschild black hole in asymptotic safety}},}\
  }\href {\doibase 10.1140/epjc/s10052-021-09644-1} {\bibfield  {journal}
  {\bibinfo  {journal} {Eur. Phys. J. C}\ }\textbf {\bibinfo {volume} {81}},\
  \bibinfo {pages} {840} (\bibinfo {year} {2021})},\ \Eprint
  {http://arxiv.org/abs/2106.03140} {arXiv:2106.03140 [gr-qc]} \BibitemShut
  {NoStop}%
\bibitem [{\citenamefont {Li}\ and\ \citenamefont
  {He}(2021{\natexlab{a}})}]{Li:2021ypw}%
  \BibitemOpen
  \bibfield  {author} {\bibinfo {author} {\bibfnamefont {Guo-Ping}\
  \bibnamefont {Li}}\ and\ \bibinfo {author} {\bibfnamefont {Ke-Jian}\
  \bibnamefont {He}},\ }\bibfield  {title} {\enquote {\bibinfo {title}
  {{Observational appearances of a f(R) global monopole black hole illuminated
  by various accretions}},}\ }\href {\doibase 10.1140/epjc/s10052-021-09817-y}
  {\bibfield  {journal} {\bibinfo  {journal} {Eur. Phys. J. C}\ }\textbf
  {\bibinfo {volume} {81}},\ \bibinfo {pages} {1018} (\bibinfo {year}
  {2021}{\natexlab{a}})}\BibitemShut {NoStop}%
\bibitem [{\citenamefont {Rahaman}\ \emph {et~al.}(2021)\citenamefont
  {Rahaman}, \citenamefont {Manna}, \citenamefont {Shaikh}, \citenamefont
  {Aktar}, \citenamefont {Mondal},\ and\ \citenamefont
  {Samanta}}]{Rahaman:2021kge}%
  \BibitemOpen
  \bibfield  {author} {\bibinfo {author} {\bibfnamefont {Farook}\ \bibnamefont
  {Rahaman}}, \bibinfo {author} {\bibfnamefont {Tuhina}\ \bibnamefont {Manna}},
  \bibinfo {author} {\bibfnamefont {Rajibul}\ \bibnamefont {Shaikh}}, \bibinfo
  {author} {\bibfnamefont {Somi}\ \bibnamefont {Aktar}}, \bibinfo {author}
  {\bibfnamefont {Monimala}\ \bibnamefont {Mondal}}, \ and\ \bibinfo {author}
  {\bibfnamefont {Bidisha}\ \bibnamefont {Samanta}},\ }\bibfield  {title}
  {\enquote {\bibinfo {title} {{Thin accretion disks around traversable
  wormholes}},}\ }\href {\doibase 10.1016/j.nuclphysb.2021.115548} {\bibfield
  {journal} {\bibinfo  {journal} {Nucl. Phys. B}\ }\textbf {\bibinfo {volume}
  {972}},\ \bibinfo {pages} {115548} (\bibinfo {year} {2021})},\ \Eprint
  {http://arxiv.org/abs/2110.09820} {arXiv:2110.09820 [gr-qc]} \BibitemShut
  {NoStop}%
\bibitem [{\citenamefont {Stashko}\ \emph {et~al.}(2021)\citenamefont
  {Stashko}, \citenamefont {Zhdanov},\ and\ \citenamefont
  {Alexandrov}}]{Stashko:2021lad}%
  \BibitemOpen
  \bibfield  {author} {\bibinfo {author} {\bibfnamefont {O.~S.}\ \bibnamefont
  {Stashko}}, \bibinfo {author} {\bibfnamefont {V.~I.}\ \bibnamefont
  {Zhdanov}}, \ and\ \bibinfo {author} {\bibfnamefont {A.~N.}\ \bibnamefont
  {Alexandrov}},\ }\bibfield  {title} {\enquote {\bibinfo {title} {{Thin
  accretion discs around spherically symmetric configurations with nonlinear
  scalar fields}},}\ }\href {\doibase 10.1103/PhysRevD.104.104055} {\bibfield
  {journal} {\bibinfo  {journal} {Phys. Rev. D}\ }\textbf {\bibinfo {volume}
  {104}},\ \bibinfo {pages} {104055} (\bibinfo {year} {2021})},\ \Eprint
  {http://arxiv.org/abs/2107.05111} {arXiv:2107.05111 [gr-qc]} \BibitemShut
  {NoStop}%
\bibitem [{\citenamefont {Liu}\ \emph {et~al.}(2022)\citenamefont {Liu},
  \citenamefont {Yang}, \citenamefont {Wu},\ and\ \citenamefont
  {Zhu}}]{Liu:2021yev}%
  \BibitemOpen
  \bibfield  {author} {\bibinfo {author} {\bibfnamefont {Cheng}\ \bibnamefont
  {Liu}}, \bibinfo {author} {\bibfnamefont {Sen}\ \bibnamefont {Yang}},
  \bibinfo {author} {\bibfnamefont {Qiang}\ \bibnamefont {Wu}}, \ and\ \bibinfo
  {author} {\bibfnamefont {Tao}\ \bibnamefont {Zhu}},\ }\bibfield  {title}
  {\enquote {\bibinfo {title} {{Thin accretion disk onto slowly rotating black
  holes in Einstein-\AE{}ther theory}},}\ }\href {\doibase
  10.1088/1475-7516/2022/02/034} {\bibfield  {journal} {\bibinfo  {journal}
  {JCAP}\ }\textbf {\bibinfo {volume} {02}},\ \bibinfo {pages} {034} (\bibinfo
  {year} {2022})},\ \Eprint {http://arxiv.org/abs/2107.04811} {arXiv:2107.04811
  [gr-qc]} \BibitemShut {NoStop}%
\bibitem [{\citenamefont {Gyulchev}\ \emph {et~al.}(2021)\citenamefont
  {Gyulchev}, \citenamefont {Nedkova}, \citenamefont {Vetsov},\ and\
  \citenamefont {Yazadjiev}}]{Gyulchev:2021dvt}%
  \BibitemOpen
  \bibfield  {author} {\bibinfo {author} {\bibfnamefont {Galin}\ \bibnamefont
  {Gyulchev}}, \bibinfo {author} {\bibfnamefont {Petya}\ \bibnamefont
  {Nedkova}}, \bibinfo {author} {\bibfnamefont {Tsvetan}\ \bibnamefont
  {Vetsov}}, \ and\ \bibinfo {author} {\bibfnamefont {Stoytcho}\ \bibnamefont
  {Yazadjiev}},\ }\bibfield  {title} {\enquote {\bibinfo {title} {{Image of the
  thin accretion disk around compact objects in the
  Einstein\textendash{}Gauss\textendash{}Bonnet gravity}},}\ }\href {\doibase
  10.1140/epjc/s10052-021-09624-5} {\bibfield  {journal} {\bibinfo  {journal}
  {Eur. Phys. J. C}\ }\textbf {\bibinfo {volume} {81}},\ \bibinfo {pages} {885}
  (\bibinfo {year} {2021})},\ \Eprint {http://arxiv.org/abs/2106.14697}
  {arXiv:2106.14697 [gr-qc]} \BibitemShut {NoStop}%
\bibitem [{\citenamefont {Guerrero}\ \emph {et~al.}(2021)\citenamefont
  {Guerrero}, \citenamefont {Olmo}, \citenamefont {Rubiera-Garcia},\ and\
  \citenamefont {G\'omez}}]{Guerrero:2021ues}%
  \BibitemOpen
  \bibfield  {author} {\bibinfo {author} {\bibfnamefont {Merce}\ \bibnamefont
  {Guerrero}}, \bibinfo {author} {\bibfnamefont {Gonzalo~J.}\ \bibnamefont
  {Olmo}}, \bibinfo {author} {\bibfnamefont {Diego}\ \bibnamefont
  {Rubiera-Garcia}}, \ and\ \bibinfo {author} {\bibfnamefont {Diego
  S\'aez-Chill\'on}\ \bibnamefont {G\'omez}},\ }\bibfield  {title} {\enquote
  {\bibinfo {title} {{Shadows and optical appearance of black bounces
  illuminated by a thin accretion disk}},}\ }\href {\doibase
  10.1088/1475-7516/2021/08/036} {\bibfield  {journal} {\bibinfo  {journal}
  {JCAP}\ }\textbf {\bibinfo {volume} {08}},\ \bibinfo {pages} {036} (\bibinfo
  {year} {2021})},\ \Eprint {http://arxiv.org/abs/2105.15073} {arXiv:2105.15073
  [gr-qc]} \BibitemShut {NoStop}%
\bibitem [{\citenamefont {Gan}\ \emph {et~al.}(2021)\citenamefont {Gan},
  \citenamefont {Wang}, \citenamefont {Wu},\ and\ \citenamefont
  {Yang}}]{Gan:2021xdl}%
  \BibitemOpen
  \bibfield  {author} {\bibinfo {author} {\bibfnamefont {Qingyu}\ \bibnamefont
  {Gan}}, \bibinfo {author} {\bibfnamefont {Peng}\ \bibnamefont {Wang}},
  \bibinfo {author} {\bibfnamefont {Houwen}\ \bibnamefont {Wu}}, \ and\
  \bibinfo {author} {\bibfnamefont {Haitang}\ \bibnamefont {Yang}},\ }\bibfield
   {title} {\enquote {\bibinfo {title} {{Photon ring and observational
  appearance of a hairy black hole}},}\ }\href {\doibase
  10.1103/PhysRevD.104.044049} {\bibfield  {journal} {\bibinfo  {journal}
  {Phys. Rev. D}\ }\textbf {\bibinfo {volume} {104}},\ \bibinfo {pages}
  {044049} (\bibinfo {year} {2021})},\ \Eprint
  {http://arxiv.org/abs/2105.11770} {arXiv:2105.11770 [gr-qc]} \BibitemShut
  {NoStop}%
\bibitem [{\citenamefont {Heydari-Fard}\ and\ \citenamefont
  {Sepangi}(2021)}]{Heydari-Fard:2020iiu}%
  \BibitemOpen
  \bibfield  {author} {\bibinfo {author} {\bibfnamefont {Mohaddese}\
  \bibnamefont {Heydari-Fard}}\ and\ \bibinfo {author} {\bibfnamefont
  {Hamid~Reza}\ \bibnamefont {Sepangi}},\ }\bibfield  {title} {\enquote
  {\bibinfo {title} {{Thin accretion disk signatures of scalarized black holes
  in Einstein-scalar-Gauss-Bonnet gravity}},}\ }\href {\doibase
  10.1016/j.physletb.2021.136276} {\bibfield  {journal} {\bibinfo  {journal}
  {Phys. Lett. B}\ }\textbf {\bibinfo {volume} {816}},\ \bibinfo {pages}
  {136276} (\bibinfo {year} {2021})},\ \Eprint
  {http://arxiv.org/abs/2009.13748} {arXiv:2009.13748 [gr-qc]} \BibitemShut
  {NoStop}%
\bibitem [{\citenamefont {Heydari-Fard}\ \emph {et~al.}(2021)\citenamefont
  {Heydari-Fard}, \citenamefont {Heydari-Fard},\ and\ \citenamefont
  {Sepangi}}]{Heydari-Fard:2021ljh}%
  \BibitemOpen
  \bibfield  {author} {\bibinfo {author} {\bibfnamefont {Mohaddese}\
  \bibnamefont {Heydari-Fard}}, \bibinfo {author} {\bibfnamefont {Malihe}\
  \bibnamefont {Heydari-Fard}}, \ and\ \bibinfo {author} {\bibfnamefont
  {Hamid~Reza}\ \bibnamefont {Sepangi}},\ }\bibfield  {title} {\enquote
  {\bibinfo {title} {{Thin accretion disks around rotating black holes in 4$D$
  Einstein\textendash{}Gauss\textendash{}Bonnet gravity}},}\ }\href {\doibase
  10.1140/epjc/s10052-021-09266-7} {\bibfield  {journal} {\bibinfo  {journal}
  {Eur. Phys. J. C}\ }\textbf {\bibinfo {volume} {81}},\ \bibinfo {pages} {473}
  (\bibinfo {year} {2021})},\ \Eprint {http://arxiv.org/abs/2105.09192}
  {arXiv:2105.09192 [gr-qc]} \BibitemShut {NoStop}%
\bibitem [{\citenamefont {Kazempour}\ \emph {et~al.}(2022)\citenamefont
  {Kazempour}, \citenamefont {Zou},\ and\ \citenamefont
  {Akbarieh}}]{Kazempour:2022asl}%
  \BibitemOpen
  \bibfield  {author} {\bibinfo {author} {\bibfnamefont {Sobhan}\ \bibnamefont
  {Kazempour}}, \bibinfo {author} {\bibfnamefont {Yuan-Chuan}\ \bibnamefont
  {Zou}}, \ and\ \bibinfo {author} {\bibfnamefont {Amin~Rezaei}\ \bibnamefont
  {Akbarieh}},\ }\bibfield  {title} {\enquote {\bibinfo {title} {{Analysis of
  accretion disk around a black hole in dRGT massive gravity}},}\ }\href
  {\doibase 10.1140/epjc/s10052-022-10153-y} {\bibfield  {journal} {\bibinfo
  {journal} {Eur. Phys. J. C}\ }\textbf {\bibinfo {volume} {82}},\ \bibinfo
  {pages} {190} (\bibinfo {year} {2022})},\ \Eprint
  {http://arxiv.org/abs/2203.05190} {arXiv:2203.05190 [gr-qc]} \BibitemShut
  {NoStop}%
\bibitem [{\citenamefont {He}\ \emph {et~al.}(2022)\citenamefont {He},
  \citenamefont {Tan},\ and\ \citenamefont {Li}}]{He:2022yse}%
  \BibitemOpen
  \bibfield  {author} {\bibinfo {author} {\bibfnamefont {Ke-Jian}\ \bibnamefont
  {He}}, \bibinfo {author} {\bibfnamefont {Shuang-Cheng}\ \bibnamefont {Tan}},
  \ and\ \bibinfo {author} {\bibfnamefont {Guo-Ping}\ \bibnamefont {Li}},\
  }\bibfield  {title} {\enquote {\bibinfo {title} {{Influence of torsion charge
  on shadow and observation signature of black hole surrounded by various
  profiles of accretions}},}\ }\href {\doibase 10.1140/epjc/s10052-022-10032-6}
  {\bibfield  {journal} {\bibinfo  {journal} {Eur. Phys. J. C}\ }\textbf
  {\bibinfo {volume} {82}},\ \bibinfo {pages} {81} (\bibinfo {year}
  {2022})}\BibitemShut {NoStop}%
\bibitem [{\citenamefont {Bisnovatyi-Kogan}\ and\ \citenamefont
  {Tsupko}(2022)}]{Bisnovatyi-Kogan:2022ujt}%
  \BibitemOpen
  \bibfield  {author} {\bibinfo {author} {\bibfnamefont {Gennady~S.}\
  \bibnamefont {Bisnovatyi-Kogan}}\ and\ \bibinfo {author} {\bibfnamefont
  {Oleg~Yu.}\ \bibnamefont {Tsupko}},\ }\bibfield  {title} {\enquote {\bibinfo
  {title} {{Analytical study of higher-order ring images of the accretion disk
  around a black hole}},}\ }\href {\doibase 10.1103/PhysRevD.105.064040}
  {\bibfield  {journal} {\bibinfo  {journal} {Phys. Rev. D}\ }\textbf {\bibinfo
  {volume} {105}},\ \bibinfo {pages} {064040} (\bibinfo {year} {2022})},\
  \Eprint {http://arxiv.org/abs/2201.01716} {arXiv:2201.01716 [gr-qc]}
  \BibitemShut {NoStop}%
\bibitem [{\citenamefont {Bauer}\ \emph {et~al.}(2022)\citenamefont {Bauer},
  \citenamefont {C\'ardenas-Avenda\~no}, \citenamefont {Gammie},\ and\
  \citenamefont {Yunes}}]{Bauer:2021atk}%
  \BibitemOpen
  \bibfield  {author} {\bibinfo {author} {\bibfnamefont {Adam~Michael}\
  \bibnamefont {Bauer}}, \bibinfo {author} {\bibfnamefont {Alejandro}\
  \bibnamefont {C\'ardenas-Avenda\~no}}, \bibinfo {author} {\bibfnamefont
  {Charles~F.}\ \bibnamefont {Gammie}}, \ and\ \bibinfo {author} {\bibfnamefont
  {Nicol\'as}\ \bibnamefont {Yunes}},\ }\bibfield  {title} {\enquote {\bibinfo
  {title} {{Spherical Accretion in Alternative Theories of Gravity}},}\ }\href
  {\doibase 10.3847/1538-4357/ac3a03} {\bibfield  {journal} {\bibinfo
  {journal} {Astrophys. J.}\ }\textbf {\bibinfo {volume} {925}},\ \bibinfo
  {pages} {119} (\bibinfo {year} {2022})},\ \Eprint
  {http://arxiv.org/abs/2111.02178} {arXiv:2111.02178 [gr-qc]} \BibitemShut
  {NoStop}%
\bibitem [{\citenamefont {Li}\ and\ \citenamefont
  {He}(2021{\natexlab{b}})}]{Li:2021riw}%
  \BibitemOpen
  \bibfield  {author} {\bibinfo {author} {\bibfnamefont {Guo-Ping}\
  \bibnamefont {Li}}\ and\ \bibinfo {author} {\bibfnamefont {Ke-Jian}\
  \bibnamefont {He}},\ }\bibfield  {title} {\enquote {\bibinfo {title}
  {{Shadows and rings of the Kehagias-Sfetsos black hole surrounded by thin
  disk accretion}},}\ }\href {\doibase 10.1088/1475-7516/2021/06/037}
  {\bibfield  {journal} {\bibinfo  {journal} {JCAP}\ }\textbf {\bibinfo
  {volume} {06}},\ \bibinfo {pages} {037} (\bibinfo {year}
  {2021}{\natexlab{b}})},\ \Eprint {http://arxiv.org/abs/2105.08521}
  {arXiv:2105.08521 [gr-qc]} \BibitemShut {NoStop}%
\bibitem [{\citenamefont {\"Ovg\"un}\ and\ \citenamefont
  {Sakall\i{}}(2020)}]{Ovgun:2020gjz}%
  \BibitemOpen
  \bibfield  {author} {\bibinfo {author} {\bibfnamefont {Ali}\ \bibnamefont
  {\"Ovg\"un}}\ and\ \bibinfo {author} {\bibfnamefont {\.Izzet}\ \bibnamefont
  {Sakall\i{}}},\ }\bibfield  {title} {\enquote {\bibinfo {title} {{Testing
  generalized Einstein\textendash{}Cartan\textendash{}Kibble\textendash{}Sciama
  gravity using weak deflection angle and shadow cast}},}\ }\href {\doibase
  10.1088/1361-6382/abb579} {\bibfield  {journal} {\bibinfo  {journal} {Class.
  Quant. Grav.}\ }\textbf {\bibinfo {volume} {37}},\ \bibinfo {pages} {225003}
  (\bibinfo {year} {2020})},\ \Eprint {http://arxiv.org/abs/2005.00982}
  {arXiv:2005.00982 [gr-qc]} \BibitemShut {NoStop}%
\bibitem [{\citenamefont {\"Ovg\"un}\ \emph {et~al.}(2020)\citenamefont
  {\"Ovg\"un}, \citenamefont {Sakall\i{}}, \citenamefont {Saavedra},\ and\
  \citenamefont {Leiva}}]{Ovgun:2019jdo}%
  \BibitemOpen
  \bibfield  {author} {\bibinfo {author} {\bibfnamefont {Ali}\ \bibnamefont
  {\"Ovg\"un}}, \bibinfo {author} {\bibfnamefont {\.Izzet}\ \bibnamefont
  {Sakall\i{}}}, \bibinfo {author} {\bibfnamefont {Joel}\ \bibnamefont
  {Saavedra}}, \ and\ \bibinfo {author} {\bibfnamefont {Carlos}\ \bibnamefont
  {Leiva}},\ }\bibfield  {title} {\enquote {\bibinfo {title} {{Shadow cast of
  noncommutative black holes in Rastall gravity}},}\ }\href {\doibase
  10.1142/S0217732320501631} {\bibfield  {journal} {\bibinfo  {journal} {Mod.
  Phys. Lett. A}\ }\textbf {\bibinfo {volume} {35}},\ \bibinfo {pages}
  {2050163} (\bibinfo {year} {2020})},\ \Eprint
  {http://arxiv.org/abs/1906.05954} {arXiv:1906.05954 [hep-th]} \BibitemShut
  {NoStop}%
\bibitem [{\citenamefont {\"Ovg\"un}\ \emph {et~al.}(2018)\citenamefont
  {\"Ovg\"un}, \citenamefont {Sakall\i{}},\ and\ \citenamefont
  {Saavedra}}]{Ovgun:2018tua}%
  \BibitemOpen
  \bibfield  {author} {\bibinfo {author} {\bibfnamefont {Ali}\ \bibnamefont
  {\"Ovg\"un}}, \bibinfo {author} {\bibfnamefont {\.Izzet}\ \bibnamefont
  {Sakall\i{}}}, \ and\ \bibinfo {author} {\bibfnamefont {Joel}\ \bibnamefont
  {Saavedra}},\ }\bibfield  {title} {\enquote {\bibinfo {title} {{Shadow cast
  and Deflection angle of Kerr-Newman-Kasuya spacetime}},}\ }\href {\doibase
  10.1088/1475-7516/2018/10/041} {\bibfield  {journal} {\bibinfo  {journal}
  {JCAP}\ }\textbf {\bibinfo {volume} {10}},\ \bibinfo {pages} {041} (\bibinfo
  {year} {2018})},\ \Eprint {http://arxiv.org/abs/1807.00388} {arXiv:1807.00388
  [gr-qc]} \BibitemShut {NoStop}%
\bibitem [{\citenamefont {\"Ovg\"un}(2021)}]{Ovgun:2021ttv}%
  \BibitemOpen
  \bibfield  {author} {\bibinfo {author} {\bibfnamefont {A.}~\bibnamefont
  {\"Ovg\"un}},\ }\bibfield  {title} {\enquote {\bibinfo {title} {{Black hole
  with confining electric potential in scalar-tensor description of regularized
  4-dimensional Einstein-Gauss-Bonnet gravity}},}\ }\href {\doibase
  10.1016/j.physletb.2021.136517} {\bibfield  {journal} {\bibinfo  {journal}
  {Phys. Lett. B}\ }\textbf {\bibinfo {volume} {820}},\ \bibinfo {pages}
  {136517} (\bibinfo {year} {2021})},\ \Eprint
  {http://arxiv.org/abs/2105.05035} {arXiv:2105.05035 [gr-qc]} \BibitemShut
  {NoStop}%
\bibitem [{\citenamefont {Ling}\ \emph {et~al.}(2021)\citenamefont {Ling},
  \citenamefont {Guo}, \citenamefont {Liu}, \citenamefont {Kuang},\ and\
  \citenamefont {Wang}}]{Ling:2021vgk}%
  \BibitemOpen
  \bibfield  {author} {\bibinfo {author} {\bibfnamefont {Ru}~\bibnamefont
  {Ling}}, \bibinfo {author} {\bibfnamefont {Hong}\ \bibnamefont {Guo}},
  \bibinfo {author} {\bibfnamefont {Hang}\ \bibnamefont {Liu}}, \bibinfo
  {author} {\bibfnamefont {Xiao-Mei}\ \bibnamefont {Kuang}}, \ and\ \bibinfo
  {author} {\bibfnamefont {Bin}\ \bibnamefont {Wang}},\ }\bibfield  {title}
  {\enquote {\bibinfo {title} {{Shadow and near-horizon characteristics of the
  acoustic charged black hole in curved spacetime}},}\ }\href {\doibase
  10.1103/PhysRevD.104.104003} {\bibfield  {journal} {\bibinfo  {journal}
  {Phys. Rev. D}\ }\textbf {\bibinfo {volume} {104}},\ \bibinfo {pages}
  {104003} (\bibinfo {year} {2021})},\ \Eprint
  {http://arxiv.org/abs/2107.05171} {arXiv:2107.05171 [gr-qc]} \BibitemShut
  {NoStop}%
\bibitem [{\citenamefont {Belhaj}\ \emph {et~al.}(2021)\citenamefont {Belhaj},
  \citenamefont {Belmahi}, \citenamefont {Benali}, \citenamefont {El~Hadri},
  \citenamefont {El~Moumni},\ and\ \citenamefont
  {Torrente-Lujan}}]{Belhaj:2020okh}%
  \BibitemOpen
  \bibfield  {author} {\bibinfo {author} {\bibfnamefont {A.}~\bibnamefont
  {Belhaj}}, \bibinfo {author} {\bibfnamefont {H.}~\bibnamefont {Belmahi}},
  \bibinfo {author} {\bibfnamefont {M.}~\bibnamefont {Benali}}, \bibinfo
  {author} {\bibfnamefont {W.}~\bibnamefont {El~Hadri}}, \bibinfo {author}
  {\bibfnamefont {H.}~\bibnamefont {El~Moumni}}, \ and\ \bibinfo {author}
  {\bibfnamefont {E.}~\bibnamefont {Torrente-Lujan}},\ }\bibfield  {title}
  {\enquote {\bibinfo {title} {{Shadows of 5D black holes from string
  theory}},}\ }\href {\doibase 10.1016/j.physletb.2020.136025} {\bibfield
  {journal} {\bibinfo  {journal} {Phys. Lett. B}\ }\textbf {\bibinfo {volume}
  {812}},\ \bibinfo {pages} {136025} (\bibinfo {year} {2021})},\ \Eprint
  {http://arxiv.org/abs/2008.13478} {arXiv:2008.13478 [hep-th]} \BibitemShut
  {NoStop}%
\bibitem [{\citenamefont {Belhaj}\ \emph {et~al.}(2020)\citenamefont {Belhaj},
  \citenamefont {Benali}, \citenamefont {El~Balali}, \citenamefont
  {El~Moumni},\ and\ \citenamefont {Ennadifi}}]{Belhaj:2020rdb}%
  \BibitemOpen
  \bibfield  {author} {\bibinfo {author} {\bibfnamefont {A.}~\bibnamefont
  {Belhaj}}, \bibinfo {author} {\bibfnamefont {M.}~\bibnamefont {Benali}},
  \bibinfo {author} {\bibfnamefont {A.}~\bibnamefont {El~Balali}}, \bibinfo
  {author} {\bibfnamefont {H.}~\bibnamefont {El~Moumni}}, \ and\ \bibinfo
  {author} {\bibfnamefont {S.~E.}\ \bibnamefont {Ennadifi}},\ }\bibfield
  {title} {\enquote {\bibinfo {title} {{Deflection angle and shadow behaviors
  of quintessential black holes in arbitrary dimensions}},}\ }\href {\doibase
  10.1088/1361-6382/abbaa9} {\bibfield  {journal} {\bibinfo  {journal} {Class.
  Quant. Grav.}\ }\textbf {\bibinfo {volume} {37}},\ \bibinfo {pages} {215004}
  (\bibinfo {year} {2020})},\ \Eprint {http://arxiv.org/abs/2006.01078}
  {arXiv:2006.01078 [gr-qc]} \BibitemShut {NoStop}%
\bibitem [{\citenamefont {Abdikamalov}\ \emph {et~al.}(2019)\citenamefont
  {Abdikamalov}, \citenamefont {Abdujabbarov}, \citenamefont {Ayzenberg},
  \citenamefont {Malafarina}, \citenamefont {Bambi},\ and\ \citenamefont
  {Ahmedov}}]{Abdikamalov:2019ztb}%
  \BibitemOpen
  \bibfield  {author} {\bibinfo {author} {\bibfnamefont {Askar~B.}\
  \bibnamefont {Abdikamalov}}, \bibinfo {author} {\bibfnamefont {Ahmadjon~A.}\
  \bibnamefont {Abdujabbarov}}, \bibinfo {author} {\bibfnamefont {Dimitry}\
  \bibnamefont {Ayzenberg}}, \bibinfo {author} {\bibfnamefont {Daniele}\
  \bibnamefont {Malafarina}}, \bibinfo {author} {\bibfnamefont {Cosimo}\
  \bibnamefont {Bambi}}, \ and\ \bibinfo {author} {\bibfnamefont {Bobomurat}\
  \bibnamefont {Ahmedov}},\ }\bibfield  {title} {\enquote {\bibinfo {title}
  {{Black hole mimicker hiding in the shadow: Optical properties of the
  $\gamma$ metric}},}\ }\href {\doibase 10.1103/PhysRevD.100.024014} {\bibfield
   {journal} {\bibinfo  {journal} {Phys. Rev. D}\ }\textbf {\bibinfo {volume}
  {100}},\ \bibinfo {pages} {024014} (\bibinfo {year} {2019})},\ \Eprint
  {http://arxiv.org/abs/1904.06207} {arXiv:1904.06207 [gr-qc]} \BibitemShut
  {NoStop}%
\bibitem [{\citenamefont {Abdujabbarov}\ \emph {et~al.}(2016)\citenamefont
  {Abdujabbarov}, \citenamefont {Juraev}, \citenamefont {Ahmedov},\ and\
  \citenamefont {Stuchl\'\i{}k}}]{Abdujabbarov:2016efm}%
  \BibitemOpen
  \bibfield  {author} {\bibinfo {author} {\bibfnamefont {Ahmadjon}\
  \bibnamefont {Abdujabbarov}}, \bibinfo {author} {\bibfnamefont {Bakhtinur}\
  \bibnamefont {Juraev}}, \bibinfo {author} {\bibfnamefont {Bobomurat}\
  \bibnamefont {Ahmedov}}, \ and\ \bibinfo {author} {\bibfnamefont
  {Zden\v{e}k}\ \bibnamefont {Stuchl\'\i{}k}},\ }\bibfield  {title} {\enquote
  {\bibinfo {title} {{Shadow of rotating wormhole in plasma environment}},}\
  }\href {\doibase 10.1007/s10509-016-2818-9} {\bibfield  {journal} {\bibinfo
  {journal} {Astrophys. Space Sci.}\ }\textbf {\bibinfo {volume} {361}},\
  \bibinfo {pages} {226} (\bibinfo {year} {2016})}\BibitemShut {NoStop}%
\bibitem [{\citenamefont {Atamurotov}\ and\ \citenamefont
  {Ahmedov}(2015)}]{Atamurotov:2015nra}%
  \BibitemOpen
  \bibfield  {author} {\bibinfo {author} {\bibfnamefont {Farruh}\ \bibnamefont
  {Atamurotov}}\ and\ \bibinfo {author} {\bibfnamefont {Bobomurat}\
  \bibnamefont {Ahmedov}},\ }\bibfield  {title} {\enquote {\bibinfo {title}
  {{Optical properties of black hole in the presence of plasma: shadow}},}\
  }\href {\doibase 10.1103/PhysRevD.92.084005} {\bibfield  {journal} {\bibinfo
  {journal} {Phys. Rev. D}\ }\textbf {\bibinfo {volume} {92}},\ \bibinfo
  {pages} {084005} (\bibinfo {year} {2015})},\ \Eprint
  {http://arxiv.org/abs/1507.08131} {arXiv:1507.08131 [gr-qc]} \BibitemShut
  {NoStop}%
\bibitem [{\citenamefont {Papnoi}\ \emph {et~al.}(2014)\citenamefont {Papnoi},
  \citenamefont {Atamurotov}, \citenamefont {Ghosh},\ and\ \citenamefont
  {Ahmedov}}]{Papnoi:2014aaa}%
  \BibitemOpen
  \bibfield  {author} {\bibinfo {author} {\bibfnamefont {Uma}\ \bibnamefont
  {Papnoi}}, \bibinfo {author} {\bibfnamefont {Farruh}\ \bibnamefont
  {Atamurotov}}, \bibinfo {author} {\bibfnamefont {Sushant~G.}\ \bibnamefont
  {Ghosh}}, \ and\ \bibinfo {author} {\bibfnamefont {Bobomurat}\ \bibnamefont
  {Ahmedov}},\ }\bibfield  {title} {\enquote {\bibinfo {title} {{Shadow of
  five-dimensional rotating Myers-Perry black hole}},}\ }\href {\doibase
  10.1103/PhysRevD.90.024073} {\bibfield  {journal} {\bibinfo  {journal} {Phys.
  Rev. D}\ }\textbf {\bibinfo {volume} {90}},\ \bibinfo {pages} {024073}
  (\bibinfo {year} {2014})},\ \Eprint {http://arxiv.org/abs/1407.0834}
  {arXiv:1407.0834 [gr-qc]} \BibitemShut {NoStop}%
\bibitem [{\citenamefont {Abdujabbarov}\ \emph {et~al.}(2013)\citenamefont
  {Abdujabbarov}, \citenamefont {Atamurotov}, \citenamefont {Kucukakca},
  \citenamefont {Ahmedov},\ and\ \citenamefont {Camci}}]{Abdujabbarov:2012bn}%
  \BibitemOpen
  \bibfield  {author} {\bibinfo {author} {\bibfnamefont {Ahmadjon}\
  \bibnamefont {Abdujabbarov}}, \bibinfo {author} {\bibfnamefont {Farruh}\
  \bibnamefont {Atamurotov}}, \bibinfo {author} {\bibfnamefont {Yusuf}\
  \bibnamefont {Kucukakca}}, \bibinfo {author} {\bibfnamefont {Bobomurat}\
  \bibnamefont {Ahmedov}}, \ and\ \bibinfo {author} {\bibfnamefont {Ugur}\
  \bibnamefont {Camci}},\ }\bibfield  {title} {\enquote {\bibinfo {title}
  {{Shadow of Kerr-Taub-NUT black hole}},}\ }\href {\doibase
  10.1007/s10509-012-1337-6} {\bibfield  {journal} {\bibinfo  {journal}
  {Astrophys. Space Sci.}\ }\textbf {\bibinfo {volume} {344}},\ \bibinfo
  {pages} {429--435} (\bibinfo {year} {2013})},\ \Eprint
  {http://arxiv.org/abs/1212.4949} {arXiv:1212.4949 [physics.gen-ph]}
  \BibitemShut {NoStop}%
\bibitem [{\citenamefont {Atamurotov}\ \emph {et~al.}(2013)\citenamefont
  {Atamurotov}, \citenamefont {Abdujabbarov},\ and\ \citenamefont
  {Ahmedov}}]{Atamurotov:2013sca}%
  \BibitemOpen
  \bibfield  {author} {\bibinfo {author} {\bibfnamefont {Farruh}\ \bibnamefont
  {Atamurotov}}, \bibinfo {author} {\bibfnamefont {Ahmadjon}\ \bibnamefont
  {Abdujabbarov}}, \ and\ \bibinfo {author} {\bibfnamefont {Bobomurat}\
  \bibnamefont {Ahmedov}},\ }\bibfield  {title} {\enquote {\bibinfo {title}
  {{Shadow of rotating non-Kerr black hole}},}\ }\href {\doibase
  10.1103/PhysRevD.88.064004} {\bibfield  {journal} {\bibinfo  {journal} {Phys.
  Rev. D}\ }\textbf {\bibinfo {volume} {88}},\ \bibinfo {pages} {064004}
  (\bibinfo {year} {2013})}\BibitemShut {NoStop}%
\bibitem [{\citenamefont {Cunha}\ and\ \citenamefont
  {Herdeiro}(2018)}]{Cunha:2018acu}%
  \BibitemOpen
  \bibfield  {author} {\bibinfo {author} {\bibfnamefont {Pedro V.~P.}\
  \bibnamefont {Cunha}}\ and\ \bibinfo {author} {\bibfnamefont {Carlos A.~R.}\
  \bibnamefont {Herdeiro}},\ }\bibfield  {title} {\enquote {\bibinfo {title}
  {{Shadows and strong gravitational lensing: a brief review}},}\ }\href
  {\doibase 10.1007/s10714-018-2361-9} {\bibfield  {journal} {\bibinfo
  {journal} {Gen. Rel. Grav.}\ }\textbf {\bibinfo {volume} {50}},\ \bibinfo
  {pages} {42} (\bibinfo {year} {2018})},\ \Eprint
  {http://arxiv.org/abs/1801.00860} {arXiv:1801.00860 [gr-qc]} \BibitemShut
  {NoStop}%
\bibitem [{\citenamefont {Perlick}\ \emph {et~al.}(2015)\citenamefont
  {Perlick}, \citenamefont {Tsupko},\ and\ \citenamefont
  {Bisnovatyi-Kogan}}]{Perlick:2015vta}%
  \BibitemOpen
  \bibfield  {author} {\bibinfo {author} {\bibfnamefont {Volker}\ \bibnamefont
  {Perlick}}, \bibinfo {author} {\bibfnamefont {Oleg~Yu.}\ \bibnamefont
  {Tsupko}}, \ and\ \bibinfo {author} {\bibfnamefont {Gennady~S.}\ \bibnamefont
  {Bisnovatyi-Kogan}},\ }\bibfield  {title} {\enquote {\bibinfo {title}
  {{Influence of a plasma on the shadow of a spherically symmetric black
  hole}},}\ }\href {\doibase 10.1103/PhysRevD.92.104031} {\bibfield  {journal}
  {\bibinfo  {journal} {Phys. Rev. D}\ }\textbf {\bibinfo {volume} {92}},\
  \bibinfo {pages} {104031} (\bibinfo {year} {2015})},\ \Eprint
  {http://arxiv.org/abs/1507.04217} {arXiv:1507.04217 [gr-qc]} \BibitemShut
  {NoStop}%
\bibitem [{\citenamefont {Nedkova}\ \emph {et~al.}(2013)\citenamefont
  {Nedkova}, \citenamefont {Tinchev},\ and\ \citenamefont
  {Yazadjiev}}]{Nedkova:2013msa}%
  \BibitemOpen
  \bibfield  {author} {\bibinfo {author} {\bibfnamefont {Petya~G.}\
  \bibnamefont {Nedkova}}, \bibinfo {author} {\bibfnamefont {Vassil~K.}\
  \bibnamefont {Tinchev}}, \ and\ \bibinfo {author} {\bibfnamefont
  {Stoytcho~S.}\ \bibnamefont {Yazadjiev}},\ }\bibfield  {title} {\enquote
  {\bibinfo {title} {{Shadow of a rotating traversable wormhole}},}\ }\href
  {\doibase 10.1103/PhysRevD.88.124019} {\bibfield  {journal} {\bibinfo
  {journal} {Phys. Rev. D}\ }\textbf {\bibinfo {volume} {88}},\ \bibinfo
  {pages} {124019} (\bibinfo {year} {2013})},\ \Eprint
  {http://arxiv.org/abs/1307.7647} {arXiv:1307.7647 [gr-qc]} \BibitemShut
  {NoStop}%
\bibitem [{\citenamefont {Li}\ and\ \citenamefont {Bambi}(2014)}]{Li:2013jra}%
  \BibitemOpen
  \bibfield  {author} {\bibinfo {author} {\bibfnamefont {Zilong}\ \bibnamefont
  {Li}}\ and\ \bibinfo {author} {\bibfnamefont {Cosimo}\ \bibnamefont
  {Bambi}},\ }\bibfield  {title} {\enquote {\bibinfo {title} {{Measuring the
  Kerr spin parameter of regular black holes from their shadow}},}\ }\href
  {\doibase 10.1088/1475-7516/2014/01/041} {\bibfield  {journal} {\bibinfo
  {journal} {JCAP}\ }\textbf {\bibinfo {volume} {01}},\ \bibinfo {pages} {041}
  (\bibinfo {year} {2014})},\ \Eprint {http://arxiv.org/abs/1309.1606}
  {arXiv:1309.1606 [gr-qc]} \BibitemShut {NoStop}%
\bibitem [{\citenamefont {Cunha}\ \emph {et~al.}(2017)\citenamefont {Cunha},
  \citenamefont {Herdeiro}, \citenamefont {Kleihaus}, \citenamefont {Kunz},\
  and\ \citenamefont {Radu}}]{Cunha:2016wzk}%
  \BibitemOpen
  \bibfield  {author} {\bibinfo {author} {\bibfnamefont {Pedro V.~P.}\
  \bibnamefont {Cunha}}, \bibinfo {author} {\bibfnamefont {Carlos A.~R.}\
  \bibnamefont {Herdeiro}}, \bibinfo {author} {\bibfnamefont {Burkhard}\
  \bibnamefont {Kleihaus}}, \bibinfo {author} {\bibfnamefont {Jutta}\
  \bibnamefont {Kunz}}, \ and\ \bibinfo {author} {\bibfnamefont {Eugen}\
  \bibnamefont {Radu}},\ }\bibfield  {title} {\enquote {\bibinfo {title}
  {{Shadows of
  Einstein\textendash{}dilaton\textendash{}Gauss\textendash{}Bonnet black
  holes}},}\ }\href {\doibase 10.1016/j.physletb.2017.03.020} {\bibfield
  {journal} {\bibinfo  {journal} {Phys. Lett. B}\ }\textbf {\bibinfo {volume}
  {768}},\ \bibinfo {pages} {373--379} (\bibinfo {year} {2017})},\ \Eprint
  {http://arxiv.org/abs/1701.00079} {arXiv:1701.00079 [gr-qc]} \BibitemShut
  {NoStop}%
\bibitem [{\citenamefont {Johannsen}\ \emph {et~al.}(2016)\citenamefont
  {Johannsen}, \citenamefont {Broderick}, \citenamefont {Plewa}, \citenamefont
  {Chatzopoulos}, \citenamefont {Doeleman}, \citenamefont {Eisenhauer},
  \citenamefont {Fish}, \citenamefont {Genzel}, \citenamefont {Gerhard},\ and\
  \citenamefont {Johnson}}]{Johannsen:2015hib}%
  \BibitemOpen
  \bibfield  {author} {\bibinfo {author} {\bibfnamefont {Tim}\ \bibnamefont
  {Johannsen}}, \bibinfo {author} {\bibfnamefont {Avery~E.}\ \bibnamefont
  {Broderick}}, \bibinfo {author} {\bibfnamefont {Philipp~M.}\ \bibnamefont
  {Plewa}}, \bibinfo {author} {\bibfnamefont {Sotiris}\ \bibnamefont
  {Chatzopoulos}}, \bibinfo {author} {\bibfnamefont {Sheperd~S.}\ \bibnamefont
  {Doeleman}}, \bibinfo {author} {\bibfnamefont {Frank}\ \bibnamefont
  {Eisenhauer}}, \bibinfo {author} {\bibfnamefont {Vincent~L.}\ \bibnamefont
  {Fish}}, \bibinfo {author} {\bibfnamefont {Reinhard}\ \bibnamefont {Genzel}},
  \bibinfo {author} {\bibfnamefont {Ortwin}\ \bibnamefont {Gerhard}}, \ and\
  \bibinfo {author} {\bibfnamefont {Michael~D.}\ \bibnamefont {Johnson}},\
  }\bibfield  {title} {\enquote {\bibinfo {title} {{Testing General Relativity
  with the Shadow Size of Sgr A*}},}\ }\href {\doibase
  10.1103/PhysRevLett.116.031101} {\bibfield  {journal} {\bibinfo  {journal}
  {Phys. Rev. Lett.}\ }\textbf {\bibinfo {volume} {116}},\ \bibinfo {pages}
  {031101} (\bibinfo {year} {2016})},\ \Eprint
  {http://arxiv.org/abs/1512.02640} {arXiv:1512.02640 [astro-ph.GA]}
  \BibitemShut {NoStop}%
\bibitem [{\citenamefont {Johannsen}(2016)}]{Johannsen:2015mdd}%
  \BibitemOpen
  \bibfield  {author} {\bibinfo {author} {\bibfnamefont {Tim}\ \bibnamefont
  {Johannsen}},\ }\bibfield  {title} {\enquote {\bibinfo {title} {{Sgr A* and
  General Relativity}},}\ }\href {\doibase 10.1088/0264-9381/33/11/113001}
  {\bibfield  {journal} {\bibinfo  {journal} {Class. Quant. Grav.}\ }\textbf
  {\bibinfo {volume} {33}},\ \bibinfo {pages} {113001} (\bibinfo {year}
  {2016})},\ \Eprint {http://arxiv.org/abs/1512.03818} {arXiv:1512.03818
  [astro-ph.GA]} \BibitemShut {NoStop}%
\bibitem [{\citenamefont {Shaikh}(2019)}]{Shaikh:2019fpu}%
  \BibitemOpen
  \bibfield  {author} {\bibinfo {author} {\bibfnamefont {Rajibul}\ \bibnamefont
  {Shaikh}},\ }\bibfield  {title} {\enquote {\bibinfo {title} {{Black hole
  shadow in a general rotating spacetime obtained through Newman-Janis
  algorithm}},}\ }\href {\doibase 10.1103/PhysRevD.100.024028} {\bibfield
  {journal} {\bibinfo  {journal} {Phys. Rev. D}\ }\textbf {\bibinfo {volume}
  {100}},\ \bibinfo {pages} {024028} (\bibinfo {year} {2019})},\ \Eprint
  {http://arxiv.org/abs/1904.08322} {arXiv:1904.08322 [gr-qc]} \BibitemShut
  {NoStop}%
\bibitem [{\citenamefont {Yumoto}\ \emph {et~al.}(2012)\citenamefont {Yumoto},
  \citenamefont {Nitta}, \citenamefont {Chiba},\ and\ \citenamefont
  {Sugiyama}}]{Yumoto:2012kz}%
  \BibitemOpen
  \bibfield  {author} {\bibinfo {author} {\bibfnamefont {Akifumi}\ \bibnamefont
  {Yumoto}}, \bibinfo {author} {\bibfnamefont {Daisuke}\ \bibnamefont {Nitta}},
  \bibinfo {author} {\bibfnamefont {Takeshi}\ \bibnamefont {Chiba}}, \ and\
  \bibinfo {author} {\bibfnamefont {Naoshi}\ \bibnamefont {Sugiyama}},\
  }\bibfield  {title} {\enquote {\bibinfo {title} {{Shadows of Multi-Black
  Holes: Analytic Exploration}},}\ }\href {\doibase 10.1103/PhysRevD.86.103001}
  {\bibfield  {journal} {\bibinfo  {journal} {Phys. Rev. D}\ }\textbf {\bibinfo
  {volume} {86}},\ \bibinfo {pages} {103001} (\bibinfo {year} {2012})},\
  \Eprint {http://arxiv.org/abs/1208.0635} {arXiv:1208.0635 [gr-qc]}
  \BibitemShut {NoStop}%
\bibitem [{\citenamefont {Cunha}\ \emph
  {et~al.}(2016{\natexlab{a}})\citenamefont {Cunha}, \citenamefont {Herdeiro},
  \citenamefont {Radu},\ and\ \citenamefont {Runarsson}}]{Cunha:2016bpi}%
  \BibitemOpen
  \bibfield  {author} {\bibinfo {author} {\bibfnamefont {Pedro V.~P.}\
  \bibnamefont {Cunha}}, \bibinfo {author} {\bibfnamefont {Carlos A.~R.}\
  \bibnamefont {Herdeiro}}, \bibinfo {author} {\bibfnamefont {Eugen}\
  \bibnamefont {Radu}}, \ and\ \bibinfo {author} {\bibfnamefont {Helgi~F.}\
  \bibnamefont {Runarsson}},\ }\bibfield  {title} {\enquote {\bibinfo {title}
  {{Shadows of Kerr black holes with and without scalar hair}},}\ }\href
  {\doibase 10.1142/S0218271816410212} {\bibfield  {journal} {\bibinfo
  {journal} {Int. J. Mod. Phys. D}\ }\textbf {\bibinfo {volume} {25}},\
  \bibinfo {pages} {1641021} (\bibinfo {year} {2016}{\natexlab{a}})},\ \Eprint
  {http://arxiv.org/abs/1605.08293} {arXiv:1605.08293 [gr-qc]} \BibitemShut
  {NoStop}%
\bibitem [{\citenamefont {Moffat}(2015)}]{Moffat:2015kva}%
  \BibitemOpen
  \bibfield  {author} {\bibinfo {author} {\bibfnamefont {J.~W.}\ \bibnamefont
  {Moffat}},\ }\bibfield  {title} {\enquote {\bibinfo {title} {{Modified
  Gravity Black Holes and their Observable Shadows}},}\ }\href {\doibase
  10.1140/epjc/s10052-015-3352-6} {\bibfield  {journal} {\bibinfo  {journal}
  {Eur. Phys. J. C}\ }\textbf {\bibinfo {volume} {75}},\ \bibinfo {pages} {130}
  (\bibinfo {year} {2015})},\ \Eprint {http://arxiv.org/abs/1502.01677}
  {arXiv:1502.01677 [gr-qc]} \BibitemShut {NoStop}%
\bibitem [{\citenamefont {Giddings}\ and\ \citenamefont
  {Psaltis}(2018)}]{Giddings:2016btb}%
  \BibitemOpen
  \bibfield  {author} {\bibinfo {author} {\bibfnamefont {Steven~B.}\
  \bibnamefont {Giddings}}\ and\ \bibinfo {author} {\bibfnamefont {Dimitrios}\
  \bibnamefont {Psaltis}},\ }\bibfield  {title} {\enquote {\bibinfo {title}
  {{Event Horizon Telescope Observations as Probes for Quantum Structure of
  Astrophysical Black Holes}},}\ }\href {\doibase 10.1103/PhysRevD.97.084035}
  {\bibfield  {journal} {\bibinfo  {journal} {Phys. Rev. D}\ }\textbf {\bibinfo
  {volume} {97}},\ \bibinfo {pages} {084035} (\bibinfo {year} {2018})},\
  \Eprint {http://arxiv.org/abs/1606.07814} {arXiv:1606.07814 [astro-ph.HE]}
  \BibitemShut {NoStop}%
\bibitem [{\citenamefont {Cunha}\ \emph
  {et~al.}(2016{\natexlab{b}})\citenamefont {Cunha}, \citenamefont {Grover},
  \citenamefont {Herdeiro}, \citenamefont {Radu}, \citenamefont {Runarsson},\
  and\ \citenamefont {Wittig}}]{Cunha:2016bjh}%
  \BibitemOpen
  \bibfield  {author} {\bibinfo {author} {\bibfnamefont {P.~V.~P.}\
  \bibnamefont {Cunha}}, \bibinfo {author} {\bibfnamefont {J.}~\bibnamefont
  {Grover}}, \bibinfo {author} {\bibfnamefont {C.}~\bibnamefont {Herdeiro}},
  \bibinfo {author} {\bibfnamefont {E.}~\bibnamefont {Radu}}, \bibinfo {author}
  {\bibfnamefont {H.}~\bibnamefont {Runarsson}}, \ and\ \bibinfo {author}
  {\bibfnamefont {A.}~\bibnamefont {Wittig}},\ }\bibfield  {title} {\enquote
  {\bibinfo {title} {{Chaotic lensing around boson stars and Kerr black holes
  with scalar hair}},}\ }\href {\doibase 10.1103/PhysRevD.94.104023} {\bibfield
   {journal} {\bibinfo  {journal} {Phys. Rev. D}\ }\textbf {\bibinfo {volume}
  {94}},\ \bibinfo {pages} {104023} (\bibinfo {year} {2016}{\natexlab{b}})},\
  \Eprint {http://arxiv.org/abs/1609.01340} {arXiv:1609.01340 [gr-qc]}
  \BibitemShut {NoStop}%
\bibitem [{\citenamefont {Zakharov}(2014)}]{Zakharov:2014lqa}%
  \BibitemOpen
  \bibfield  {author} {\bibinfo {author} {\bibfnamefont {Alexander~F.}\
  \bibnamefont {Zakharov}},\ }\bibfield  {title} {\enquote {\bibinfo {title}
  {{Constraints on a charge in the Reissner-Nordstr\"om metric for the black
  hole at the Galactic Center}},}\ }\href {\doibase 10.1103/PhysRevD.90.062007}
  {\bibfield  {journal} {\bibinfo  {journal} {Phys. Rev. D}\ }\textbf {\bibinfo
  {volume} {90}},\ \bibinfo {pages} {062007} (\bibinfo {year} {2014})},\
  \Eprint {http://arxiv.org/abs/1407.7457} {arXiv:1407.7457 [gr-qc]}
  \BibitemShut {NoStop}%
\bibitem [{\citenamefont {Tsukamoto}(2018)}]{Tsukamoto:2017fxq}%
  \BibitemOpen
  \bibfield  {author} {\bibinfo {author} {\bibfnamefont {Naoki}\ \bibnamefont
  {Tsukamoto}},\ }\bibfield  {title} {\enquote {\bibinfo {title} {{Black hole
  shadow in an asymptotically-flat, stationary, and axisymmetric spacetime: The
  Kerr-Newman and rotating regular black holes}},}\ }\href {\doibase
  10.1103/PhysRevD.97.064021} {\bibfield  {journal} {\bibinfo  {journal} {Phys.
  Rev. D}\ }\textbf {\bibinfo {volume} {97}},\ \bibinfo {pages} {064021}
  (\bibinfo {year} {2018})},\ \Eprint {http://arxiv.org/abs/1708.07427}
  {arXiv:1708.07427 [gr-qc]} \BibitemShut {NoStop}%
\bibitem [{\citenamefont {Hennigar}\ \emph {et~al.}(2018)\citenamefont
  {Hennigar}, \citenamefont {Poshteh},\ and\ \citenamefont
  {Mann}}]{Hennigar:2018hza}%
  \BibitemOpen
  \bibfield  {author} {\bibinfo {author} {\bibfnamefont {Robie~A.}\
  \bibnamefont {Hennigar}}, \bibinfo {author} {\bibfnamefont {Mohammad
  Bagher~Jahani}\ \bibnamefont {Poshteh}}, \ and\ \bibinfo {author}
  {\bibfnamefont {Robert~B.}\ \bibnamefont {Mann}},\ }\bibfield  {title}
  {\enquote {\bibinfo {title} {{Shadows, Signals, and Stability in Einsteinian
  Cubic Gravity}},}\ }\href {\doibase 10.1103/PhysRevD.97.064041} {\bibfield
  {journal} {\bibinfo  {journal} {Phys. Rev. D}\ }\textbf {\bibinfo {volume}
  {97}},\ \bibinfo {pages} {064041} (\bibinfo {year} {2018})},\ \Eprint
  {http://arxiv.org/abs/1801.03223} {arXiv:1801.03223 [gr-qc]} \BibitemShut
  {NoStop}%
\bibitem [{\citenamefont {Kumar}\ \emph {et~al.}(2020)\citenamefont {Kumar},
  \citenamefont {Ghosh},\ and\ \citenamefont {Wang}}]{Kumar:2020hgm}%
  \BibitemOpen
  \bibfield  {author} {\bibinfo {author} {\bibfnamefont {Rahul}\ \bibnamefont
  {Kumar}}, \bibinfo {author} {\bibfnamefont {Sushant~G.}\ \bibnamefont
  {Ghosh}}, \ and\ \bibinfo {author} {\bibfnamefont {Anzhong}\ \bibnamefont
  {Wang}},\ }\bibfield  {title} {\enquote {\bibinfo {title} {{Gravitational
  deflection of light and shadow cast by rotating Kalb-Ramond black holes}},}\
  }\href {\doibase 10.1103/PhysRevD.101.104001} {\bibfield  {journal} {\bibinfo
   {journal} {Phys. Rev. D}\ }\textbf {\bibinfo {volume} {101}},\ \bibinfo
  {pages} {104001} (\bibinfo {year} {2020})}\BibitemShut {NoStop}%
\bibitem [{\citenamefont {Li}\ \emph {et~al.}(2020{\natexlab{a}})\citenamefont
  {Li}, \citenamefont {Guo},\ and\ \citenamefont {Chen}}]{Li2020}%
  \BibitemOpen
  \bibfield  {author} {\bibinfo {author} {\bibfnamefont {Peng-Cheng}\
  \bibnamefont {Li}}, \bibinfo {author} {\bibfnamefont {Minyong}\ \bibnamefont
  {Guo}}, \ and\ \bibinfo {author} {\bibfnamefont {Bin}\ \bibnamefont {Chen}},\
  }\bibfield  {title} {\enquote {\bibinfo {title} {{Shadow of a spinning black
  hole in an expanding universe}},}\ }\href {\doibase
  10.1103/physrevd.101.084041} {\bibfield  {journal} {\bibinfo  {journal}
  {Phys. Rev. D}\ }\textbf {\bibinfo {volume} {101}},\ \bibinfo {pages} {1--26}
  (\bibinfo {year} {2020}{\natexlab{a}})}\BibitemShut {NoStop}%
\bibitem [{\citenamefont {\c{C}imdiker}\ \emph {et~al.}(2021)\citenamefont
  {\c{C}imdiker}, \citenamefont {Demir},\ and\ \citenamefont
  {\"Ovg\"un}}]{Cimdiker:2021cpz}%
  \BibitemOpen
  \bibfield  {author} {\bibinfo {author} {\bibfnamefont {\.Irfan}\ \bibnamefont
  {\c{C}imdiker}}, \bibinfo {author} {\bibfnamefont {Durmu\c{s}}\ \bibnamefont
  {Demir}}, \ and\ \bibinfo {author} {\bibfnamefont {Ali}\ \bibnamefont
  {\"Ovg\"un}},\ }\bibfield  {title} {\enquote {\bibinfo {title} {{Black hole
  shadow in symmergent gravity}},}\ }\href {\doibase
  10.1016/j.dark.2021.100900} {\bibfield  {journal} {\bibinfo  {journal} {Phys.
  Dark Univ.}\ }\textbf {\bibinfo {volume} {34}},\ \bibinfo {pages} {100900}
  (\bibinfo {year} {2021})},\ \Eprint {http://arxiv.org/abs/2110.11904}
  {arXiv:2110.11904 [gr-qc]} \BibitemShut {NoStop}%
\bibitem [{\citenamefont {Hu}\ \emph {et~al.}(2021)\citenamefont {Hu},
  \citenamefont {Zhong}, \citenamefont {Li}, \citenamefont {Guo},\ and\
  \citenamefont {Chen}}]{Hu:2020usx}%
  \BibitemOpen
  \bibfield  {author} {\bibinfo {author} {\bibfnamefont {Zezhou}\ \bibnamefont
  {Hu}}, \bibinfo {author} {\bibfnamefont {Zhen}\ \bibnamefont {Zhong}},
  \bibinfo {author} {\bibfnamefont {Peng-Cheng}\ \bibnamefont {Li}}, \bibinfo
  {author} {\bibfnamefont {Minyong}\ \bibnamefont {Guo}}, \ and\ \bibinfo
  {author} {\bibfnamefont {Bin}\ \bibnamefont {Chen}},\ }\bibfield  {title}
  {\enquote {\bibinfo {title} {{QED effect on a black hole shadow}},}\ }\href
  {\doibase 10.1103/PhysRevD.103.044057} {\bibfield  {journal} {\bibinfo
  {journal} {Phys. Rev. D}\ }\textbf {\bibinfo {volume} {103}},\ \bibinfo
  {pages} {044057} (\bibinfo {year} {2021})},\ \Eprint
  {http://arxiv.org/abs/2012.07022} {arXiv:2012.07022 [gr-qc]} \BibitemShut
  {NoStop}%
\bibitem [{\citenamefont {Zhong}\ \emph {et~al.}(2021)\citenamefont {Zhong},
  \citenamefont {Hu}, \citenamefont {Yan}, \citenamefont {Guo},\ and\
  \citenamefont {Chen}}]{Zhong:2021mty}%
  \BibitemOpen
  \bibfield  {author} {\bibinfo {author} {\bibfnamefont {Zhen}\ \bibnamefont
  {Zhong}}, \bibinfo {author} {\bibfnamefont {Zezhou}\ \bibnamefont {Hu}},
  \bibinfo {author} {\bibfnamefont {Haopeng}\ \bibnamefont {Yan}}, \bibinfo
  {author} {\bibfnamefont {Minyong}\ \bibnamefont {Guo}}, \ and\ \bibinfo
  {author} {\bibfnamefont {Bin}\ \bibnamefont {Chen}},\ }\bibfield  {title}
  {\enquote {\bibinfo {title} {{QED effects on Kerr black hole shadows immersed
  in uniform magnetic fields}},}\ }\href {\doibase 10.1103/PhysRevD.104.104028}
  {\bibfield  {journal} {\bibinfo  {journal} {Phys. Rev. D}\ }\textbf {\bibinfo
  {volume} {104}},\ \bibinfo {pages} {104028} (\bibinfo {year} {2021})},\
  \Eprint {http://arxiv.org/abs/2108.06140} {arXiv:2108.06140 [gr-qc]}
  \BibitemShut {NoStop}%
\bibitem [{\citenamefont {Luminet}(1979)}]{Luminet:1979nyg}%
  \BibitemOpen
  \bibfield  {author} {\bibinfo {author} {\bibfnamefont {J.~P.}\ \bibnamefont
  {Luminet}},\ }\bibfield  {title} {\enquote {\bibinfo {title} {{Image of a
  spherical black hole with thin accretion disk}},}\ }\href@noop {} {\bibfield
  {journal} {\bibinfo  {journal} {Astron. Astrophys.}\ }\textbf {\bibinfo
  {volume} {75}},\ \bibinfo {pages} {228--235} (\bibinfo {year}
  {1979})}\BibitemShut {NoStop}%
\bibitem [{\citenamefont {{Cunningham}}\ and\ \citenamefont
  {{Bardeen}}(1973)}]{1973Apx}%
  \BibitemOpen
  \bibfield  {author} {\bibinfo {author} {\bibfnamefont {C.~T.}\ \bibnamefont
  {{Cunningham}}}\ and\ \bibinfo {author} {\bibfnamefont {James~M.}\
  \bibnamefont {{Bardeen}}},\ }\bibfield  {title} {\enquote {\bibinfo {title}
  {{The Optical Appearance of a Star Orbiting an Extreme Kerr Black Hole}},}\
  }\href {\doibase 10.1086/152223} {\bibfield  {journal} {\bibinfo  {journal}
  {\apj}\ }\textbf {\bibinfo {volume} {183}},\ \bibinfo {pages} {237--264}
  (\bibinfo {year} {1973})}\BibitemShut {NoStop}%
\bibitem [{\citenamefont {Shakura}\ and\ \citenamefont
  {Sunyaev}(1973)}]{1973A}%
  \BibitemOpen
  \bibfield  {author} {\bibinfo {author} {\bibfnamefont {N.~I.}\ \bibnamefont
  {Shakura}}\ and\ \bibinfo {author} {\bibfnamefont {R.~A.}\ \bibnamefont
  {Sunyaev}},\ }\bibfield  {title} {\enquote {\bibinfo {title} {{Black holes in
  binary systems. Observational appearance.}}}\ }\href@noop {} {\bibfield
  {journal} {\bibinfo  {journal} {Astron.Astrophys.}\ }\textbf {\bibinfo
  {volume} {24}},\ \bibinfo {pages} {337--355} (\bibinfo {year}
  {1973})}\BibitemShut {NoStop}%
\bibitem [{\citenamefont {DeWitt}\ and\ \citenamefont
  {DeWitt}(1973)}]{NovikovThorne}%
  \BibitemOpen
  \bibinfo {editor} {\bibfnamefont {C\'ecile}\ \bibnamefont {DeWitt}}\ and\
  \bibinfo {editor} {\bibfnamefont {Bryce~Seligman}\ \bibnamefont {DeWitt}},\
  eds.,\ \href@noop {} {\emph {\bibinfo {title} {{Novikov, I. D. and Thorne, K.
  S. 1973, in Black Holes}: {Les Houches, France, August, 1972}}}},\ \bibinfo
  {series} {Les Houches Summer School}, Vol.~\bibinfo {volume} {23}\ (\bibinfo
  {publisher} {Gordon and Breach},\ \bibinfo {address} {New York, NY},\
  \bibinfo {year} {1973})\BibitemShut {NoStop}%
\bibitem [{\citenamefont {{Page}}\ and\ \citenamefont
  {{Thorne}}(1974)}]{1974ApJ...191..499P}%
  \BibitemOpen
  \bibfield  {author} {\bibinfo {author} {\bibfnamefont {Don~N.}\ \bibnamefont
  {{Page}}}\ and\ \bibinfo {author} {\bibfnamefont {Kip~S.}\ \bibnamefont
  {{Thorne}}},\ }\bibfield  {title} {\enquote {\bibinfo {title}
  {{Disk-Accretion onto a Black Hole. Time-Averaged Structure of Accretion
  Disk}},}\ }\href {\doibase 10.1086/152990} {\bibfield  {journal} {\bibinfo
  {journal} {\apj}\ }\textbf {\bibinfo {volume} {191}},\ \bibinfo {pages}
  {499--506} (\bibinfo {year} {1974})}\BibitemShut {NoStop}%
\bibitem [{\citenamefont {Virbhadra}\ and\ \citenamefont
  {Ellis}(2000)}]{Virbhadra:1999nm}%
  \BibitemOpen
  \bibfield  {author} {\bibinfo {author} {\bibfnamefont {K.~S.}\ \bibnamefont
  {Virbhadra}}\ and\ \bibinfo {author} {\bibfnamefont {George F.~R.}\
  \bibnamefont {Ellis}},\ }\bibfield  {title} {\enquote {\bibinfo {title}
  {{Schwarzschild black hole lensing}},}\ }\href {\doibase
  10.1103/PhysRevD.62.084003} {\bibfield  {journal} {\bibinfo  {journal} {Phys.
  Rev. D}\ }\textbf {\bibinfo {volume} {62}},\ \bibinfo {pages} {084003}
  (\bibinfo {year} {2000})},\ \Eprint {http://arxiv.org/abs/astro-ph/9904193}
  {arXiv:astro-ph/9904193} \BibitemShut {NoStop}%
\bibitem [{\citenamefont {Virbhadra}\ and\ \citenamefont
  {Ellis}(2002)}]{Virbhadra:2002ju}%
  \BibitemOpen
  \bibfield  {author} {\bibinfo {author} {\bibfnamefont {K.~S.}\ \bibnamefont
  {Virbhadra}}\ and\ \bibinfo {author} {\bibfnamefont {G.~F.~R.}\ \bibnamefont
  {Ellis}},\ }\bibfield  {title} {\enquote {\bibinfo {title} {{Gravitational
  lensing by naked singularities}},}\ }\href {\doibase
  10.1103/PhysRevD.65.103004} {\bibfield  {journal} {\bibinfo  {journal} {Phys.
  Rev. D}\ }\textbf {\bibinfo {volume} {65}},\ \bibinfo {pages} {103004}
  (\bibinfo {year} {2002})}\BibitemShut {NoStop}%
\bibitem [{\citenamefont {Virbhadra}\ \emph {et~al.}(1998)\citenamefont
  {Virbhadra}, \citenamefont {Narasimha},\ and\ \citenamefont
  {Chitre}}]{Virbhadra:1998dy}%
  \BibitemOpen
  \bibfield  {author} {\bibinfo {author} {\bibfnamefont {K.~S.}\ \bibnamefont
  {Virbhadra}}, \bibinfo {author} {\bibfnamefont {D.}~\bibnamefont
  {Narasimha}}, \ and\ \bibinfo {author} {\bibfnamefont {S.~M.}\ \bibnamefont
  {Chitre}},\ }\bibfield  {title} {\enquote {\bibinfo {title} {{Role of the
  scalar field in gravitational lensing}},}\ }\href@noop {} {\bibfield
  {journal} {\bibinfo  {journal} {Astron. Astrophys.}\ }\textbf {\bibinfo
  {volume} {337}},\ \bibinfo {pages} {1--8} (\bibinfo {year} {1998})},\ \Eprint
  {http://arxiv.org/abs/astro-ph/9801174} {arXiv:astro-ph/9801174} \BibitemShut
  {NoStop}%
\bibitem [{\citenamefont {Virbhadra}\ and\ \citenamefont
  {Keeton}(2008)}]{Virbhadra:2007kw}%
  \BibitemOpen
  \bibfield  {author} {\bibinfo {author} {\bibfnamefont {K.~S.}\ \bibnamefont
  {Virbhadra}}\ and\ \bibinfo {author} {\bibfnamefont {C.~R.}\ \bibnamefont
  {Keeton}},\ }\bibfield  {title} {\enquote {\bibinfo {title} {{Time delay and
  magnification centroid due to gravitational lensing by black holes and naked
  singularities}},}\ }\href {\doibase 10.1103/PhysRevD.77.124014} {\bibfield
  {journal} {\bibinfo  {journal} {Phys. Rev. D}\ }\textbf {\bibinfo {volume}
  {77}},\ \bibinfo {pages} {124014} (\bibinfo {year} {2008})},\ \Eprint
  {http://arxiv.org/abs/0710.2333} {arXiv:0710.2333 [gr-qc]} \BibitemShut
  {NoStop}%
\bibitem [{\citenamefont {Virbhadra}(2009)}]{Virbhadra:2008ws}%
  \BibitemOpen
  \bibfield  {author} {\bibinfo {author} {\bibfnamefont {K.~S.}\ \bibnamefont
  {Virbhadra}},\ }\bibfield  {title} {\enquote {\bibinfo {title} {{Relativistic
  images of Schwarzschild black hole lensing}},}\ }\href {\doibase
  10.1103/PhysRevD.79.083004} {\bibfield  {journal} {\bibinfo  {journal} {Phys.
  Rev. D}\ }\textbf {\bibinfo {volume} {79}},\ \bibinfo {pages} {083004}
  (\bibinfo {year} {2009})},\ \Eprint {http://arxiv.org/abs/0810.2109}
  {arXiv:0810.2109 [gr-qc]} \BibitemShut {NoStop}%
\bibitem [{\citenamefont {Adler}\ and\ \citenamefont
  {Virbhadra}(2022)}]{Adler:2022qtb}%
  \BibitemOpen
  \bibfield  {author} {\bibinfo {author} {\bibfnamefont {Stephen~L.}\
  \bibnamefont {Adler}}\ and\ \bibinfo {author} {\bibfnamefont {K.~S.}\
  \bibnamefont {Virbhadra}},\ }\bibfield  {title} {\enquote {\bibinfo {title}
  {{Cosmological constant corrections to the photon sphere and black hole
  shadow radii}},}\ }\href@noop {} {\  (\bibinfo {year} {2022})},\ \Eprint
  {http://arxiv.org/abs/2205.04628} {arXiv:2205.04628 [gr-qc]} \BibitemShut
  {NoStop}%
\bibitem [{\citenamefont {Virbhadra}(2022{\natexlab{a}})}]{Virbhadra:2022ybp}%
  \BibitemOpen
  \bibfield  {author} {\bibinfo {author} {\bibfnamefont {K.~S.}\ \bibnamefont
  {Virbhadra}},\ }\bibfield  {title} {\enquote {\bibinfo {title} {{Compactness
  of supermassive dark objects at galactic centers}},}\ }\href@noop {} {\
  (\bibinfo {year} {2022}{\natexlab{a}})},\ \Eprint
  {http://arxiv.org/abs/2204.01792} {arXiv:2204.01792 [gr-qc]} \BibitemShut
  {NoStop}%
\bibitem [{\citenamefont {Virbhadra}(2022{\natexlab{b}})}]{Virbhadra:2022iiy}%
  \BibitemOpen
  \bibfield  {author} {\bibinfo {author} {\bibfnamefont {K.~S.}\ \bibnamefont
  {Virbhadra}},\ }\bibfield  {title} {\enquote {\bibinfo {title} {{Distortions
  of images of Schwarzschild lensing}},}\ }\href@noop {} {\  (\bibinfo {year}
  {2022}{\natexlab{b}})},\ \Eprint {http://arxiv.org/abs/2204.01879}
  {arXiv:2204.01879 [gr-qc]} \BibitemShut {NoStop}%
\bibitem [{\citenamefont {Bozza}\ \emph {et~al.}(2001)\citenamefont {Bozza},
  \citenamefont {Capozziello}, \citenamefont {Iovane},\ and\ \citenamefont
  {Scarpetta}}]{Bozza:2001xd}%
  \BibitemOpen
  \bibfield  {author} {\bibinfo {author} {\bibfnamefont {V.}~\bibnamefont
  {Bozza}}, \bibinfo {author} {\bibfnamefont {S.}~\bibnamefont {Capozziello}},
  \bibinfo {author} {\bibfnamefont {G.}~\bibnamefont {Iovane}}, \ and\ \bibinfo
  {author} {\bibfnamefont {G.}~\bibnamefont {Scarpetta}},\ }\bibfield  {title}
  {\enquote {\bibinfo {title} {{Strong field limit of black hole gravitational
  lensing}},}\ }\href {\doibase 10.1023/A:1012292927358} {\bibfield  {journal}
  {\bibinfo  {journal} {Gen. Rel. Grav.}\ }\textbf {\bibinfo {volume} {33}},\
  \bibinfo {pages} {1535--1548} (\bibinfo {year} {2001})},\ \Eprint
  {http://arxiv.org/abs/gr-qc/0102068} {arXiv:gr-qc/0102068} \BibitemShut
  {NoStop}%
\bibitem [{\citenamefont {Bozza}(2002)}]{Bozza:2002zj}%
  \BibitemOpen
  \bibfield  {author} {\bibinfo {author} {\bibfnamefont {V.}~\bibnamefont
  {Bozza}},\ }\bibfield  {title} {\enquote {\bibinfo {title} {{Gravitational
  lensing in the strong field limit}},}\ }\href {\doibase
  10.1103/PhysRevD.66.103001} {\bibfield  {journal} {\bibinfo  {journal} {Phys.
  Rev. D}\ }\textbf {\bibinfo {volume} {66}},\ \bibinfo {pages} {103001}
  (\bibinfo {year} {2002})},\ \Eprint {http://arxiv.org/abs/gr-qc/0208075}
  {arXiv:gr-qc/0208075} \BibitemShut {NoStop}%
\bibitem [{\citenamefont {Hasse}\ and\ \citenamefont
  {Perlick}(2002)}]{Hasse:2001by}%
  \BibitemOpen
  \bibfield  {author} {\bibinfo {author} {\bibfnamefont {Wolfgang}\
  \bibnamefont {Hasse}}\ and\ \bibinfo {author} {\bibfnamefont {Volker}\
  \bibnamefont {Perlick}},\ }\bibfield  {title} {\enquote {\bibinfo {title}
  {{Gravitational lensing in spherically symmetric static space-times with
  centrifugal force reversal}},}\ }\href {\doibase 10.1023/A:1015384604371}
  {\bibfield  {journal} {\bibinfo  {journal} {Gen. Rel. Grav.}\ }\textbf
  {\bibinfo {volume} {34}},\ \bibinfo {pages} {415--433} (\bibinfo {year}
  {2002})},\ \Eprint {http://arxiv.org/abs/gr-qc/0108002} {arXiv:gr-qc/0108002}
  \BibitemShut {NoStop}%
\bibitem [{\citenamefont {Perlick}(2004)}]{Perlick:2003vg}%
  \BibitemOpen
  \bibfield  {author} {\bibinfo {author} {\bibfnamefont {Volker}\ \bibnamefont
  {Perlick}},\ }\bibfield  {title} {\enquote {\bibinfo {title} {{On the Exact
  gravitational lens equation in spherically symmetric and static
  space-times}},}\ }\href {\doibase 10.1103/PhysRevD.69.064017} {\bibfield
  {journal} {\bibinfo  {journal} {Phys. Rev. D}\ }\textbf {\bibinfo {volume}
  {69}},\ \bibinfo {pages} {064017} (\bibinfo {year} {2004})},\ \Eprint
  {http://arxiv.org/abs/gr-qc/0307072} {arXiv:gr-qc/0307072} \BibitemShut
  {NoStop}%
\bibitem [{\citenamefont {He}\ \emph {et~al.}(2020)\citenamefont {He},
  \citenamefont {Zhou}, \citenamefont {Feng}, \citenamefont {Mu}, \citenamefont
  {Wang}, \citenamefont {Li}, \citenamefont {Pan},\ and\ \citenamefont
  {Lin}}]{He:2020eah}%
  \BibitemOpen
  \bibfield  {author} {\bibinfo {author} {\bibfnamefont {Guansheng}\
  \bibnamefont {He}}, \bibinfo {author} {\bibfnamefont {Xia}\ \bibnamefont
  {Zhou}}, \bibinfo {author} {\bibfnamefont {Zhongwen}\ \bibnamefont {Feng}},
  \bibinfo {author} {\bibfnamefont {Xueling}\ \bibnamefont {Mu}}, \bibinfo
  {author} {\bibfnamefont {Hui}\ \bibnamefont {Wang}}, \bibinfo {author}
  {\bibfnamefont {Weijun}\ \bibnamefont {Li}}, \bibinfo {author} {\bibfnamefont
  {Chaohong}\ \bibnamefont {Pan}}, \ and\ \bibinfo {author} {\bibfnamefont
  {Wenbin}\ \bibnamefont {Lin}},\ }\bibfield  {title} {\enquote {\bibinfo
  {title} {{Gravitational deflection of massive particles in Schwarzschild-de
  Sitter spacetime}},}\ }\href {\doibase 10.1140/epjc/s10052-020-8382-z}
  {\bibfield  {journal} {\bibinfo  {journal} {Eur. Phys. J. C}\ }\textbf
  {\bibinfo {volume} {80}},\ \bibinfo {pages} {835} (\bibinfo {year}
  {2020})}\BibitemShut {NoStop}%
\bibitem [{\citenamefont {Gibbons}\ and\ \citenamefont
  {Werner}(2008)}]{Gibbons:2008rj}%
  \BibitemOpen
  \bibfield  {author} {\bibinfo {author} {\bibfnamefont {G.~W.}\ \bibnamefont
  {Gibbons}}\ and\ \bibinfo {author} {\bibfnamefont {M.~C.}\ \bibnamefont
  {Werner}},\ }\bibfield  {title} {\enquote {\bibinfo {title} {{Applications of
  the Gauss-Bonnet theorem to gravitational lensing}},}\ }\href {\doibase
  10.1088/0264-9381/25/23/235009} {\bibfield  {journal} {\bibinfo  {journal}
  {Class. Quant. Grav.}\ }\textbf {\bibinfo {volume} {25}},\ \bibinfo {pages}
  {235009} (\bibinfo {year} {2008})},\ \Eprint {http://arxiv.org/abs/0807.0854}
  {arXiv:0807.0854 [gr-qc]} \BibitemShut {NoStop}%
\bibitem [{\citenamefont {Werner}(2012)}]{Werner_2012}%
  \BibitemOpen
  \bibfield  {author} {\bibinfo {author} {\bibfnamefont {M.~C.}\ \bibnamefont
  {Werner}},\ }\bibfield  {title} {\enquote {\bibinfo {title} {Gravitational
  lensing in the kerr-randers optical geometry},}\ }\href {\doibase
  10.1007/s10714-012-1458-9} {\bibfield  {journal} {\bibinfo  {journal} {Gen.
  Relativ. Gravit.}\ }\textbf {\bibinfo {volume} {44}},\ \bibinfo {pages}
  {3047} (\bibinfo {year} {2012})}\BibitemShut {NoStop}%
\bibitem [{\citenamefont {\"Ovg\"un}(2018)}]{Ovgun:2018fnk}%
  \BibitemOpen
  \bibfield  {author} {\bibinfo {author} {\bibfnamefont {Ali}\ \bibnamefont
  {\"Ovg\"un}},\ }\bibfield  {title} {\enquote {\bibinfo {title} {{Light
  deflection by Damour-Solodukhin wormholes and Gauss-Bonnet theorem}},}\
  }\href {\doibase 10.1103/PhysRevD.98.044033} {\bibfield  {journal} {\bibinfo
  {journal} {Phys. Rev. D}\ }\textbf {\bibinfo {volume} {98}},\ \bibinfo
  {pages} {044033} (\bibinfo {year} {2018})},\ \Eprint
  {http://arxiv.org/abs/1805.06296} {arXiv:1805.06296 [gr-qc]} \BibitemShut
  {NoStop}%
\bibitem [{\citenamefont {\"Ovg\"un}(2019{\natexlab{a}})}]{Ovgun:2019wej}%
  \BibitemOpen
  \bibfield  {author} {\bibinfo {author} {\bibfnamefont {A.}~\bibnamefont
  {\"Ovg\"un}},\ }\bibfield  {title} {\enquote {\bibinfo {title} {{Weak field
  deflection angle by regular black holes with cosmic strings using the
  Gauss-Bonnet theorem}},}\ }\href {\doibase 10.1103/PhysRevD.99.104075}
  {\bibfield  {journal} {\bibinfo  {journal} {Phys. Rev. D}\ }\textbf {\bibinfo
  {volume} {99}},\ \bibinfo {pages} {104075} (\bibinfo {year}
  {2019}{\natexlab{a}})},\ \Eprint {http://arxiv.org/abs/1902.04411}
  {arXiv:1902.04411 [gr-qc]} \BibitemShut {NoStop}%
\bibitem [{\citenamefont {\"Ovg\"un}(2019{\natexlab{b}})}]{Ovgun:2018oxk}%
  \BibitemOpen
  \bibfield  {author} {\bibinfo {author} {\bibfnamefont {Ali}\ \bibnamefont
  {\"Ovg\"un}},\ }\bibfield  {title} {\enquote {\bibinfo {title} {{Deflection
  Angle of Photons through Dark Matter by Black Holes and Wormholes Using
  Gauss\textendash{}Bonnet Theorem}},}\ }\href {\doibase
  10.3390/universe5050115} {\bibfield  {journal} {\bibinfo  {journal}
  {Universe}\ }\textbf {\bibinfo {volume} {5}},\ \bibinfo {pages} {115}
  (\bibinfo {year} {2019}{\natexlab{b}})},\ \Eprint
  {http://arxiv.org/abs/1806.05549} {arXiv:1806.05549 [physics.gen-ph]}
  \BibitemShut {NoStop}%
\bibitem [{\citenamefont {Javed}\ \emph
  {et~al.}(2019{\natexlab{a}})\citenamefont {Javed}, \citenamefont {Abbas},\
  and\ \citenamefont {\"Ovg\"un}}]{Javed:2019kon}%
  \BibitemOpen
  \bibfield  {author} {\bibinfo {author} {\bibfnamefont {Wajiha}\ \bibnamefont
  {Javed}}, \bibinfo {author} {\bibfnamefont {Jameela}\ \bibnamefont {Abbas}},
  \ and\ \bibinfo {author} {\bibfnamefont {Ali}\ \bibnamefont {\"Ovg\"un}},\
  }\bibfield  {title} {\enquote {\bibinfo {title} {{Deflection angle of photon
  from magnetized black hole and effect of nonlinear electrodynamics}},}\
  }\href {\doibase 10.1140/epjc/s10052-019-7208-3} {\bibfield  {journal}
  {\bibinfo  {journal} {Eur. Phys. J. C}\ }\textbf {\bibinfo {volume} {79}},\
  \bibinfo {pages} {694} (\bibinfo {year} {2019}{\natexlab{a}})},\ \Eprint
  {http://arxiv.org/abs/1908.09632} {arXiv:1908.09632 [physics.gen-ph]}
  \BibitemShut {NoStop}%
\bibitem [{\citenamefont {Javed}\ \emph
  {et~al.}(2019{\natexlab{b}})\citenamefont {Javed}, \citenamefont {Abbas},\
  and\ \citenamefont {\"Ovg\"un}}]{Javed:2019rrg}%
  \BibitemOpen
  \bibfield  {author} {\bibinfo {author} {\bibfnamefont {Wajiha}\ \bibnamefont
  {Javed}}, \bibinfo {author} {\bibfnamefont {jameela}\ \bibnamefont {Abbas}},
  \ and\ \bibinfo {author} {\bibfnamefont {Ali}\ \bibnamefont {\"Ovg\"un}},\
  }\bibfield  {title} {\enquote {\bibinfo {title} {{Effect of the Hair on
  Deflection Angle by Asymptotically Flat Black Holes in
  Einstein-Maxwell-Dilaton Theory}},}\ }\href {\doibase
  10.20944/preprints201906.0101.v1} {\bibfield  {journal} {\bibinfo  {journal}
  {Phys. Rev. D}\ }\textbf {\bibinfo {volume} {100}},\ \bibinfo {pages}
  {044052} (\bibinfo {year} {2019}{\natexlab{b}})},\ \Eprint
  {http://arxiv.org/abs/1908.05241} {arXiv:1908.05241 [gr-qc]} \BibitemShut
  {NoStop}%
\bibitem [{\citenamefont {Javed}\ \emph
  {et~al.}(2019{\natexlab{c}})\citenamefont {Javed}, \citenamefont {Babar},\
  and\ \citenamefont {\"Ovg\"un}}]{Javed:2019ynm}%
  \BibitemOpen
  \bibfield  {author} {\bibinfo {author} {\bibfnamefont {Wajiha}\ \bibnamefont
  {Javed}}, \bibinfo {author} {\bibfnamefont {Rimsha}\ \bibnamefont {Babar}}, \
  and\ \bibinfo {author} {\bibfnamefont {Al\"\i{}}\ \bibnamefont {\"Ovg\"un}},\
  }\bibfield  {title} {\enquote {\bibinfo {title} {{Effect of the dilaton field
  and plasma medium on deflection angle by black holes in
  Einstein-Maxwell-dilaton-axion theory}},}\ }\href {\doibase
  10.1103/PhysRevD.100.104032} {\bibfield  {journal} {\bibinfo  {journal}
  {Phys. Rev. D}\ }\textbf {\bibinfo {volume} {100}},\ \bibinfo {pages}
  {104032} (\bibinfo {year} {2019}{\natexlab{c}})},\ \Eprint
  {http://arxiv.org/abs/1910.11697} {arXiv:1910.11697 [gr-qc]} \BibitemShut
  {NoStop}%
\bibitem [{\citenamefont {Javed}\ \emph
  {et~al.}(2020{\natexlab{a}})\citenamefont {Javed}, \citenamefont {Hamza},\
  and\ \citenamefont {\"Ovg\"un}}]{Javed:2020lsg}%
  \BibitemOpen
  \bibfield  {author} {\bibinfo {author} {\bibfnamefont {Wajiha}\ \bibnamefont
  {Javed}}, \bibinfo {author} {\bibfnamefont {Ali}\ \bibnamefont {Hamza}}, \
  and\ \bibinfo {author} {\bibfnamefont {Ali}\ \bibnamefont {\"Ovg\"un}},\
  }\bibfield  {title} {\enquote {\bibinfo {title} {{Effect of nonlinear
  electrodynamics on the weak field deflection angle by a black hole}},}\
  }\href {\doibase 10.20944/preprints201911.0142.v1} {\bibfield  {journal}
  {\bibinfo  {journal} {Phys. Rev. D}\ }\textbf {\bibinfo {volume} {101}},\
  \bibinfo {pages} {103521} (\bibinfo {year} {2020}{\natexlab{a}})},\ \Eprint
  {http://arxiv.org/abs/2005.09464} {arXiv:2005.09464 [gr-qc]} \BibitemShut
  {NoStop}%
\bibitem [{\citenamefont {Javed}\ \emph
  {et~al.}(2019{\natexlab{d}})\citenamefont {Javed}, \citenamefont {Babar},\
  and\ \citenamefont {\"Ovg\"un}}]{Javed:2019qyg}%
  \BibitemOpen
  \bibfield  {author} {\bibinfo {author} {\bibfnamefont {Wajiha}\ \bibnamefont
  {Javed}}, \bibinfo {author} {\bibfnamefont {Rimsha}\ \bibnamefont {Babar}}, \
  and\ \bibinfo {author} {\bibfnamefont {Ali}\ \bibnamefont {\"Ovg\"un}},\
  }\bibfield  {title} {\enquote {\bibinfo {title} {{The effect of the
  Brane-Dicke coupling parameter on weak gravitational lensing by wormholes and
  naked singularities}},}\ }\href {\doibase 10.1103/PhysRevD.99.084012}
  {\bibfield  {journal} {\bibinfo  {journal} {Phys. Rev. D}\ }\textbf {\bibinfo
  {volume} {99}},\ \bibinfo {pages} {084012} (\bibinfo {year}
  {2019}{\natexlab{d}})},\ \Eprint {http://arxiv.org/abs/1903.11657}
  {arXiv:1903.11657 [gr-qc]} \BibitemShut {NoStop}%
\bibitem [{\citenamefont {\"Ovg\"un}\ \emph {et~al.}(2019)\citenamefont
  {\"Ovg\"un}, \citenamefont {Sakall\i{}},\ and\ \citenamefont
  {Saavedra}}]{Ovgun:2018fte}%
  \BibitemOpen
  \bibfield  {author} {\bibinfo {author} {\bibfnamefont {Ali}\ \bibnamefont
  {\"Ovg\"un}}, \bibinfo {author} {\bibfnamefont {\.Izzet}\ \bibnamefont
  {Sakall\i{}}}, \ and\ \bibinfo {author} {\bibfnamefont {Joel}\ \bibnamefont
  {Saavedra}},\ }\bibfield  {title} {\enquote {\bibinfo {title} {{Weak
  gravitational lensing by Kerr-MOG black hole and Gauss\textendash{}Bonnet
  theorem}},}\ }\href {\doibase 10.1016/j.aop.2019.167978} {\bibfield
  {journal} {\bibinfo  {journal} {Annals Phys.}\ }\textbf {\bibinfo {volume}
  {411}},\ \bibinfo {pages} {167978} (\bibinfo {year} {2019})},\ \Eprint
  {http://arxiv.org/abs/1806.06453} {arXiv:1806.06453 [gr-qc]} \BibitemShut
  {NoStop}%
\bibitem [{\citenamefont {Javed}\ \emph
  {et~al.}(2020{\natexlab{b}})\citenamefont {Javed}, \citenamefont {Abbas},\
  and\ \citenamefont {\"Ovg\"un}}]{Javed:2019jag}%
  \BibitemOpen
  \bibfield  {author} {\bibinfo {author} {\bibfnamefont {W.}~\bibnamefont
  {Javed}}, \bibinfo {author} {\bibfnamefont {J.}~\bibnamefont {Abbas}}, \ and\
  \bibinfo {author} {\bibfnamefont {A.}~\bibnamefont {\"Ovg\"un}},\ }\bibfield
  {title} {\enquote {\bibinfo {title} {{Effect of the Quintessential Dark
  Energy on Weak Deflection Angle by Kerr-Newmann Black Hole}},}\ }\href
  {\doibase 10.20944/preprints201906.0124.v1} {\bibfield  {journal} {\bibinfo
  {journal} {Annals Phys.}\ }\textbf {\bibinfo {volume} {418}},\ \bibinfo
  {pages} {168183} (\bibinfo {year} {2020}{\natexlab{b}})},\ \Eprint
  {http://arxiv.org/abs/2007.16027} {arXiv:2007.16027 [gr-qc]} \BibitemShut
  {NoStop}%
\bibitem [{\citenamefont {Ishihara}\ \emph {et~al.}(2016)\citenamefont
  {Ishihara}, \citenamefont {Suzuki}, \citenamefont {Ono}, \citenamefont
  {Kitamura},\ and\ \citenamefont {Asada}}]{Ishihara:2016vdc}%
  \BibitemOpen
  \bibfield  {author} {\bibinfo {author} {\bibfnamefont {Asahi}\ \bibnamefont
  {Ishihara}}, \bibinfo {author} {\bibfnamefont {Yusuke}\ \bibnamefont
  {Suzuki}}, \bibinfo {author} {\bibfnamefont {Toshiaki}\ \bibnamefont {Ono}},
  \bibinfo {author} {\bibfnamefont {Takao}\ \bibnamefont {Kitamura}}, \ and\
  \bibinfo {author} {\bibfnamefont {Hideki}\ \bibnamefont {Asada}},\ }\bibfield
   {title} {\enquote {\bibinfo {title} {{Gravitational bending angle of light
  for finite distance and the Gauss-Bonnet theorem}},}\ }\href {\doibase
  10.1103/PhysRevD.94.084015} {\bibfield  {journal} {\bibinfo  {journal} {Phys.
  Rev. D}\ }\textbf {\bibinfo {volume} {94}},\ \bibinfo {pages} {084015}
  (\bibinfo {year} {2016})},\ \Eprint {http://arxiv.org/abs/1604.08308}
  {arXiv:1604.08308 [gr-qc]} \BibitemShut {NoStop}%
\bibitem [{\citenamefont {Takizawa}\ \emph {et~al.}(2020)\citenamefont
  {Takizawa}, \citenamefont {Ono},\ and\ \citenamefont
  {Asada}}]{Takizawa:2020egm}%
  \BibitemOpen
  \bibfield  {author} {\bibinfo {author} {\bibfnamefont {Keita}\ \bibnamefont
  {Takizawa}}, \bibinfo {author} {\bibfnamefont {Toshiaki}\ \bibnamefont
  {Ono}}, \ and\ \bibinfo {author} {\bibfnamefont {Hideki}\ \bibnamefont
  {Asada}},\ }\bibfield  {title} {\enquote {\bibinfo {title} {{Gravitational
  deflection angle of light: Definition by an observer and its application to
  an asymptotically nonflat spacetime}},}\ }\href {\doibase
  10.1103/PhysRevD.101.104032} {\bibfield  {journal} {\bibinfo  {journal}
  {Phys. Rev. D}\ }\textbf {\bibinfo {volume} {101}},\ \bibinfo {pages}
  {104032} (\bibinfo {year} {2020})},\ \Eprint
  {http://arxiv.org/abs/2001.03290} {arXiv:2001.03290 [gr-qc]} \BibitemShut
  {NoStop}%
\bibitem [{\citenamefont {Ono}\ and\ \citenamefont
  {Asada}(2019)}]{Ono:2019hkw}%
  \BibitemOpen
  \bibfield  {author} {\bibinfo {author} {\bibfnamefont {Toshiaki}\
  \bibnamefont {Ono}}\ and\ \bibinfo {author} {\bibfnamefont {Hideki}\
  \bibnamefont {Asada}},\ }\bibfield  {title} {\enquote {\bibinfo {title} {{The
  effects of finite distance on the gravitational deflection angle of
  light}},}\ }\href {\doibase 10.3390/universe5110218} {\bibfield  {journal}
  {\bibinfo  {journal} {Universe}\ }\textbf {\bibinfo {volume} {5}},\ \bibinfo
  {pages} {218} (\bibinfo {year} {2019})},\ \Eprint
  {http://arxiv.org/abs/1906.02414} {arXiv:1906.02414 [gr-qc]} \BibitemShut
  {NoStop}%
\bibitem [{\citenamefont {Ishihara}\ \emph {et~al.}(2017)\citenamefont
  {Ishihara}, \citenamefont {Suzuki}, \citenamefont {Ono},\ and\ \citenamefont
  {Asada}}]{Ishihara:2016sfv}%
  \BibitemOpen
  \bibfield  {author} {\bibinfo {author} {\bibfnamefont {Asahi}\ \bibnamefont
  {Ishihara}}, \bibinfo {author} {\bibfnamefont {Yusuke}\ \bibnamefont
  {Suzuki}}, \bibinfo {author} {\bibfnamefont {Toshiaki}\ \bibnamefont {Ono}},
  \ and\ \bibinfo {author} {\bibfnamefont {Hideki}\ \bibnamefont {Asada}},\
  }\bibfield  {title} {\enquote {\bibinfo {title} {{Finite-distance corrections
  to the gravitational bending angle of light in the strong deflection
  limit}},}\ }\href {\doibase 10.1103/PhysRevD.95.044017} {\bibfield  {journal}
  {\bibinfo  {journal} {Phys. Rev. D}\ }\textbf {\bibinfo {volume} {95}},\
  \bibinfo {pages} {044017} (\bibinfo {year} {2017})}\BibitemShut {NoStop}%
\bibitem [{\citenamefont {Ono}\ \emph {et~al.}(2017)\citenamefont {Ono},
  \citenamefont {Ishihara},\ and\ \citenamefont {Asada}}]{Ono:2017pie}%
  \BibitemOpen
  \bibfield  {author} {\bibinfo {author} {\bibfnamefont {Toshiaki}\
  \bibnamefont {Ono}}, \bibinfo {author} {\bibfnamefont {Asahi}\ \bibnamefont
  {Ishihara}}, \ and\ \bibinfo {author} {\bibfnamefont {Hideki}\ \bibnamefont
  {Asada}},\ }\bibfield  {title} {\enquote {\bibinfo {title} {{Gravitomagnetic
  bending angle of light with finite-distance corrections in stationary
  axisymmetric spacetimes}},}\ }\href {\doibase 10.1103/PhysRevD.96.104037}
  {\bibfield  {journal} {\bibinfo  {journal} {Phys. Rev. D}\ }\textbf {\bibinfo
  {volume} {96}},\ \bibinfo {pages} {104037} (\bibinfo {year}
  {2017})}\BibitemShut {NoStop}%
\bibitem [{\citenamefont {Li}\ and\ \citenamefont
  {\"Ovg\"un}(2020)}]{Li:2020dln}%
  \BibitemOpen
  \bibfield  {author} {\bibinfo {author} {\bibfnamefont {Zonghai}\ \bibnamefont
  {Li}}\ and\ \bibinfo {author} {\bibfnamefont {Ali}\ \bibnamefont
  {\"Ovg\"un}},\ }\bibfield  {title} {\enquote {\bibinfo {title}
  {{Finite-distance gravitational deflection of massive particles by a
  Kerr-like black hole in the bumblebee gravity model}},}\ }\href {\doibase
  10.1103/PhysRevD.101.024040} {\bibfield  {journal} {\bibinfo  {journal}
  {Phys. Rev. D}\ }\textbf {\bibinfo {volume} {101}},\ \bibinfo {pages}
  {024040} (\bibinfo {year} {2020})}\BibitemShut {NoStop}%
\bibitem [{\citenamefont {Li}\ \emph {et~al.}(2020{\natexlab{b}})\citenamefont
  {Li}, \citenamefont {Zhang},\ and\ \citenamefont {\"Ovg\"un}}]{Li:2020wvn}%
  \BibitemOpen
  \bibfield  {author} {\bibinfo {author} {\bibfnamefont {Zonghai}\ \bibnamefont
  {Li}}, \bibinfo {author} {\bibfnamefont {Guodong}\ \bibnamefont {Zhang}}, \
  and\ \bibinfo {author} {\bibfnamefont {Ali}\ \bibnamefont {\"Ovg\"un}},\
  }\bibfield  {title} {\enquote {\bibinfo {title} {{Circular Orbit of a
  Particle and Weak Gravitational Lensing}},}\ }\href {\doibase
  10.1103/PhysRevD.101.124058} {\bibfield  {journal} {\bibinfo  {journal}
  {Phys. Rev. D}\ }\textbf {\bibinfo {volume} {101}},\ \bibinfo {pages}
  {124058} (\bibinfo {year} {2020}{\natexlab{b}})}\BibitemShut {NoStop}%
\bibitem [{\citenamefont {Javed}\ \emph {et~al.}(2022)\citenamefont {Javed},
  \citenamefont {Riaz}, \citenamefont {Pantig},\ and\ \citenamefont
  {\"Ovg\"un}}]{Javed:2022fsn}%
  \BibitemOpen
  \bibfield  {author} {\bibinfo {author} {\bibfnamefont {Wajiha}\ \bibnamefont
  {Javed}}, \bibinfo {author} {\bibfnamefont {Sibgha}\ \bibnamefont {Riaz}},
  \bibinfo {author} {\bibfnamefont {Reggie~C.}\ \bibnamefont {Pantig}}, \ and\
  \bibinfo {author} {\bibfnamefont {Ali}\ \bibnamefont {\"Ovg\"un}},\
  }\bibfield  {title} {\enquote {\bibinfo {title} {{Weak gravitational lensing
  in dark matter and plasma mediums for wormhole-like static aether
  solution}},}\ }\href {\doibase 10.1140/epjc/s10052-022-11030-4} {\bibfield
  {journal} {\bibinfo  {journal} {Eur. Phys. J. C}\ }\textbf {\bibinfo {volume}
  {82}},\ \bibinfo {pages} {1057} (\bibinfo {year} {2022})},\ \Eprint
  {http://arxiv.org/abs/2212.00804} {arXiv:2212.00804 [gr-qc]} \BibitemShut
  {NoStop}%
\bibitem [{\citenamefont {Javed}\ \emph {et~al.}(2023)\citenamefont {Javed},
  \citenamefont {Atique}, \citenamefont {Pantig},\ and\ \citenamefont
  {\"{O}vg\"{u}n}}]{doi:10.1142/S0219887823500408}%
  \BibitemOpen
  \bibfield  {author} {\bibinfo {author} {\bibfnamefont {Wajiha}\ \bibnamefont
  {Javed}}, \bibinfo {author} {\bibfnamefont {Mehak}\ \bibnamefont {Atique}},
  \bibinfo {author} {\bibfnamefont {Reggie~C.}\ \bibnamefont {Pantig}}, \ and\
  \bibinfo {author} {\bibfnamefont {Ali}\ \bibnamefont {\"{O}vg\"{u}n}},\
  }\bibfield  {title} {\enquote {\bibinfo {title} {Weak lensing, hawking
  radiation and greybody factor bound by a charged black holes with non-linear
  electrodynamics corrections},}\ }\href {\doibase 10.1142/S0219887823500408}
  {\bibfield  {journal} {\bibinfo  {journal} {International Journal of
  Geometric Methods in Modern Physics}\ }\textbf {\bibinfo {volume} {0}},\
  \bibinfo {pages} {2350040} (\bibinfo {year} {2023})},\ \Eprint
  {http://arxiv.org/abs/https://doi.org/10.1142/S0219887823500408}
  {https://doi.org/10.1142/S0219887823500408} \BibitemShut {NoStop}%
\bibitem [{\citenamefont {Tucker}\ \emph {et~al.}(2018)\citenamefont {Tucker},
  \citenamefont {Shappee}, \citenamefont {Holoien}, \citenamefont {Auchettl},
  \citenamefont {Strader}, \citenamefont {Stanek}, \citenamefont {Kochanek},
  \citenamefont {Bahramian}, \citenamefont {Dong}, \citenamefont {Prieto},
  \citenamefont {Shields}, \citenamefont {Thompson}, \citenamefont {Beacom},
  \citenamefont {Chomiuk}, \citenamefont {Denneau}, \citenamefont {Flewelling},
  \citenamefont {Heinze}, \citenamefont {Smith}, \citenamefont {Stalder},
  \citenamefont {Tonry}, \citenamefont {Weiland}, \citenamefont {Rest},
  \citenamefont {Huber}, \citenamefont {Rowan}, \citenamefont {Dage},\ and\
  \citenamefont {and}}]{Tucker_2018}%
  \BibitemOpen
  \bibfield  {author} {\bibinfo {author} {\bibfnamefont {M.~A.}\ \bibnamefont
  {Tucker}}, \bibinfo {author} {\bibfnamefont {B.~J.}\ \bibnamefont {Shappee}},
  \bibinfo {author} {\bibfnamefont {T.~W.-S.}\ \bibnamefont {Holoien}},
  \bibinfo {author} {\bibfnamefont {K.}~\bibnamefont {Auchettl}}, \bibinfo
  {author} {\bibfnamefont {J.}~\bibnamefont {Strader}}, \bibinfo {author}
  {\bibfnamefont {K.~Z.}\ \bibnamefont {Stanek}}, \bibinfo {author}
  {\bibfnamefont {C.~S.}\ \bibnamefont {Kochanek}}, \bibinfo {author}
  {\bibfnamefont {A.}~\bibnamefont {Bahramian}}, \bibinfo {author}
  {\bibfnamefont {Subo}\ \bibnamefont {Dong}}, \bibinfo {author} {\bibfnamefont
  {J.~L.}\ \bibnamefont {Prieto}}, \bibinfo {author} {\bibfnamefont
  {J.}~\bibnamefont {Shields}}, \bibinfo {author} {\bibfnamefont {Todd~A.}\
  \bibnamefont {Thompson}}, \bibinfo {author} {\bibfnamefont {John~F.}\
  \bibnamefont {Beacom}}, \bibinfo {author} {\bibfnamefont {L.}~\bibnamefont
  {Chomiuk}}, \bibinfo {author} {\bibfnamefont {L.}~\bibnamefont {Denneau}},
  \bibinfo {author} {\bibfnamefont {H.}~\bibnamefont {Flewelling}}, \bibinfo
  {author} {\bibfnamefont {A.~N.}\ \bibnamefont {Heinze}}, \bibinfo {author}
  {\bibfnamefont {K.~W.}\ \bibnamefont {Smith}}, \bibinfo {author}
  {\bibfnamefont {B.}~\bibnamefont {Stalder}}, \bibinfo {author} {\bibfnamefont
  {J.~L.}\ \bibnamefont {Tonry}}, \bibinfo {author} {\bibfnamefont
  {H.}~\bibnamefont {Weiland}}, \bibinfo {author} {\bibfnamefont
  {A.}~\bibnamefont {Rest}}, \bibinfo {author} {\bibfnamefont {M.~E.}\
  \bibnamefont {Huber}}, \bibinfo {author} {\bibfnamefont {D.~M.}\ \bibnamefont
  {Rowan}}, \bibinfo {author} {\bibfnamefont {K.}~\bibnamefont {Dage}}, \ and\
  \bibinfo {author} {\bibnamefont {and}},\ }\bibfield  {title} {\enquote
  {\bibinfo {title} {{ASASSN}-18ey: The rise of a new black hole x-ray
  binary},}\ }\href {\doibase 10.3847/2041-8213/aae88a} {\bibfield  {journal}
  {\bibinfo  {journal} {The Astrophysical Journal}\ }\textbf {\bibinfo {volume}
  {867}},\ \bibinfo {pages} {L9} (\bibinfo {year} {2018})}\BibitemShut
  {NoStop}%
\bibitem [{\citenamefont {Corral-Santana}\ \emph {et~al.}(2016)\citenamefont
  {Corral-Santana}, \citenamefont {Casares}, \citenamefont {Munoz-Darias},
  \citenamefont {Bauer}, \citenamefont {Martinez-Pais},\ and\ \citenamefont
  {Russell}}]{Corral-Santana:2015fud}%
  \BibitemOpen
  \bibfield  {author} {\bibinfo {author} {\bibfnamefont {Jesus~M.}\
  \bibnamefont {Corral-Santana}}, \bibinfo {author} {\bibfnamefont {Jorge}\
  \bibnamefont {Casares}}, \bibinfo {author} {\bibfnamefont {Teo}\ \bibnamefont
  {Munoz-Darias}}, \bibinfo {author} {\bibfnamefont {Franz~E.}\ \bibnamefont
  {Bauer}}, \bibinfo {author} {\bibfnamefont {Ignacio~G.}\ \bibnamefont
  {Martinez-Pais}}, \ and\ \bibinfo {author} {\bibfnamefont {David~M.}\
  \bibnamefont {Russell}},\ }\bibfield  {title} {\enquote {\bibinfo {title}
  {{BlackCAT: A catalogue of stellar-mass black holes in X-ray transients}},}\
  }\href {\doibase 10.1051/0004-6361/201527130} {\bibfield  {journal} {\bibinfo
   {journal} {Astron. Astrophys.}\ }\textbf {\bibinfo {volume} {587}},\
  \bibinfo {pages} {A61} (\bibinfo {year} {2016})},\ \Eprint
  {http://arxiv.org/abs/1510.08869} {arXiv:1510.08869 [astro-ph.HE]}
  \BibitemShut {NoStop}%
\bibitem [{\citenamefont {Perez}\ \emph {et~al.}(2013)\citenamefont {Perez},
  \citenamefont {Romero},\ and\ \citenamefont {Bergliaffa}}]{Perez:2012bx}%
  \BibitemOpen
  \bibfield  {author} {\bibinfo {author} {\bibfnamefont {Daniela}\ \bibnamefont
  {Perez}}, \bibinfo {author} {\bibfnamefont {Gustavo~E.}\ \bibnamefont
  {Romero}}, \ and\ \bibinfo {author} {\bibfnamefont {Santiago E.~Perez}\
  \bibnamefont {Bergliaffa}},\ }\bibfield  {title} {\enquote {\bibinfo {title}
  {{Accretion disks around black holes in modified strong gravity}},}\ }\href
  {\doibase 10.1051/0004-6361/201220378} {\bibfield  {journal} {\bibinfo
  {journal} {Astron. Astrophys.}\ }\textbf {\bibinfo {volume} {551}},\ \bibinfo
  {pages} {A4} (\bibinfo {year} {2013})},\ \Eprint
  {http://arxiv.org/abs/1212.2640} {arXiv:1212.2640 [astro-ph.CO]} \BibitemShut
  {NoStop}%
\bibitem [{\citenamefont {Harko}\ \emph {et~al.}(2009)\citenamefont {Harko},
  \citenamefont {Kovacs},\ and\ \citenamefont {Lobo}}]{Harko:2009rp}%
  \BibitemOpen
  \bibfield  {author} {\bibinfo {author} {\bibfnamefont {Tiberiu}\ \bibnamefont
  {Harko}}, \bibinfo {author} {\bibfnamefont {Zoltan}\ \bibnamefont {Kovacs}},
  \ and\ \bibinfo {author} {\bibfnamefont {Francisco S.~N.}\ \bibnamefont
  {Lobo}},\ }\bibfield  {title} {\enquote {\bibinfo {title} {{Testing
  Ho\v{r}ava-Lifshitz gravity using thin accretion disk properties}},}\ }\href
  {\doibase 10.1103/PhysRevD.80.044021} {\bibfield  {journal} {\bibinfo
  {journal} {Phys. Rev. D}\ }\textbf {\bibinfo {volume} {80}},\ \bibinfo
  {pages} {044021} (\bibinfo {year} {2009})},\ \Eprint
  {http://arxiv.org/abs/0907.1449} {arXiv:0907.1449 [gr-qc]} \BibitemShut
  {NoStop}%
\bibitem [{\citenamefont {Pun}\ \emph {et~al.}(2008)\citenamefont {Pun},
  \citenamefont {Kovacs},\ and\ \citenamefont {Harko}}]{Pun:2008ua}%
  \BibitemOpen
  \bibfield  {author} {\bibinfo {author} {\bibfnamefont {C.~S.~J.}\
  \bibnamefont {Pun}}, \bibinfo {author} {\bibfnamefont {Z.}~\bibnamefont
  {Kovacs}}, \ and\ \bibinfo {author} {\bibfnamefont {T.}~\bibnamefont
  {Harko}},\ }\bibfield  {title} {\enquote {\bibinfo {title} {{Thin accretion
  disks onto brane world black holes}},}\ }\href {\doibase
  10.1103/PhysRevD.78.084015} {\bibfield  {journal} {\bibinfo  {journal} {Phys.
  Rev. D}\ }\textbf {\bibinfo {volume} {78}},\ \bibinfo {pages} {084015}
  (\bibinfo {year} {2008})},\ \Eprint {http://arxiv.org/abs/0809.1284}
  {arXiv:0809.1284 [gr-qc]} \BibitemShut {NoStop}%
\bibitem [{\citenamefont {Heydari-Fard}(2010)}]{Heydari-Fard:2010agr}%
  \BibitemOpen
  \bibfield  {author} {\bibinfo {author} {\bibfnamefont {Malihe}\ \bibnamefont
  {Heydari-Fard}},\ }\bibfield  {title} {\enquote {\bibinfo {title} {{Black
  hole accretion disks in brane gravity via a confining potential}},}\ }\href
  {\doibase 10.1088/0264-9381/27/23/235004} {\bibfield  {journal} {\bibinfo
  {journal} {Class. Quant. Grav.}\ }\textbf {\bibinfo {volume} {27}},\ \bibinfo
  {pages} {235004} (\bibinfo {year} {2010})}\BibitemShut {NoStop}%
\bibitem [{\citenamefont {Bambi}(2013{\natexlab{a}})}]{Bambi:2013jda}%
  \BibitemOpen
  \bibfield  {author} {\bibinfo {author} {\bibfnamefont {Cosimo}\ \bibnamefont
  {Bambi}},\ }\bibfield  {title} {\enquote {\bibinfo {title} {{Broad
  K\ensuremath{\alpha} iron line from accretion disks around traversable
  wormholes}},}\ }\href {\doibase 10.1103/PhysRevD.87.084039} {\bibfield
  {journal} {\bibinfo  {journal} {Phys. Rev. D}\ }\textbf {\bibinfo {volume}
  {87}},\ \bibinfo {pages} {084039} (\bibinfo {year} {2013}{\natexlab{a}})},\
  \Eprint {http://arxiv.org/abs/1303.0624} {arXiv:1303.0624 [gr-qc]}
  \BibitemShut {NoStop}%
\bibitem [{\citenamefont {Bambi}\ \emph {et~al.}(2016)\citenamefont {Bambi},
  \citenamefont {Jiang},\ and\ \citenamefont {Steiner}}]{Bambi:2015ldr}%
  \BibitemOpen
  \bibfield  {author} {\bibinfo {author} {\bibfnamefont {Cosimo}\ \bibnamefont
  {Bambi}}, \bibinfo {author} {\bibfnamefont {Jiachen}\ \bibnamefont {Jiang}},
  \ and\ \bibinfo {author} {\bibfnamefont {James~F.}\ \bibnamefont {Steiner}},\
  }\bibfield  {title} {\enquote {\bibinfo {title} {{Testing the no-hair theorem
  with the continuum-fitting and the iron line methods: a short review}},}\
  }\href {\doibase 10.1088/0264-9381/33/6/064001} {\bibfield  {journal}
  {\bibinfo  {journal} {Class. Quant. Grav.}\ }\textbf {\bibinfo {volume}
  {33}},\ \bibinfo {pages} {064001} (\bibinfo {year} {2016})},\ \Eprint
  {http://arxiv.org/abs/1511.07587} {arXiv:1511.07587 [gr-qc]} \BibitemShut
  {NoStop}%
\bibitem [{\citenamefont {Jiang}\ \emph {et~al.}(2016)\citenamefont {Jiang},
  \citenamefont {Bambi},\ and\ \citenamefont {Steiner}}]{Jiang:2016bdj}%
  \BibitemOpen
  \bibfield  {author} {\bibinfo {author} {\bibfnamefont {Jiachen}\ \bibnamefont
  {Jiang}}, \bibinfo {author} {\bibfnamefont {Cosimo}\ \bibnamefont {Bambi}}, \
  and\ \bibinfo {author} {\bibfnamefont {James~F.}\ \bibnamefont {Steiner}},\
  }\bibfield  {title} {\enquote {\bibinfo {title} {{Testing the Kerr nature of
  black hole candidates using iron line reverberation mapping in the
  Cardoso-Pani-Rico framework}},}\ }\href {\doibase 10.1103/PhysRevD.93.123008}
  {\bibfield  {journal} {\bibinfo  {journal} {Phys. Rev. D}\ }\textbf {\bibinfo
  {volume} {93}},\ \bibinfo {pages} {123008} (\bibinfo {year} {2016})},\
  \Eprint {http://arxiv.org/abs/1601.00838} {arXiv:1601.00838 [gr-qc]}
  \BibitemShut {NoStop}%
\bibitem [{\citenamefont {Kruglov}(2015{\natexlab{a}})}]{Kruglov_2015}%
  \BibitemOpen
  \bibfield  {author} {\bibinfo {author} {\bibfnamefont {S.~I.}\ \bibnamefont
  {Kruglov}},\ }\bibfield  {title} {\enquote {\bibinfo {title} {Nonlinear
  electrodynamics and black holes},}\ }\href {\doibase
  10.1142/s0219887815500735} {\bibfield  {journal} {\bibinfo  {journal}
  {International Journal of Geometric Methods in Modern Physics}\ }\textbf
  {\bibinfo {volume} {12}},\ \bibinfo {pages} {1550073} (\bibinfo {year}
  {2015}{\natexlab{a}})}\BibitemShut {NoStop}%
\bibitem [{\citenamefont {Born}(1934)}]{Born1934410ï12437}%
  \BibitemOpen
  \bibfield  {author} {\bibinfo {author} {\bibfnamefont {Max}\ \bibnamefont
  {Born}},\ }\bibfield  {title} {\enquote {\bibinfo {title} {ï¿1/2on the
  quantum theory of the electromagnetic fieldï¿1/2},}\ }\href
  {https://www.scopus.com/inward/record.uri?eid=2-s2.0-85093265662&partnerID=40&md5=19ade31392ad436c29c628c057bcb0b6}
  {\bibfield  {journal} {\bibinfo  {journal} {Proceedings of the Royal Society
  of London A: Mathematical, Physical and Engineering Sciences}\ }\textbf
  {\bibinfo {volume} {143}},\ \bibinfo {pages} {410ï12437} (\bibinfo {year}
  {1934})},\ \bibinfo {note} {cited by: 0}\BibitemShut {NoStop}%
\bibitem [{\citenamefont {Kruglov}(2015{\natexlab{b}})}]{kruglov2015nonlinear}%
  \BibitemOpen
  \bibfield  {author} {\bibinfo {author} {\bibfnamefont {SI}~\bibnamefont
  {Kruglov}},\ }\bibfield  {title} {\enquote {\bibinfo {title} {Nonlinear
  electrodynamics with birefringence},}\ }\href@noop {} {\bibfield  {journal}
  {\bibinfo  {journal} {Physics Letters A}\ }\textbf {\bibinfo {volume}
  {379}},\ \bibinfo {pages} {623--625} (\bibinfo {year}
  {2015}{\natexlab{b}})}\BibitemShut {NoStop}%
\bibitem [{\citenamefont {Mizuno}\ \emph {et~al.}(2018)\citenamefont {Mizuno},
  \citenamefont {Younsi}, \citenamefont {Fromm}, \citenamefont {Porth},
  \citenamefont {De~Laurentis}, \citenamefont {Olivares}, \citenamefont
  {Falcke}, \citenamefont {Kramer},\ and\ \citenamefont
  {Rezzolla}}]{Mizuno:2018lxz}%
  \BibitemOpen
  \bibfield  {author} {\bibinfo {author} {\bibfnamefont {Yosuke}\ \bibnamefont
  {Mizuno}}, \bibinfo {author} {\bibfnamefont {Ziri}\ \bibnamefont {Younsi}},
  \bibinfo {author} {\bibfnamefont {Christian~M.}\ \bibnamefont {Fromm}},
  \bibinfo {author} {\bibfnamefont {Oliver}\ \bibnamefont {Porth}}, \bibinfo
  {author} {\bibfnamefont {Mariafelicia}\ \bibnamefont {De~Laurentis}},
  \bibinfo {author} {\bibfnamefont {Hector}\ \bibnamefont {Olivares}}, \bibinfo
  {author} {\bibfnamefont {Heino}\ \bibnamefont {Falcke}}, \bibinfo {author}
  {\bibfnamefont {Michael}\ \bibnamefont {Kramer}}, \ and\ \bibinfo {author}
  {\bibfnamefont {Luciano}\ \bibnamefont {Rezzolla}},\ }\bibfield  {title}
  {\enquote {\bibinfo {title} {{The Current Ability to Test Theories of Gravity
  with Black Hole Shadows}},}\ }\href {\doibase 10.1038/s41550-018-0449-5}
  {\bibfield  {journal} {\bibinfo  {journal} {Nature Astron.}\ }\textbf
  {\bibinfo {volume} {2}},\ \bibinfo {pages} {585--590} (\bibinfo {year}
  {2018})},\ \Eprint {http://arxiv.org/abs/1804.05812} {arXiv:1804.05812
  [astro-ph.GA]} \BibitemShut {NoStop}%
\bibitem [{\citenamefont {Wang}(2022)}]{Wang2022}%
  \BibitemOpen
  \bibfield  {author} {\bibinfo {author} {\bibfnamefont {Deng}\ \bibnamefont
  {Wang}},\ }\bibfield  {title} {\enquote {\bibinfo {title} {{Shaving the Hair
  of Black Hole with Sagittarius A$^*$ from Event Horizon Telescope}},}\
  }\href@noop {} {\  (\bibinfo {year} {2022})},\ \Eprint
  {http://arxiv.org/abs/2205.08026} {arXiv:2205.08026 [gr-qc]} \BibitemShut
  {NoStop}%
\bibitem [{\citenamefont {Bambi}\ \emph {et~al.}(2019)\citenamefont {Bambi},
  \citenamefont {Freese}, \citenamefont {Vagnozzi},\ and\ \citenamefont
  {Visinelli}}]{Bambi:2019tjh}%
  \BibitemOpen
  \bibfield  {author} {\bibinfo {author} {\bibfnamefont {Cosimo}\ \bibnamefont
  {Bambi}}, \bibinfo {author} {\bibfnamefont {Katherine}\ \bibnamefont
  {Freese}}, \bibinfo {author} {\bibfnamefont {Sunny}\ \bibnamefont
  {Vagnozzi}}, \ and\ \bibinfo {author} {\bibfnamefont {Luca}\ \bibnamefont
  {Visinelli}},\ }\bibfield  {title} {\enquote {\bibinfo {title} {{Testing the
  rotational nature of the supermassive object M87* from the circularity and
  size of its first image}},}\ }\href {\doibase 10.1103/PhysRevD.100.044057}
  {\bibfield  {journal} {\bibinfo  {journal} {Phys. Rev. D}\ }\textbf {\bibinfo
  {volume} {100}},\ \bibinfo {pages} {044057} (\bibinfo {year} {2019})},\
  \Eprint {http://arxiv.org/abs/1904.12983} {arXiv:1904.12983 [gr-qc]}
  \BibitemShut {NoStop}%
\bibitem [{\citenamefont {Do~Carmo}(2016)}]{Carmo2016}%
  \BibitemOpen
  \bibfield  {author} {\bibinfo {author} {\bibfnamefont {Manfredo~P}\
  \bibnamefont {Do~Carmo}},\ }\href@noop {} {\emph {\bibinfo {title}
  {Differential geometry of curves and surfaces: revised and updated second
  edition}}}\ (\bibinfo  {publisher} {Courier Dover Publications},\ \bibinfo
  {year} {2016})\BibitemShut {NoStop}%
\bibitem [{\citenamefont {Klingenberg}(2013)}]{Klingenberg2013}%
  \BibitemOpen
  \bibfield  {author} {\bibinfo {author} {\bibfnamefont {Wilhelm}\ \bibnamefont
  {Klingenberg}},\ }\href@noop {} {\emph {\bibinfo {title} {A course in
  differential geometry}}},\ Vol.~\bibinfo {volume} {51}\ (\bibinfo
  {publisher} {Springer Science \& Business Media},\ \bibinfo {year}
  {2013})\BibitemShut {NoStop}%
\bibitem [{\citenamefont {Joshi}\ \emph {et~al.}(2014)\citenamefont {Joshi},
  \citenamefont {Malafarina},\ and\ \citenamefont {Narayan}}]{Joshi:2013dva}%
  \BibitemOpen
  \bibfield  {author} {\bibinfo {author} {\bibfnamefont {Pankaj~S.}\
  \bibnamefont {Joshi}}, \bibinfo {author} {\bibfnamefont {Daniele}\
  \bibnamefont {Malafarina}}, \ and\ \bibinfo {author} {\bibfnamefont {Ramesh}\
  \bibnamefont {Narayan}},\ }\bibfield  {title} {\enquote {\bibinfo {title}
  {{Distinguishing black holes from naked singularities through their accretion
  disc properties}},}\ }\href@noop {} {\bibfield  {journal} {\bibinfo
  {journal} {Class. Quant. Grav.}\ }\textbf {\bibinfo {volume} {31}},\ \bibinfo
  {pages} {015002} (\bibinfo {year} {2014})},\ \Eprint
  {http://arxiv.org/abs/1304.7331} {arXiv:1304.7331 [gr-qc]} \BibitemShut
  {NoStop}%
\bibitem [{\citenamefont {Gralla}\ \emph {et~al.}(2019)\citenamefont {Gralla},
  \citenamefont {Holz},\ and\ \citenamefont {Wald}}]{Gralla:2019xty}%
  \BibitemOpen
  \bibfield  {author} {\bibinfo {author} {\bibfnamefont {Samuel~E.}\
  \bibnamefont {Gralla}}, \bibinfo {author} {\bibfnamefont {Daniel~E.}\
  \bibnamefont {Holz}}, \ and\ \bibinfo {author} {\bibfnamefont {Robert~M.}\
  \bibnamefont {Wald}},\ }\bibfield  {title} {\enquote {\bibinfo {title}
  {{Black Hole Shadows, Photon Rings, and Lensing Rings}},}\ }\href {\doibase
  10.1103/PhysRevD.100.024018} {\bibfield  {journal} {\bibinfo  {journal}
  {Phys. Rev. D}\ }\textbf {\bibinfo {volume} {100}},\ \bibinfo {pages}
  {024018} (\bibinfo {year} {2019})},\ \Eprint
  {http://arxiv.org/abs/1906.00873} {arXiv:1906.00873 [astro-ph.HE]}
  \BibitemShut {NoStop}%
\bibitem [{\citenamefont {Chakhchi}\ \emph {et~al.}(2022)\citenamefont
  {Chakhchi}, \citenamefont {El~Moumni},\ and\ \citenamefont
  {Masmar}}]{Chakhchi:2022fls}%
  \BibitemOpen
  \bibfield  {author} {\bibinfo {author} {\bibfnamefont {L.}~\bibnamefont
  {Chakhchi}}, \bibinfo {author} {\bibfnamefont {H.}~\bibnamefont {El~Moumni}},
  \ and\ \bibinfo {author} {\bibfnamefont {K.}~\bibnamefont {Masmar}},\
  }\bibfield  {title} {\enquote {\bibinfo {title} {{Shadows and optical
  appearance of a power-Yang-Mills black hole surrounded by different accretion
  disk profiles}},}\ }\href {\doibase 10.1103/PhysRevD.105.064031} {\bibfield
  {journal} {\bibinfo  {journal} {Phys. Rev. D}\ }\textbf {\bibinfo {volume}
  {105}},\ \bibinfo {pages} {064031} (\bibinfo {year} {2022})}\BibitemShut
  {NoStop}%
\bibitem [{\citenamefont {Zeng}\ and\ \citenamefont
  {Zhang}(2020)}]{Zeng:2020vsj}%
  \BibitemOpen
  \bibfield  {author} {\bibinfo {author} {\bibfnamefont {Xiao-Xiong}\
  \bibnamefont {Zeng}}\ and\ \bibinfo {author} {\bibfnamefont {Hai-Qing}\
  \bibnamefont {Zhang}},\ }\bibfield  {title} {\enquote {\bibinfo {title}
  {{Influence of quintessence dark energy on the shadow of black hole}},}\
  }\href {\doibase 10.1140/epjc/s10052-020-08656-7} {\bibfield  {journal}
  {\bibinfo  {journal} {Eur. Phys. J. C}\ }\textbf {\bibinfo {volume} {80}},\
  \bibinfo {pages} {1058} (\bibinfo {year} {2020})},\ \Eprint
  {http://arxiv.org/abs/2007.06333} {arXiv:2007.06333 [gr-qc]} \BibitemShut
  {NoStop}%
\bibitem [{\citenamefont {Bambi}(2013{\natexlab{b}})}]{Bambi:2013nla}%
  \BibitemOpen
  \bibfield  {author} {\bibinfo {author} {\bibfnamefont {Cosimo}\ \bibnamefont
  {Bambi}},\ }\bibfield  {title} {\enquote {\bibinfo {title} {{Can the
  supermassive objects at the centers of galaxies be traversable wormholes? The
  first test of strong gravity for mm/sub-mm very long baseline interferometry
  facilities}},}\ }\href {\doibase 10.1103/PhysRevD.87.107501} {\bibfield
  {journal} {\bibinfo  {journal} {Phys. Rev. D}\ }\textbf {\bibinfo {volume}
  {87}},\ \bibinfo {pages} {107501} (\bibinfo {year} {2013}{\natexlab{b}})},\
  \Eprint {http://arxiv.org/abs/1304.5691} {arXiv:1304.5691 [gr-qc]}
  \BibitemShut {NoStop}%
\bibitem [{\citenamefont {Liu}\ and\ \citenamefont
  {Prokopec}(2017)}]{Liu_2017}%
  \BibitemOpen
  \bibfield  {author} {\bibinfo {author} {\bibfnamefont {L.}~\bibnamefont
  {Liu}}\ and\ \bibinfo {author} {\bibfnamefont {L}~\bibnamefont {Prokopec}},\
  }\bibfield  {title} {\enquote {\bibinfo {title} {Gravitational microlensing
  in verlinde’s emergent gravity},}\ }\href {\doibase
  10.1016/j.physletb.2017.03.061} {\bibfield  {journal} {\bibinfo  {journal}
  {Phys. Lett. B}\ }\textbf {\bibinfo {volume} {769}},\ \bibinfo {pages} {281}
  (\bibinfo {year} {2017})}\BibitemShut {NoStop}%
\bibitem [{\citenamefont {Kardashev}\ \emph {et~al.}(2013)\citenamefont
  {Kardashev}, \citenamefont {Khartov}, \citenamefont {Abramov}, \citenamefont
  {Avdeev} \emph {et~al.}}]{Kardashev2013}%
  \BibitemOpen
  \bibfield  {author} {\bibinfo {author} {\bibfnamefont {N.~S.}\ \bibnamefont
  {Kardashev}}, \bibinfo {author} {\bibfnamefont {V.~V.}\ \bibnamefont
  {Khartov}}, \bibinfo {author} {\bibfnamefont {V.~V.}\ \bibnamefont
  {Abramov}}, \bibinfo {author} {\bibfnamefont {V.~Yu.}\ \bibnamefont
  {Avdeev}},  \emph {et~al.},\ }\bibfield  {title} {\enquote {\bibinfo {title}
  {“radioastron”-a telescope with a size of 300 000 km: Main parameters and
  first observational results},}\ }\href {\doibase 10.1134/s1063772913030025}
  {\bibfield  {journal} {\bibinfo  {journal} {Astron. Rep.}\ }\textbf {\bibinfo
  {volume} {57}},\ \bibinfo {pages} {153–194} (\bibinfo {year}
  {2013})}\BibitemShut {NoStop}%
\end{thebibliography}%
\end{document}